\documentclass[11pt,onecolumn,nofootinbib,floatfix]{revtex4}

\usepackage{amsmath}
\usepackage{epsfig}
\usepackage{color}			
\usepackage{colordvi}			
\usepackage{slashbox}			
\usepackage{nolbreaks}			
\usepackage{overpic}			
\usepackage{subfig}			
\usepackage{setspace}   		
\usepackage{xspace}
\usepackage{relsize}
\usepackage{titlesec}
\titlespacing*{\section}{0pt}{20pt}{10pt}
\titlespacing*{\subsection}{0pt}{10pt}{10pt}

\def\babar{\mbox{\slshape B\kern-0.1em{\smaller A}\kern-0.1em
    B\kern-0.1em{\smaller A\kern-0.2em R}}}

\def\Bbar    {\kern 0.18em\overline{\kern -0.18em B}{}\xspace}

\def\BB      {\ensuremath{B\Bbar}\xspace} 
\def\Bz      {\ensuremath{B^0}\xspace}
\def\Bzb     {\ensuremath{\Bbar^0}\xspace}
\def\BzBzb   {\ensuremath{\Bz {\kern -0.16em \Bzb}}\xspace}
\def\Bu      {\ensuremath{B^+}\xspace}
\def\Bub     {\ensuremath{B^-}\xspace}
\def\Bp      {\ensuremath{\Bu}\xspace}
\def\Bm      {\ensuremath{\Bub}\xspace}
\def\BpBm    {\ensuremath{\Bu {\kern -0.16em \Bub}}\xspace}
\def\Dbar    {\kern 0.2em\overline{\kern -0.2em D}{}\xspace}

\def\Dz      {\ensuremath{D^0}\xspace}
\def\Dzb     {\ensuremath{\Dbar^0}\xspace}
\def\Dmaybezb{\ensuremath{^{^(}\kern -0.54em \Dbar^{{^)}0}}\xspace}
\def\Dp      {\ensuremath{D^+}\xspace}
\def\Dm      {\ensuremath{D^-}\xspace}
\def\Dstar   {\ensuremath{D^\ast}\xspace}

\def\Dstarz  {\ensuremath{D^{\ast0}}\xspace}

\def\Dstarp  {\ensuremath{D^{\ast+}}\xspace}
\def\Dstarm  {\ensuremath{D^{\ast-}}\xspace}
\def\Kp      {\ensuremath{K^+}\xspace}
\def\Km      {\ensuremath{K^-}\xspace}
\def\piz     {\ensuremath{\pi^0}\xspace}
\def\pip     {\ensuremath{\pi^+}\xspace}
\def\pim     {\ensuremath{\pi^-}\xspace}
\def\FourS   {\ensuremath{\Upsilon{(4S)}}\xspace}
\def\epem    {\ensuremath{e^+e^-}\xspace}
\def\mes     {\mbox{$m_{\smaller\rm ES}$}\xspace}
\def\DeltaE  {\mbox{$\Delta E$}\xspace}
\def\invfb   {\ensuremath{\mbox{\,fb}^{-1}}\xspace}
\def\pep2{PEP-II} 
\def\gev     {\ensuremath{\mathrm{\,Ge\kern -0.1em V}}\xspace}
\def\mev     {\ensuremath{\mathrm{\,Me\kern -0.1em V}}\xspace}
\def\gevc    {\ensuremath{{\mathrm{\,Ge\kern -0.1em V\!/}c}}\xspace}
\def\mevc    {\ensuremath{{\mathrm{\,Me\kern -0.1em V\!/}c}}\xspace}
\def\gevcc   {\ensuremath{{\mathrm{\,Ge\kern -0.1em V\!/}c^2}}\xspace}
\def\mevcc   {\ensuremath{{\mathrm{\,Me\kern -0.1em V\!/}c^2}}\xspace}
\def\gevccsq {\ensuremath{{\mathrm{\,Ge\kern -0.1em V^2\!/}c^4}}\xspace}
\def\mevccsq {\ensuremath{{\mathrm{\,Me\kern -0.1em V^2\!/}c^4}}\xspace}
\def\gevnc   {\ensuremath{{\mathrm{\,Ge\kern -0.1em V\!/}c^n}}\xspace}
\def\mevnc   {\ensuremath{{\mathrm{\,Me\kern -0.1em V\!/}c^n}}\xspace}

\def\Dmaybestar {\ensuremath{D^{(\!\ast\!)}}}
\def\Dmaybestarz{\ensuremath{D^{(\!\ast\!)0}}}
\def\Dmaybestarp{\ensuremath{D^{(\!\ast\!)+}}}
\def\Dmaybestarm{\ensuremath{D^{(\!\ast\!)-}}}
\def\Wm   {\ensuremath{W^-}}

\def\ubar {\ensuremath{\kern 0.10em \overline{\kern-0.10em u}\kern 0.05em{}\xspace}}
\def\dbar {\ensuremath{\kern 0.15em \overline{\kern-0.15em d}\kern 0.05em{}\xspace}}
\def\sbar {\ensuremath{\kern 0.10em \overline{\kern-0.10em s}\kern 0.05em{}\xspace}}
\def\cbar {\ensuremath{\kern 0.10em \overline{\kern-0.10em c}\kern 0.05em{}\xspace}}
\def\qbar {\ensuremath{\kern 0.15em \overline{\kern-0.15em q}\kern 0.05em{}\xspace}}
\def\nbar {\ensuremath{\kern 0.10em \overline{\kern-0.10em n}\kern 0.05em{}\xspace}}
\def\pbar {\ensuremath{\kern 0.10em \overline{\kern-0.10em p}\kern 0.05em{}\xspace}}
\def\pmaybebar{\ensuremath{^{\mbox{\fontsize{1}{2}\selectfont$^($}}
	\kern-0.15em \pbar^{\mbox{\fontsize{1}{2}\selectfont$^)$}}
	}\xspace}
\def\ppbar{\ensuremath{p\pbar}}
\def\Nbar {\ensuremath{\kern 0.25em \overline{\kern-0.25em N}\kern 0.05em{}\xspace}}
\def\NNbar{\nolbreaks{\ensuremath{N\Nbar^{\prime}}}}
\def\To{\ensuremath{\!\rightarrow\!}}
\def\eq{\!\,\,=\!\,\,}
\def\dbline{\noalign{\vskip 0.10truecm\hrule\vskip 0.05truecm\hrule\vskip 0.10truecm}}
\def\sgline{\noalign{\vskip 0.10truecm\hrule\vskip 0.10truecm}}

\def\percent{\%}

\newcommand{\BABARPubYear}    {09}
\newcommand{\BABARConfNumber} {002}
\newcommand{\SLACPubNumber}   {13755}
\newcommand{\LANLNumber}      {}

\long\def\inst#1{\par\nobreak\kern 4pt\nobreak
    {\it #1}\par\vskip 10pt plus 3pt minus 3pt}

\begin{document}
\singlespacing

\thispagestyle{empty}

\begin{flushright}
BABAR-CONF-\BABARPubYear/\BABARConfNumber \\
SLAC-PUB-\SLACPubNumber \\
hep-ex/\LANLNumber \\
August 2009 \\
\end{flushright}

\par\vskip 4cm

\begin{center}
\Large \bf \boldmath Observation and study of baryonic $B$ decays:\\
$B\To\Dmaybestar\ppbar$, $\Dmaybestar\ppbar\pi$, and $\Dmaybestar\ppbar\pi\pi$ 
\end{center}
\bigskip

\begin{center}
\large The \babar\ Collaboration\\
\mbox{ }\\
\today
\end{center}
\bigskip \bigskip

\begin{center}
\bf ABSTRACT
\end{center}
\noindent 
We present a study of ten $B$-meson decays to a ${D}^{(\!\ast\!)}$, a
proton-antiproton pair, and a system of up to two pions using \babar's
data set of $455\times 10^6$ $B{\kern 0.18em\overline{\kern -0.18em B}}$
pairs.  Four of the modes 
(\nolbreaks{${\kern 0.18em\overline{\kern -0.18em B}}^0\!\rightarrow\!{D}^0{p}\overline{p}$},
 \nolbreaks{${\kern 0.18em\overline{\kern -0.18em B}}^0\!\rightarrow\!{D}^{\ast0}{p}\overline{p}$},
 \nolbreaks{${\kern 0.18em\overline{\kern -0.18em B}}^0\!\rightarrow\!{D}^+{p}\overline{p}\pi^-$},
 \nolbreaks{${\kern 0.18em\overline{\kern -0.18em B}}^0\!\rightarrow\!{D}^{ast+}{p}\overline{p}\pi^-$})
are studied with improved statistics compared to previous
measurements; six of the modes
(\nolbreaks{$B^-\!\rightarrow\!{D}^0{p}\overline{p}\pi^-$},
 \nolbreaks{$B^-\!\rightarrow\!{D}^{\ast0}{p}\overline{p}\pi^-$},
 \nolbreaks{${\kern 0.18em\overline{\kern -0.18em B}}^0\!\rightarrow\!{D}^0{p}\overline{p}\pi^-\pi^+$},
 \nolbreaks{${\kern 0.18em\overline{\kern -0.18em B}}^0\!\rightarrow\!{D}^{\ast0}{p}\overline{p}\pi^-\pi^+$},
 \nolbreaks{$B^-\!\rightarrow\!{D}^+{p}\overline{p}\pi^-\pi^-$},
 \nolbreaks{$B^-\!\rightarrow\!{D}^{ast+}{p}\overline{p}\pi^-\pi^-$})
are first observations.  The branching fractions for 3- and 5-body
decays are suppressed compared to 4-body decays.  Kinematic
distributions for 3-body decays show non-overlapping threshold
enhancements in $m({p}\overline{p})$ and $m({D}^{(\!\ast\!)0}{p})$ in
the Dalitz plots.  For 4-body decays, $m(p\pi^-)$ mass projections show
a narrow peak with mass and full width of
($1497.4\pm 3.0\pm 0.9$)$\mathrm{\,Me\kern -0.1em V\!/}c^2$ and
($47\pm 12\pm 4$)$\mathrm{\,Me\kern -0.1em V\!/}c^2$, respectively,
where the first (second) errors are statistical (systematic). For
5-body decays, mass projections are similar to phase space
expectations.  All results are preliminary. 

\\

\vfill
\begin{center}
Submitted to the XXIV International Symposium on Lepton Photon \\
Interactions at High Energies, \\ August 17--22, 2009, Hamburg,
Germany.
\end{center}

\vspace{1.0cm}
\begin{center}
{\em SLAC National Accelerator Laboratory, Stanford University,
  Stanford, CA 94309} \\
\vspace{0.1cm}\hrule\vspace{0.1cm}
Work supported in part by Department of Energy contract
DE-AC02-76SF00515.
\end{center}

\newpage
\pagestyle{plain}

\begin{center}
\small

The \babar\ Collaboration,
\bigskip

%
{B.~Aubert,}
{Y.~Karyotakis,}
{J.~P.~Lees,}
{V.~Poireau,}
{E.~Prencipe,}
{X.~Prudent,}
{V.~Tisserand}
\inst{Laboratoire d'Annecy-le-Vieux de Physique des Particules (LAPP), Universit\'e de Savoie, CNRS/IN2P3,  F-74941 Annecy-Le-Vieux, France}
{J.~Garra~Tico,}
{E.~Grauges}
\inst{Universitat de Barcelona, Facultat de Fisica, Departament ECM, E-08028 Barcelona, Spain }
{M.~Martinelli$^{ab}$,}
{A.~Palano$^{ab}$,}
{M.~Pappagallo$^{ab}$ }
\inst{INFN Sezione di Bari$^{a}$; Dipartimento di Fisica, Universit\`a di Bari$^{b}$, I-70126 Bari, Italy }
% ...

\end{center}\newpage

\section{INTRODUCTION}
\label{sec:introduction}

The decays\footnote{Charge conjugation of particles and decays are
implied throughout this document unless otherwise stated.} of $B$
mesons to final states with baryons have been explored much less
systematically than decays to meson-only final states.  Such decays
have their own distinctive features; in particular, the suppression of
the rate for two-body decays and the rate enhancement for low masses
of the baryon-antibaryon system in multi-body decays \cite{Chen:2008jy,
  Aubert:2007qea,
  Wang:2007as,
  Medvedeva:2007zz,
  Wei:2007fg,
  Aubert:2005gw,
  Lee:2004mg}.
The first observed exclusive decays were the CLEO measurements of
\nolbreaks{$B\To\Lambda_c^+\,\pbar\pi(\!\pi\!)$} \cite{Fu:1996qt}
followed by \nolbreaks{$\Bzb\To\Dstarp{p}\pbar\pim$} and
\nolbreaks{$\Bz\To\Dstarm{p}\nbar\,$}\footnote{The charge conjugate
$\Bzb\To\Dstarp\pbar{n}$ is not included since CLEO only detects
antineutrons $\nbar$.} \cite{Anderson:2000tz} supporting a
prediction~\cite{Dunietz:1998uz} that the decays with $\Lambda_c$ are
not the only significant contributions to the baryonic $B$ decay rate
and that \nolbreaks{$\Dmaybestar\NNbar\!+\!\textrm{anything}$} is also
important, where $N$ and $N^{\prime}$ are nucleons. Belle observed
\nolbreaks{$\Bzb\To\Dmaybestarz{p}\pbar$} \cite{Abe:2002tw} obtaining
a branching fraction about five times smaller than for
$\Dstarp{p}\pbar\pim$.

\begin{figure}[b!]
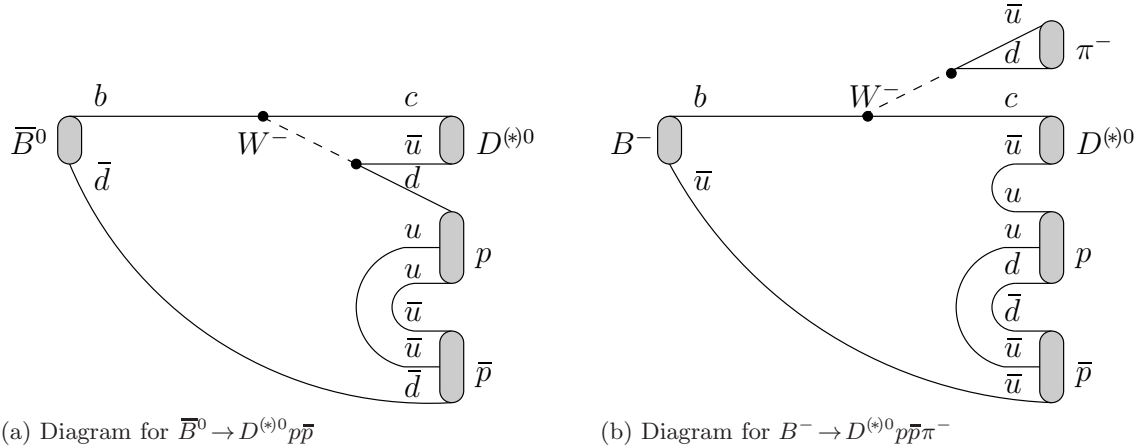

\centering
\subfloat[Diagram for $\Bzb\To\Dmaybestarz\ppbar$]{
\label{fig:feyn1}
\input{feyn_3bodyNt.pstex_t}
}%
\hspace{0.5in}
\subfloat[Diagram for $\Bm\To\Dmaybestarz\ppbar\pim$]{
\label{fig:feyn2}
\input{feyn_4bodyCh.pstex_t}
}
\caption{Typical valence-quark diagrams for (a) 3- and (b) 4-body $B$
  decays.} 
\label{fig:feyn}
\end{figure}

We report the branching fractions and kinematic distributions for
ten baryonic $B$ decays:
\begin{eqnarray}
\begin{array}{clllllll}
\textrm{3-body decays}&\Bzb\To\Dz{p}\pbar         &\textrm{and}&\Dstarz{p}\pbar,          \\
\textrm{4-body decays}&\Bzb\To\Dp{p}\pbar\pim     &\textrm{and}&\Dstarp{p}\pbar\pim,      \\
''                    &\Bm\To\Dz{p}\pbar\pim      &\textrm{and}&\Dstarz{p}\pbar\pim,      \\ 
\textrm{5-body decays}&\Bzb\To\Dz{p}\pbar\pim\pip &\textrm{and}&\Dstarz{p}\pbar\pim\pip,  \\
''                    &\Bm\To\Dp{p}\pbar\pim\pim  &\textrm{and}&\Dstarp{p}\pbar\pim\pim,
\end{array}
\label{eqn:decays}
\end{eqnarray}
where the latter six in the list are first
observations.\footnote{\nolbreaks{$\Bm\To\Dmaybestarm\ppbar$} are
suppressed by $\lambda^2$ \cite{Wolfenstein:1983yz} with respect to
Fig.~\ref{fig:feyn1} for \nolbreaks{$b\To{u}\Dmaybestarm$} and are
beyond our sensitivity.} The $\Dstar$ are reconstructed as $\Dz\pip$
and $\Dz\piz$; $D$ as $\Km\pip$, $\Km\pip\piz$, $\Km\pip\pim\pip$, and
$\Km\pip\pip$.  This makes up 26 reconstructed decay chains (3 for
each of eight $B$ decays with $\Dz$ and 1 for two with \Dp).  This
study uses $455\times10^6$ \BB\ pairs and supersedes the previous
\babar\ publication of \nolbreaks{$\Bzb\To\Dmaybestarz\ppbar$} and
\nolbreaks{$\Dmaybestarp\ppbar\pim$} \cite{Aubert:2006qx} using
$232\times10^6$ \BB\ pairs.  Figure~\ref{fig:feyn} shows the typical
valence-quark diagrams for 3- and 4-body decays.  

Until recently, interest has focused on the dynamical features of
baryonic decays, on studying rare modes (both $b\To{u}$ and $b\To{s}$
transitions), and on the role these modes play in accounting for the
overall production of charm in $B$ decays.  Many theoretical studies
have appeared
\cite{Dunietz:1998uz,
   Hou:2000bz,
   Chua:2001vh,
   Chua:2002wp,
   Kerbikov:2004gs,
   Cheng:2001tr,
   Cheng:2001ub,
   Cheng:2006nm,
   Cheng:2003fq,
   Cheng:2002sa,
   Luo:2003pv,
   Rosner:2003bm,
   Rosner:2003ia,
   Suzuki:2005iq,
   Cheng:2006bn}
using the decays as a playground for phenomenological models including
some with $\ppbar$ bound states \cite{Rosner:2003bm} and multi-quark
intermediate resonances \cite{Rosner:2003ia}.  We hope to shed light
on these models by studying the relatively unexplored territory of
$b\To{c}$ baryonic $B$ decays involving a $\Dmaybestar$ meson.

\section{\boldmath THE \babar\ DETECTOR AND DATA SET}
\label{sec:babar}

The \babar\ detector is described in detail
elsewhere~\cite{Aubert:2001tu}.  Exclusive $B$-meson decays are
simulated with the Monte Carlo (MC) event generator
E{\sc vt}G{\sc en}~\cite{Lange:2001uf} and hadronization is simulated
with J\textsc{etset~7.4}~\cite{Sjostrand:1993yb}.  We use
GEANT4~\cite{Agostinelli:2002hh} to model interactions of
particles traversing the detector, taking into account the varying
detector conditions and beam backgrounds. 

\begin{figure}[b!]
\centering
  \begin{overpic}[width=1\textwidth]{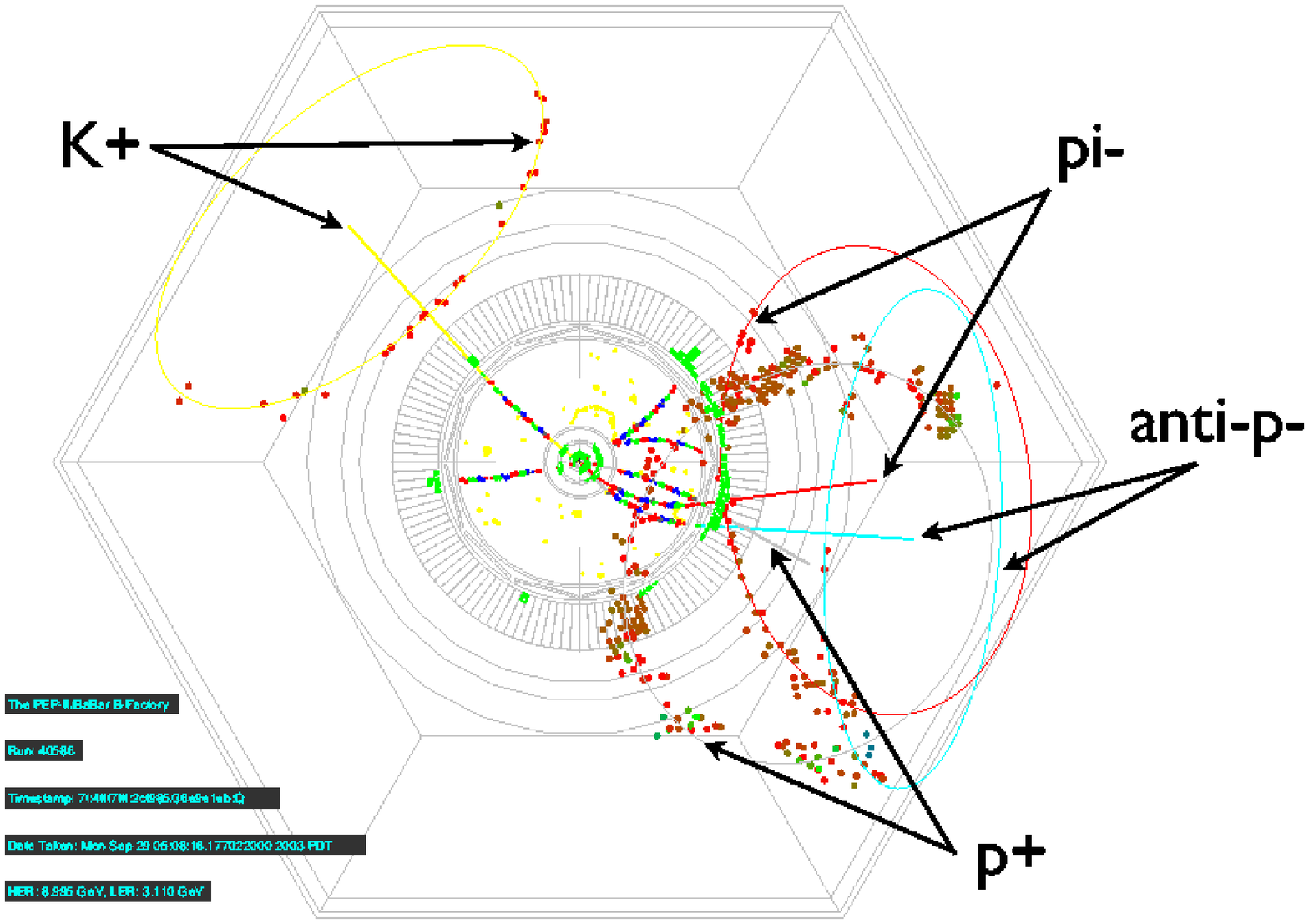}
    \put(2,56){\colorbox{white}{\Huge\Km}}
    \put(79,56){\colorbox{white}{\Huge\pim}}
    \put(84,37){\colorbox{white}{\Huge\quad\ \pbar\ \ }}
    \put(72,5){\colorbox{white}{\Huge$p$\quad}}
  \end{overpic}
\caption{Event display of $\Bz\To\Dzb\ppbar$,
  $\Dzb\To\Kp\pim$ transverse to the $e^-$ beam direction.
}
\label{fig:evtdisplay}
\end{figure}

This study uses the data set of $414\invfb$ collected with the \babar\
detector at the PEP-II asymmetric-energy \epem\ collider at the
SLAC National Accelerator Laboratory.  Figure~\ref{fig:evtdisplay} is a
data event display of $\Bz\To\Dzb\ppbar$ followed by $\Dzb\To\Kp\pim$.
For each daughter particle, we can see the drift chamber hits matching
up with projected Cherenkov light cones in the ring-imaging detector.

\section{ANALYSIS METHOD}
\label{sec:analysis}

\subsection{Event selection}
\label{sec:selection}

We select events with a $B$ candidate reconstructed in one of 26 decay
chains in Eqn.~\ref{eqn:decays} in two steps:

First, the \babar\ data set is reduced by requiring the presence of a
$p$, $\pbar$, and $D$. The average momentum of a proton produced in a
typical $B$ decay listed in Eqn.~\ref{eqn:decays} is $1\gevc$.
Protons are identified with a likelihood-based selector using
information from the silicon vertex tracker, drift chamber, and
ring-imaging Cherenkov detector, which has a $98\percent$ selection
efficiency and a $1\percent$ kaon fake rate.  The $D$ mesons decay to
charged tracks and $\piz$s. Charged tracks are required to be in the
fiducial volume and have a distance of closest approach to the beam
spot $<\!1.5$ cm.  Neutral pions are formed from two well-separated
photons with \nolbreaks{$115\!<\!m_{\gamma\gamma}\!<\!150\mevcc$} or
from two unseparated photons using the second moment of the
electromagnetic calorimeter energy distribution.  The same
requirements apply to the non-composite daughters of $B$ and $\Dstar$
decays.

Second, the data set is further reduced by the explicit reconstruction
of $B$ candidates. $B$ candidates are formed with a $D$ mass
\cite{Amsler:2008zzb} constraint using a Kalman
fitter~\cite{Hulsbergen:2005pu} with its vertex $\chi^2$ probability
$>\!0.1\percent$.  To suppress continuum $\epem\To{q}\qbar$
($q\!=\!u,d,s,c$) events, we compute the angle between the thrust axes
of the $B$ candidate and the rest of the event \cite{Aubert:2001xs}.
For modes with a $\Dstar$, the mass difference
$\Delta{m}{\eq}m(\Dz\pi)\!-\!m(\Dz)$ is required to be within
$3\sigma$ of the nominal value, where the resolution is around
$0.8\mevcc$.  For $\Bzb\To\Dz\ppbar\pim\pip$, we require
$\Delta{m}\!>\!160\mevcc$ on the $\Dz\pip$ system.  For all decay
modes, $D$ candidates are required to be within $3\sigma$ around the
nominal value, where the mass resolution for $K\pi$, $K\pi\piz$,
$K\pi\pi\pi$, and $K\pi\pi$ is around $6$, $10$, $5$, and $5$ \mevcc,
respectively.  Furthermore, the average momentum of a kaon produced in
a typical $D$ decay is $0.9\gevc$ and is found with the
above-mentioned likehood technique, which has an $85\percent$ selection
efficiency and a $2\percent$ pion fake rate.  For $K\pi\piz$ decays, we use
the squared decay amplitude based on a Dalitz plot model
\cite{Frabetti:1994di}.  Lastly, requirements are optimized by
maximizing the ratio of the squared expected signal yield and its sum
with expected backgrounds.  For cut values with broad maxima, they are
chosen to be uniform across related decay modes.

After events are filtered through the two steps, we are left with
$\mathcal{O}(10^5)$ for all modes.  The average number of $B$
candidates per event that pass all requirements ranges from
\nolbreaks{$1.0$--$1.7$}, increasing with the multiplicity of the
decay.  If there are multiple such candidates, then we choose one with
the $\Dmaybestar$ mass closest to the nominal value; furthermore, if
candidates share a $\Dmaybestar$, we choose one at random.  The
reconstruction efficiencies are found using MC containing the desired
$B$ decay.  In general, efficiency decreases with particle
multiplicity: $\Dz\ppbar$, $K\pi$ is highest at $19\percent$ and
\nolbreaks{$\Dstar\ppbar\pi\pi$, $K\pi\pi\pi$} is lowest at
$\mathcal{O}(1\percent)$; they are given later in Table~\ref{tab:bfchain}.

\subsection{Fit method and yields}
\label{sec:fit}

\begin{figure}[b!]
\centering
\subfloat[$\Bm\To\Dz\ppbar\pim$]{
  \hspace{-0.03\textwidth}%
  \begin{overpic}[width=0.35\textwidth]
    {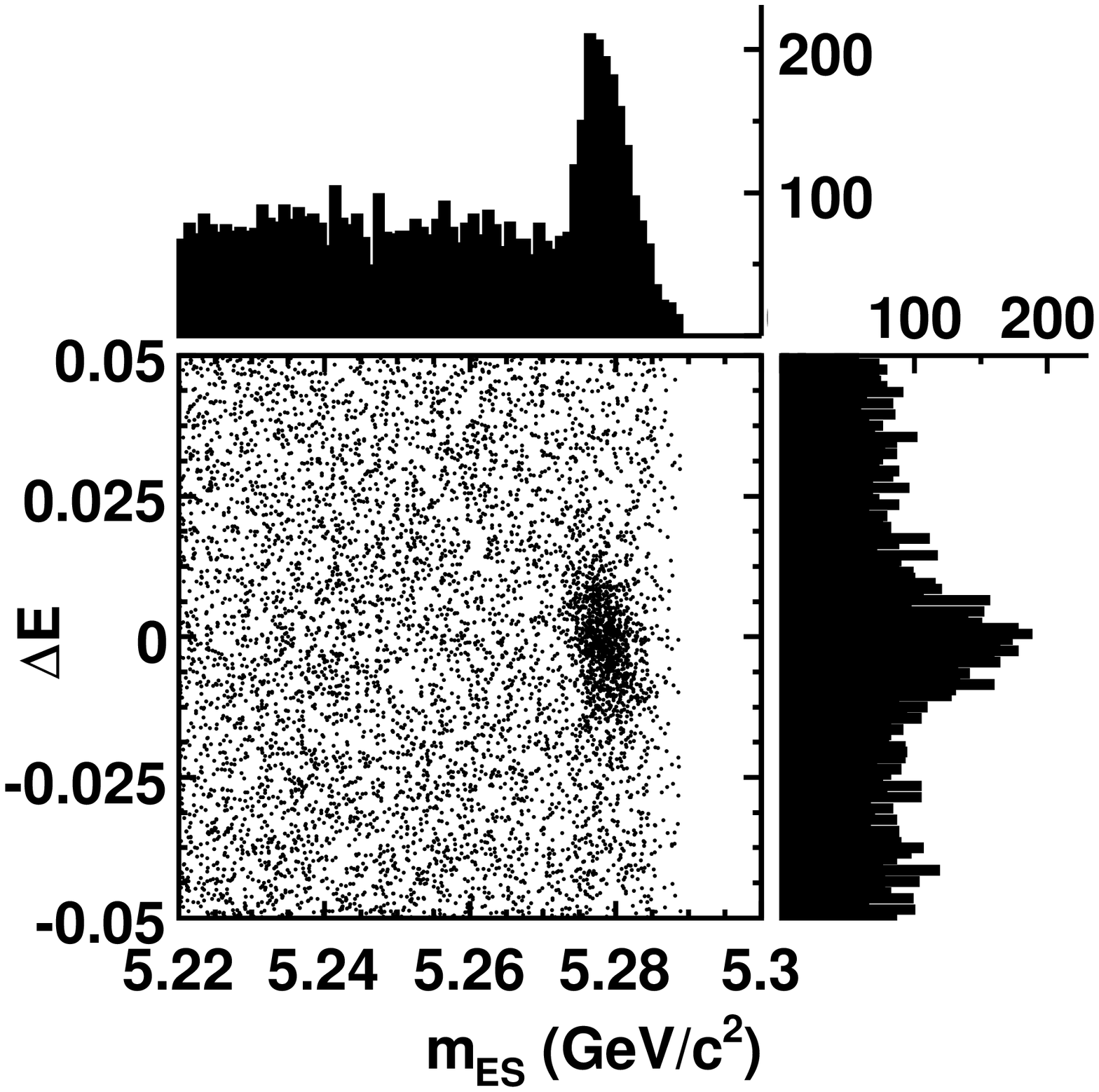}
    \put(20,88){\small{\babar}}
    \put(20,82){\small{prelim.}}
  \end{overpic}
}%
\subfloat[$\Bzb\To\Dz\ppbar\pim\pip$]{
  \hspace{-0.03\textwidth}%
  \begin{overpic}[width=0.35\textwidth]
    {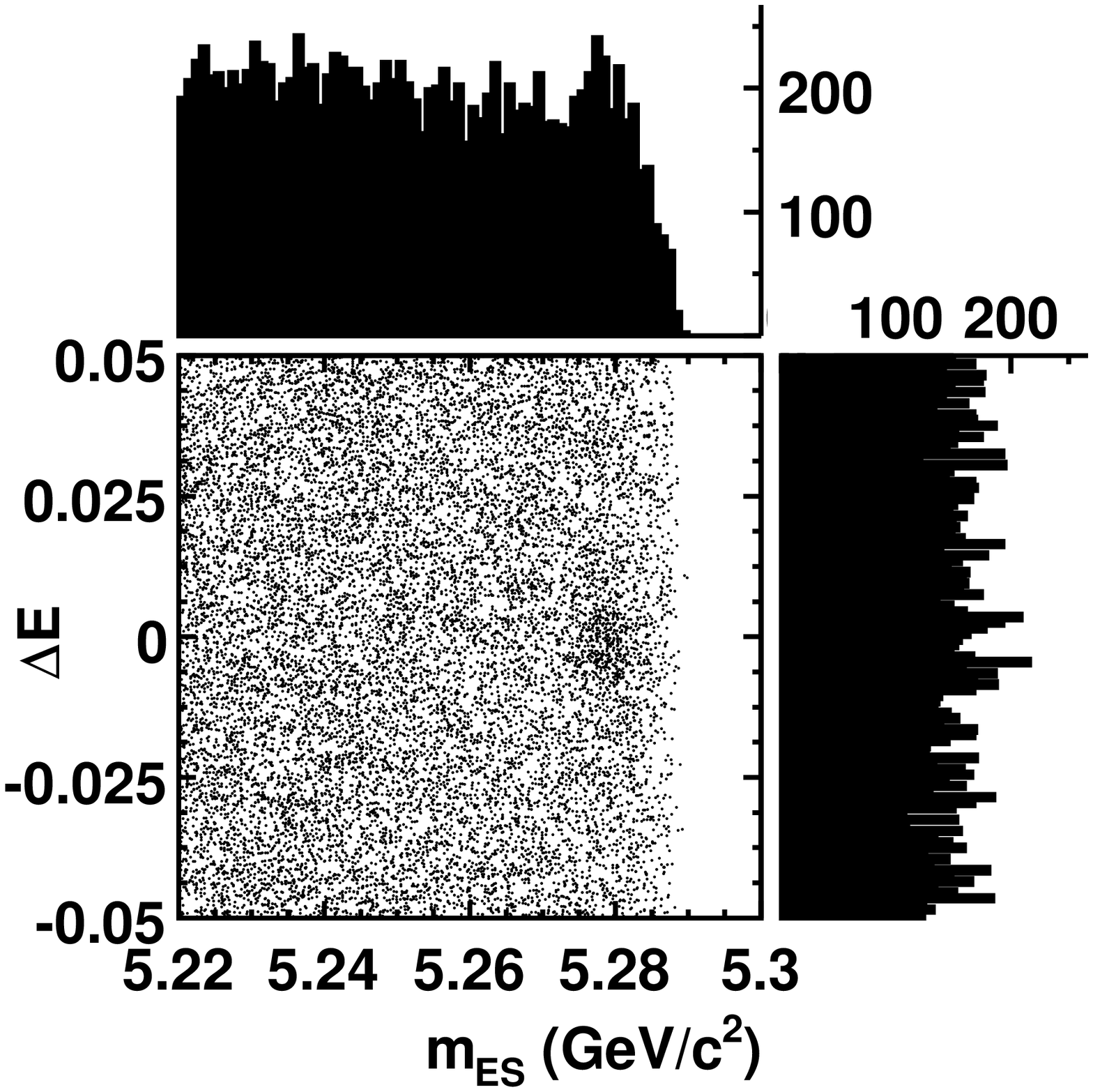}
    \put(20,75){\small\White{\babar}}
    \put(20,69){\small\White{prelim.}}
\end{overpic}
}%
\subfloat[$\Bm\To\Dp\ppbar\pim\pim$]{
  \hspace{-0.03\textwidth}%
  \begin{overpic}[width=0.35\textwidth]
    {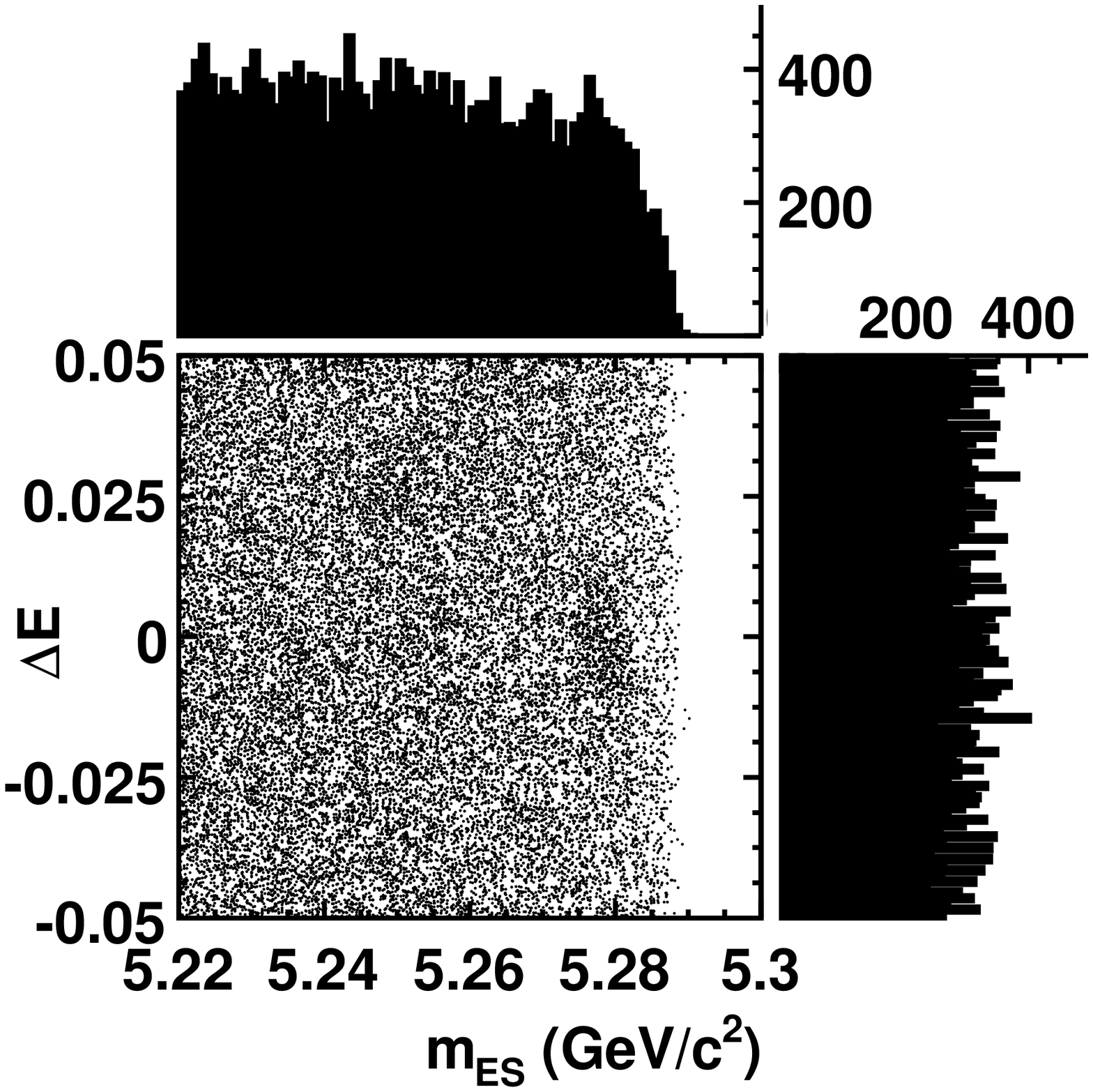}
    \put(20,75){\small\White{\babar}}
    \put(20,69){\small\White{prelim.}}
    \end{overpic}
}%
\\
\subfloat[$\Bm\To\Dstarz\ppbar\pim$]{
  \hspace{-0.03\textwidth}%
  \begin{overpic}[width=0.35\textwidth]
    {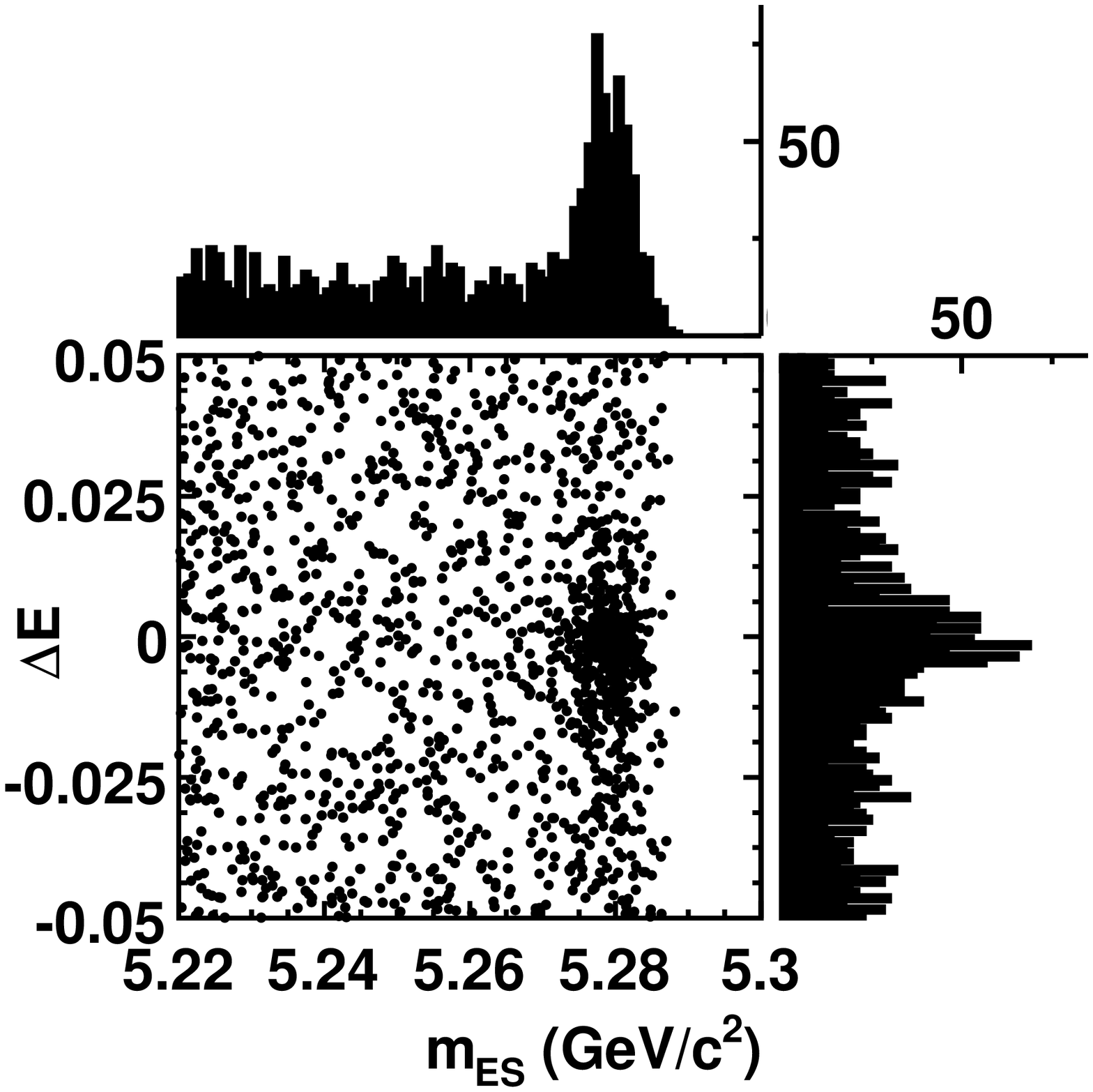}
    \put(20,86){\small{\babar}}
    \put(20,80){\small{prelim.}}
    \end{overpic}
}%
\subfloat[$\Bzb\To\Dstarz\ppbar\pim\pip$]{
  \hspace{-0.03\textwidth}%
  \begin{overpic}[width=0.35\textwidth]
    {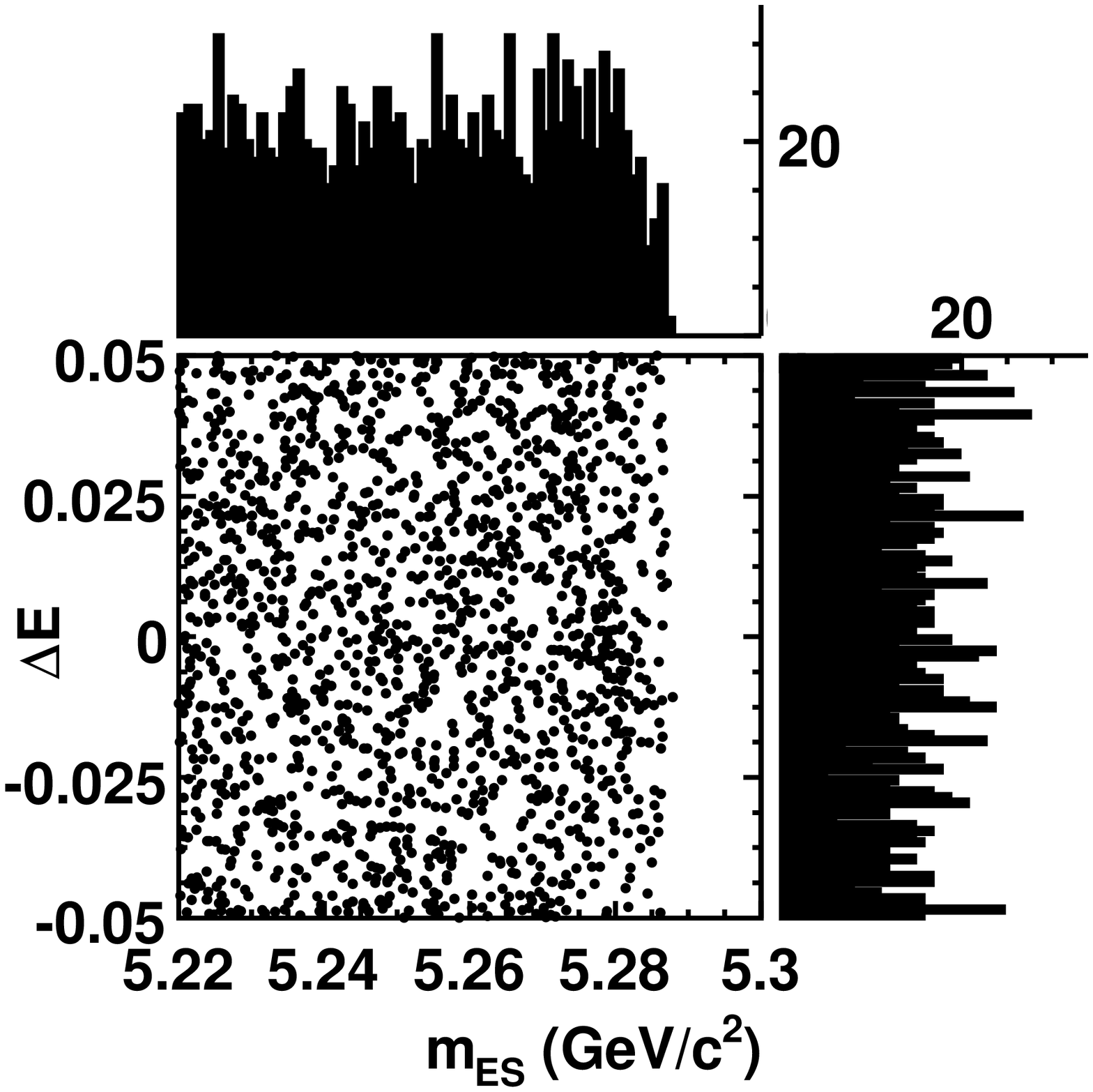}
    \put(20,73){\small\White{\babar}}
    \put(20,67){\small\White{prelim.}}
  \end{overpic}
}%
\subfloat[$\Bm\To\Dstarp\ppbar\pim\pim$]{
  \hspace{-0.03\textwidth}%
  \begin{overpic}[width=0.35\textwidth]
    {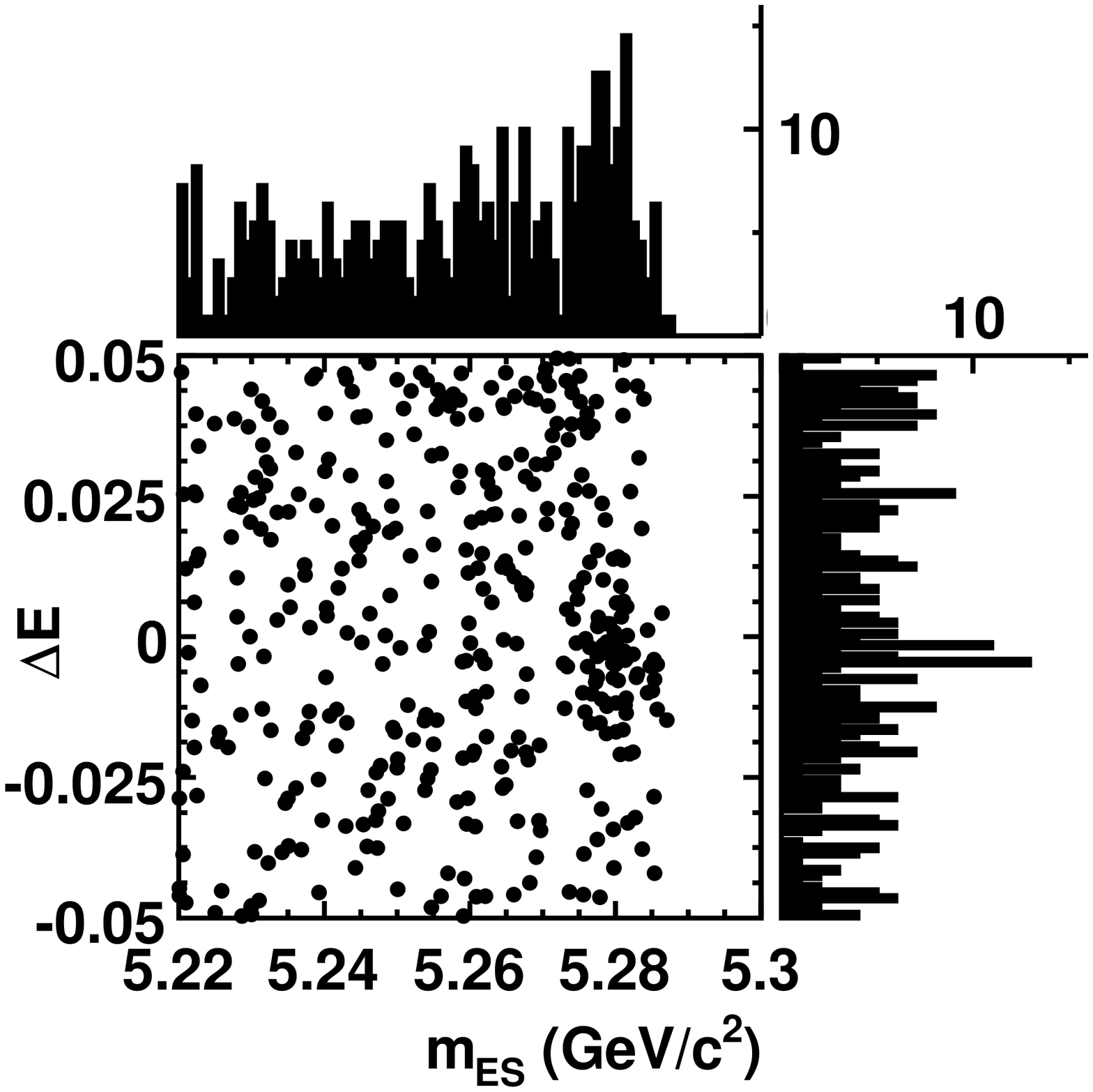}
    \put(20,88){\small{\babar}}
    \put(20,82){\small{prelim.}}
\end{overpic}
}%
\caption{\mes-\DeltaE\ scatter plots for new observations of
  $B$ decays:
    (a) $\Bm\To\Dz\ppbar\pi$, 
    (b) $\Bzb\To\Dz\ppbar\pi\pi$, 
    (c) $\Bm\To\Dp\ppbar\pi\pi$, 
    (d) $\Bm\To\Dstarz\ppbar\pi$, 
    (e) $\Bzb\To\Dstarz\ppbar\pi\pi$, and
    (f) $\Bm\To\Dstarp\ppbar\pi\pi$. 
  We show the cleanest mode with $D^{0,+}\To\Km\pim$,
  $\Km\pip\pip$, resp.  Histograms on top (right) are \mes\ (\DeltaE)
  projections in $1$ \mevcc\ (\mev) bins; no cut is made on the
  complementary variable.}
\label{fig:run16_2dscatter}
\end{figure}

The signal $B$ yield is extracted by fitting the joint distribution of
\begin{eqnarray}
  \mes    = \sqrt{\frac{s}{4} - (\mathbf{P}_B)^2} \quad\textrm{ and }\quad
  \DeltaE =  E_B - \sqrt{\frac{s}{4}},
\end{eqnarray}
where $\sqrt{s}$ is the \epem\ center-of-mass (cms) energy and
$\mathbf{P}_B$ ($E_B$) is the $B$ candidate momentum (energy) in the
cms.  Figure~\ref{fig:run16_2dscatter} gives \mes-\DeltaE\ scatter plots
for the six $B$ decays that are first observations.  A concentration
of correctly reconstructed $B$ candidates is visible in the region of
$\mes$ and $\DeltaE$ near the nominal $B$ mass of $5.28\gevcc$ and
zero, respectively.  The uniform distribution of dots over the entire
plane is indicative of the general smoothness of the background
events.

A 2-dimensional (2d) probability distribution function (pdf) is used
to fit $\mes$-$\DeltaE$ via the unbinned extended maximum likelihood
technique \cite{Barlow:1990vc}.  The 2d pdf is a sum of two
components for all $B$ decay modes except for $\Dstarz\ppbar\pi$,
which requires three.  These are denoted as $P_S$, $P_B$, and $P_P$
respectively, for the signal, background, and peaking background
events.  The motivation and description for $P_P$ is given later and
continued in Sec.~\ref{sec:syst}.  The parameters (yields) associated
with the components are $\Omega_S$, $\Omega_B$, and $\Omega_P$ ($n_S$,
$n_B$, and $n_P$), respectively.  Since the fit variables are
uncorrelated to a good approximation, each 2d pdf component is written
as product of two 1d pdfs.  $P_S$ is the product of two functions with
a Gaussian core and a power-law tail \cite{CBshape} written as
$P_g(\mes;\Omega_S)P_g(\DeltaE;\Omega_S^\prime)$.  $P_B$ is the
product of a threshold function vanishing at the nominal $B$ mass for
\mes\ and a $2^\textrm{nd}$-order Chebyshev polynomial of the first
kind for \DeltaE\ written as $P_t(\mes;\Omega_B)P_c(\DeltaE;\Omega_B^\prime)$.
$P_P$ for $\Dstarz\ppbar\pi$ is written as
$P_g(\mes;\Omega_P)P_g(\DeltaE;\Omega_P^\prime)$.

\begin{figure}[bp!]
\centering
\subfloat[$-18.8\!<\!\DeltaE\!<\!14.7$]{
  \hspace{-0.03\textwidth}%
  \begin{overpic}[width=0.260\textwidth]
    {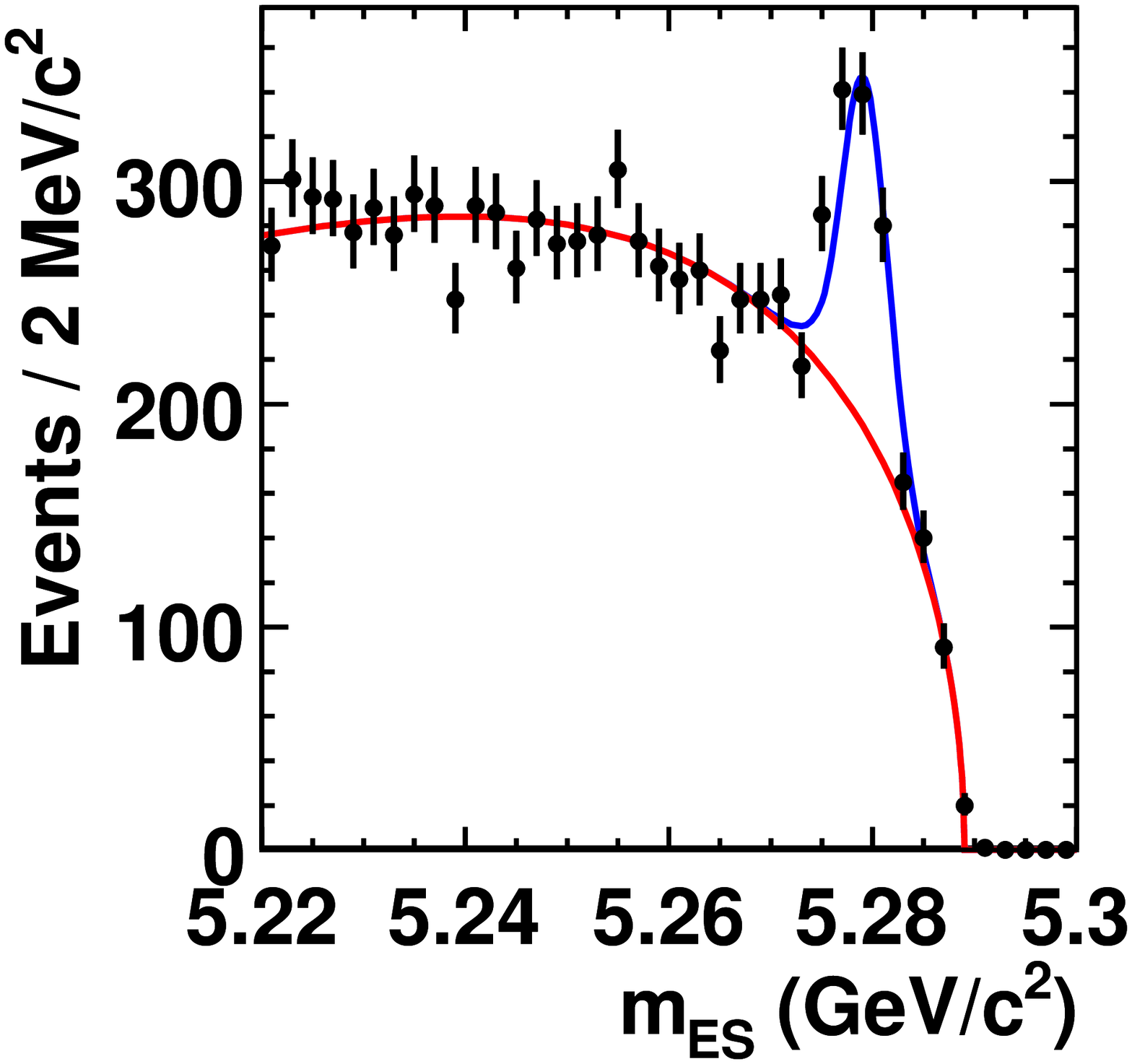}
    \put(30,36){\small{\babar}}
    \put(30,28){\small{prelim.}}
  \end{overpic}
}%
\subfloat[$5273.6\!<\!\mes\!<\!5284.8$]{
  \hspace{-0.03\textwidth}%
  \begin{overpic}[width=0.260\textwidth]
    {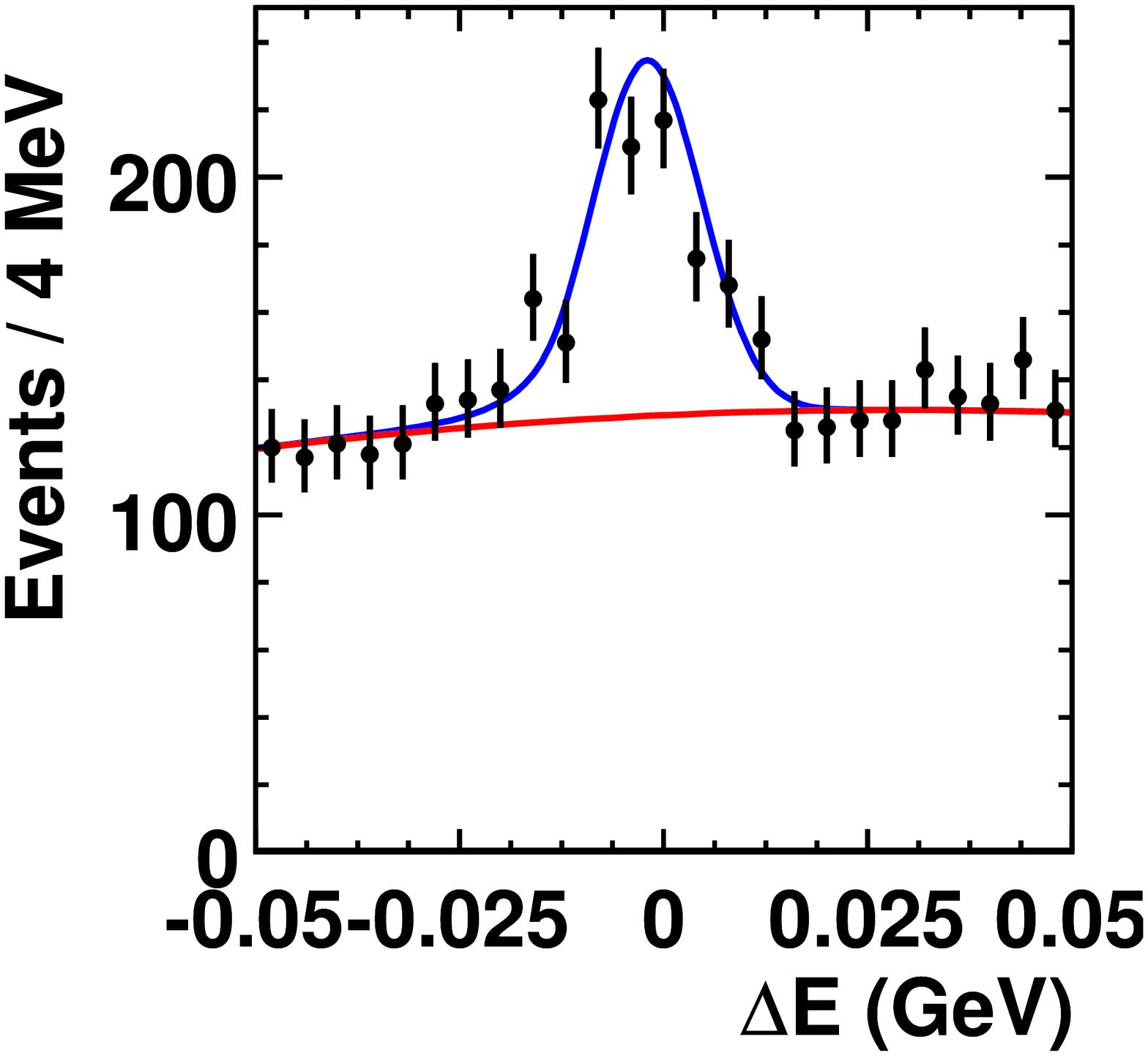}
    \put(30,36){\small{\babar}}
    \put(30,28){\small{prelim.}}
  \end{overpic}
}%
\subfloat[$\DeltaE\!<\!-28.9$ or $\!>24.8$]{
  \hspace{-0.03\textwidth}%
  \begin{overpic}[width=0.260\textwidth]
    {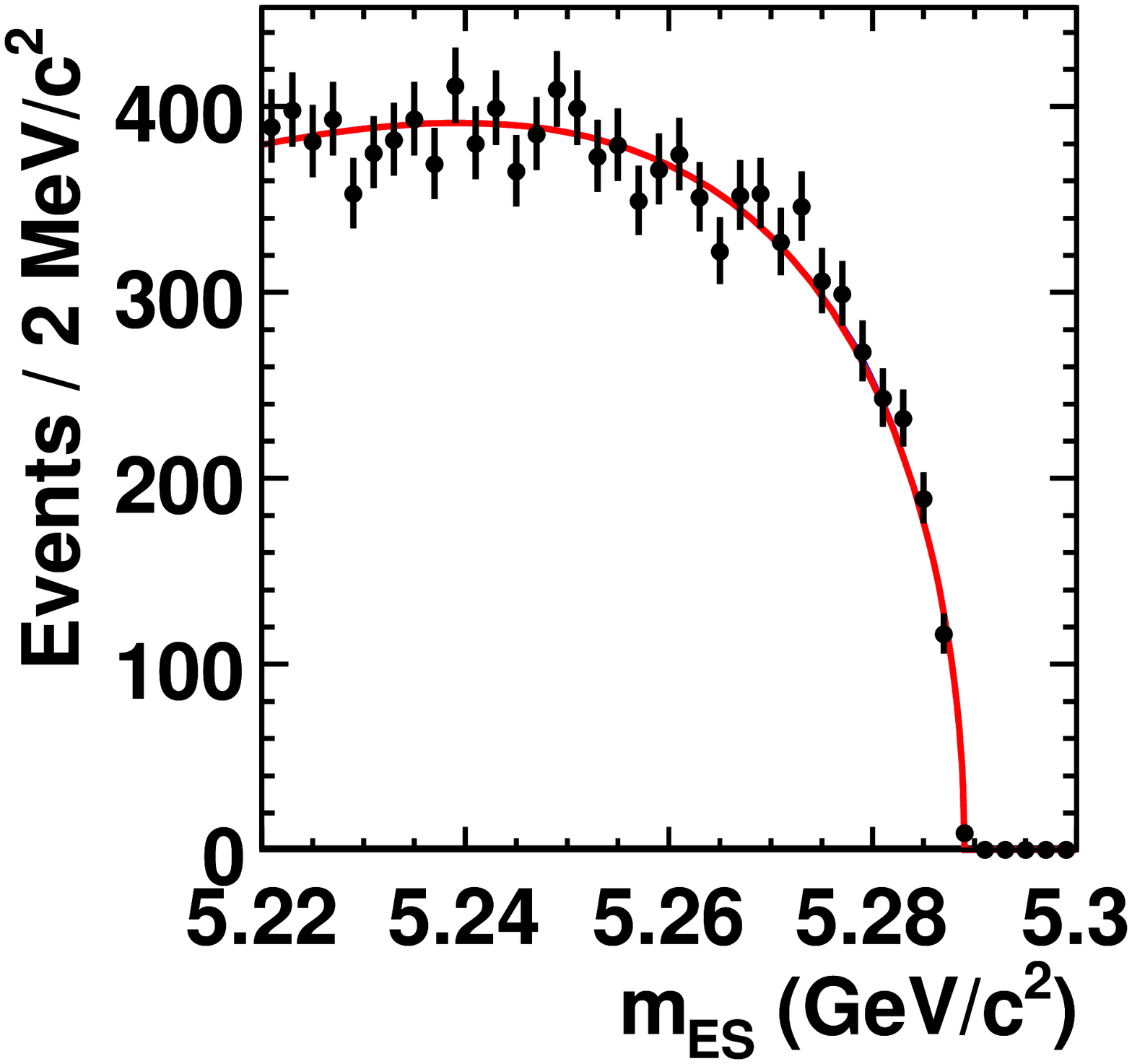}
    \put(30,36){\small{\babar}}
    \put(30,28){\small{prelim.}}
  \end{overpic}
}%
\subfloat[$\mes\!<\!-5270.3$]{
  \hspace{-0.03\textwidth}%
  \begin{overpic}[width=0.260\textwidth]
    {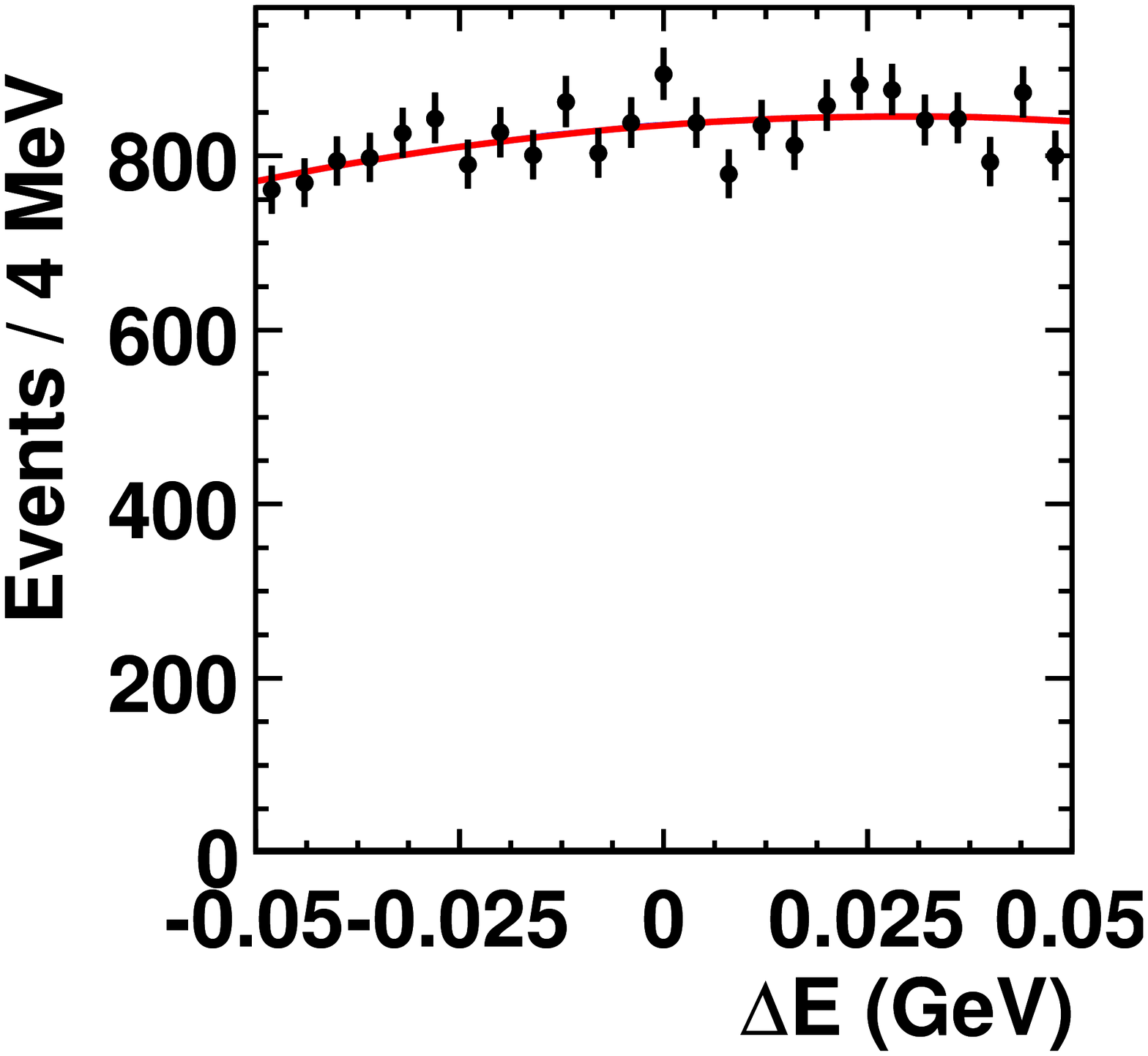}
    \put(30,36){\small{\babar}}
    \put(30,28){\small{prelim.}}
  \end{overpic}
}%
\caption{The \mes-\DeltaE\ fit projections for
  $\Bm\To\Dp\ppbar\pi\pi$, $\Dp\To{K}\pi\pi$ in (ab) signal and (cd)
  sideband regions with requirements denoted in the caption in \mevcc\
  (\mev) for \mes\ (\DeltaE).  For signal slices in (ab), the top blue
  curve is the signal pdf $P_S$; bottom red is the background pdf
  $P_B$.  For sideband slices in (cd), no part of the signal pdf is
  present in this region and the background pdf is identical to the
  corresponding plot in (ab).}
\label{fig:2dfits_example}
\end{figure}

The fit window of $5.22\!<\!\mes\!<\!5.30\gevcc$ and
$|\DeltaE|\!<\!50\mev$ provides ample sideband regions because the
signal resolution of \mes\ and \DeltaE\ are relatively narrow at
$2.2$--$2.5$\mevcc and $8$--$10$\mev, respectively.  Signal slice
projection plots for \mes\ are
defined to be within $2.5\sigma$ of the mean value of \DeltaE\ and
sideband to be outside $4\sigma$ and vice versa for \DeltaE\ plots. As
an illustration of the fit projection over the entire window,
Fig.~\ref{fig:2dfits_example}ab(cd) gives projections of the 2d pdf in
the signal (sideband) slices. For $\Bm\To\Dm\ppbar\pi\pi$,
corresponding to the scatter plot in Fig.~\ref{fig:run16_2dscatter}c,
the good description of background in the sideband slices gives us
confidence that it is well modeled in the signal box.

The likelihood for $N$ data events is defined as
\begin{equation}
  \mathcal{L} = 
  \frac{e^{-(n_S+n_B)}}{N!}
  \,\prod^{N\textrm{ events}}_{i=1} 
  \big[
    n_S\underbrace{\phantom{\sum}\!\!\!\!\!\!\!\!
      \,{P_g}(\mes_{i};\Omega_S)
	\,{P_g}(\DeltaE_{i};\Omega_S^\prime)
    }_{\textrm{signal pdf $P_S$}}
    +
    n_B\underbrace{\phantom{\sum}\!\!\!\!\!\!\!\!
      \,{P_t}(\mes_{i};\Omega_B)
      \,{P_c}(\DeltaE_{i};\Omega_B^\prime)
    }_{\textrm{background pdf $P_B$}}
  \big],
\label{eqn:pdf}
\end{equation}
where $\mes_{i}$ and $\DeltaE_{i}$ are the values for the
$i^\textrm{th}$ event.  For $\Dstarz\ppbar\pi$, we add a third
component in Eqn.~\ref{eqn:pdf} for peaking background.  The quantity
$-\ln\mathcal{L}$ is minimized with respect to the parameters and
yields using M\textsc{inuit} \cite{James:1975dr} via the
R\textsc{ooFit} toolkit \cite{Verkerke:2003ir} in ROOT
\cite{Brun:1997pa}.  We fix some parameters to values obtained in fits
to MC distributions.  For $P_S$, these are the Gaussian width and
power-law tail parameters; for $P_B$, the end-point of the threshold
pdf for $\mes$; and for $P_P$, all parameters including the the yield
$n_P$ using the branching fraction and acceptance found for
$\Dstarp\ppbar\pi$.  To check for possible pathological behaviors in
the fit, we perform one thousand mock experiments where events are
drawn from the pdf and subsequently fit with the same pdf.  All means
(widths) of signal yield pull distributions are consistent with zero
(unity).

\begin{figure}[bp!]
\centering
\subfloat[$\Bzb\To\Dz\ppbar$, $K\pi$]{
  \begin{overpic}[width=0.260\textwidth]
    {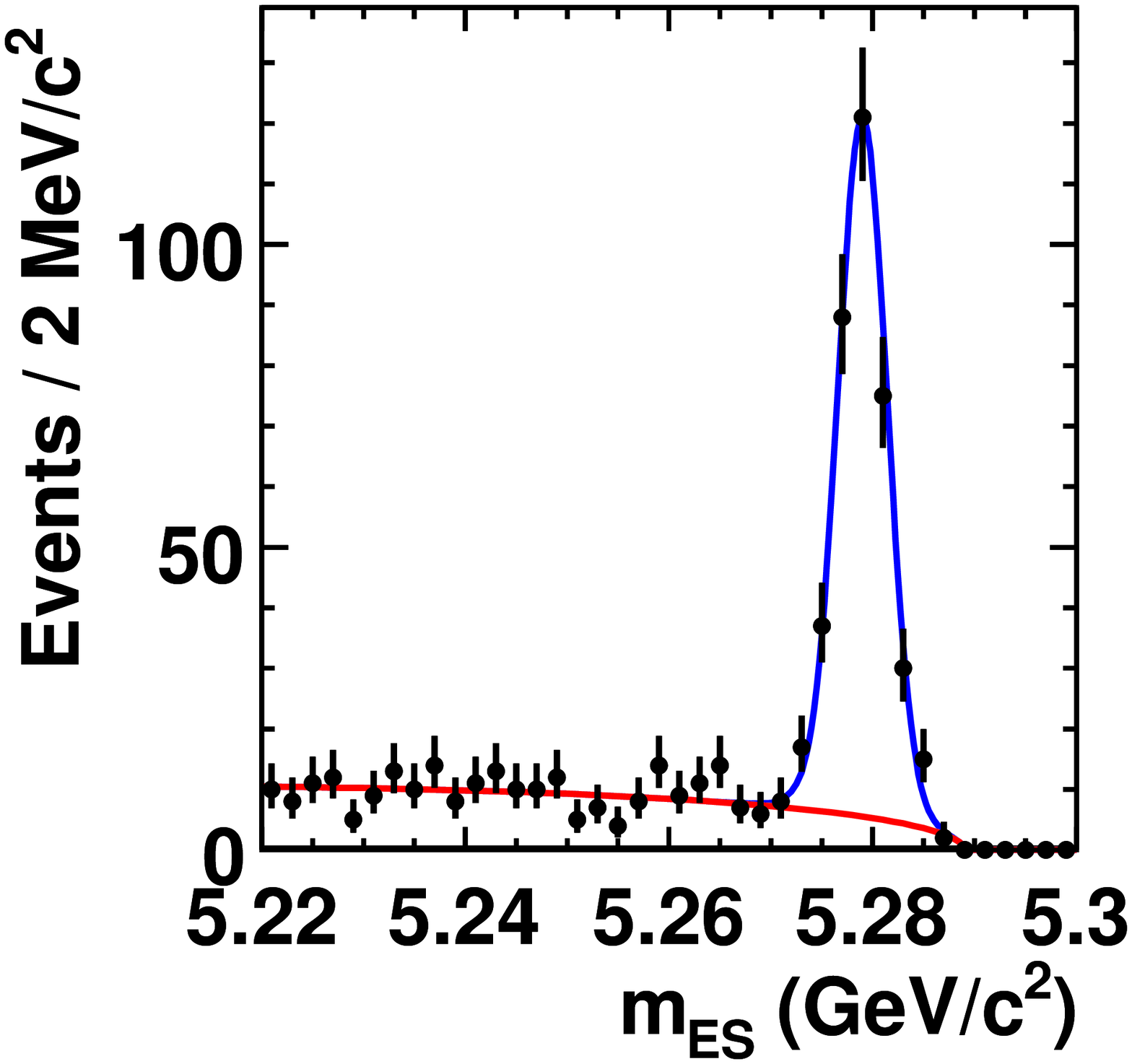}
    \put(30,80){\small{\babar}}
    \put(30,72){\small{prelim.}}
  \end{overpic}
}%
\hspace{0.02\textwidth}%
\subfloat[$\Bzb\To\Dz\ppbar$, $K\pi\piz$]{
  \begin{overpic}[width=0.260\textwidth]
    {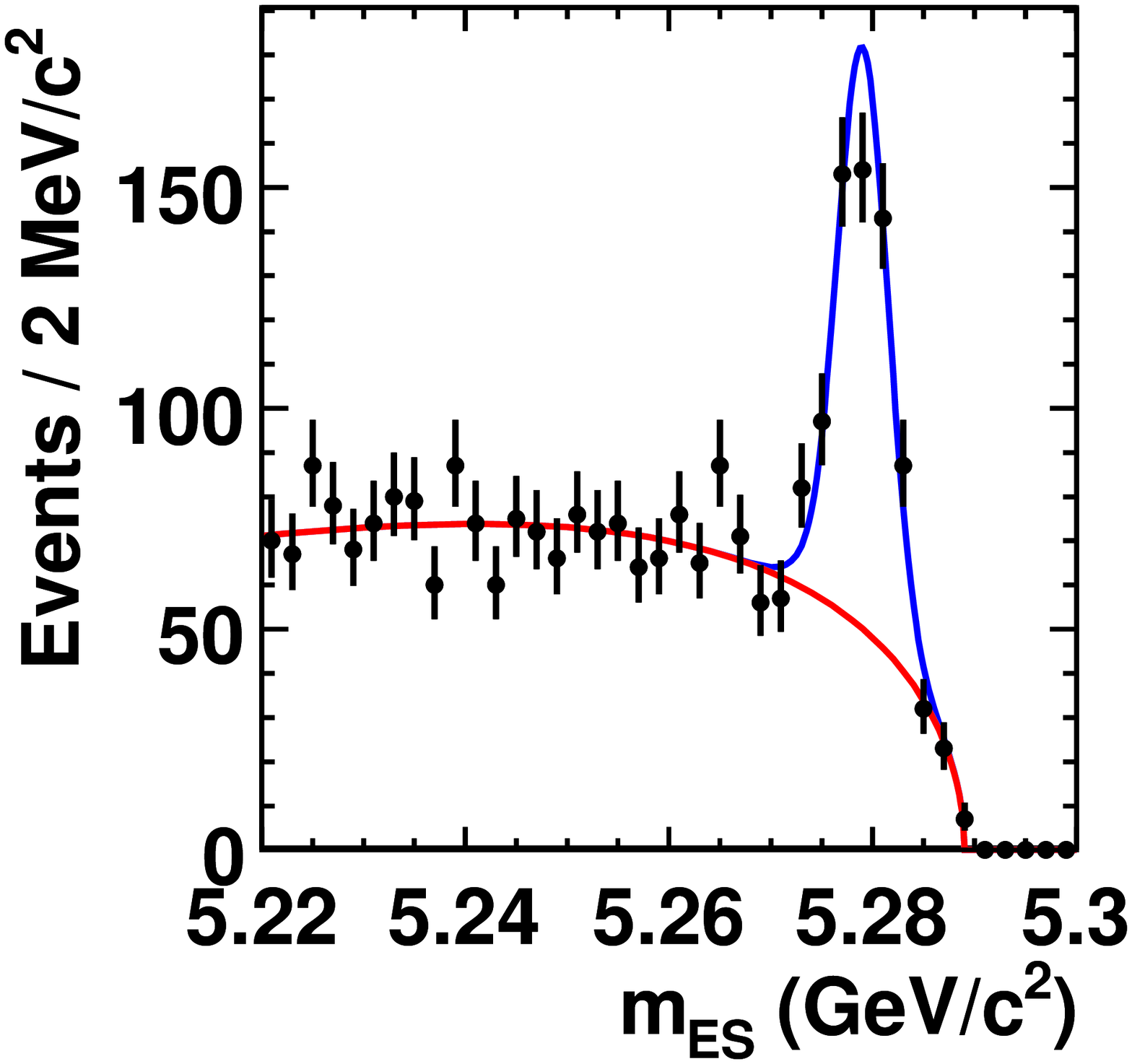}
    \put(30,80){\small{\babar}}
    \put(30,72){\small{prelim.}}
  \end{overpic}
}%
\hspace{0.02\textwidth}%
\subfloat[$\Bzb\To\Dz\ppbar$, $K\pi\pi\pi$]{
  \begin{overpic}[width=0.260\textwidth]
    {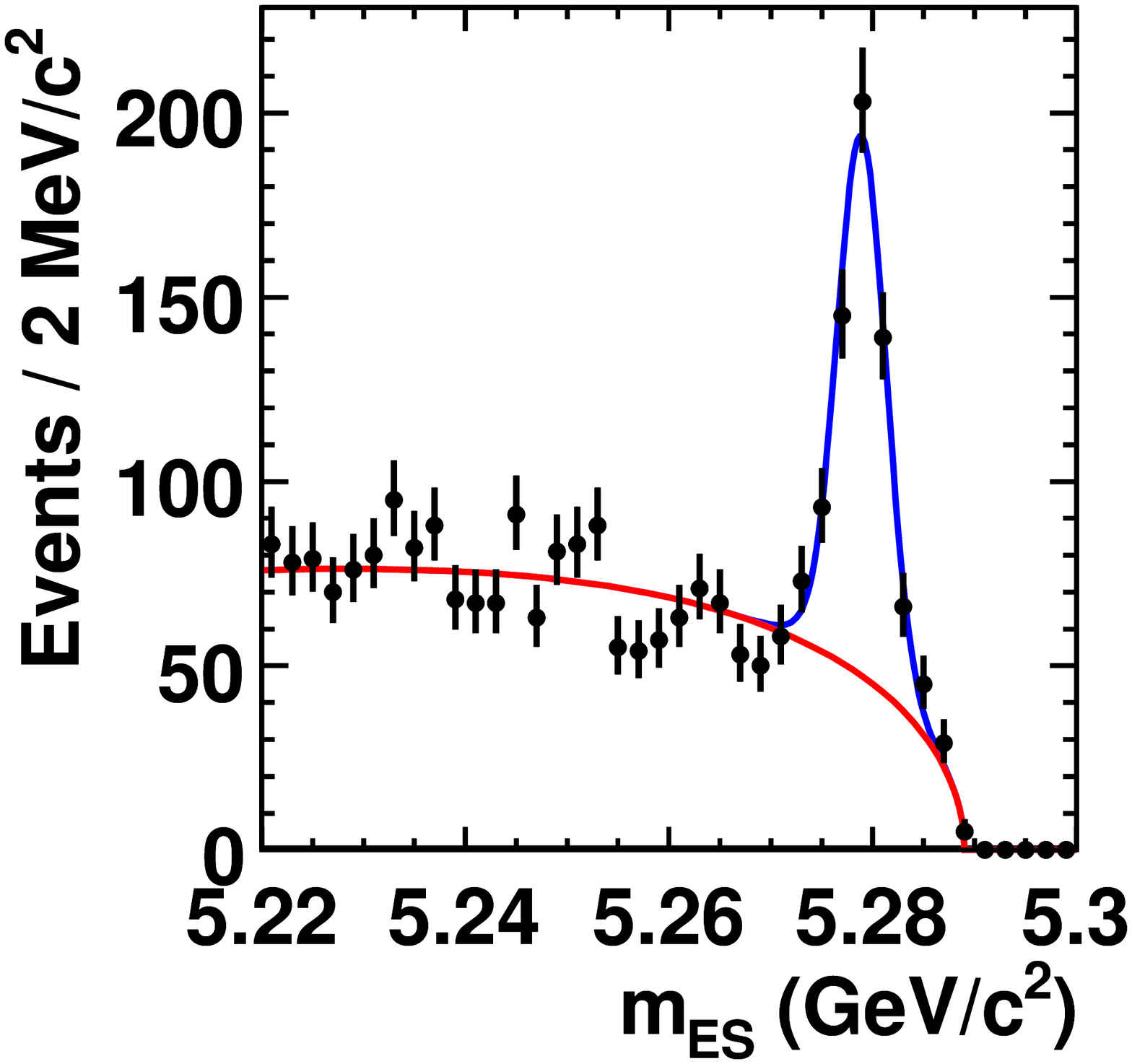}
    \put(30,80){\small{\babar}}
    \put(30,72){\small{prelim.}}
  \end{overpic}
}%
\\
\subfloat[$\Bzb\To\Dstarz\ppbar$, $K\pi$]{
  \begin{overpic}[width=0.260\textwidth]
    {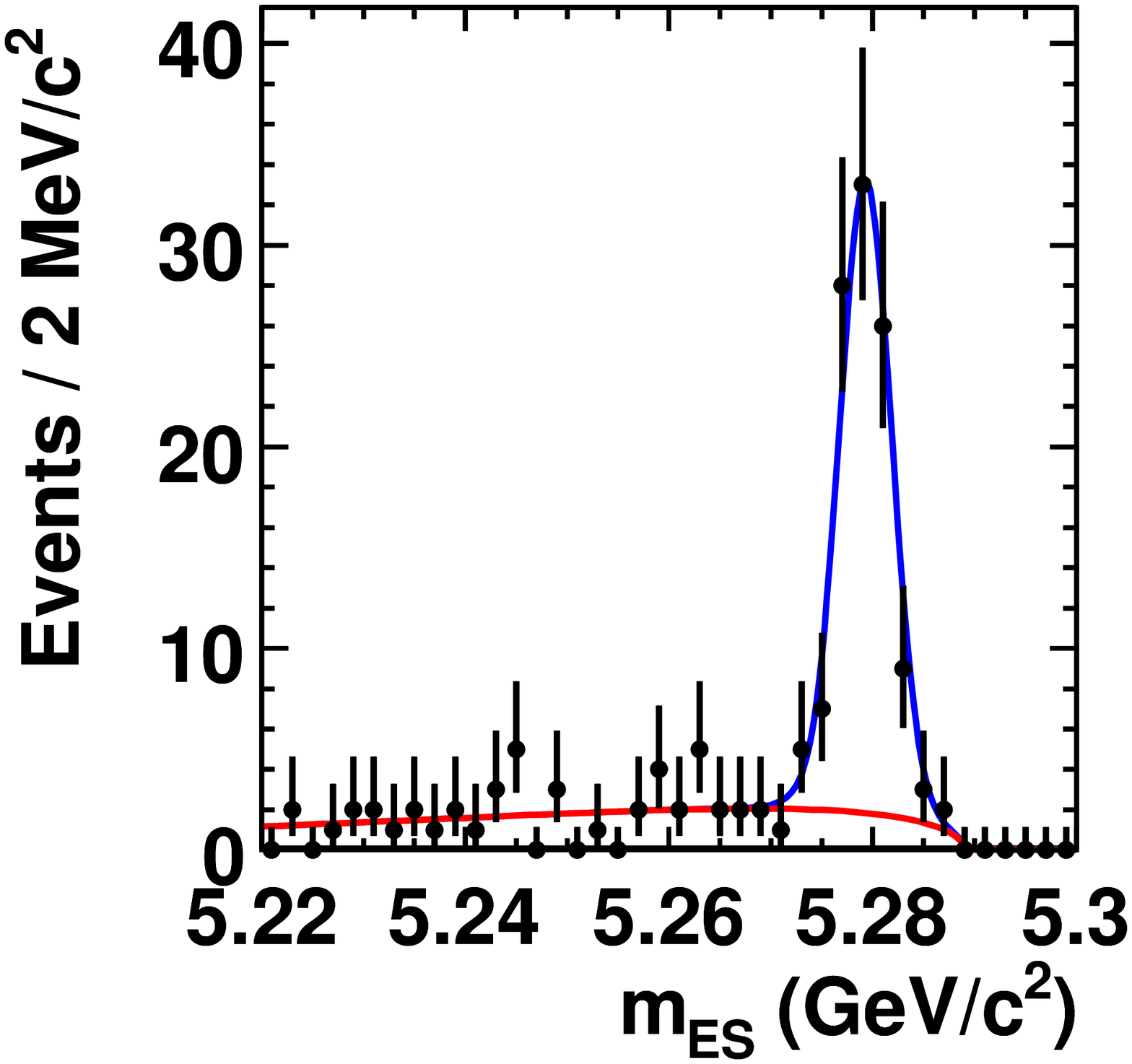}
    \put(30,80){\small{\babar}}
    \put(30,72){\small{prelim.}}
  \end{overpic}
}%
\hspace{0.02\textwidth}%
\subfloat[$\Bzb\To\Dstarz\ppbar$, $K\pi\piz$]{
  \begin{overpic}[width=0.260\textwidth]
    {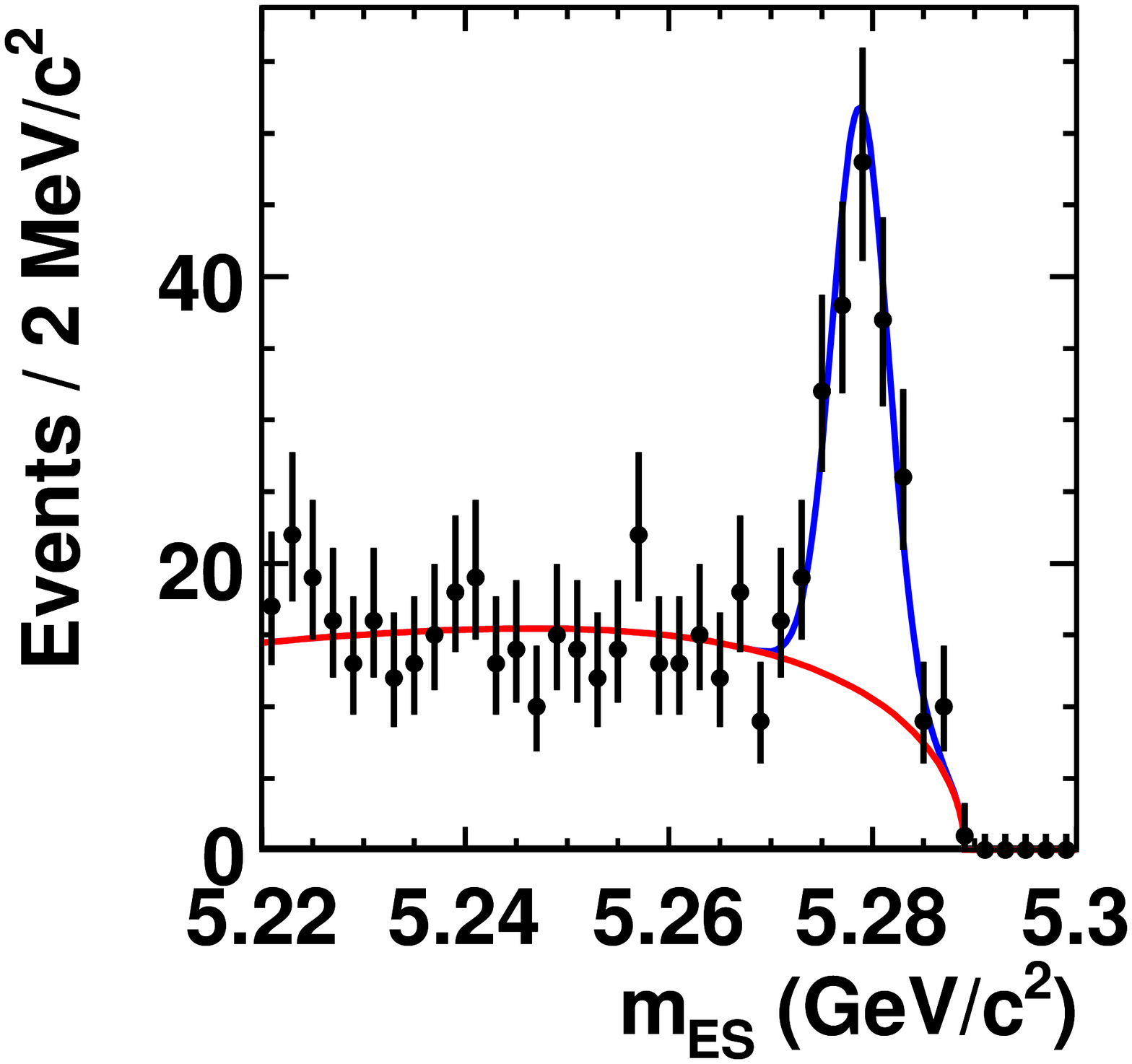}
    \put(30,80){\small{\babar}}
    \put(30,72){\small{prelim.}}
  \end{overpic}
}%
\hspace{0.02\textwidth}%
\subfloat[$\Bzb\To\Dstarz\ppbar$, $K\pi\pi\pi$]{
  \begin{overpic}[width=0.260\textwidth]
    {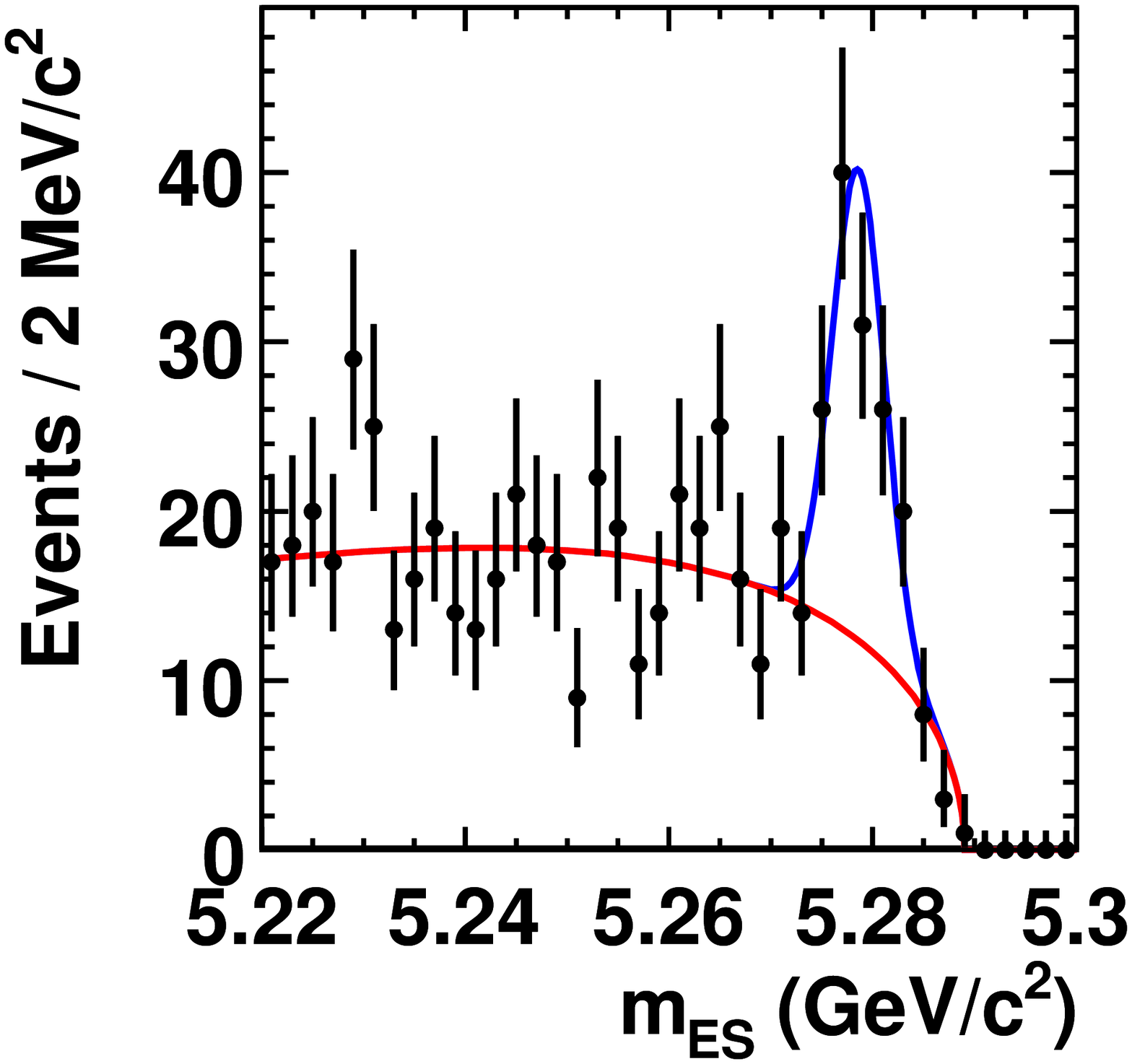}
    \put(30,80){\small{\babar}}
    \put(30,72){\small{prelim.}}
  \end{overpic}
}%
\caption{Fit projections of \mes\ for 3-body decays (abc) $\Bzb\To\Dz\ppbar$
  and (def) $\Bzb\To\Dstarz\ppbar$ for each $D$ decay chain.  The data
  sample is a selection of events within $2.5\sigma$ of \DeltaE\ mean.
  The pdf is integrated over the said range and the components from
  the top are $P_S$ in blue and $P_B$ in red.
}
\label{fig:2dfits_3body}
\end{figure}

\begin{figure}[p!]
\centering
\subfloat[$\Bzb\To\Dp\ppbar\pi$, $K\pi\pi$]{
  \begin{overpic}[width=0.260\textwidth]
    {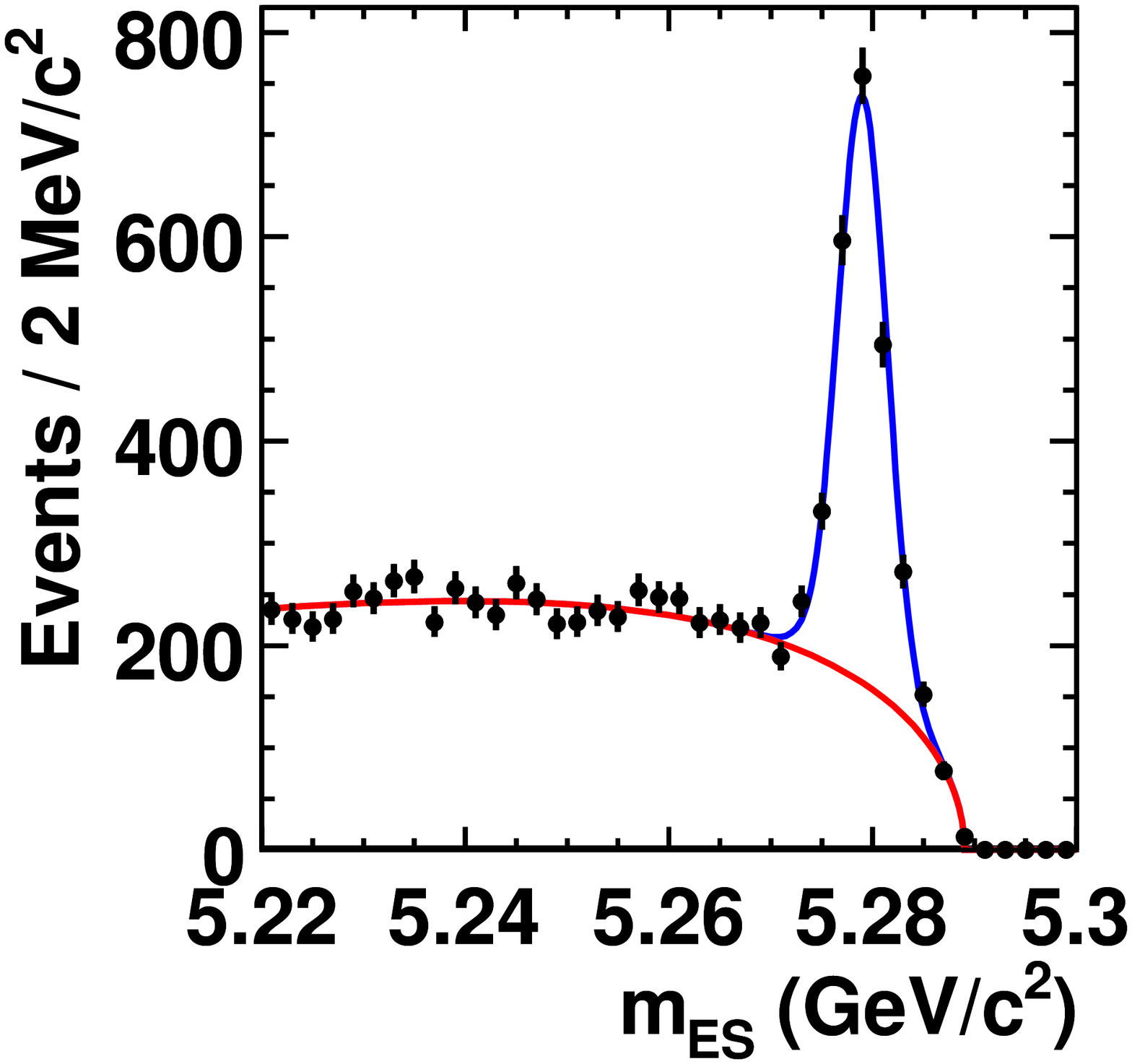}
    \put(30,80){\small{\babar}}
    \put(30,72){\small{prelim.}}
  \end{overpic}
}%
\\
\subfloat[$\Bzb\To\Dstarp\ppbar\pi$, $K\pi$]{
  \begin{overpic}[width=0.260\textwidth]
    {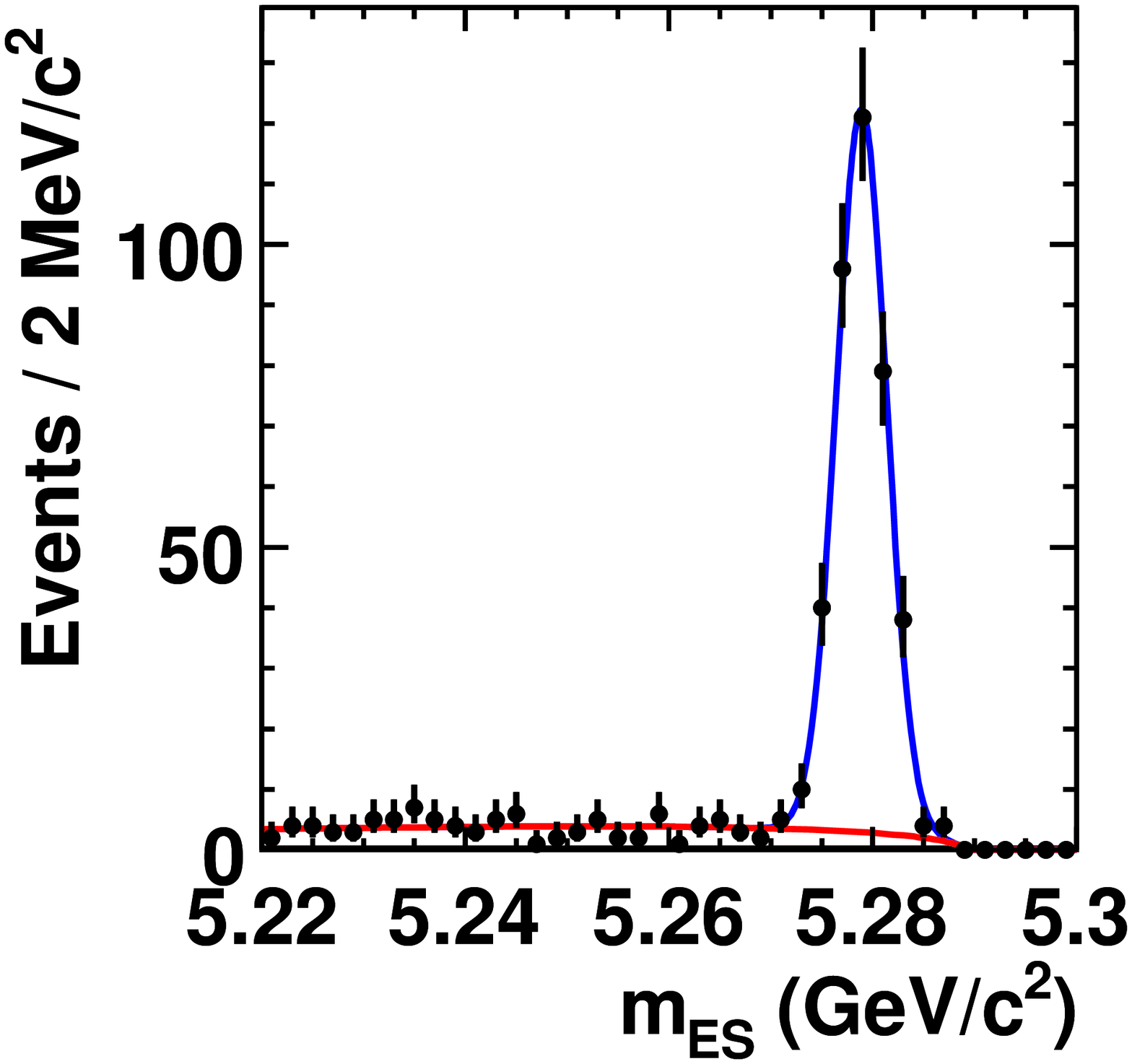}
    \put(30,80){\small{\babar}}
    \put(30,72){\small{prelim.}}
  \end{overpic}
}%
\hspace{0.02\textwidth}%
\subfloat[$\Bzb\To\Dstarp\ppbar\pi$, $K\pi\piz$]{
  \begin{overpic}[width=0.260\textwidth]
    {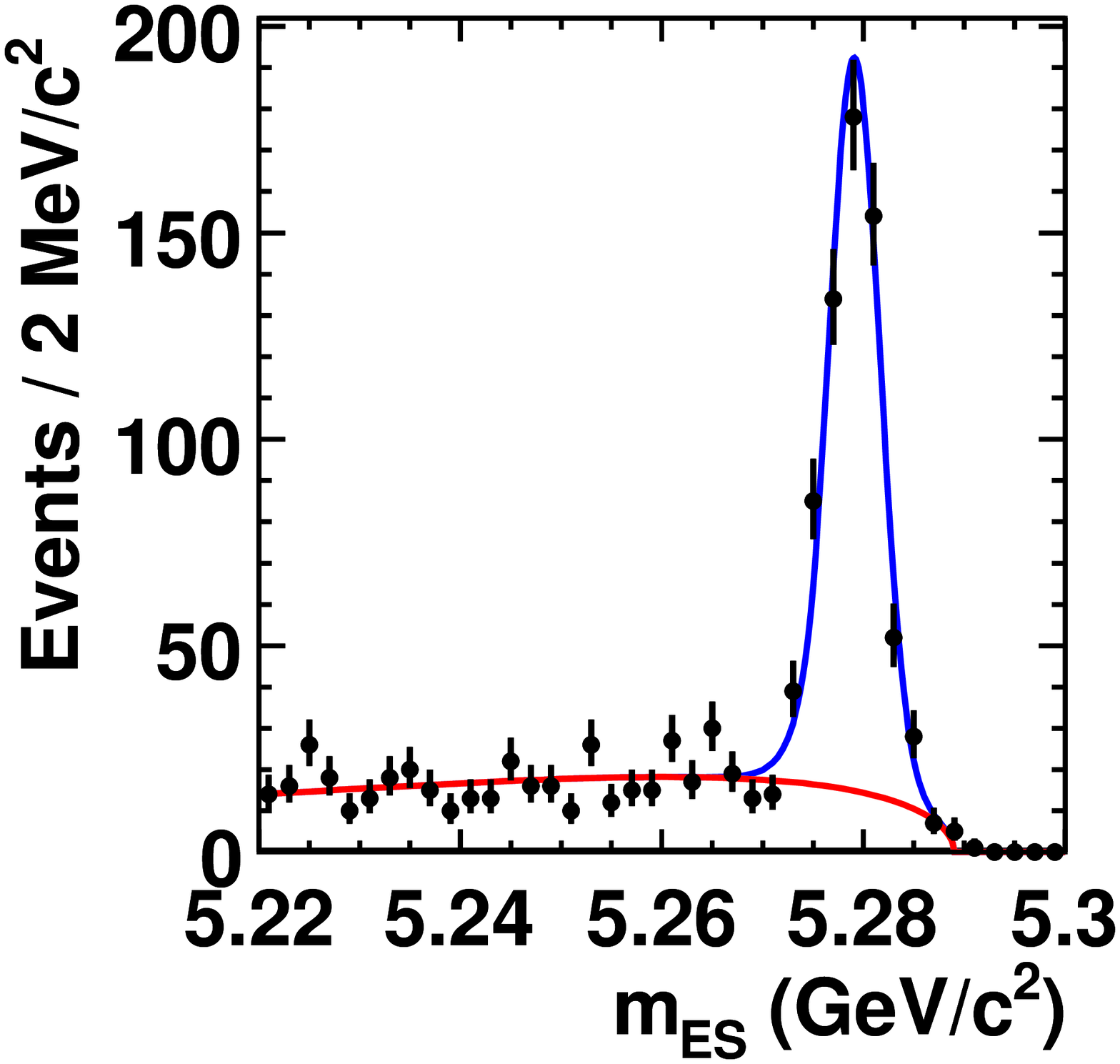}
    \put(30,80){\small{\babar}}
    \put(30,72){\small{prelim.}}
  \end{overpic}
}%
\hspace{0.02\textwidth}%
\subfloat[$\Bzb\To\Dstarp\ppbar\pi$, $K\pi\pi\pi$]{
  \begin{overpic}[width=0.260\textwidth]
    {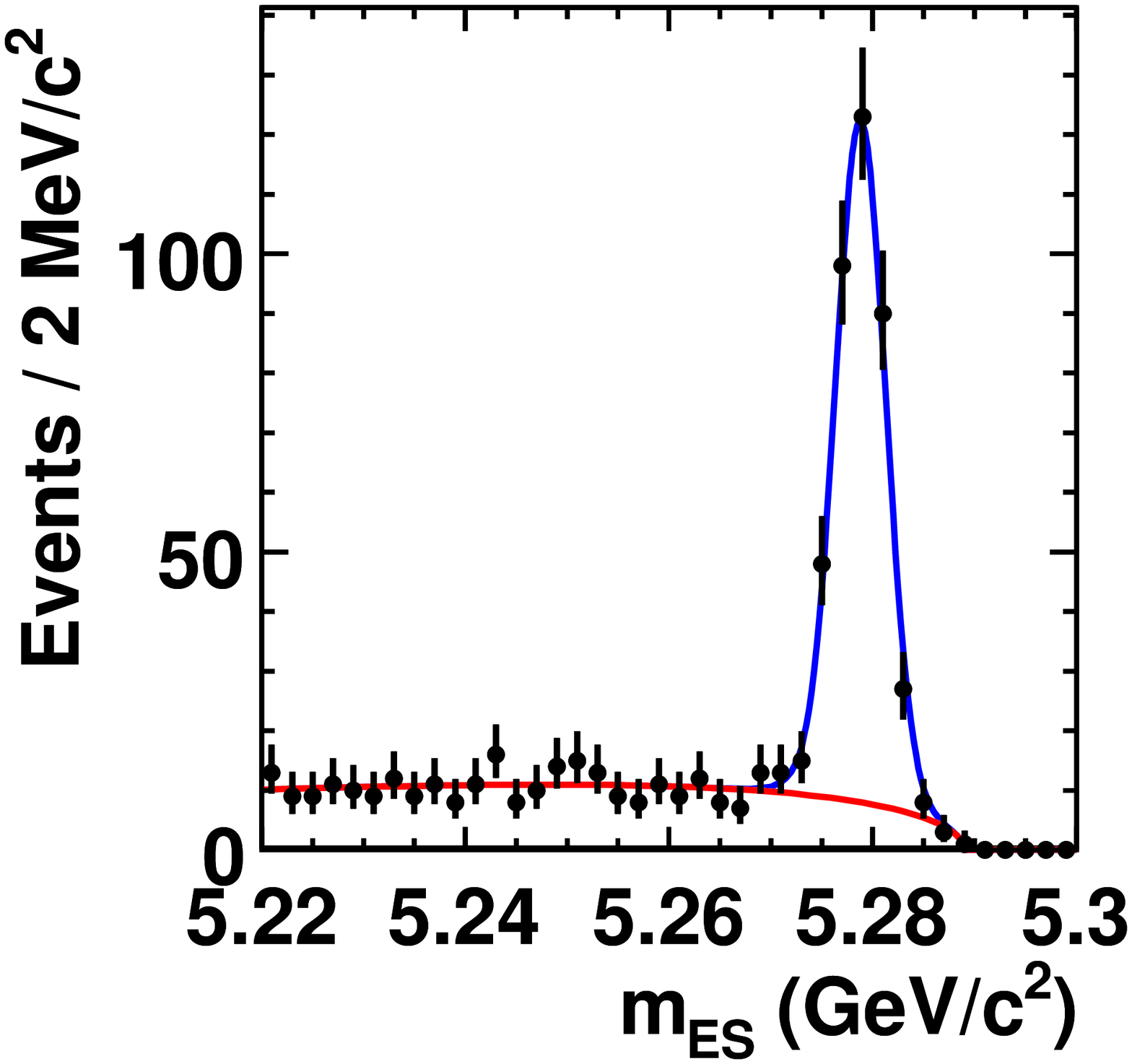}
    \put(30,80){\small{\babar}}
    \put(30,72){\small{prelim.}}
  \end{overpic}
}%
\\
\subfloat[$\Bm\To\Dz\ppbar\pi$, $K\pi$]{
  \begin{overpic}[width=0.260\textwidth]
    {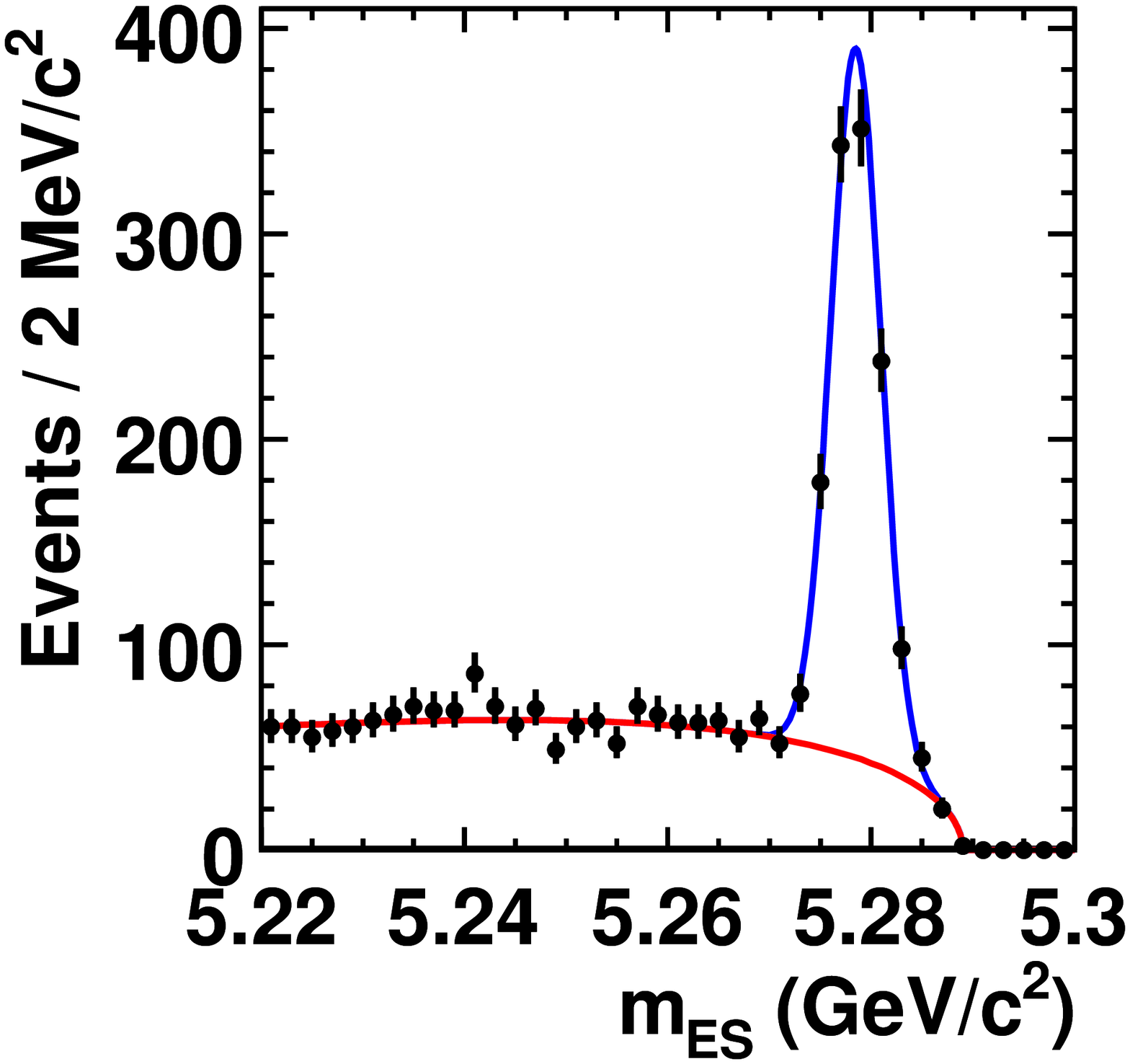}
    \put(30,80){\small{\babar}}
    \put(30,72){\small{prelim.}}
  \end{overpic}
}%
\hspace{0.02\textwidth}%
\subfloat[$\Bm\To\Dz\ppbar\pi$, $K\pi\piz$]{
  \begin{overpic}[width=0.260\textwidth]
    {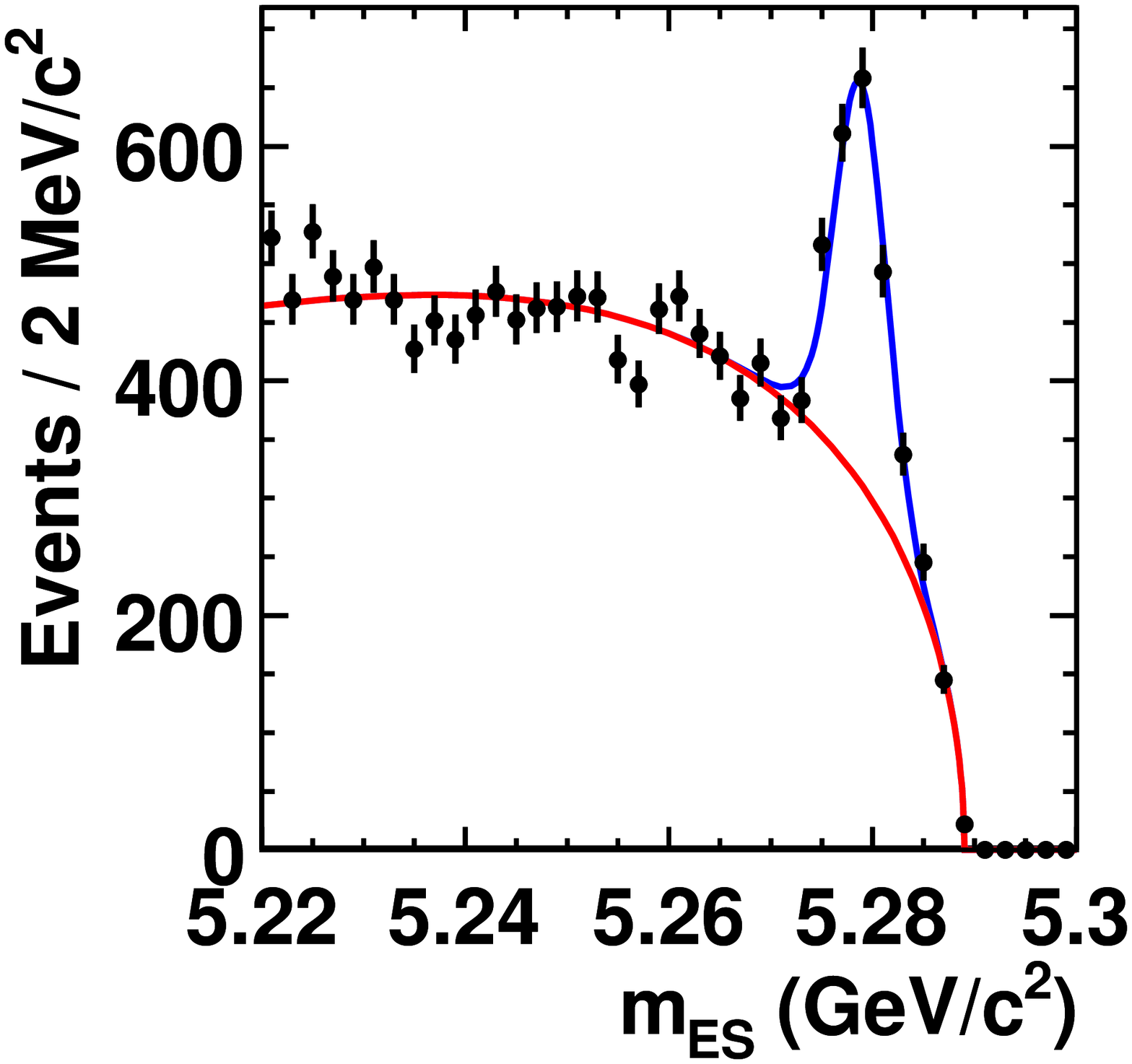}
    \put(30,36){\small{\babar}}
    \put(30,28){\small{prelim.}}
  \end{overpic}
}%
\hspace{0.02\textwidth}%
\subfloat[$\Bm\To\Dz\ppbar\pi$, $K\pi\pi\pi$]{
  \begin{overpic}[width=0.260\textwidth]
    {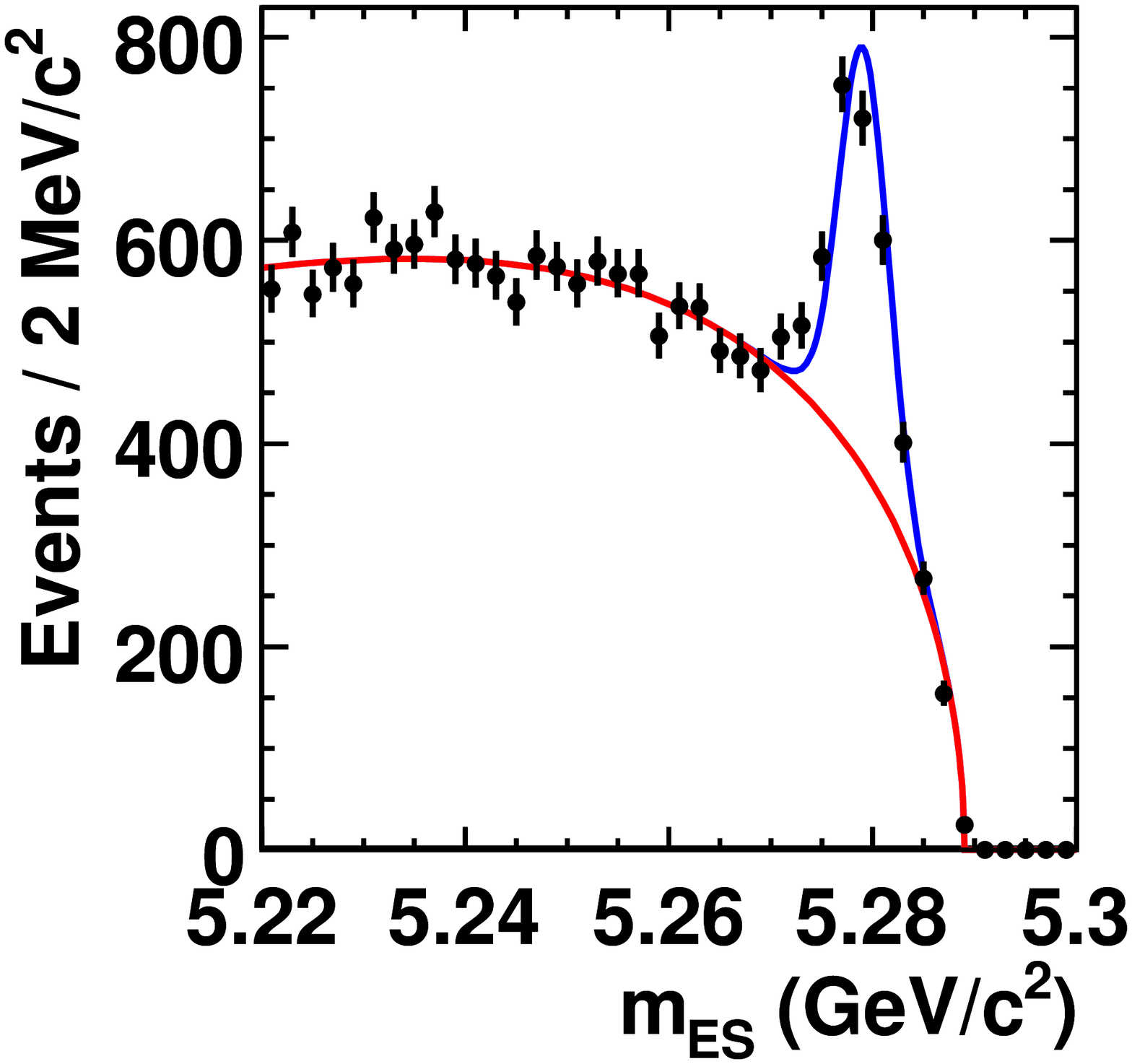}
    \put(30,36){\small{\babar}}
    \put(30,28){\small{prelim.}}
  \end{overpic}
}%
\\
\subfloat[$\Bm\To\Dstarz\ppbar\pi$, $K\pi$]{
  \begin{overpic}[width=0.260\textwidth]
    {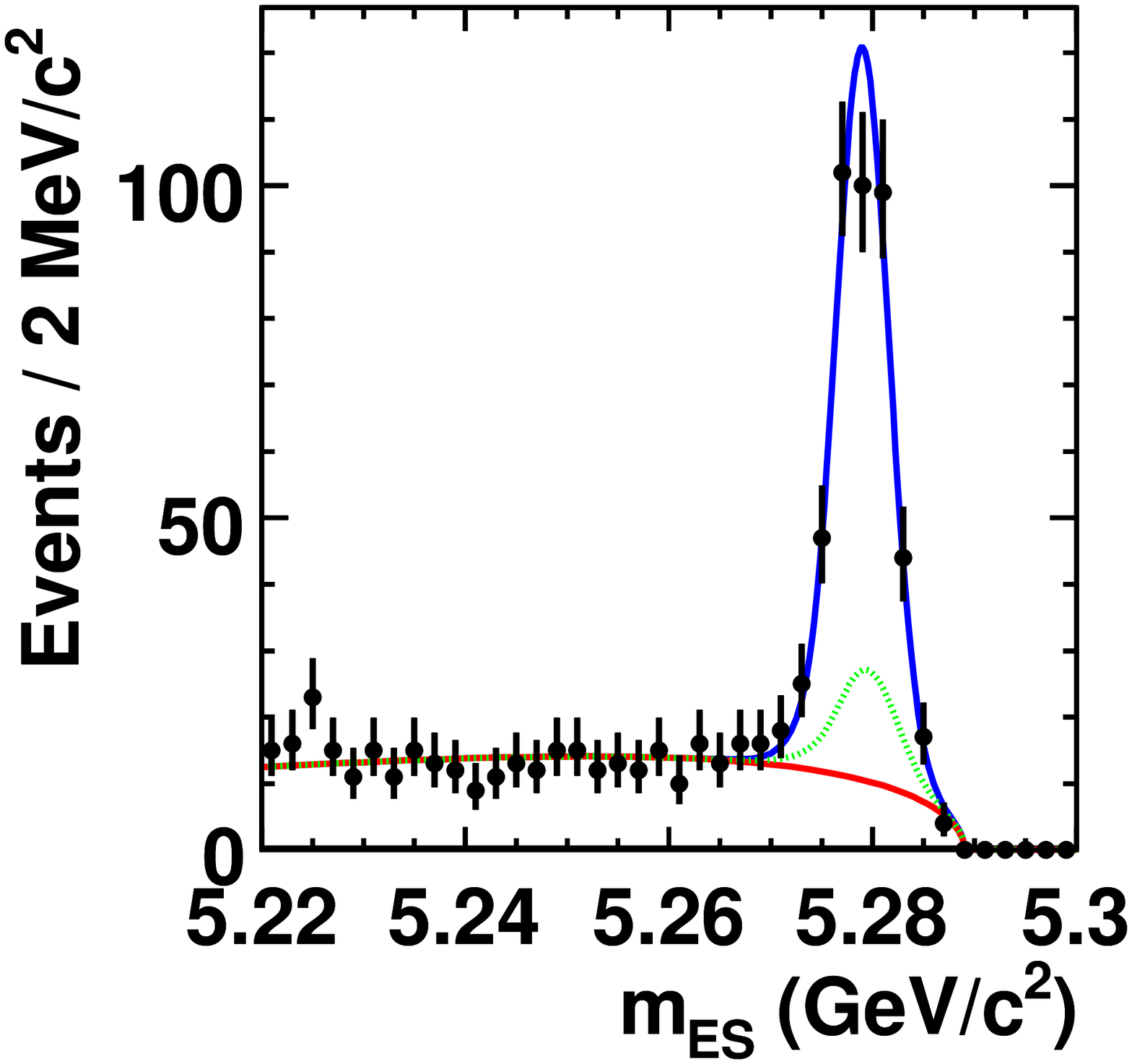}
    \put(30,80){\small{\babar}}
    \put(30,72){\small{prelim.}}
  \end{overpic}
}%
\hspace{0.02\textwidth}%
\subfloat[$\Bm\To\Dstarz\ppbar\pi$, $K\pi\piz$]{
  \begin{overpic}[width=0.260\textwidth]
    {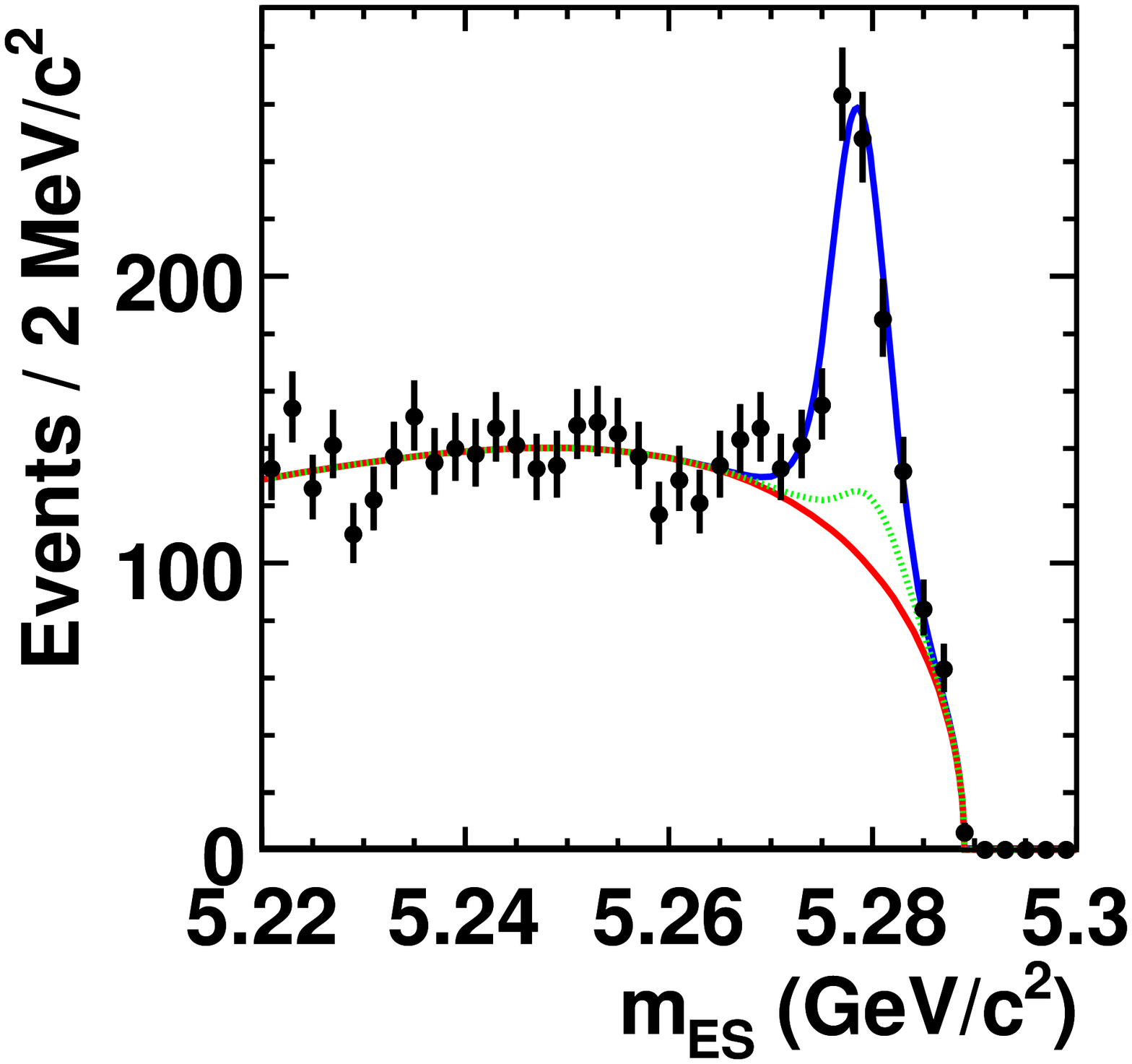}
    \put(30,36){\small{\babar}}
    \put(30,28){\small{prelim.}}
  \end{overpic}
}%
\hspace{0.02\textwidth}%
\subfloat[$\Bm\To\Dstarz\ppbar\pi$, $K\pi\pi\pi$]{
  \begin{overpic}[width=0.260\textwidth]
    {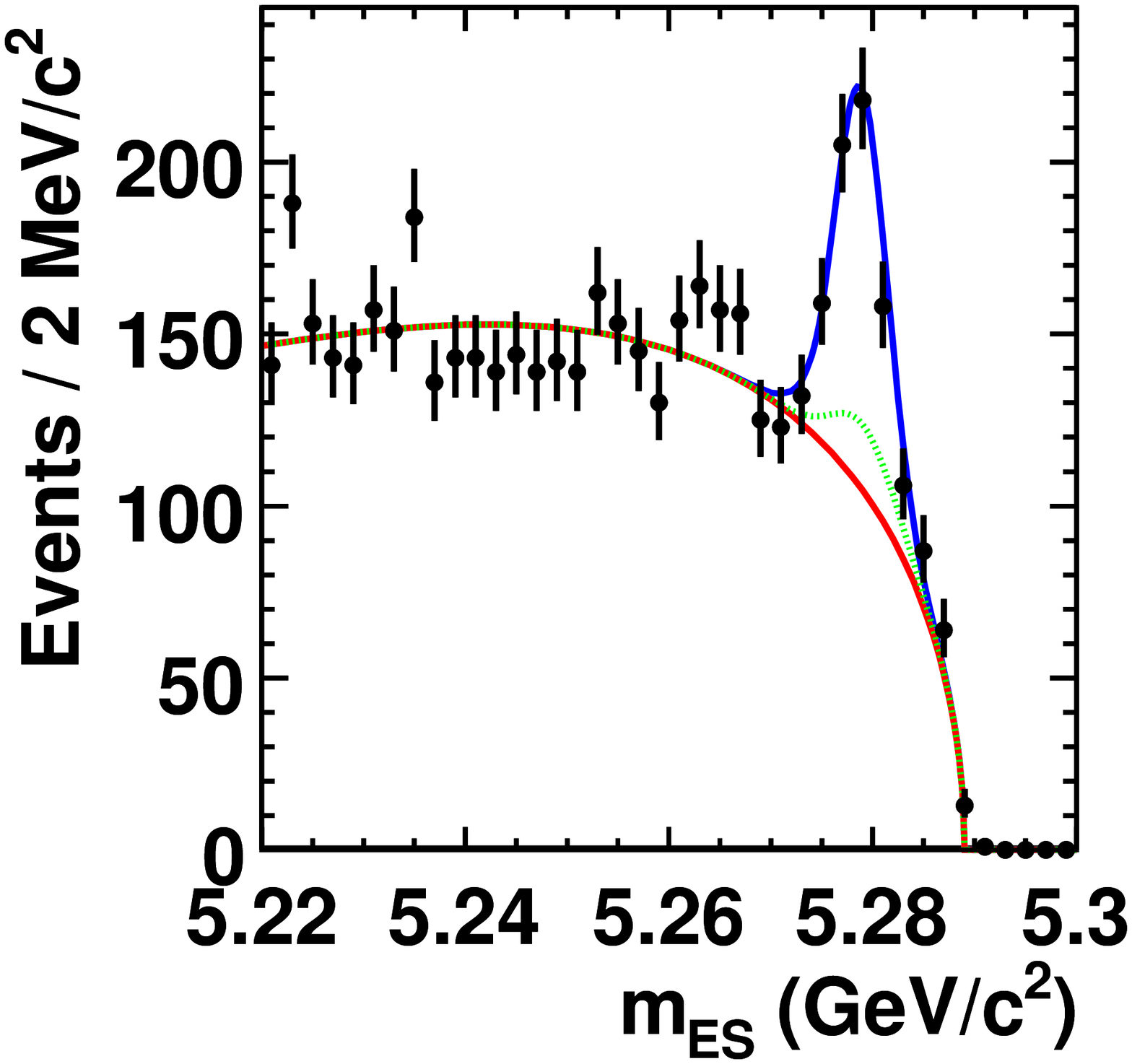}
    \put(30,36){\small{\babar}}
    \put(30,28){\small{prelim.}}
  \end{overpic}
}
\caption{Fit projections of \mes\ for 4-body decays
  (a) $\Bzb\To\Dp\ppbar\pim$,
  (bcd) $\Bzb\To\Dstarp\ppbar\pim$,
  (efg) $\Bm\To\Dz\ppbar\pim$, and
  (hij) $\Bm\To\Dstarz\ppbar\pim$
  for each $D$ decay chain.
  The data
  sample is a selection of events within $2.5\sigma$ of \DeltaE\ mean.
  The pdf is integrated over the said range and the components from
  the top are $P_S$ in blue and $P_B$ in red.  For (hij), $P_P$ is the
  middle pdf in green.
}
\label{fig:2dfits_4body}
\end{figure}

\begin{figure}[p!]
\centering
\subfloat[$\Bzb\To\Dz\ppbar\pi\pi$, $K\pi$]{
  \begin{overpic}[width=0.260\textwidth]
    {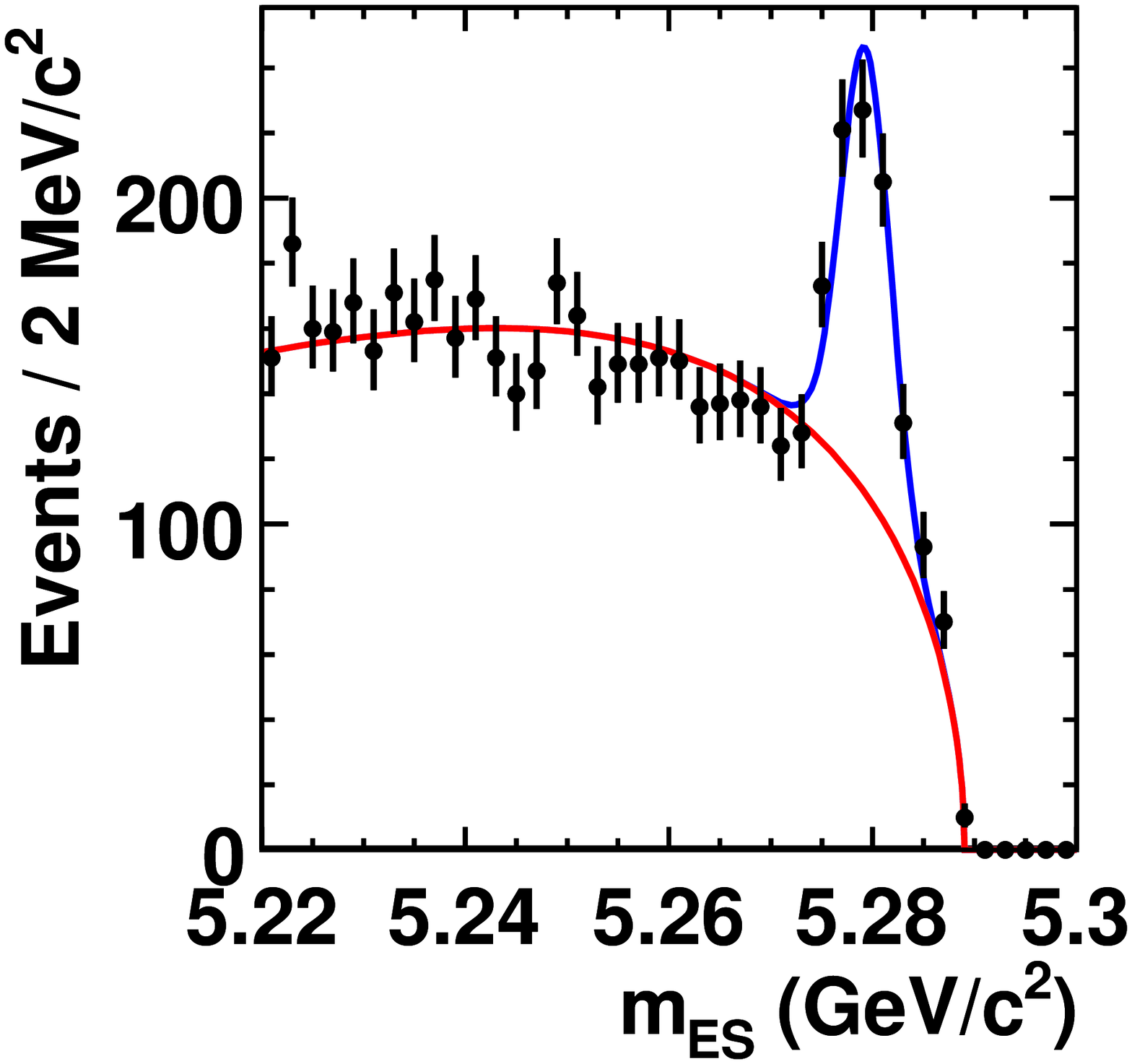}
    \put(30,36){\small{\babar}}
    \put(30,28){\small{prelim.}}
  \end{overpic}
}%
\hspace{0.02\textwidth}%
\subfloat[$\Bzb\To\Dz\ppbar\pi\pi$, $K\pi\piz$]{
  \begin{overpic}[width=0.260\textwidth]
    {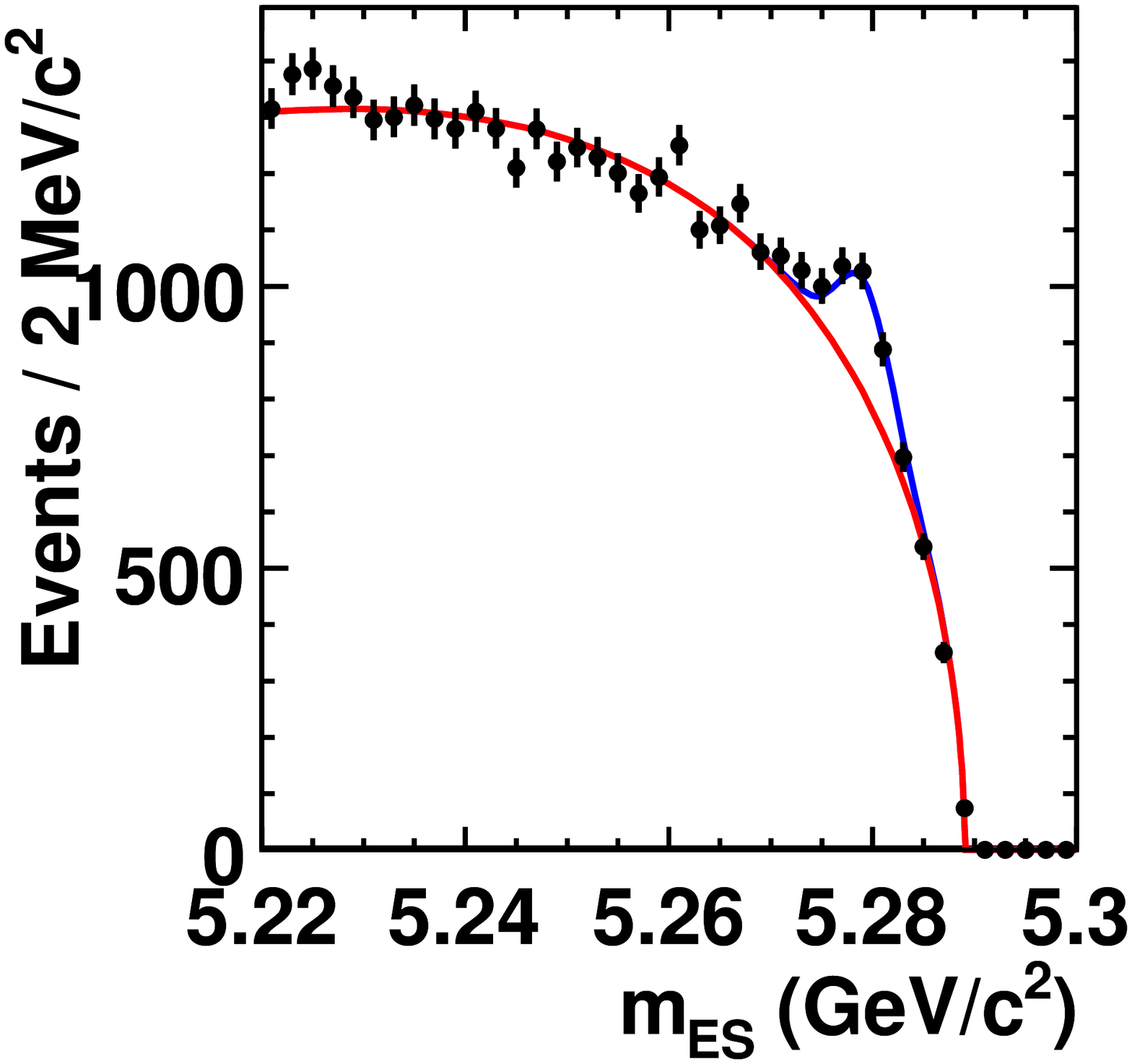}
    \put(30,36){\small{\babar}}
    \put(30,28){\small{prelim.}}
  \end{overpic}
}%
\hspace{0.02\textwidth}%
\subfloat[$\Bzb\To\Dz\ppbar\pi\pi$, $K\pi\pi\pi$]{
  \begin{overpic}[width=0.260\textwidth]
    {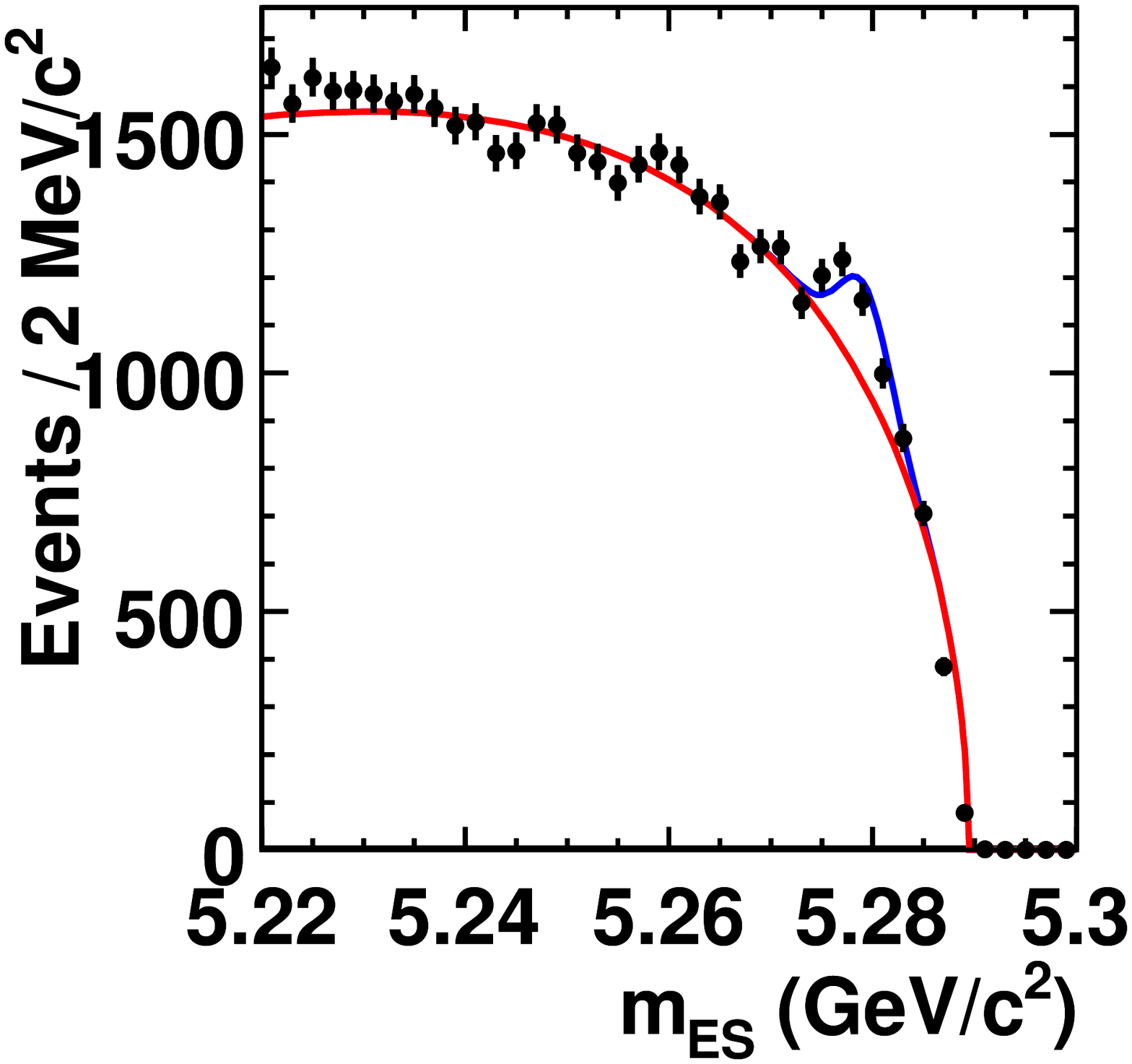}
    \put(30,36){\small{\babar}}
    \put(30,28){\small{prelim.}}
  \end{overpic}
}%
\\
\subfloat[$\Bzb\To\To\Dstarz\ppbar\pi\pi$, $K\pi$]{
  \begin{overpic}[width=0.260\textwidth]
    {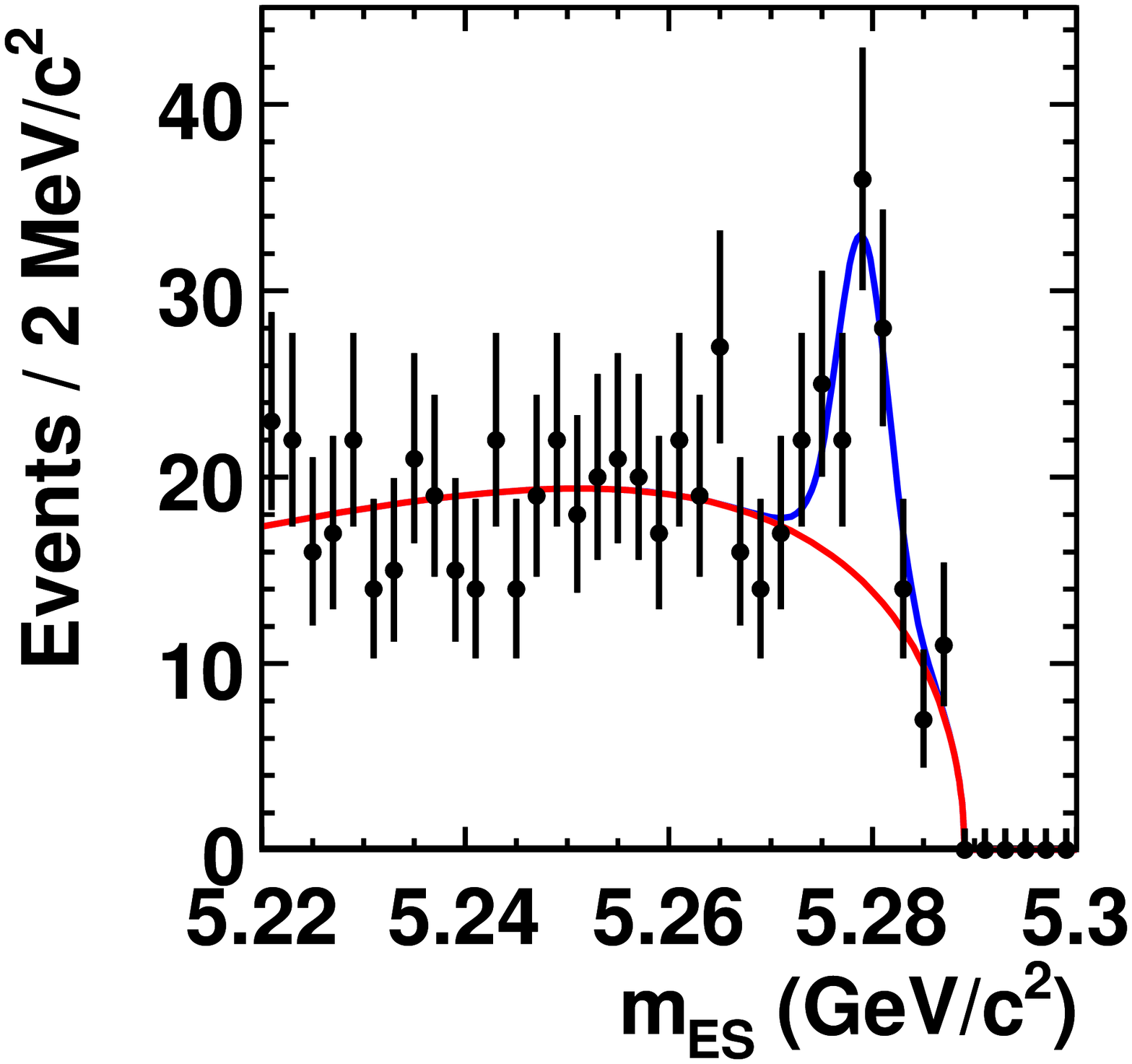}
    \put(30,80){\small{\babar}}
    \put(30,72){\small{prelim.}}
  \end{overpic}
}%
\hspace{0.02\textwidth}%
\subfloat[$\Bzb\To\Dstarz\ppbar\pi\pi$, $K\pi\piz$]{
  \begin{overpic}[width=0.260\textwidth]
    {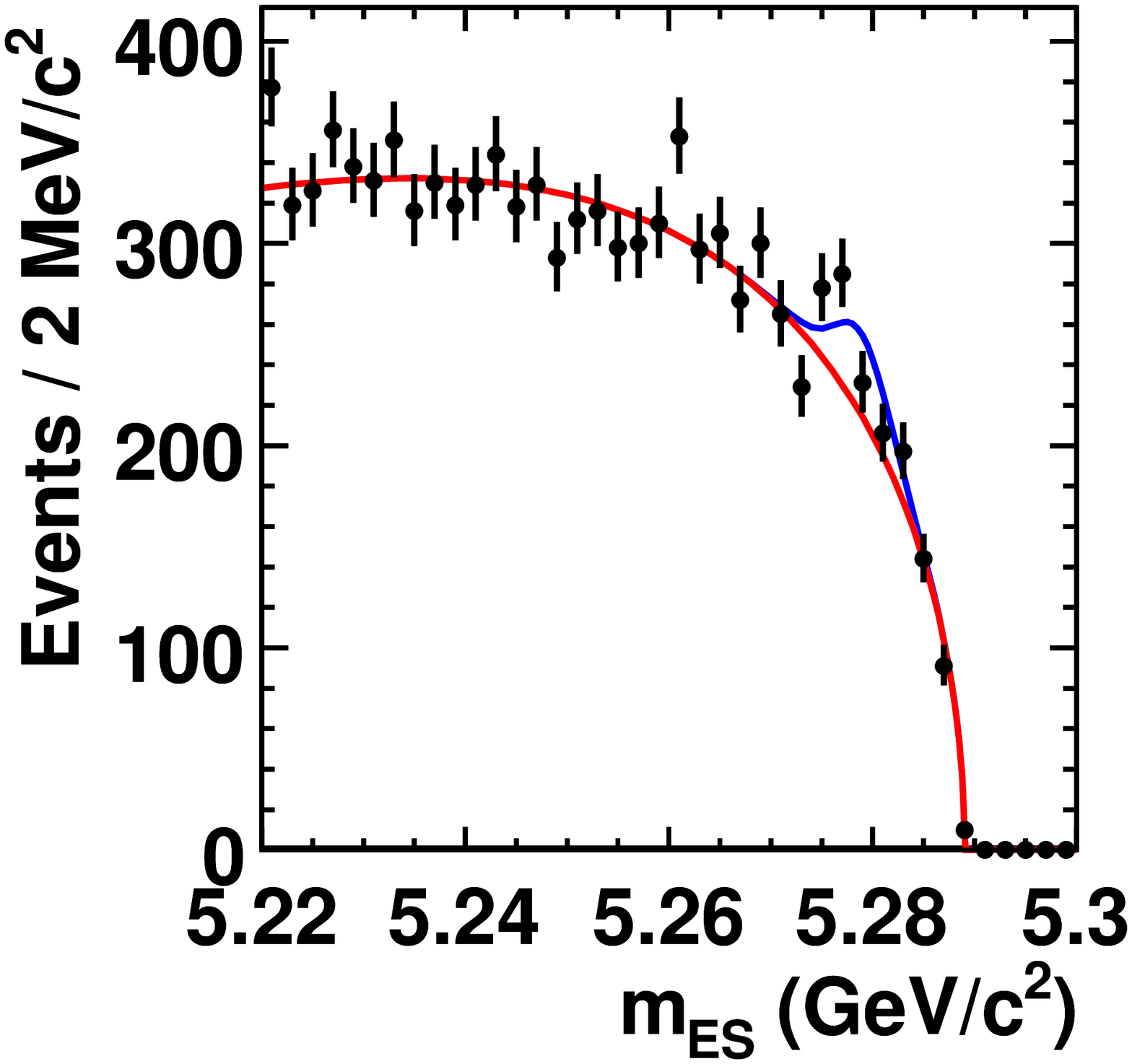}
    \put(30,36){\small{\babar}}
    \put(30,28){\small{prelim.}}
  \end{overpic}
}%
\hspace{0.02\textwidth}%
\subfloat[$\Bzb\To\Dstarz\ppbar\pi\pi$, $K\pi\pi\pi$]{
  \begin{overpic}[width=0.260\textwidth]
    {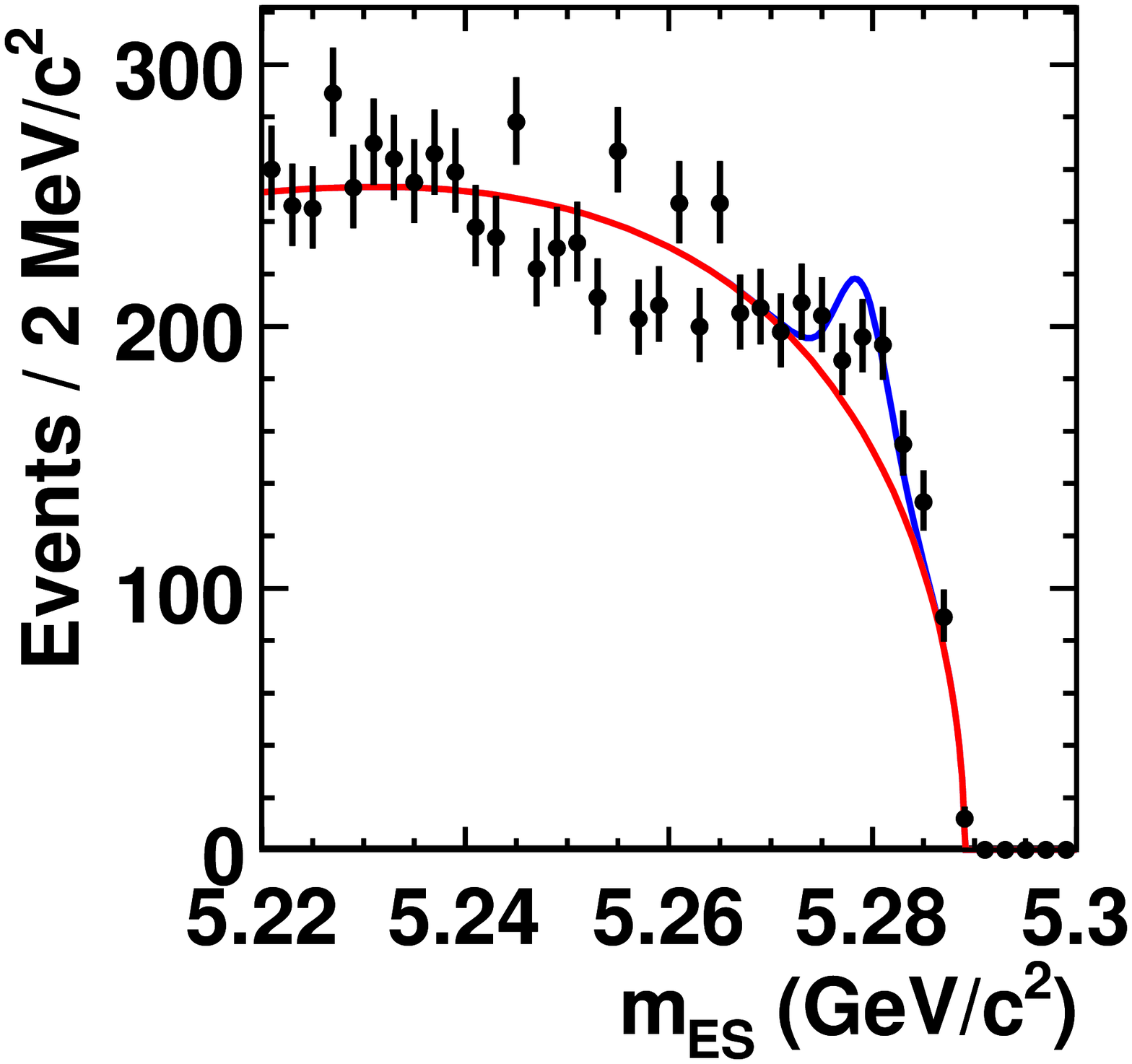}
    \put(30,36){\small{\babar}}
    \put(30,28){\small{prelim.}}
  \end{overpic}
}%
\\
\subfloat[$\Bm\To\Dp\ppbar\pi\pi$, $K\pi\pi$]{
  \begin{overpic}[width=0.260\textwidth]
    {BToDPPbarX-Run16-OnPeak-R22d-all-90011_mesplot_signalsliceonly.eps}
    \put(30,36){\small{\babar}}
    \put(30,28){\small{prelim.}}
  \end{overpic}
}%
\\
\subfloat[$\Bm\To\Dstarp\ppbar\pi\pi$, $K\pi$]{
  \begin{overpic}[width=0.260\textwidth]
    {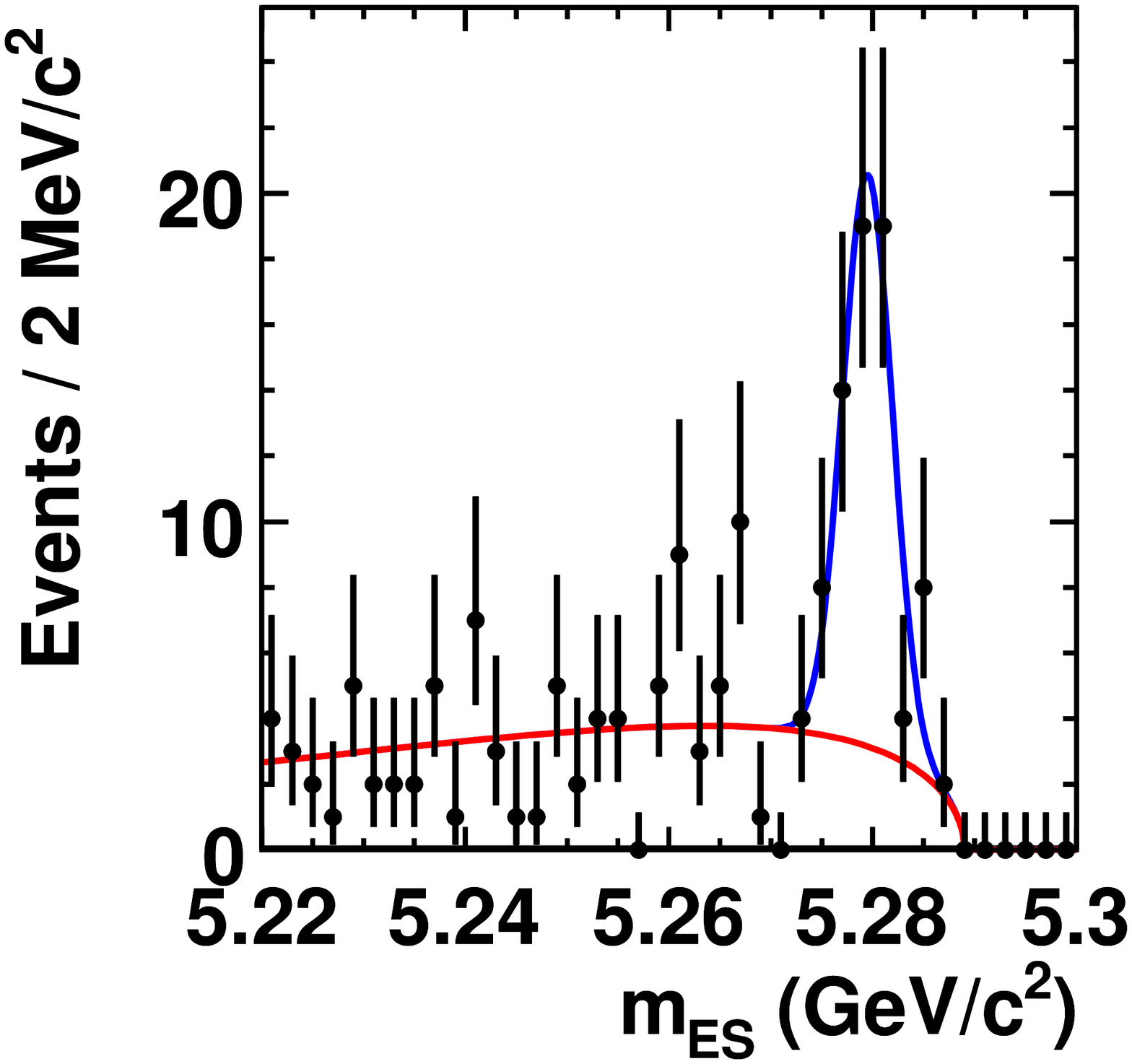}
    \put(30,80){\small{\babar}}
    \put(30,72){\small{prelim.}}
  \end{overpic}
}%
\hspace{0.02\textwidth}%
\subfloat[$\Bm\To\Dstarp\ppbar\pi\pi$, $K\pi\piz$]{
  \begin{overpic}[width=0.260\textwidth]
    {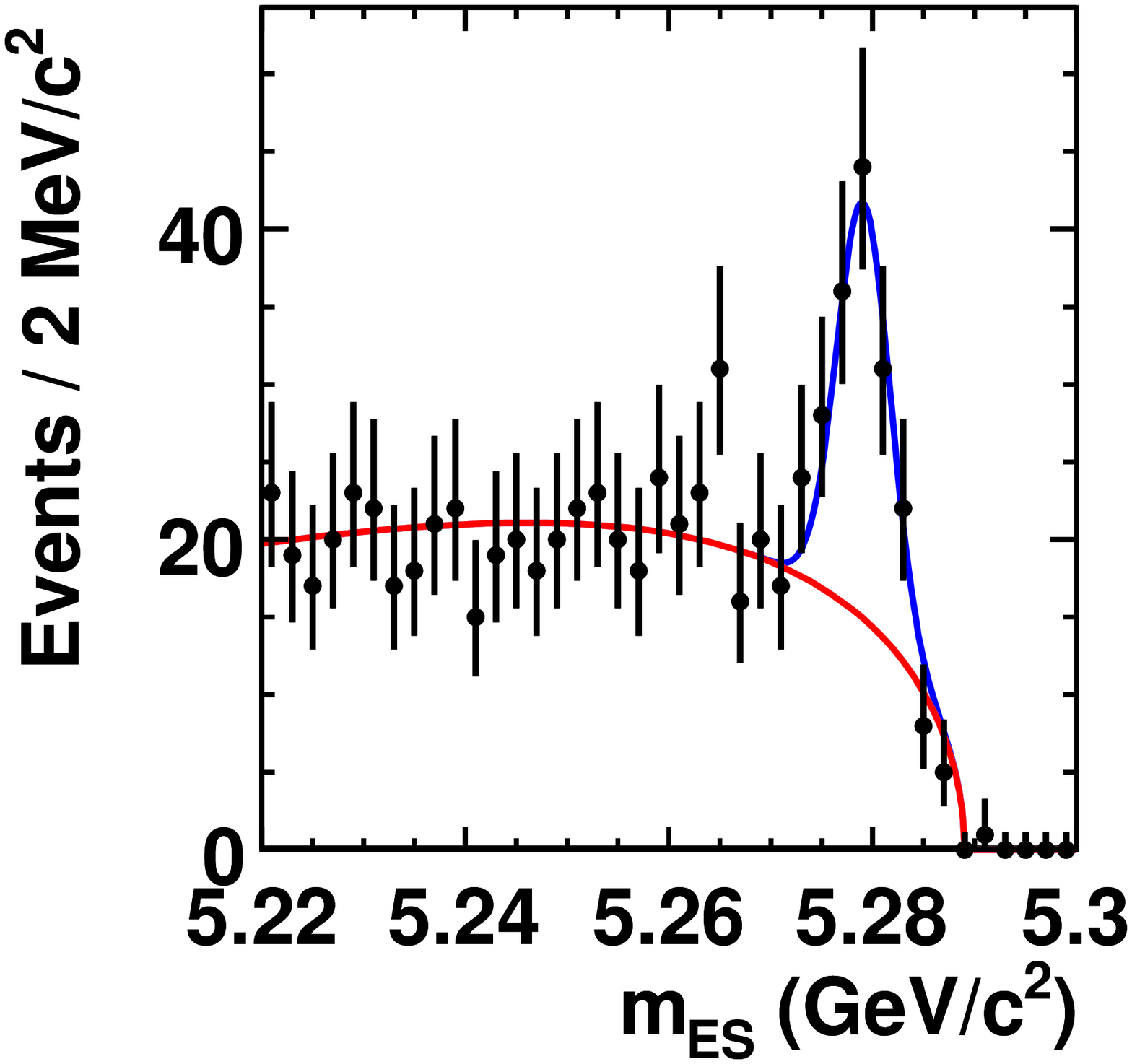}
    \put(30,80){\small{\babar}}
    \put(30,72){\small{prelim.}}
  \end{overpic}
}%
\hspace{0.02\textwidth}%
\subfloat[$\Bm\To\Dstarp\ppbar\pi\pi$, $K\pi\pi\pi$]{
  \begin{overpic}[width=0.260\textwidth]
    {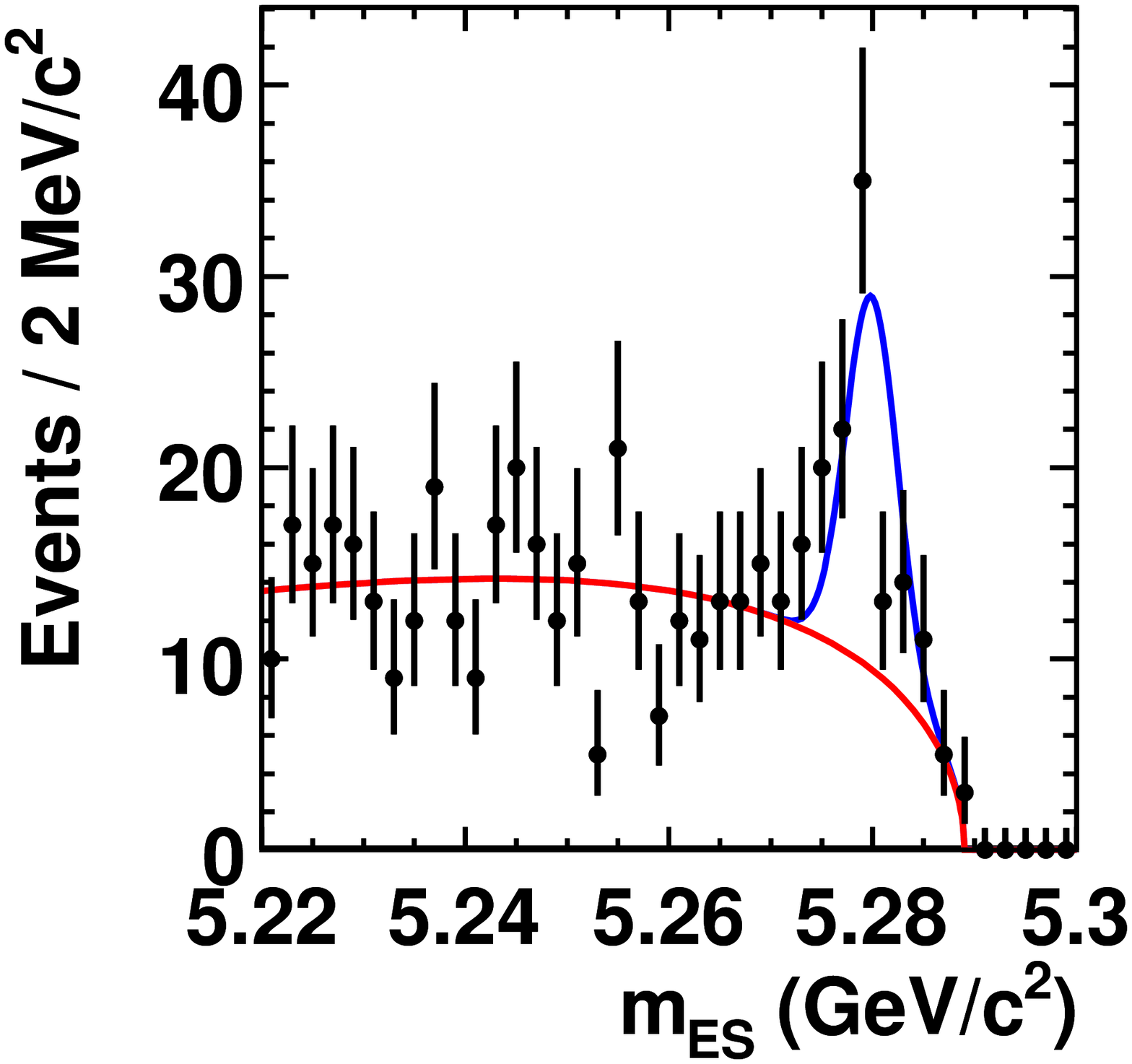}
    \put(30,80){\small{\babar}}
    \put(30,72){\small{prelim.}}
  \end{overpic}
}
\caption{Fit projections of \mes\ for 5-body decays 
  (abc) $\Bzb\To\Dz\ppbar\pim\pip$,
  (def) $\Bzb\To\Dstarp\ppbar\pim$,
  (g)   $\Bm\To\Dz\ppbar\pim\pim$, and
  (hij) $\Bm\To\Dstarz\ppbar\pim\pim$
  for each $D$ decay chain.
  The data
  sample is a selection of events within $2.5\sigma$ of \DeltaE\ mean.
  The pdf is integrated over the said range and the components from
  the top are $P_S$ in blue and $P_B$ in red.
}
\label{fig:2dfits_5body}
\end{figure}

Figs.~\ref{fig:2dfits_3body}, \ref{fig:2dfits_4body}, and
\ref{fig:2dfits_5body} give the collection of \mes\ sliced projections
in the \DeltaE\ signal region for 3-, 4-, and 5-body $B$ decays,
respectively.  Signal yields range from \nolbreaks{$50$--$3500$}
events for the ten $B$ decays; they are given later in
Table~\ref{tab:bfchain}.  Lastly, additional work is done for four
$\Bzb$ decay chains $\Dmaybestarz\ppbar\pi\pi$, $K\pi\piz$\ and
$\Dmaybestarz\ppbar\pi\pi$, $K\pi\pi\pi$ shown in
Fig.~\ref{fig:2dfits_5body}bcef because the fits converged to
non-physical Gaussian mean values for $P_g(\mes)$ and $P_g(\DeltaE)$
due to the low signal-to-background ratio.  For these, the means are
fixed to the values found for the $K\pi$ counterpart and the end-point
for $P_t(\mes)$ is floated.  We note that these measurements do not
contribute significantly to the branching fraction averaged over $\Dz$
decay modes because of large systematic uncertainties compared to the
corresponding $K\pi$ measurement; this will be discussed later in
Sec.~\ref{sec:syst}.

\subsection{Systematic studies}
\label{sec:syst}

Table~\ref{tab:syst_summary} lists the individual sources of
systematic uncertainties for the branching fraction measurement in
five categories (I--V): $B$ counting, assumed branching fractions for
$\FourS$ and $\Dmaybestar$, acceptance, pdf, and peaking background
estimate.

\begin{table}[b!]
\caption{Systematic uncertainties for measurements of $B$ branching
  fractions, $\mathcal{B}$.  }
\centering
{\small
\subfloat[List of individual systematic uncertainties in five
  categories (I--V).]{
\label{tab:syst_summary}
\begin{tabular}{rll}
\dbline
Item & Source description & \percent of $\mathcal{B}$ \\
\sgline
I\phantom{.0}&  $\BB$ counting 			& $1.1$ \\
II.1  & Branching fraction of $\FourS$ 		& $1.6$ \\
II.2  & Branching fraction of $D\To{K}\pi$,
        $K\pi\piz$, $K\pi\pi\pi$, $K\pi\pi$     & $1.8$, $4.4$, $3.2$, $3.6$, resp. \\
II.3  & Branching fraction of 
        $\Dstar\To\Dz\piz$, $\Dz\pip$           & $4.7$, $0.7$, resp. \\
III.1 & Efficiency of finding charged tracks 
        not including soft pions                & $0.5$ per track \\
III.2 & Efficiency of finding soft charged
        pions from $\Dstarp$                    & $3.1$ per $\pi$ \\
III.3 & Efficiency of finding neutral pions 	& $3.0$ per $\piz$ \\
III.4 & Efficiency of finding $B$ decays in 
        bins of $m^2(\ppbar)$ and 
	$m^2(\Dmaybestar{p})$                   & $0.8$ to $9.7$ \\
III.5 & Efficiency of finding kaons in $B$ 
        decays                                  & $0.5$ per $K$	\\
III.6 & Efficiency of finding protons in $B$ 
        decays                                  & $1.0$ per $p$ \\
III.7 & Efficiency of particle identifcation 
        based on data control samples           & $1.5$ to $2.5$ \\
IV.1  & Pdf parameter variation for modes with
        $D\To{K}\pi$, $K\pi\piz$, 
        $K\pi\pi\pi$, $K\pi\pi$                 & $1.3$, $2.8$, $5.7$, $3.4$, resp. \\
IV.2  & Pdf choice for signal events 		& $0.6$ \\
IV.3  & Pdf choice for background events for
        $D\To{K}\pi$, $K\pi\piz$, 
        $K\pi\pi\pi$, $K\pi\pi$                 & $0.8$, $4.5$, $1.3$, $2.0$, resp. \\
IV.4  & Pdf yield bias by fitting mock 
        experiments embedded with MC            & $0.4$ to $2.2$ \\
V.1   & Peaking background in \mes-\DeltaE 	& $8.0$ for $\Dstarz\ppbar\pi$ 	\\
V.2   & Peaking background in \mes\ only 
        ($\Dmaybestarz\ppbar\pi\pi$, $K\pi\piz$ and $\Dmaybestarz\ppbar\pi\pi$, $K\pi\pi\pi$) 
						& $0.0$ to $14.5$ ($77$ to $85$)\\
V.3   & Peaking background from identical final
        states without a $D$ meson              & $0.5$ to $13.5$ \\
\dbline
\end{tabular}
}%
\\
\subfloat[Uncorrelated sources (\percent of $\mathcal{B}$)]{
\label{tab:syst_tot_uncorr}
\begin{tabular}{clrrrr}
\dbline
 $N$-body&\backslashbox{$B$ decay}{$D$ decay}
& $K\pi$& $K\pi\piz$& $K\pi\pi\pi$& $K\pi\pi$ \\
\sgline
  3-body &$\Bzb\To\Dz\ppbar$ 		&  $2.7$  & $5.5$  & $4.7$  & -      \\
  $''$   &$\Bzb\To\Dstarz\ppbar$ 	&  $2.2$  & $8.6$  & $8.7$  & -      \\
  4-body &$\Bzb\To\Dp\ppbar\pim$ 	&  -      &-       &-       & $5.7$  \\
  $''$   &$\Bzb\To\Dstarp\ppbar\pim$ 	&  $4.2$  & $7.6$  & $9.9$  & -      \\
  $''$   &$\Bm\To\Dz\ppbar\pim$ 	&  $14.5$ & $81.2$ & $77.3$ & -      \\
  $''$   &$\Bm\To\Dstarz\ppbar\pim$ 	&  $13.8$ & $86.3$ & $85.4$ & -      \\
  5-body &$\Bzb\To\Dz\ppbar\pim\pip$ 	&  $4.4$  & $8.8$  & $11.6$ & -      \\
  $''$   &$\Bzb\To\Dstarz\ppbar\pim\pip$&  $10.4$ & $11.6$ & $13.2$ & -      \\
  $''$   &$\Bm\To\Dp\ppbar\pim\pim$ 	&  -      &-       &-       & $15.0$ \\
  $''$   &$\Bm\To\Dstarp\ppbar\pim\pim$ &  $5.9$  & $14.9$ & $19.0$ & -      \\
\dbline
\end{tabular}
}
\subfloat[Correlated sources excluding II.2, II.3 (\percent of $\mathcal{B}$)]{
\label{tab:syst_tot_corr}
\begin{tabular}{lrrrr}
\dbline
  
\backslashbox{$B$ decay}{$D$ decay}
	& $K\pi$& $K\pi\piz$& $K\pi\pi\pi$& $K\pi\pi$ \\
\sgline
 $\Bzb\To\Dz\ppbar$ 			& $3.5$ & $6.9$ & $7.0$ & -     \\
 $\Bzb\To\Dstarz\ppbar$ 		& $4.6$ & $8.6$ & $7.7$ & -     \\
 $\Bzb\To\Dp\ppbar\pim$ 		& -     & -     & -     & $5.6$ \\
 $\Bzb\To\Dstarp\ppbar\pim$ 		& $6.3$ & $8.6$ & $8.9$ & -     \\
 $\Bm\To\Dz\ppbar\pim$ 			& $3.9$ & $7.0$ & $7.0$ & -     \\
 $\Bm\To\Dstarz\ppbar\pim$ 		& $4.9$ & $8.8$ & $7.7$ & -     \\
 $\Bzb\To\Dz\ppbar\pim\pip$ 		& $4.2$ & $7.2$ & $7.3$ & -     \\
 $\Bzb\To\Dstarz\ppbar\pim\pip$ 	& $5.2$ & $8.9$ & $7.9$ & -     \\
 $\Bm\To\Dp\ppbar\pim\pim$ 		& -     & -     & -     & $5.9$ \\
 $\Bm\To\Dstarp\ppbar\pim\pim$ 		& $6.8$ & $9.0$ & $9.2$ & -     \\
\dbline
\end{tabular}
}
}
\end{table}

I. \BB\ is counted by subtracting the total number of hadronic events
taken at $\sqrt{s}$ of $10.58\gev$ by the expected number due to
continuum events, which is estimated using data taken at $10.54\gev$.
The total uncertainty is $1.1\percent$ and the largest component comes from
the comparison of detection efficiencies of hadronic events in data
and MC.

II. We assume equal production of $\FourS\To\Bz\Bzb$ and $\Bp\Bm$ as
well as nominal branching fractions for $\Dstar$ and $D$. (1) For
$\FourS$, we take the absolute difference between the nominal value
and $50\percent$, which gives $1.6\percent$. (2) For $\Dstarz$ ($\Dstarp$), the
uncertainties are $4.7\percent$ ($0.7\percent$).  For $D$, the uncertainties are
is $1.3\percent$, $3.7\percent$, $2.5\percent$, and $2.3\percent$ for $K\pi$, $K\pi\piz$,
$K\pi\pi\pi$, and $K\pi\pi$, respectively.

III. The reconstruction efficiency has seven contributing sources.
(1) Charged track identification uncertainty of $0.5\percent$ per track is
obtained by comparing data and simulations with $\epem\To\tau^+\tau^-$
events where one tau decays leptonically and the other tau decays
hadronically. (2) Because soft charged pions from $\Dstarp$ decays
leave hits only in the silicon vertex tracker, they have an additional
uncertainty of $3.1\percent$, which is obtained from using the helicity
angle distribution of the decay daughters \cite{Long:2001}.  (3)
Neutral pions identification of $3.0\percent$ is obtained by comparing data
and MC $\tau^+\tau^-$ events where one tau decays leptonically and the
other decay includes a $\piz$.  (4) Acceptance of signal $B$ decays
computed using the MC assuming the uniform phase space decay model is
compared with the data distribution in bins of $m^2(\ppbar)$ and
$m^2(\Dmaybestar{p})$ where the most variation is seen.  We obtain
uncertainties ranging from $0.8$--$9.7\percent$ with modes containing a
\Dstarp\ at the higher end of the spectrum.  (5) Kaon identification
uncertainty of $0.5\percent$ is obtained by comparing data control samples
of $\Dz\To{K}\pi$ from inclusively produced $\Dstar\To\Dz\pi$ and from
similar $B$ decays.  (6) Proton identification uncertainty of $1\percent$ is
obtained by comparing various data control samples of inclusively
produced $\Lambda\To{p}\pim$ with decay topologies of high
charged-particle multiplicities that mimic our $B$ decay environment.
(7) Corrections made for particle identification contribute additional
statistical uncertainties that range from $1.5$--$2.5\percent$.

IV. The fit pdf has four contributing sources. (1) As described in
Sec.~\ref{sec:analysis}, some parameters are fixed to values
determined by the MC sample.  These are varied by the uncertainties
obtained when fitting the MC and the quadrature sum of the typical
yield changes are $1.3$, $2.8$, $5.7$, and $3.4$\percent for modes with
$D\To{K}\pi$, $K\pi\piz$, $K\pi\pi\pi$, and $K\pi\pi$, respectively.
(2) A more general signal pdf is chosen and the typical yield change
is $0.6\percent$. (3) A more general background pdf is chosen and the
typical yield changes are $0.8$, $4.5$, $1.3$, and $2.0$\percent for modes
with $D\To{K}\pi$, $K\pi\piz$, $K\pi\pi\pi$, and $K\pi\pi$,
respectively.  (4) Sec.~\ref{sec:fit} described a thousand mock
experiments to check for possible pathologies of the pdf.  We
generalize this study by embedding the MC with signal $B$ decays into
background events drawn from the background pdf.  Signal yield bias
ranges from \nolbreaks{$0.4$--$2.2\percent$}, which we assign as a
systematic uncertainty.

V. Peaking background sources have three contributing sources. (1)
Possible sources of events that peak in the signal box around
$\mes{\eq}5.28\gevcc$ and $\DeltaE{\eq}0\mev$ have been studied
extensively using a sample of $\mathcal{O}(10^{9})$ MC events with $B$
decays hadronized using J{\sc etset~7.4}~\cite{Sjostrand:1993yb} as
well as a sample of $\mathcal{O}(10^{8})$ MC events dedicated to a set
of $B$ decays to a \Dmaybestar, a \ppbar\ pair, and a system of up to
two charged or neutral pions followed by a comprehensive set of
$\Dmaybestar$ decays including those not reconstructed in our study.
We find only one scenario meeting our requirement: events generated as
\nolbreaks{$\Bzb\To\Dstarp\ppbar\pim$} reconstructed as
\nolbreaks{$\Bm\To\Dstarz\ppbar\pim$}.  In this case, we fit the MC
and add the pdf component in the fit to data in Eqn.~\ref{eqn:pdf} as
described in Sec.~\ref{sec:fit}.  The normalization is the product of
the acceptance and branching fraction for $\Dstarp\ppbar\pim$.  The
total uncertainty on the latter quantity is $8\percent$ and is assigned as a
systematic.  (2) Possible sources of events that peak either in
$\mes{\eq}5.28\gevcc$ or in $\DeltaE{\eq}0\mev$ have been studied
using the above-mentioned set of MC.  We find that only a few samples
exhibit a peak in \mes\ that occur when a related decay is
misreconstructed by losing or gaining either a \piz\ or a $\gamma$
from the other $B$ decay in the event; we find no cases of events
peaking only in \DeltaE.  In the \mes-peaking cases, the distributions
peak broadly with a Gaussian width of around $10\mevcc$ and has a
smooth variation in \DeltaE.  One such distribution is fit and the
component is added to our nominal fit in Eqn.~\ref{eqn:pdf} as a
systematic study.  The signal yield changes varied from $0$--$15\percent$
with four exceptions: in the decays
\nolbreaks{$\Dmaybestar\ppbar\pi\pi$,} $K\pi\piz$ and
\nolbreaks{$\Dmaybestar\ppbar\pi\pi$,} $K\pi\pi\pi$ shown in
Fig.~\ref{fig:2dfits_4body}bc (\ref{fig:2dfits_4body}ef), the signal
yields varied by $77$--$85\percent$.  This is not surprising because we saw
in Sec.~\ref{sec:fit} that the fit was not stable against our initial
strategy and had to be forced to find a peak at the desired location
by fixing the means of $P_S$.  For these four $B$ modes, the $K\pi$
measurement is heavily weighted due to these systematic errors with
respect to both $K\pi\piz$ and $K\pi\pi\pi$ when we average the $B$
branching fraction over $D$ decay modes.  (3) Possible sources of
events that share the final states with desired decays are studied by
using $m(D)$ sidebands in data.  One such example scenario would be
\nolbreaks{$B\To\Lambda_c\,\pbar\piz$},
\nolbreaks{$\Lambda_c\To{p}K\pi$} misreconstructed as
\nolbreaks{$\Bzb\To\Dz\ppbar$}, \nolbreaks{$\Dz\To{K}\pi\piz$}, which
would live under the $D$ mass peak in the $m(D)$ distribution.  To
correct for such a bias, we scale the $B$ signal yield in the $m(D)$
sideband region to the amount of $D$ background in the $m(D)$ signal
region.  This is given later as a correction factor $n_b$ in
Table~\ref{tab:bfchain} and the uncertainties range from
\nolbreaks{$0.5$--$13.5\percent$}.

Combining systematic uncertainties requires us to divide uncorrelated
sources from those that are not.  Table~\ref{tab:syst_tot_uncorr}
gives the uncorrelated items III.4, III.7, IV.4, V.1, V.2, and V.3
summed in quadrature.  Table~\ref{tab:syst_tot_corr} gives the
correlated sources, which are computed in two steps.  Correlated
sources for the individual decay chains---III.1, 2, 3, and 6---are
summed linearly; Correlated sources for $B$ decays---I; II.1; III.5
and 6; and IV.1, 2, and 3---are summed in quadrature.  Lastly, the
error matrix, $\mathbf{V}$, is constructed.  For decays with a $\Dz$,
$\mathbf{V}$ is a $3\!\times\!3$ matrix spanned by the three decay
modes ($K\pi,K\pi\piz,K\pi\pi\pi$) labeled by $\alpha$ and $\beta$;
for those with a $\Dp$, it is a $1\times1$ matrix for $K\pi\pi$:
\begin{equation} 
\begin{array}{lll}
\mathbf{V}
&\!=
\mathbf{V}_\textrm{stat}
+
\overbrace{
\mathbf{V}_\textrm{unc} +
\mathbf{V}_\textrm{cor}
}^{\mathbf{V}_\textrm{syst}}
&\textrm{is the sum of statistical and systematic
  components}\\
\mathbf{V}_\textrm{stat} 
&\!=\textrm{diag}[\sigma_{S,\alpha}^2]
&\textrm{for statistical errors from fitting for the signal yield} \\
\mathbf{V}_\textrm{unc} 
&\!=\textrm{diag}[\sigma_{\textrm{unc},\alpha}^2]
&\textrm{for uncorrelated systematic errors} \\
\mathbf{V}_\textrm{cor}
&\!=\textrm{diag}[\sigma_{\textrm{cor},\alpha}^2]
+
\rho_{\alpha\beta}
\,\sigma_{\textrm{cor},\alpha}
\,\sigma_{\textrm{cor},\beta}
&\textrm{for correlated systematic errors,}
\end{array}
\label{eqn:V}
\end{equation}
where $\rho_{\alpha\beta}$ is the correlation coefficient and
$\sigma_{S}$ is given later in Table~\ref{tab:bfchain}.  The
$\rho_{\alpha\beta}$ for $\Dz$ branching fraction are given in
\cite{Amsler:2008zzb} and all other coefficients are assumed to be
unity.

\section{BRANCHING FRACTIONS}
\label{sec:bf}

\subsection{Weighted average}

Branching fractions are found in two steps: they are computed for 26
decay chains, then averaged over $D$ decay modes.  The branching
fraction for a $B$ decay followed by $D\To{K}\pi$ denoted by $\alpha$,
is 
\begin{equation}
\mathcal{B}_{\alpha} =
\frac{1}{N_{\BB}}\
\frac{1}{\mathcal{B}(\Dstar\To\Dz\pi)}\
\frac{1}{\mathcal{B}(D\To\alpha)}\
\frac{1}{\epsilon}\
(n_S - n_b),
\label{eqn:bf}
\end{equation}
where $N_{\BB}{\eq}455\times10^6$, $\mathcal{B}(\Dmaybestar)$ is the
$\Dmaybestar$ branching fraction, $\epsilon$ is the acceptance, $n_S$
is the signal yield, and $n_b$ is the $D$-mass-sideband peaking
background correction.  Branching fractions for eight $B$
decays with a $\Dz$ form a vector
$\vec{\mathcal{B}}{\eq}(\mathcal{B}_{K\pi},\mathcal{B}_{K\pi\piz},\mathcal{B}_{K\pi\pi\pi})^T$;
for two with a $\Dp$,
$\vec{\mathcal{B}}{\eq}(\mathcal{B}_{K\pi\pi})^T$.

\begin{table}[bp!]
\caption{$B$ branching fractions, $\mathcal{B}$, (a) for 26 decay
  chains and (b) averaged over $D$ modes.}
\label{tab:bf}
\centering
{\small
\subfloat[$\mathcal{B}$ and its ingredients. See Eqn.~\ref{eqn:bf}.]{
\label{tab:bfchain}
\begin{tabular*}{0.95\textwidth}{clr@{$\pm$} rrrrc r@{$\pm$}p{0.75in}}
\dbline
\multicolumn{1}{l}{$N$-body}
& \multicolumn{1}{l}{$B$ decay chain}
& $n_S$
& $\sigma_S$
& $n_b$\ \
& $\epsilon$ (\percent)
& $\mathcal{B}_{D}$(\percent)
& $\mathcal{B}_{\Dstar}$(\percent)
& $\mathcal{B}$ 
& $\sigma_\textrm{stat}$ ($10^{-4}$)
\\
\sgline
3-body & $\Bzb\To\Dz\ppbar$, $K\pi$			&$351   $&${20}      $&$7.6  $&$19.04$&$3.89 $& -     &$1.02 $&$0.06$ \\
$''$ & $\Bzb\To\Dz\ppbar$, $K\pi\piz$			&$431   $&$_{27}^{28}$&$23.7 $&$7.01 $&$13.5\phantom{0}$& -     &$0.95 $&$0.06$ \\
$''$ & $\Bzb\To\Dz\ppbar$, $K\pi\pi\pi$			&$448   $&$_{26}^{27}$&$10.1 $&$9.85 $&$8.10 $& -     &$1.21 $&$0.07$ \\
$''$ & $\Bzb\To\Dstarz\ppbar$, $\Dz\pi$, $K\pi$   	&$110   $&$_{11}^{12}$&$-1.4 $&$9.40 $&$3.89 $&$61.9 $&$1.08 $&$0.12$ \\
$''$ & $\Bzb\To\Dstarz\ppbar$, $\Dz\pi$, $K\pi\piz$	&$148   $&${15}      $&$3.9  $&$3.24 $&$13.5\phantom{0}$&$61.9 $&$1.17 $&$0.12$ \\
$''$ & $\Bzb\To\Dstarz\ppbar$, $\Dz\pi$, $K\pi\pi\pi$	&$95    $&$_{13}^{14}$&$5.5  $&$5.15 $&$8.10 $&$61.9 $&$0.76 $&$0.12$ \\
4-body & $\Bzb\To\Dp\ppbar\pi$, $K\pi\pi$		&$1816  $&$_{52}^{53}$&$55.2 $&$12.64$&$9.22 $& -     &$3.32 $&$0.10$ \\
$''$ & $\Bzb\To\Dstarp\ppbar\pi$, $\Dz\pi$, $K\pi$	&$392   $&$_{20}^{21}$&$2.3  $&$6.79 $&$3.89 $&$67.7 $&$4.79 $&$0.26$ \\
$''$ & $\Bzb\To\Dstarp\ppbar\pi$, $\Dz\pi$, $K\pi\piz$	&$601   $&$_{27}^{28}$&$20.7 $&$3.08 $&$13.5\phantom{0}$&$67.7 $&$4.53 $&$0.22$ \\
$''$ & $\Bzb\To\Dstarp\ppbar\pi$, $\Dz\pi$, $K\pi\pi\pi$&$378   $&$_{21}^{22}$&$19.9 $&$3.66 $&$8.10 $&$67.7 $&$3.92 $&$0.24$ \\
$''$ & $\Bm\To\Dz\ppbar\pi$, $K\pi$			&$1078  $&$_{37}^{38}$&$13.1 $&$15.89$&$3.89 $& -     &$3.79 $&$0.14$ \\
$''$ & $\Bm\To\Dz\ppbar\pi$, $K\pi\piz$			&$1176  $&$_{53}^{54}$&$41.1 $&$5.53 $&$13.5\phantom{0}$& -     &$3.34 $&$0.16$ \\
$''$ & $\Bm\To\Dz\ppbar\pi$, $K\pi\pi\pi$		&$1296  $&$_{56}^{57}$&$33.0 $&$7.82 $&$8.10 $& -     &$4.38 $&$0.20$ \\
$''$ & $\Bm\To\Dstarz\ppbar\pi$, $\Dz\pi$, $K\pi$	&$328   $&${22}      $&$2.1  $&$7.71 $&$3.89 $&$61.9 $&$3.86 $&$0.26$ \\
$''$ & $\Bm\To\Dstarz\ppbar\pi$, $\Dz\pi$, $K\pi\piz$	&$482   $&$_{34}^{35}$&$46.5 $&$2.87 $&$13.5\phantom{0}$&$61.9 $&$3.99 $&$0.32$ \\
$''$ & $\Bm\To\Dstarz\ppbar\pi$, $\Dz\pi$, $K\pi\pi\pi$	&$343   $&$_{30}^{31}$&$32.4 $&$4.04 $&$8.10 $&$61.9 $&$3.37 $&$0.34$ \\
5-body & $\Bzb\To\Dz\ppbar\pi\pi$, $K\pi$		&$438   $&$_{32}^{32}$&$7.7  $&$8.19 $&$3.89 $& -     &$2.97 $&$0.22$ \\
$''$ & $\Bzb\To\Dz\ppbar\pi\pi$, $K\pi\piz$		&$663   $&$_{64}^{65}$&$155.2$&$2.92 $&$13.5\phantom{0}$& -     &$2.83 $&$0.36$ \\
$''$ & $\Bzb\To\Dz\ppbar\pi\pi$, $K\pi\pi\pi$		&$770   $&$_{67}^{68}$&$39.7 $&$3.75 $&$8.10 $& -     &$5.28 $&$0.48$ \\
$''$ & $\Bzb\To\Dstarz\ppbar\pi\pi$, $\Dz\pi$, $K\pi$	&$61    $&${12}      $&$1.8  $&$2.89 $&$3.89 $&$61.9 $&$1.87 $&$0.38$ \\
$''$ & $\Bzb\To\Dstarz\ppbar\pi\pi$, $\Dz\pi$, $K\pi\piz$&$142  $&$_{31}^{33}$&$36.7 $&$1.27 $&$13.5\phantom{0}$&$61.9 $&$2.19 $&$0.66$ \\
$''$ & $\Bzb\To\Dstarz\ppbar\pi\pi$, $\Dz\pi$, $K\pi\pi\pi$&$163$&$_{29}^{30}$&$12.8 $&$1.33 $&$8.10 $&$61.9 $&$4.93 $&$0.99$ \\
$''$ & $\Bm\To\Dp\ppbar\pi\pi$, $K\pi\pi$		&$475   $&$_{36}^{37}$&$6.6  $&$6.74 $&$9.22 $& -     &$1.66 $&$0.13$ \\
$''$ & $\Bm\To\Dstarp\ppbar\pi\pi$, $\Dz\pi$, $K\pi$    &$57    $&${9}       $&$-12.4$&$2.93 $&$3.89 $&$67.7 $&$1.98 $&$0.26$ \\
$''$ & $\Bm\To\Dstarp\ppbar\pi\pi$, $\Dz\pi$, $K\pi\piz$&$94    $&$_{13}^{14}$&$-0.6 $&$1.25 $&$13.5\phantom{0}$&$67.7 $&$1.82 $&$0.27$ \\
$''$ & $\Bm\To\Dstarp\ppbar\pi\pi$, $\Dz\pi$, $K\pi\pi\pi$&$66  $&$_{11}^{12}$&$4.8  $&$1.52 $&$8.10 $&$67.7 $&$1.61 $&$0.32$ \\
\dbline
\end{tabular*}
}\\
\subfloat[$\mathcal{B}$ averaged over $D$ decay modes and previous
measurements \cite{Aubert:2006qx,Anderson:2000tz,Abe:2002tw} in units
of ($10^{-4}$).  See Eqn.~\ref{eqn:bfavg}.] {
\label{tab:bfsummary}
\begin{tabular*}{0.95\textwidth}{cl r@{$\pm$}l @{$\pm$}l cc
  r@{$\pm$}l @{$\pm$}l
  r@{$\pm$}l @{$\pm$}l
}
\dbline
$N$-body
& $B$ decay
& $\mathcal{B}$
& $\sigma_\textrm{stat}$ 
& $\sigma_\textrm{syst}$
& $\chi^2$
&  $P(\chi^2)$
& \multicolumn{3}{c}{Refs. \cite{Anderson:2000tz,Abe:2002tw}}
& \multicolumn{3}{c}{Ref. \cite{Aubert:2006qx}}
\\
\sgline
3-body & $\Bzb\To\Dz\ppbar$ 		& $1.02$ &  $0.04$ & $0.05$ & $4.3$ & $12\percent$ & $1.18$ & $0.15$ & $0.16$ \cite{Abe:2002tw} & $1.13$ & $0.06$ & $0.08$ \\
$''$   & $\Bzb\To\Dstarz\ppbar$ 	& $0.97$ &  $0.07$ & $0.09$ & $4.1$ & $13\percent$ & $1.20$ & $^{0.33}_{0.29}$ & $0.21$ \cite{Abe:2002tw} & $1.01$ & $0.10$ & $0.09$ \\
4-body & $\Bzb\To\Dp\ppbar\pim$ 	& $3.32$ &  $0.10$ & $0.27$ & -     & -      & \multicolumn{3}{l}{\quad -}& $3.38$ & $0.14$ & $0.29$ \\
$''$   & $\Bzb\To\Dstarp\ppbar\pim$ 	& $4.55$ &  $0.16$ & $0.37$ & $1.2$ & $54\percent$ 
					& $6.5\phantom{0}$ & $^{1.3}_{1.2}\phantom{0}$
					& $1.0\phantom{0}$ \cite{Anderson:2000tz}& $4.81$ & $0.22$ & $0.44$ \\
$''$   & $\Bm\To\Dz\ppbar\pim$ 		& $3.72$ &  $0.11$ & $0.23$ & $3.4$ & $19\percent$ & \multicolumn{3}{l}{\quad -}& \multicolumn{3}{l}{\quad -}\\
$''$   & $\Bm\To\Dstarz\ppbar\pim$ 	& $3.73$ &  $0.17$ & $0.39$ & $0.5$ & $79\percent$ & \multicolumn{3}{l}{\quad -}& \multicolumn{3}{l}{\quad -}\\
5-body & $\Bzb\To\Dz\ppbar\pim\pip$ 	& $2.99$ &  $0.21$ & $0.44$ & $0.3$ & $85\percent$ & \multicolumn{3}{l}{\quad -}& \multicolumn{3}{l}{\quad -}\\
$''$   & $\Bzb\To\Dstarz\ppbar\pim\pip$ & $1.91$ &  $0.36$ & $0.29$ & $0.5$ & $78\percent$ & \multicolumn{3}{l}{\quad -}& \multicolumn{3}{l}{\quad -}\\
$''$   & $\Bm\To\Dp\ppbar\pim\pim$ 	& $1.66$ &  $0.13$ & $0.27$ & -     & -      & \multicolumn{3}{l}{\quad -}& \multicolumn{3}{l}{\quad -}\\
$''$   & $\Bm\To\Dstarp\ppbar\pim\pim$ 	& $1.86$ &  $0.16$ & $0.18$ & $0.2$ & $91\percent$ & \multicolumn{3}{l}{\quad -}& \multicolumn{3}{l}{\quad -}\\
\dbline
\end{tabular*}
}
}
\end{table}

Table~\ref{tab:bf} presents the branching fractions and their
ingredients for the 26 decay chains and the
combined values,
$\mathcal{B}_\textrm{avg}{\eq}\vec{\mathcal{B}}\cdot\vec{w}$, which
are averaged over \Dz\ decay modes using a method \cite{Lyons:1988rp}
to compute the weights from error matrices in Eqn.~\ref{eqn:V}
\begin{eqnarray}
\vec{w} = \frac{\mathbf{V}^{-1}\cdot\vec{u}}{\vec{u}^T\cdot\mathbf{V}^{-1}\cdot\vec{u}},
\label{eqn:bfavg}
\end{eqnarray}
where $\vec{u}{\eq}(1\textrm{,}1\textrm{,}1)^T$.  The squared combined statistical
uncertainty is
\nolbreaks{$\sigma_\textrm{stat}^2{\eq}\vec{w}^T\!\cdot\!\mathbf{V}_\textrm{stat}\!\cdot\!\vec{w}$}
and similarly computed for systematics.  All measurements are
statistically significant.

For $B$ decays with a $\Dz$ meson, consistency between the three
measurements is evaluated by
\nolbreaks{$\chi^2{\eq}(\vec{\mathcal{B}}-\mathcal{B}_\textrm{avg}\vec{u})^T\cdot\mathbf{V}^{-1}\cdot(\vec{\mathcal{B}}-\mathcal{B}_\textrm{avg}\vec{u})$}
and the associated probability
\nolbreaks{$P(\chi^2){\eq}\exp[-\chi^2/2]/2$}.  For all cases, 
$P(\chi^2)$ is greater than $10\percent$.

\subsection{Ratios}

Tables~\ref{tab:ratios} gives ratios of branching fractions that show
two general patterns.  First, the ratios are of order unity for modes
related by replacing $D$ mesons.  The unit ratio for
$D\!\leftrightarrow\!\!{\Dstar}$ replacement suggests that the
additional $\Dstar$ polarization degrees of freedom does not increase
production.  The unit ratio for
$\Dmaybestarp\!\!\leftrightarrow\!\!\Dmaybestarz$ suggests that the
production can be described by simple isospin relations.  Second, the
ratios for modes related by addition of pions imply the hierarchy of
\nolbreaks{$\mathcal{B}_\textrm{3-body} < \mathcal{B}_\textrm{5-body}
  < \mathcal{B}_\textrm{4-body}$}.

 \begin{table}[b!]
 \centering
 \caption{Ratios of branching fractions, $r$.  The 
   $\sigma_\textrm{tot}$ is the quadrature sum of
     $\sigma_\textrm{stat}$ and $\sigma_\textrm{syst}$.
 }
 \label{tab:ratios}
{\small
 \subfloat[Modes related by 
 $s$- and $d$-wave $\Dmaybestar$ mesons
]{
 \begin{tabular}{p{0.7in} @{/}p{1.1in}  r @{$\pm$}c@{ (}c@{)}}
 \dbline
 \multicolumn{1}{l}{Quantity} & & $r$ & $\sigma_\textrm{tot}$ & $\sigma_\textrm{stat}$ \\
 \sgline
 $\mathcal{B}_{\Dstarz\ppbar}      $&$\mathcal{B}_{\Dz\ppbar}$ 		& $0.95$ & $0.13$ & $0.08$ \\
 $\mathcal{B}_{\Dstarp\ppbar\pi}   $&$\mathcal{B}_{\Dp\ppbar\pi}$ 	& $1.37$ & $0.17$ & $0.06$ \\
 $\mathcal{B}_{\Dstarz\ppbar\pi}   $&$\mathcal{B}_{\Dz\ppbar\pi}$ 	& $1.00$ & $0.13$ & $0.05$ \\
 $\mathcal{B}_{\Dstarz\ppbar\pi\pi}$&$\mathcal{B}_{\Dz\ppbar\pi\pi}$ 	& $0.64$ & $0.19$ & $0.13$ \\
 $\mathcal{B}_{\Dstarp\ppbar\pi\pi}$&$\mathcal{B}_{\Dp\ppbar\pi\pi}$ 	& $1.12$ & $0.25$ & $0.13$ \\
 \multicolumn{2}{l}{Average with $\chi^2$ of $8.9/4$}			& $1.01$ & \multicolumn{2}{l}{\!$0.07$} \\
 \dbline
 \end{tabular}
 }
 \subfloat[Modes related by 
charged and neutral $\Dmaybestar$ mesons
]{
 \begin{tabular}{p{0.7in} @{/}p{1.1in} r @{$\pm$}c@{ (}c@{)}}
 \dbline
 \multicolumn{1}{l}{Quantity} & & $r$ & $\sigma_\textrm{tot}$ & $\sigma_\textrm{stat}$ \\
 \sgline
 $\mathcal{B}_{\Dz\ppbar\pi}       $&$\mathcal{B}_{\Dp\ppbar\pi}       $& $1.12$ & $0.12$ & $0.05$ \\
 $\mathcal{B}_{\Dstarz\ppbar\pi}   $&$\mathcal{B}_{\Dstarp\ppbar\pi}   $& $0.82$ & $0.12$ & $0.05$ \\
 $\mathcal{B}_{\Dz\ppbar\pi\pi}    $&$\mathcal{B}_{\Dp\ppbar\pi\pi}    $& $1.80$ & $0.44$ & $0.19$ \\
 $\mathcal{B}_{\Dstarz\ppbar\pi\pi}$&$\mathcal{B}_{\Dstarp\ppbar\pi\pi}$& $1.03$ & $0.28$ & $0.21$ \\
 \multicolumn{2}{l}{Average with $\chi^2$ of $6.6/3$}			& $1.00$ & \multicolumn{2}{l}{\!$0.08$} \\  
 \dbline
 \end{tabular}
 }\\
 \subfloat[Comparison of 3- to 4-body $B$ decay]{
 \begin{tabular}{p{0.7in} @{/}p{1.1in} r @{$\pm$}c@{ (}c@{)}}
 \dbline
 \multicolumn{1}{l}{Quantity} & & $r$ & $\sigma_\textrm{tot}$ & $\sigma_\textrm{stat}$ \\
 \sgline
 $\mathcal{B}_{\Dstarz\ppbar\pi}   $&$\mathcal{B}_{\Dstarz\ppbar}      $& $3.66$ & $0.48$  & $0.22$ \\
 $\mathcal{B}_{\Dz\ppbar\pi}       $&$\mathcal{B}_{\Dz\ppbar} 	       $& $3.08$ & $0.62$  & $0.31$ \\
 \multicolumn{2}{l}{Average with $\chi^2$ of $0.5/1$ }			& $3.44$ & \multicolumn{2}{l}{\!$0.38$} \\
 \dbline
 \end{tabular}
 }
 \subfloat[Comparison of 4- to 5-body $B$ decay]{
 \begin{tabular}{p{0.7in} @{/}p{1.1in} r @{$\pm$}c@{ (}c@{)}}
 \dbline
 \multicolumn{1}{l}{Quantity} & & $r$ & $\sigma_\textrm{tot}$ & $\sigma_\textrm{stat}$ \\
 \sgline
 $\mathcal{B}_{\Dp\ppbar\pi\pi}       $&$\mathcal{B}_{\Dp\ppbar\pi}    $& $0.50$ & $0.10$ & $0.04$ \\
 $\mathcal{B}_{\Dstarp\ppbar\pi\pi}   $&$\mathcal{B}_{\Dstarp\ppbar\pi}$& $0.41$ & $0.06$ & $0.04$ \\
 $\mathcal{B}_{\Dz\ppbar\pi\pi}       $&$\mathcal{B}_{\Dz\ppbar\pi}    $& $0.80$ & $0.14$ & $0.06$ \\
 $\mathcal{B}_{\Dstarz\ppbar\pi\pi}   $&$\mathcal{B}_{\Dstarz\ppbar\pi}$& $0.51$ & $0.14$ & $0.10$ \\
 \multicolumn{2}{l}{Average with $\chi^2$ of $6.5/3$}			& $0.49$ & \multicolumn{2}{l}{\!$0.05$} \\
 \dbline
 \end{tabular}
 }
}
\end{table} 

\section{DECAY DYNAMICS}
\label{sec:dynamics}

\subsection{\boldmath Three-body decays $B\To\Dmaybestar\ppbar$}

\begin{figure}[b!]
\centering
\subfloat[$\Bzb\To\Dz\ppbar$ in \mes-\DeltaE\ signal box]{
  \begin{overpic}[width=0.40\textwidth]{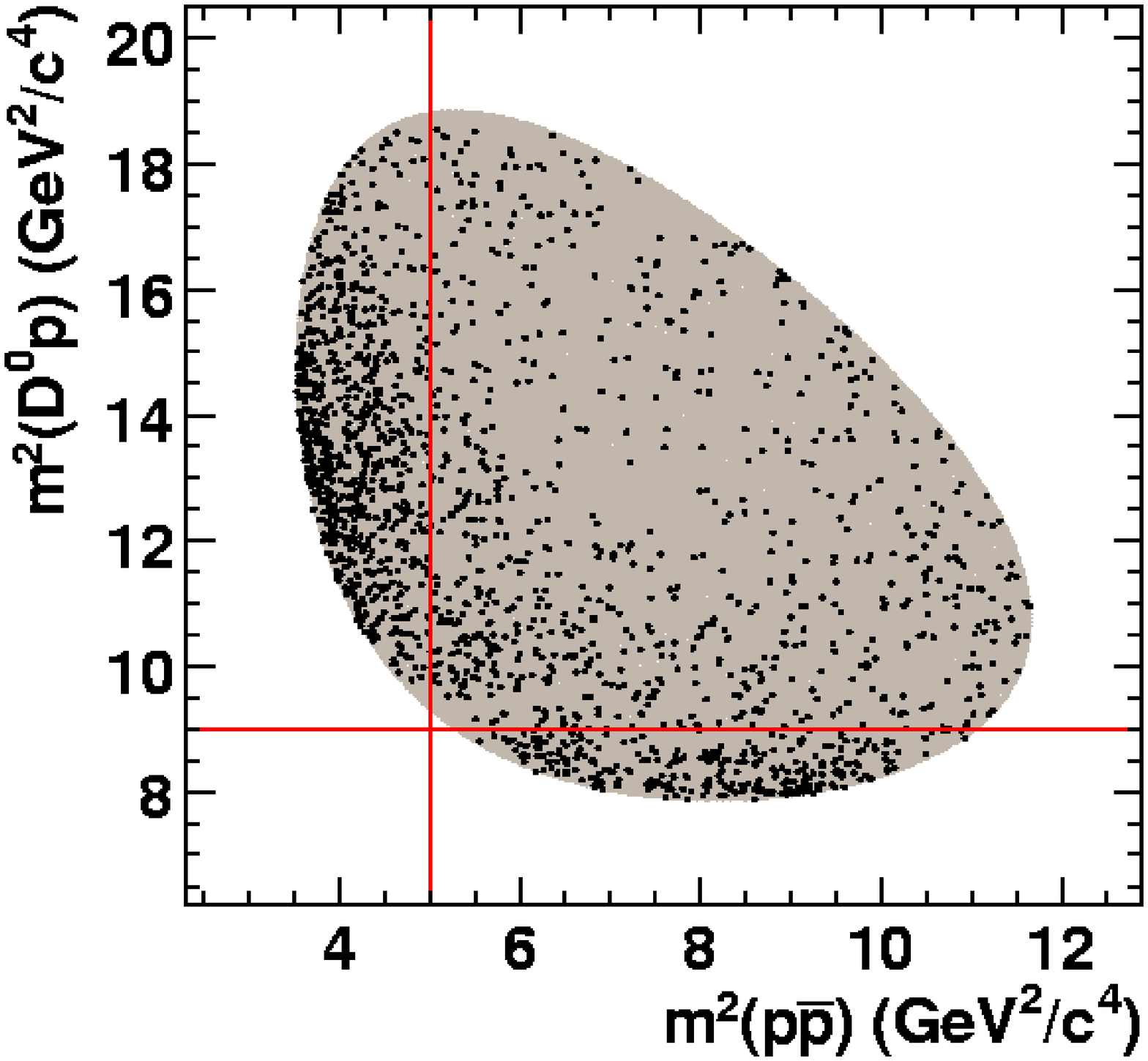}
    \put(70,82){\small{\babar}}
    \put(70,74){\small{prelim.}}
    \end{overpic}
}%
\hspace{0.5in}
\subfloat[$\Bzb\To\Dz\ppbar$ scaled \mes-\DeltaE\ sideband]{
  \begin{overpic}[width=0.40\textwidth]{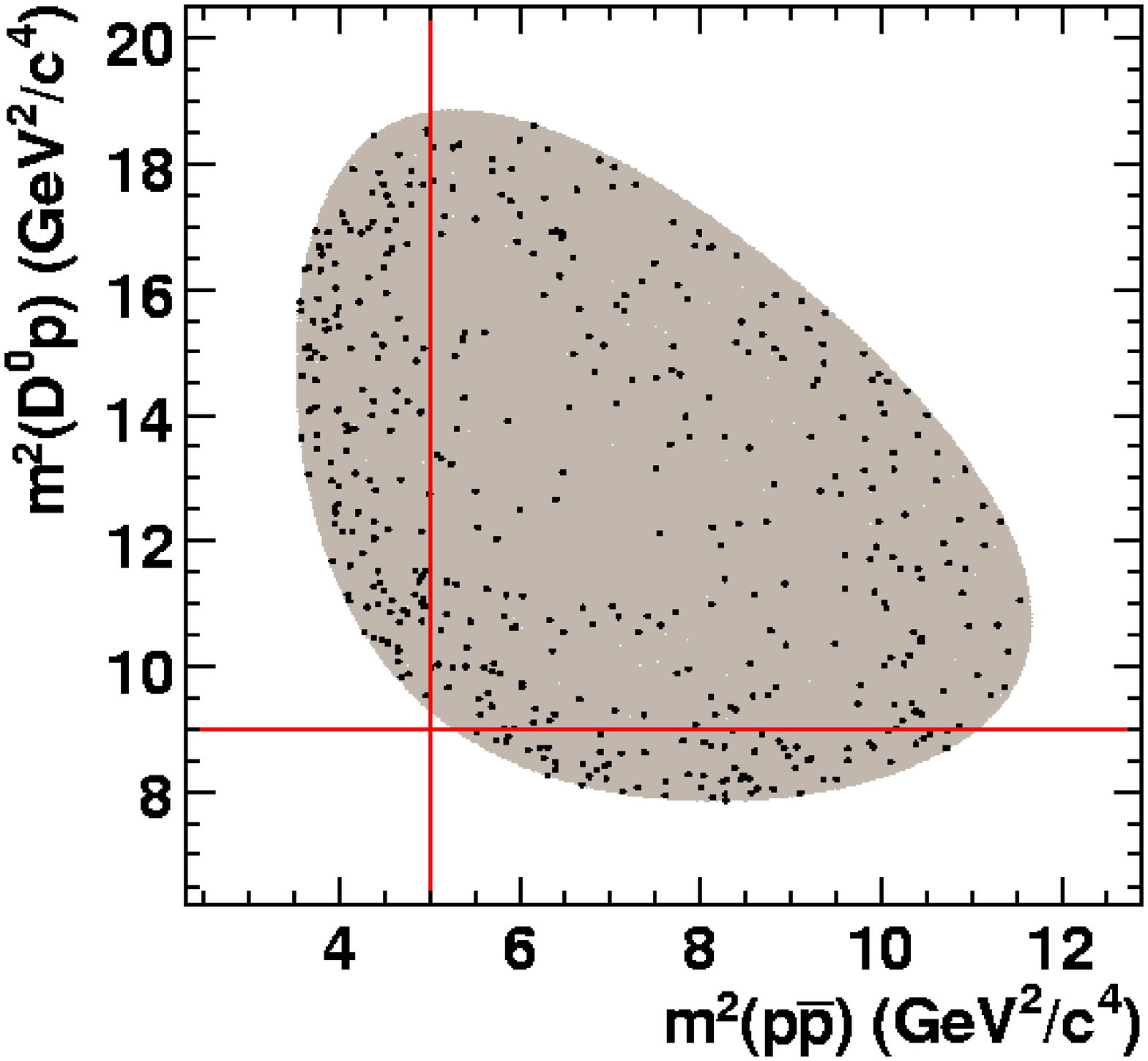}
    \put(70,82){\small{\babar}}
    \put(70,74){\small{prelim.}}
\end{overpic}
}%
\\
\subfloat[$\Bzb\To\Dstarz\ppbar$ in \mes-\DeltaE\ signal box]{
  \begin{overpic}[width=0.40\textwidth]{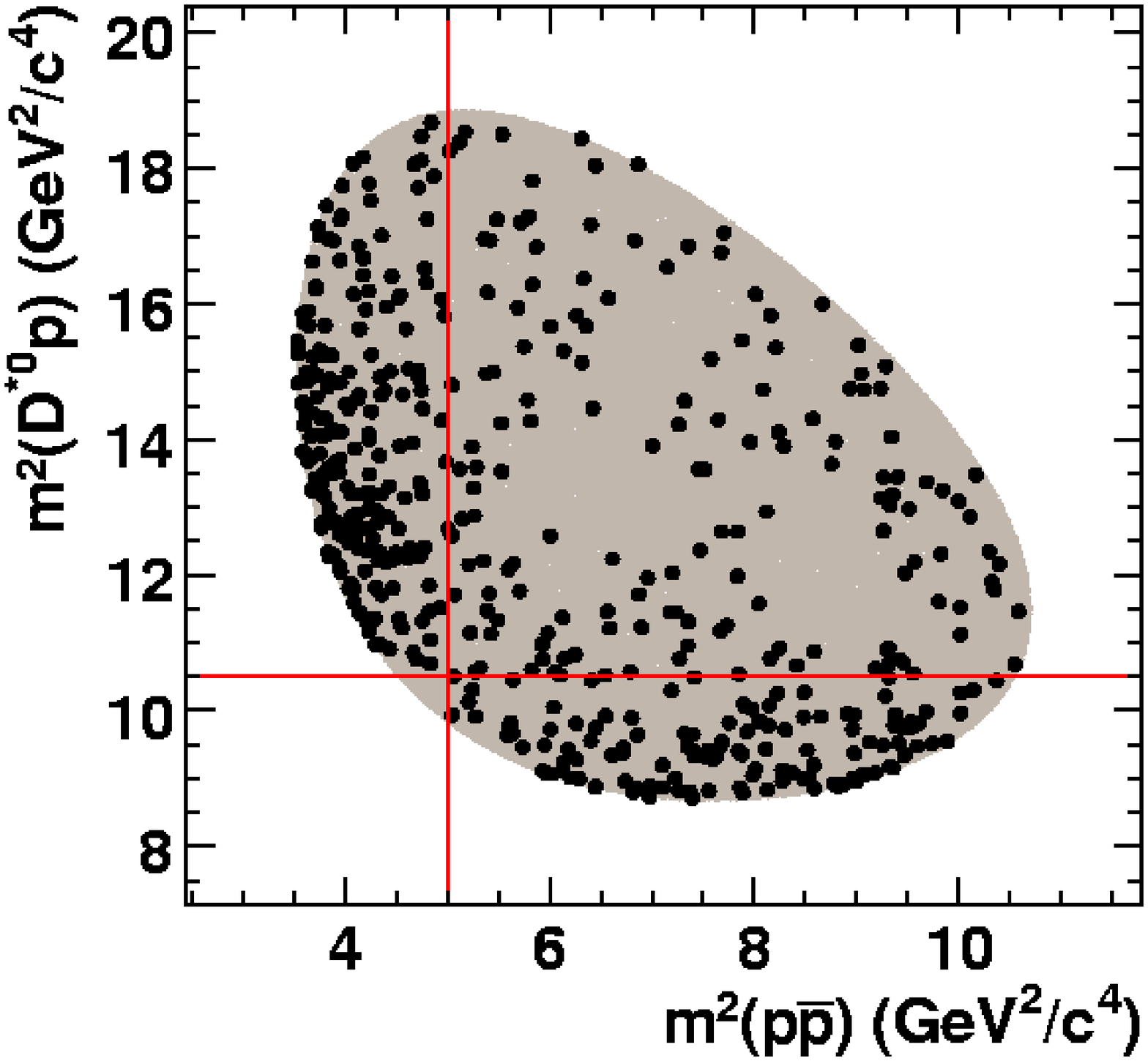}
    \put(70,82){\small{\babar}}
    \put(70,74){\small{prelim.}}
  \end{overpic}
}%
\hspace{0.5in}
\subfloat[$\Bzb\To\Dstarz\ppbar$ scaled \mes-\DeltaE\ sideband]{
  \begin{overpic}[width=0.40\textwidth]{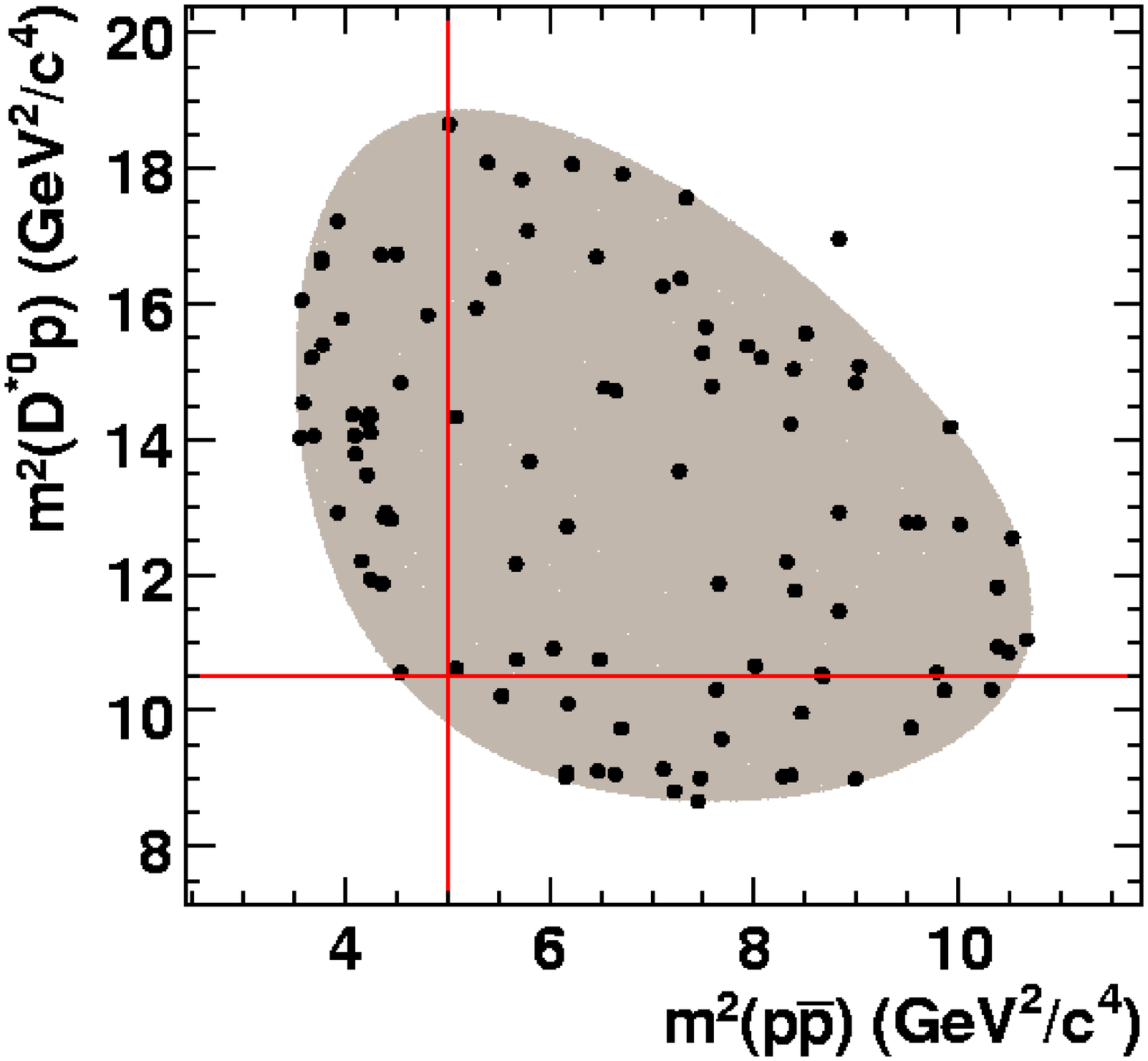}
    \put(70,82){\small{\babar}}
    \put(70,74){\small{prelim.}}
  \end{overpic}
}
\caption{Dalitz plots of $m^2(\ppbar)$ vs.  $m^2(\Dmaybestar{p})$ for
  two 3-body $B$ decays.  Plot in (a) shows $\Bzb\To\Dz\ppbar$ in the
  \mes-\DeltaE\ signal box and (b) the sideband region with the number
  of events corresponding to the background yield in (a).
  Corresponding plots for $\Bzb\To\Dstarz\ppbar$ are given in (c) and
  (d), respectively.  Events live in the shaded allowed kinematic
  region except for one outlier in (d), which failed the re-fit
  procedure described in footnote d.  Red lines are drawn at $(x,y)$
  of $(5\gevccsq,9\gevccsq)$ for (ab) and $(5\gevccsq,11.5\gevccsq)$
  for (cd).
}
\label{fig:dalitz}
\end{figure}

Figure~\ref{fig:dalitz} gives the Dalitz plot for 3-body $B$ decay
modes, where we sum over $\Dz$ decays.  Each point in the
$m^2(\ppbar)$-$m^2(\Dmaybestarz{p})$ plane represents an event and we
observe threshold enhancements in both variables.\footnote{$B$
candidates are refit with the $B$-mass constraint so that events lie
in the region allowed by kinematic limits.} Red lines are drawn as
visual guides: the threshold effect in $\ppbar$ ($\Dmaybestarz{p}$) is
on the left (bottom) of the vertical (horizontal) line at $5$ ($9$)
\gevccsq.  The two regions do not overlap.

We investigate the enhancements with mass projection plots using the
background subtraction technique \cite{Pivk:2004ty} by weighting pdf
components and associated yields, which were described in
Eqn.~\ref{eqn:pdf} and given in
Table~\ref{tab:bfchain}.  The weight for the
$i^\textrm{th}$ event, where \nolbreaks{$y_i{\eq}(\mes_{i},\DeltaE_{i})$}, is
\begin{eqnarray}
W_i =
\frac{C_{SS}P_S(y_i)+C_{SB}P_B(y_i)}
     {n_S P_S(y_i)+n_B P_B(y_i)}
\textrm{\quad and \quad}
(C_{S\Lambda})^{-1}{\eq}\!\!
\sum_{j=1}^{N\textrm{ events}}
\frac{P_S(y_j)P_\Lambda(y_j)}
     {\big(n_S P_S(y_j)+n_B P_B(y_j)\big)^2},
\label{eqn:w}
\end{eqnarray}
where $\Lambda{\eq}S,B$ and $C_{SB}$ quantifies the correlation between the
signal and background yields.

\begin{figure}[b!]
\centering
\subfloat[$\Dz\ppbar$, $m^2(\ppbar)\!<\!5$]{
  \begin{overpic}[width=0.25\textwidth]{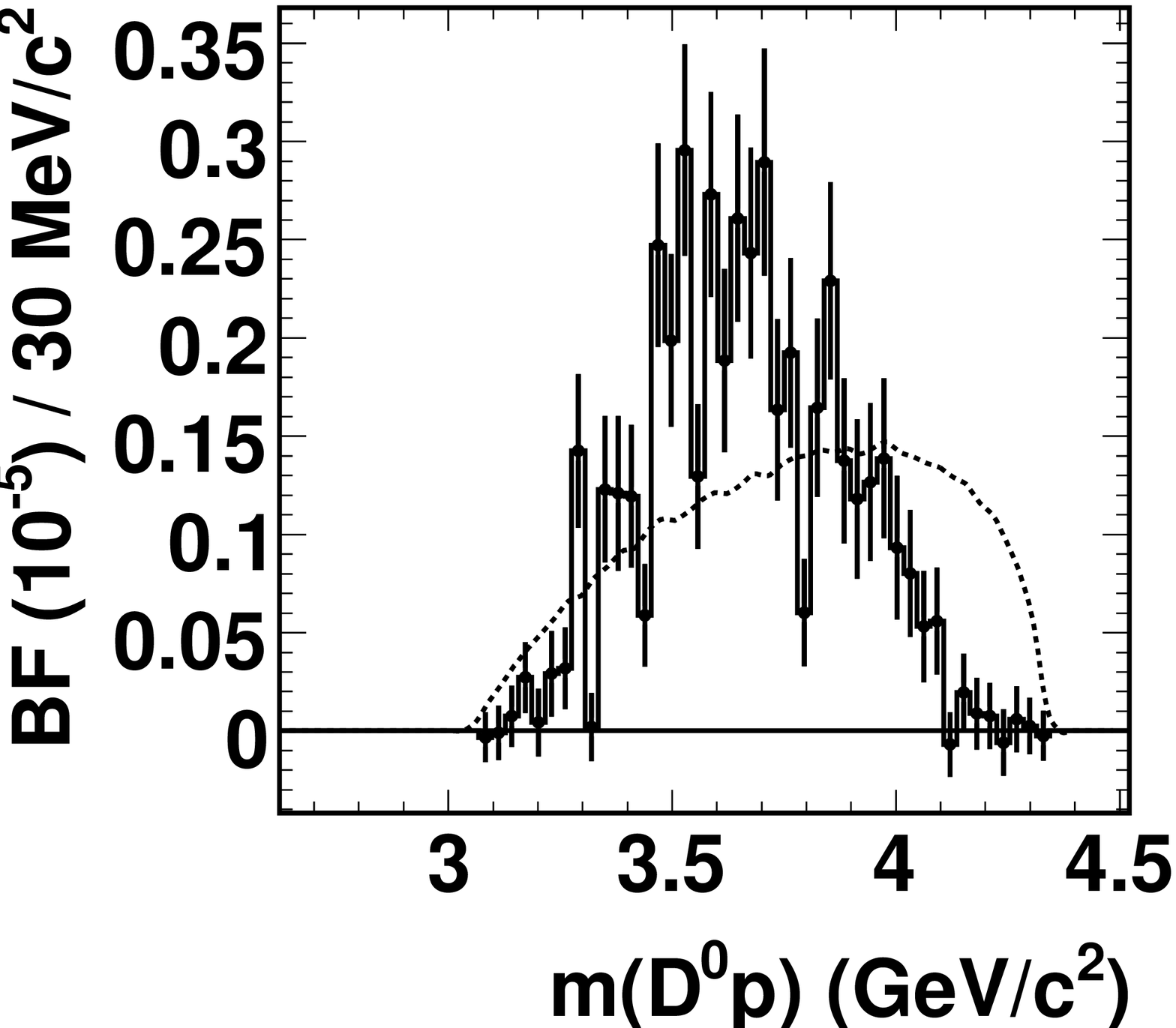}
    \put(28,76){\small{\babar}}
    \put(28,68){\small{prelim.}}
  \end{overpic}
}%
\subfloat[$\Dz\ppbar$, $m^2(\ppbar)\!>\!5$]{
  \hspace{-0.017\textwidth}%
  \begin{overpic}[width=0.25\textwidth]{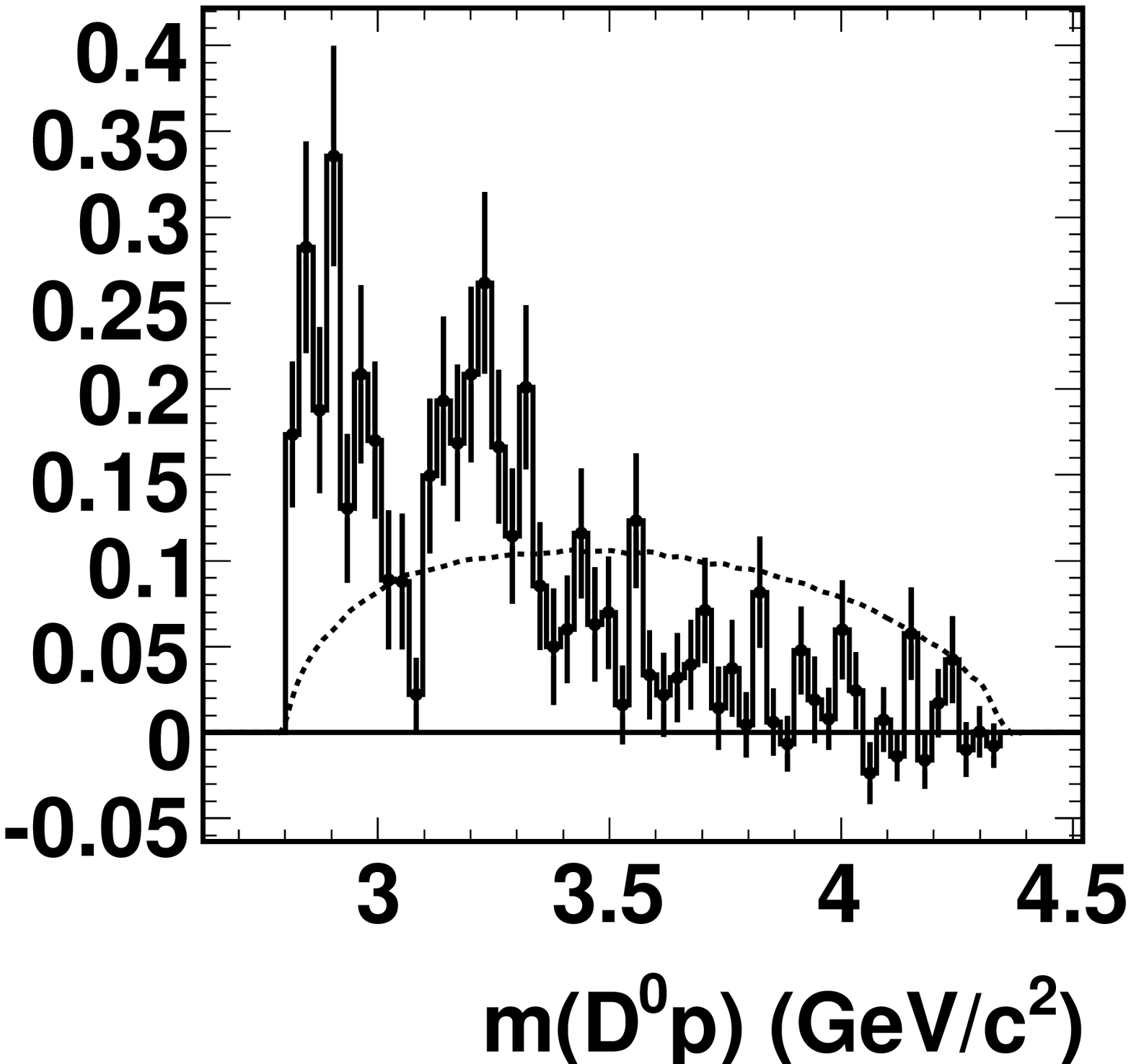}
    \put(66,76){\small{\babar}}
    \put(66,68){\small{prelim.}}
  \end{overpic}
}%
\subfloat[$\Dz\ppbar$, $m^2(\Dz{p})\!<\!9$]{
  \hspace{-0.017\textwidth}%
  \begin{overpic}[width=0.25\textwidth]{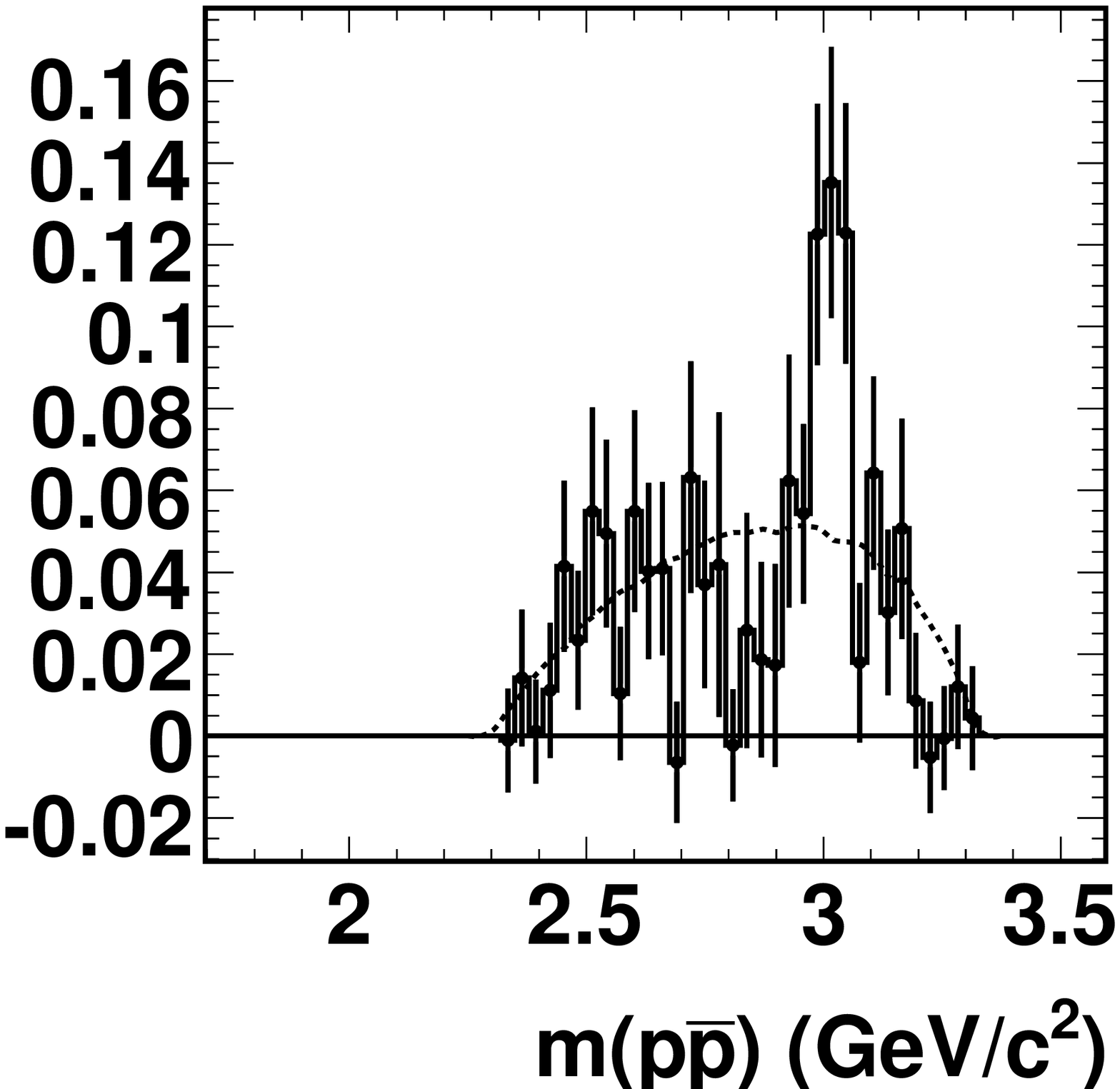}
    \put(28,76){\small{\babar}}
    \put(28,68){\small{prelim.}}
  \end{overpic}
}%
\subfloat[$\Dz\ppbar$, $m^2(\Dz{p})\!>\!9$]{
  \hspace{-0.017\textwidth}%
  \begin{overpic}[width=0.25\textwidth]{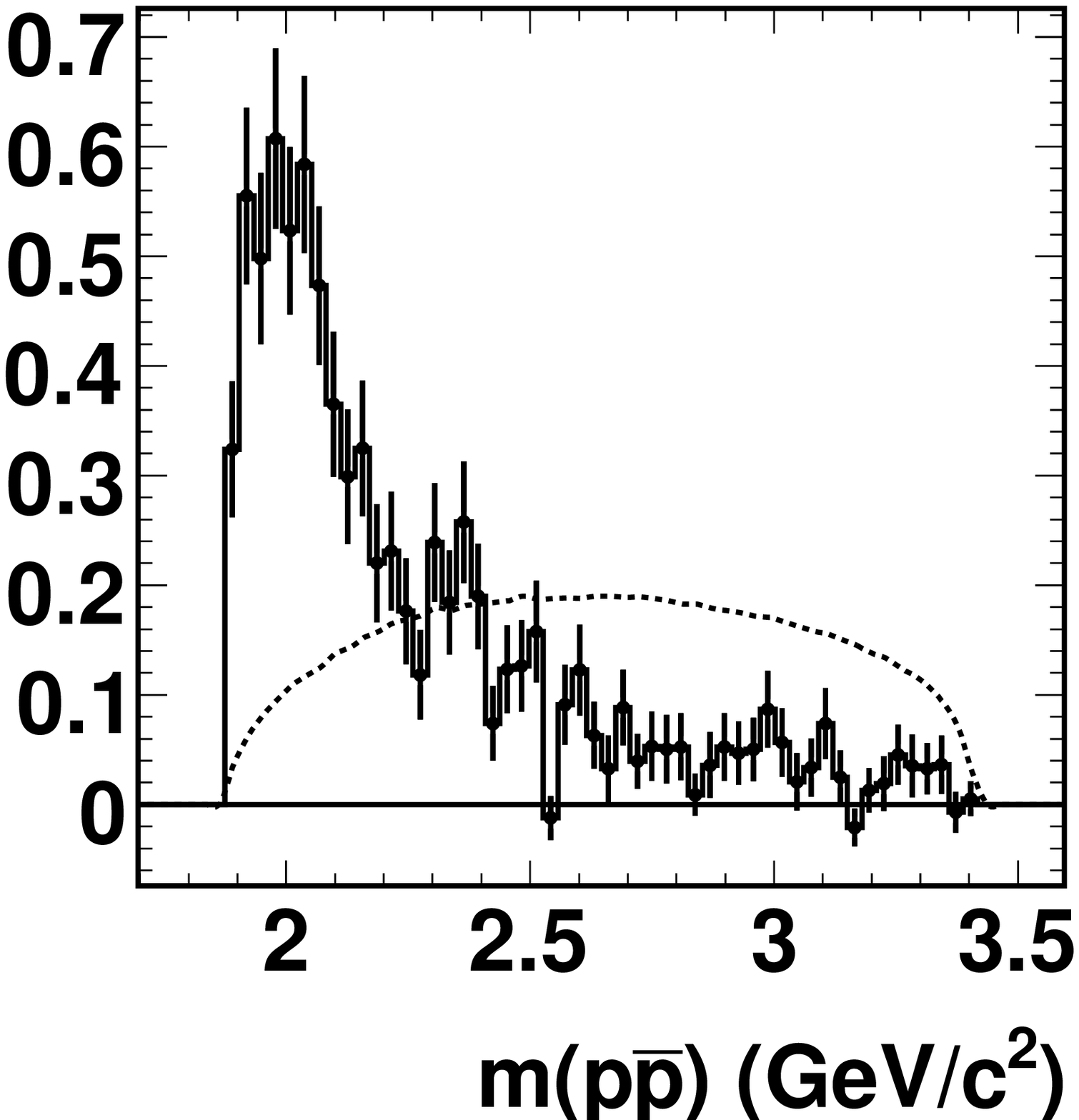}
    \put(66,76){\small{\babar}}
    \put(66,68){\small{prelim.}}
  \end{overpic}
}%
\\
\subfloat[$\Dstarz\ppbar$,\,$m^2(\ppbar)\!<\!5$]{
  \begin{overpic}[width=0.25\textwidth]{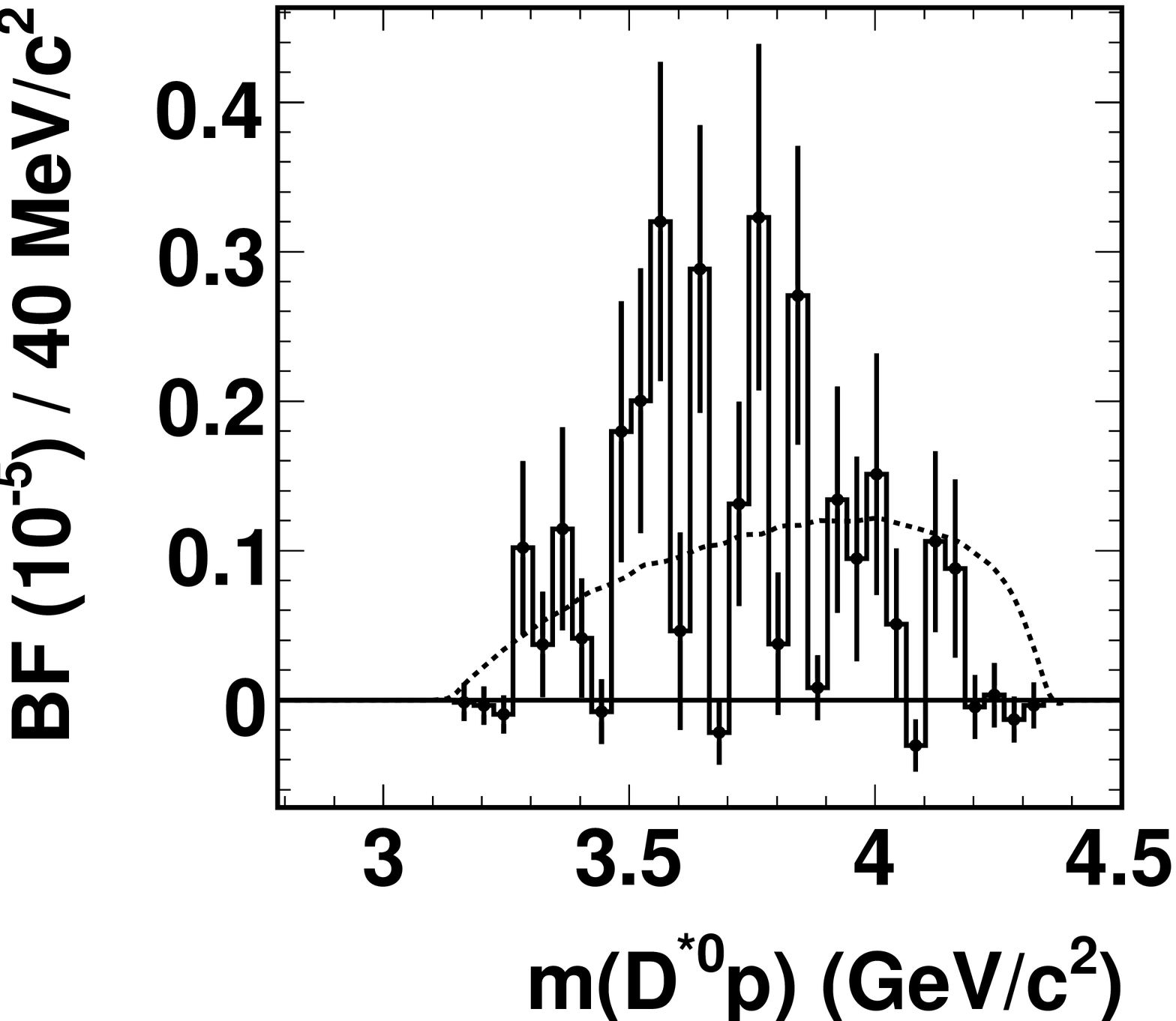}
    \put(28,76){\small{\babar}}
    \put(28,68){\small{prelim.}}
  \end{overpic}
}%
\subfloat[$\Dstarz\ppbar$,\,$m^2(\ppbar)\!>\!5$]{
  \hspace{-0.017\textwidth}%
  \begin{overpic}[width=0.25\textwidth]{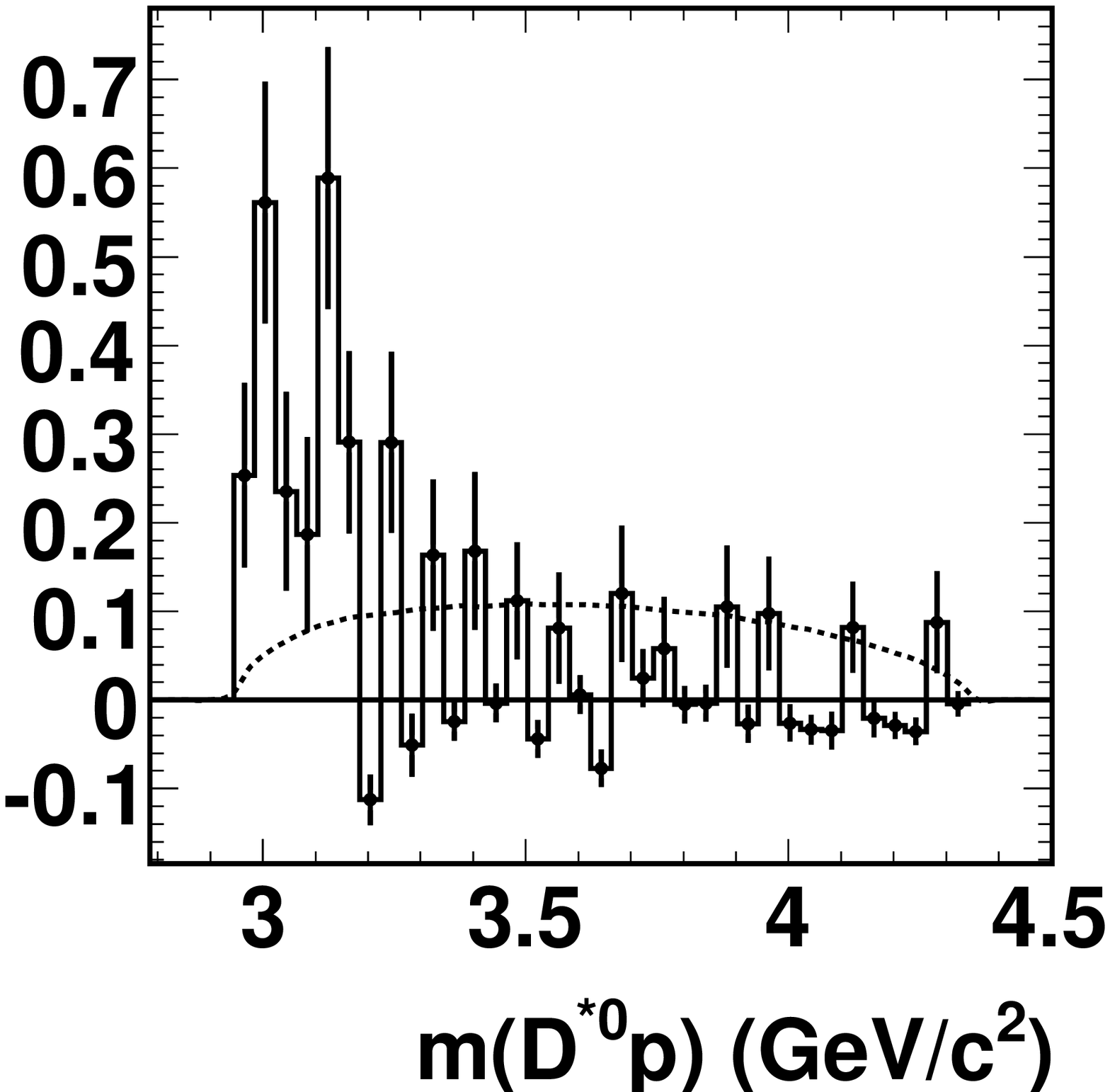}
    \put(66,76){\small{\babar}}
    \put(66,68){\small{prelim.}}
  \end{overpic}
}%
\subfloat[$\Dstarz\ppbar$,\,$m^2(\Dstarz{p})\!<\!11.5$]{
  \hspace{-0.017\textwidth}%
  \begin{overpic}[width=0.25\textwidth]{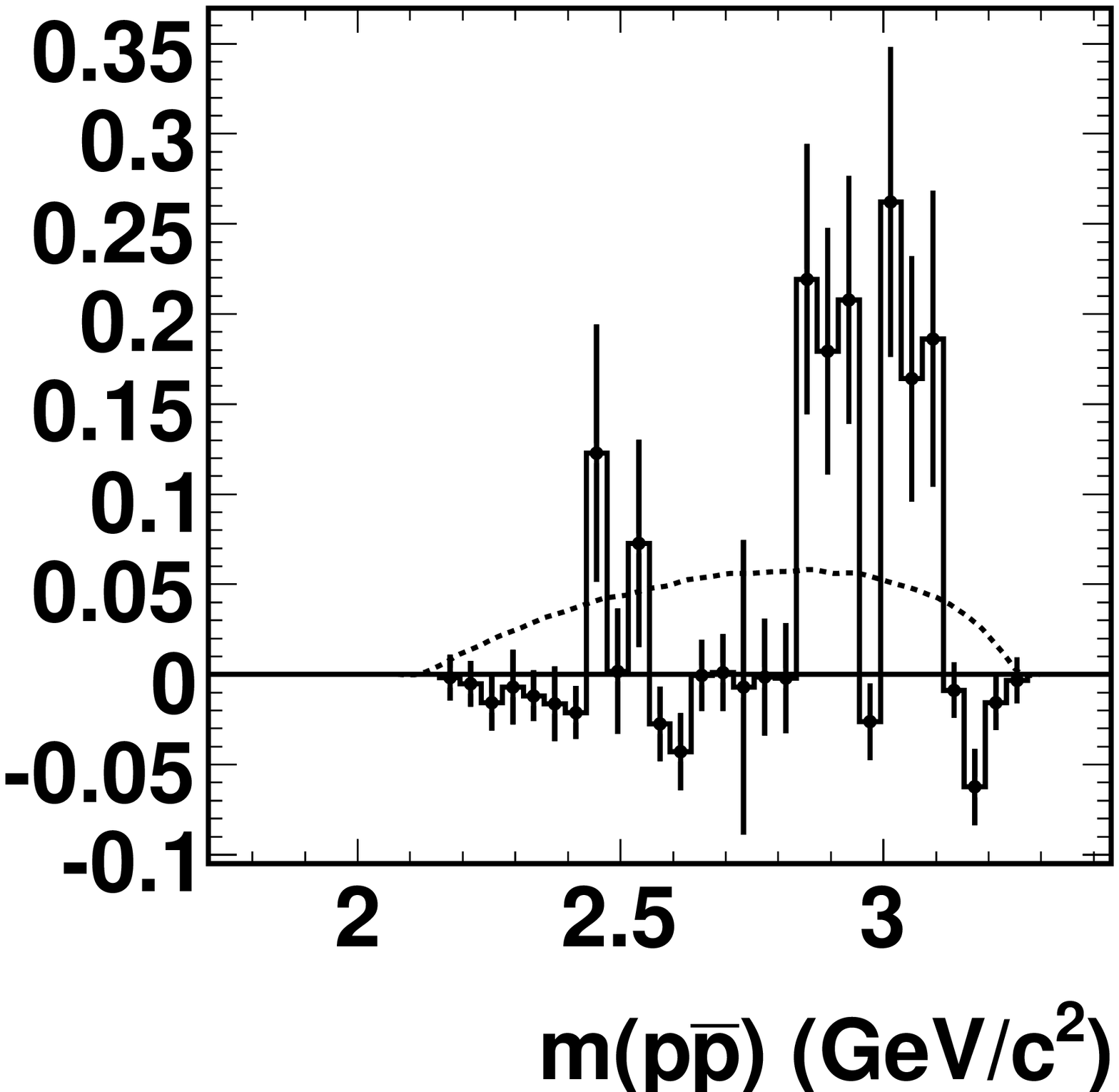}
    \put(28,76){\small{\babar}}
    \put(28,68){\small{prelim.}}
  \end{overpic}
}%
\subfloat[$\Dstarz\ppbar$,\,$m^2(\Dstarz{p})\!>\!11.5$]{
  \hspace{-0.017\textwidth}%
  \begin{overpic}[width=0.25\textwidth]{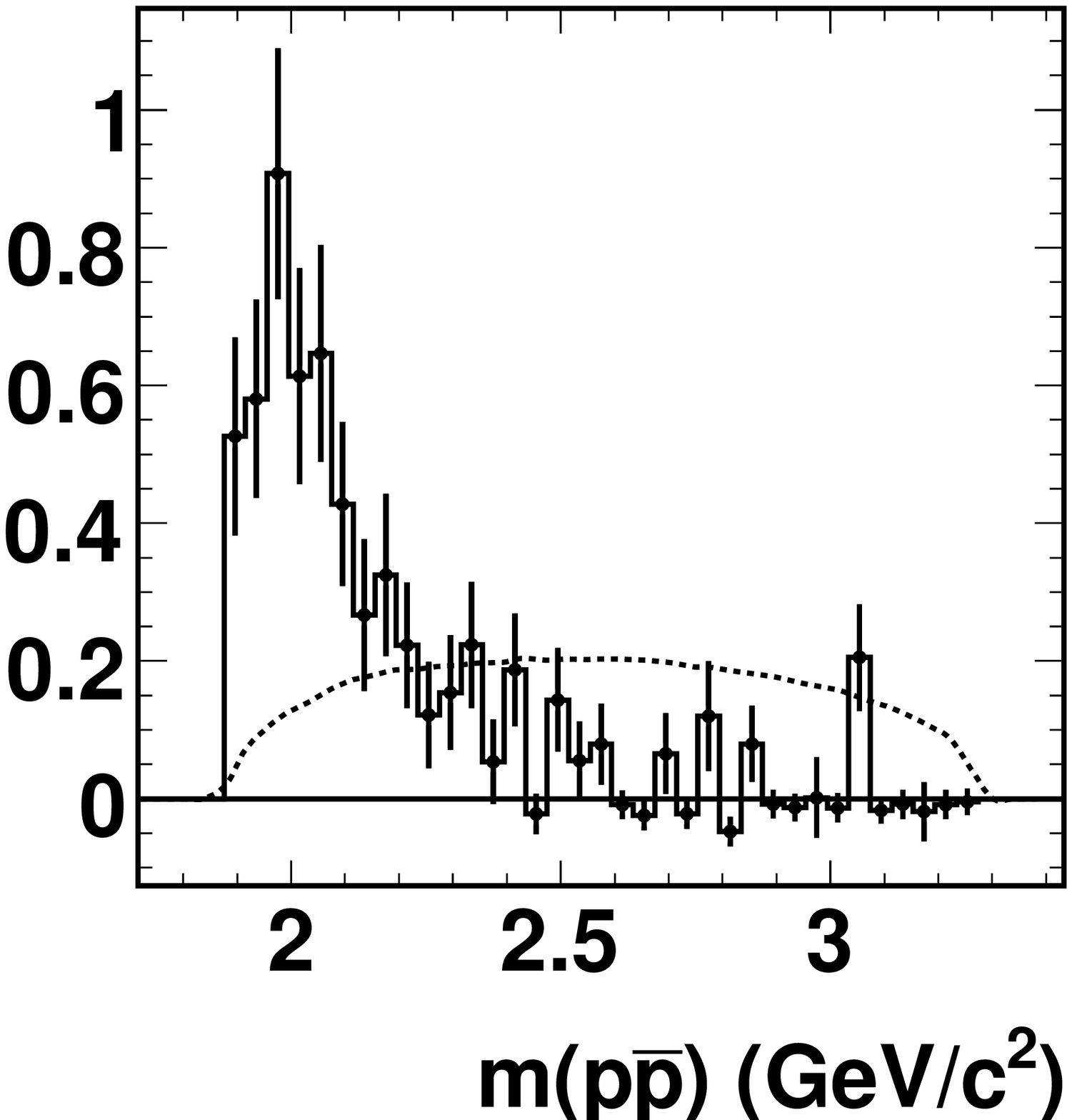}
    \put(66,76){\small{\babar}}
    \put(66,68){\small{prelim.}}
  \end{overpic}
}
\caption{3-body $B$ decay mass projection plots for (abcd)
  $\Bzb\To\Dz\ppbar$ and (efgh) $\Bzb\To\Dstarz\ppbar$.  Fig. (abef)
  gives plots of $m(\Dmaybestar{p})$ in the regions to the left and
  right of the vertical red line in Fig.~\ref{fig:dalitz}ab; Fig.
  (cdgh) plots of $m(\ppbar)$ above and below the horizontal red line.
  The curve represents decays following the uniform phase space
  model.  All requirements are given in \gevccsq.
}
\label{fig:3body_mass}
\end{figure}

Figure~\ref{fig:3body_mass} shows the differential branching fractions
for $\ppbar$ and $\Dmaybestarz{p}$ in different regions of the
complementary variable, which are found by substituting $W_i$ for
$n_S-n_b$ in Eqn.~\ref{eqn:bf} and computing $\epsilon$ for the
appropriate region binned in the two variables.  In general, the
distributions for $\Dz\ppbar$ and $\Dstarz\ppbar$ are similar.  One
exception is the comparison of $\Dmaybestar{p}$ in the region
$m^2(\ppbar)>5\gevccsq$ shown in Fig.~\ref{fig:3body_mass}b and f,
where two broad peaks around $100\mevcc$ wide are seen for $\Dz\ppbar$
at $2.9$ and $3.2$ \gevcc, but only one is seen in $\Dstarz\ppbar$
near threshold.

Lastly, we consider the possibility that the $\Dz{p}$ threshold effect
in $\Bzb\To\Dz\ppbar$ may be due to one or both of two $\Lambda_c$
states observed \cite{Aubert:2006sp} at $2.94$ and $2.88$ \gevcc\ with
a full width of $18$ and $6$ \mevcc, respectively.  These states would
span $1$--$2$ bins in Fig.~\ref{fig:3body_mass}b and are not
sufficient to describe the $150$--$200\mevcc$ wide enhancement.

\subsection{\boldmath Four-body decays $B\To\Dmaybestar\ppbar\pi$}

\begin{figure}[bp!]
\centering
\subfloat[$\Dp\ppbar\pi$, $m(\ppbar)$]{
  \begin{overpic}[width=0.25\textwidth]{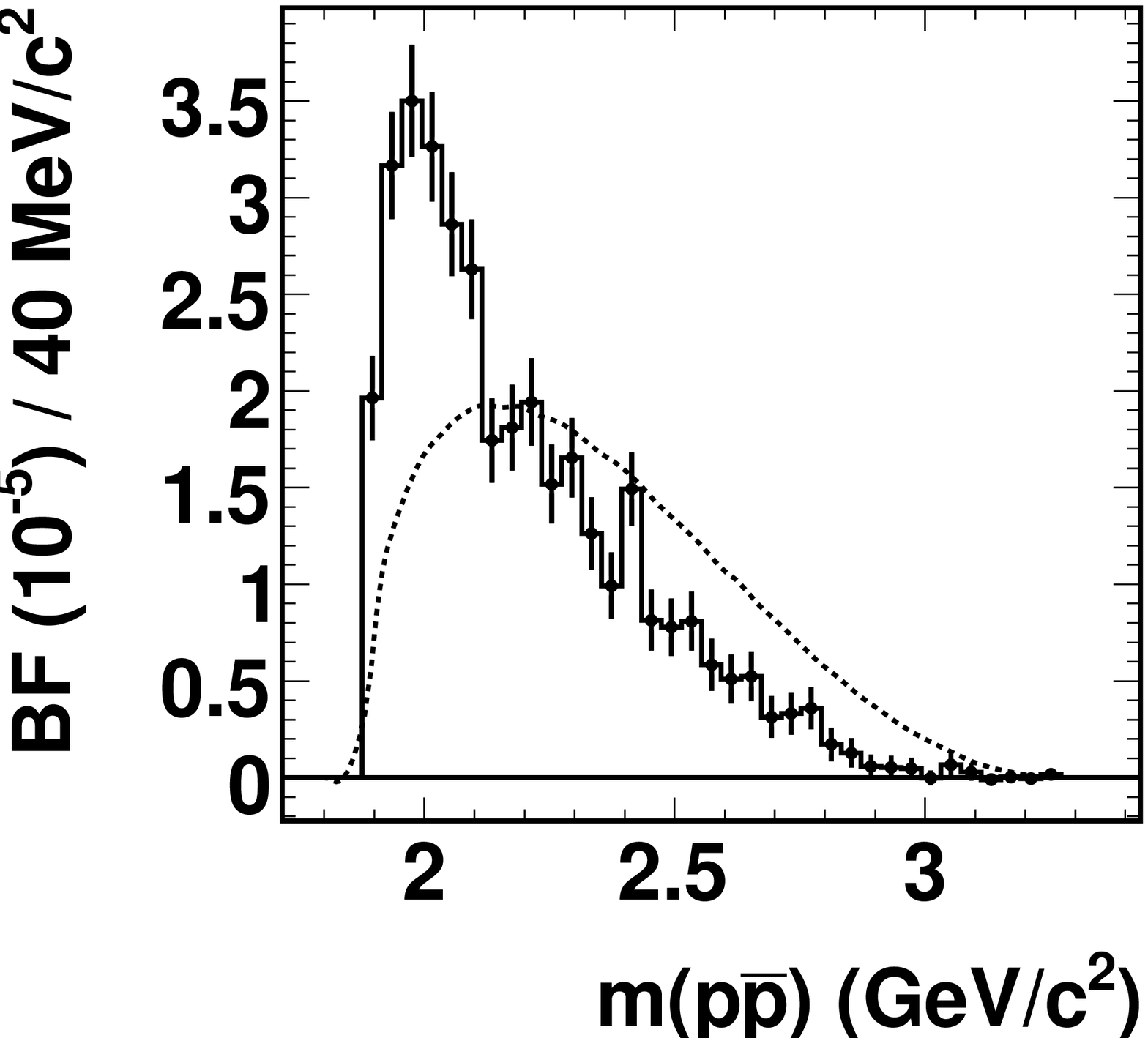}
    \put(66,74){\small{\babar}}
    \put(66,66){\small{prelim.}}
  \end{overpic}
}%
\subfloat[$\Dp\ppbar\pi$, $m(\Dp\pbar)$]{
  \hspace{-0.017\textwidth}%
  \begin{overpic}[width=0.25\textwidth]{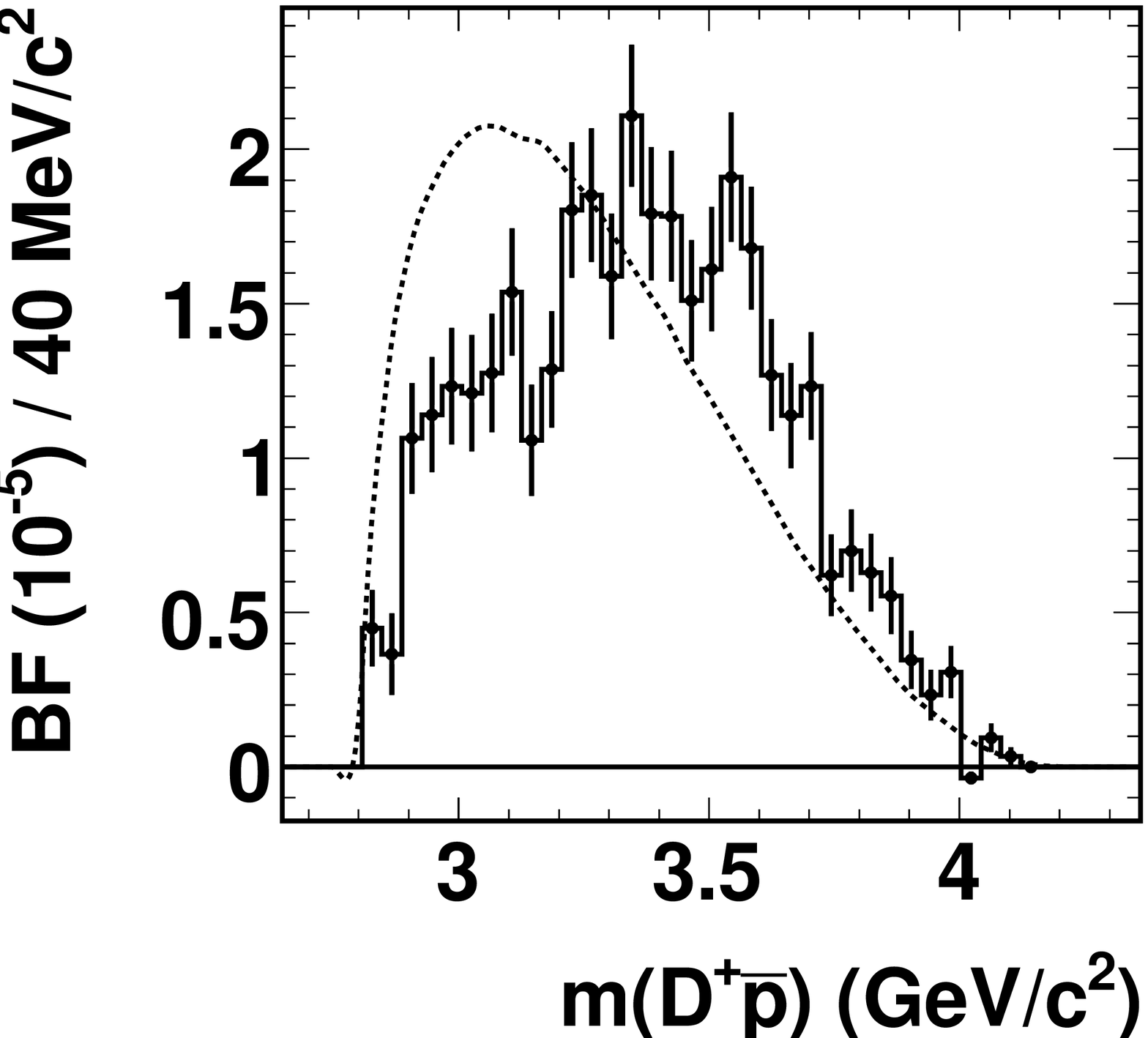}
    \put(66,74){\small{\babar}}
    \put(66,66){\small{prelim.}}
  \end{overpic}
}%
\subfloat[$\Dp\ppbar\pi$, $m(\Dp{p})$]{
  \hspace{-0.017\textwidth}%
  \begin{overpic}[width=0.25\textwidth]{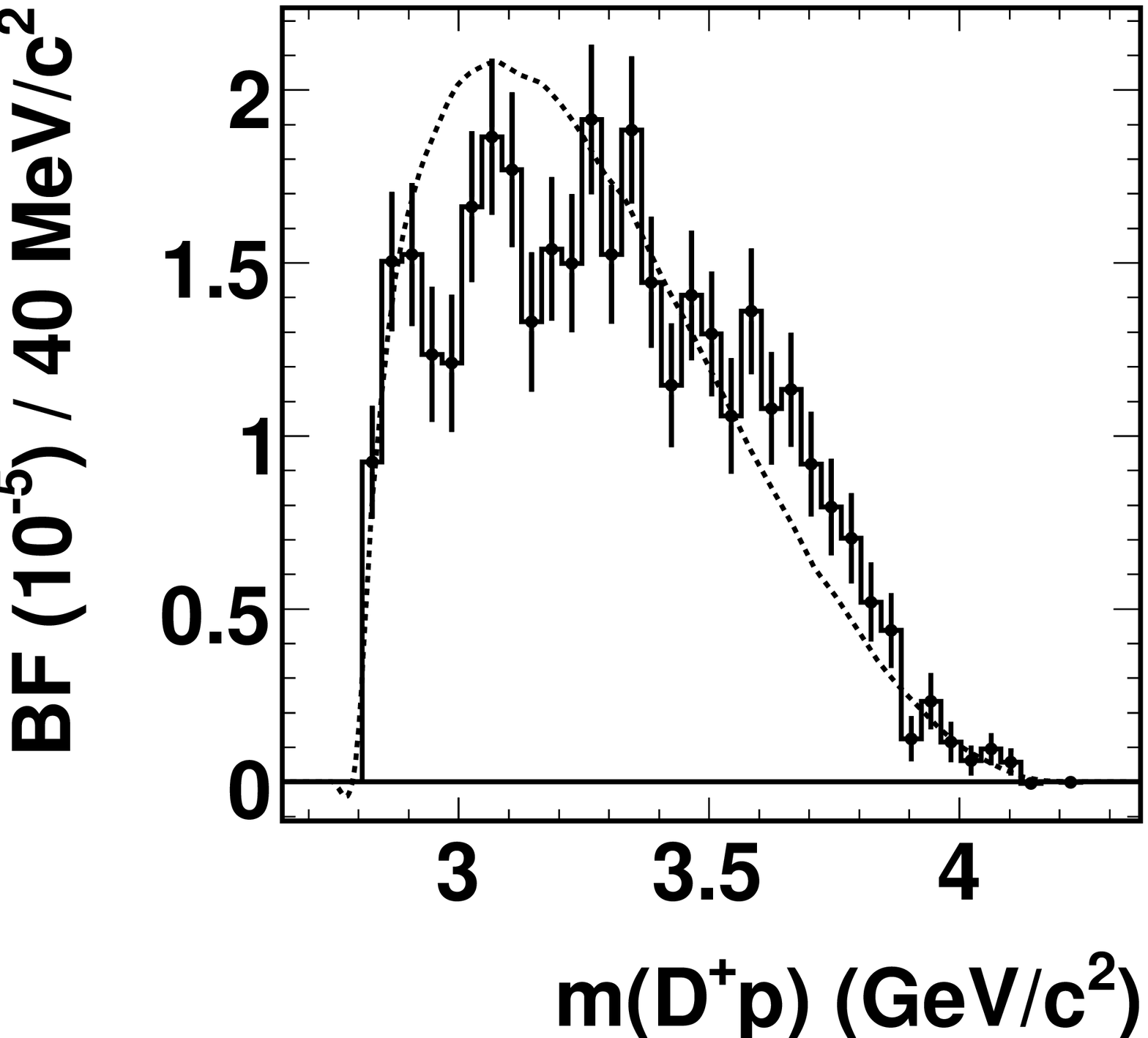}
    \put(66,74){\small{\babar}}
    \put(66,66){\small{prelim.}}
  \end{overpic}
}%
\subfloat[$\Dp\ppbar\pi$, $m(p\pim)$]{
  \hspace{-0.017\textwidth}%
  \begin{overpic}[width=0.25\textwidth]{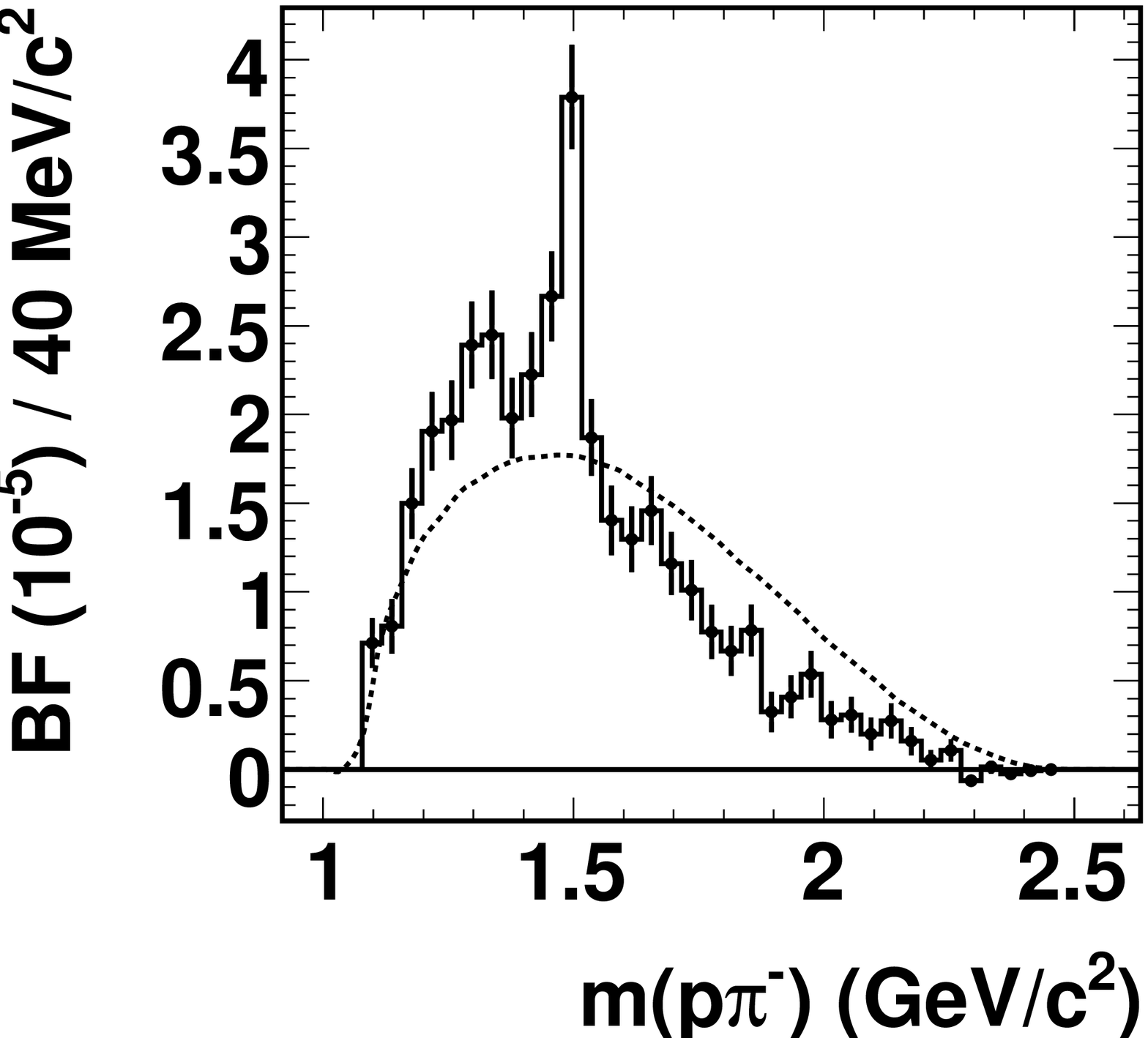}
    \put(66,74){\small{\babar}}
    \put(66,66){\small{prelim.}}
  \end{overpic}
}%
\\
\subfloat[$\Dstarp\ppbar\pi$, $m(\ppbar)$]{
  \begin{overpic}[width=0.25\textwidth]{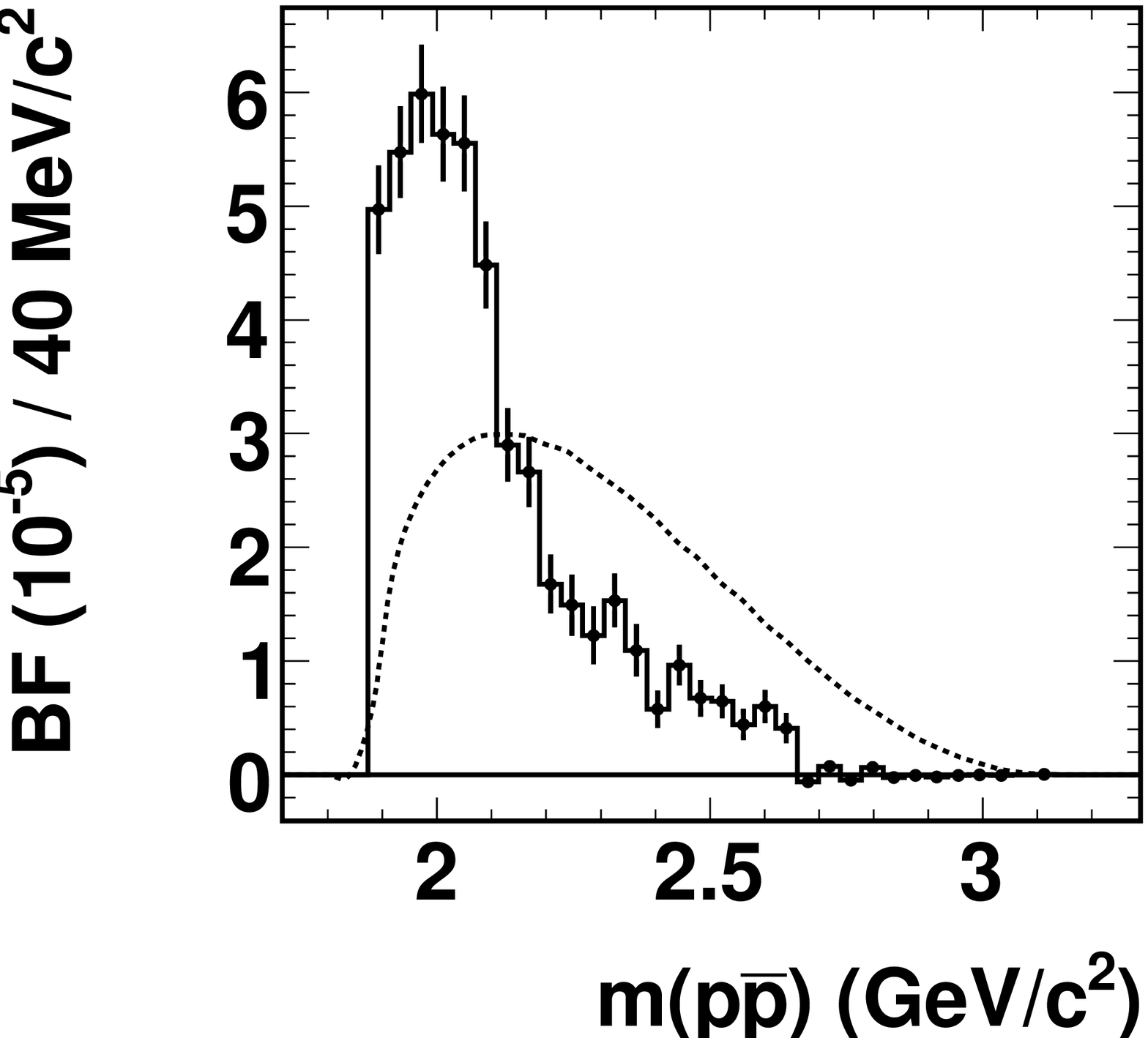}
    \put(66,74){\small{\babar}}
    \put(66,66){\small{prelim.}}
  \end{overpic}
}%
\subfloat[$\Dstarp\ppbar\pi$, $m(\Dp\pbar)$]{
  \hspace{-0.017\textwidth}%
  \begin{overpic}[width=0.25\textwidth]{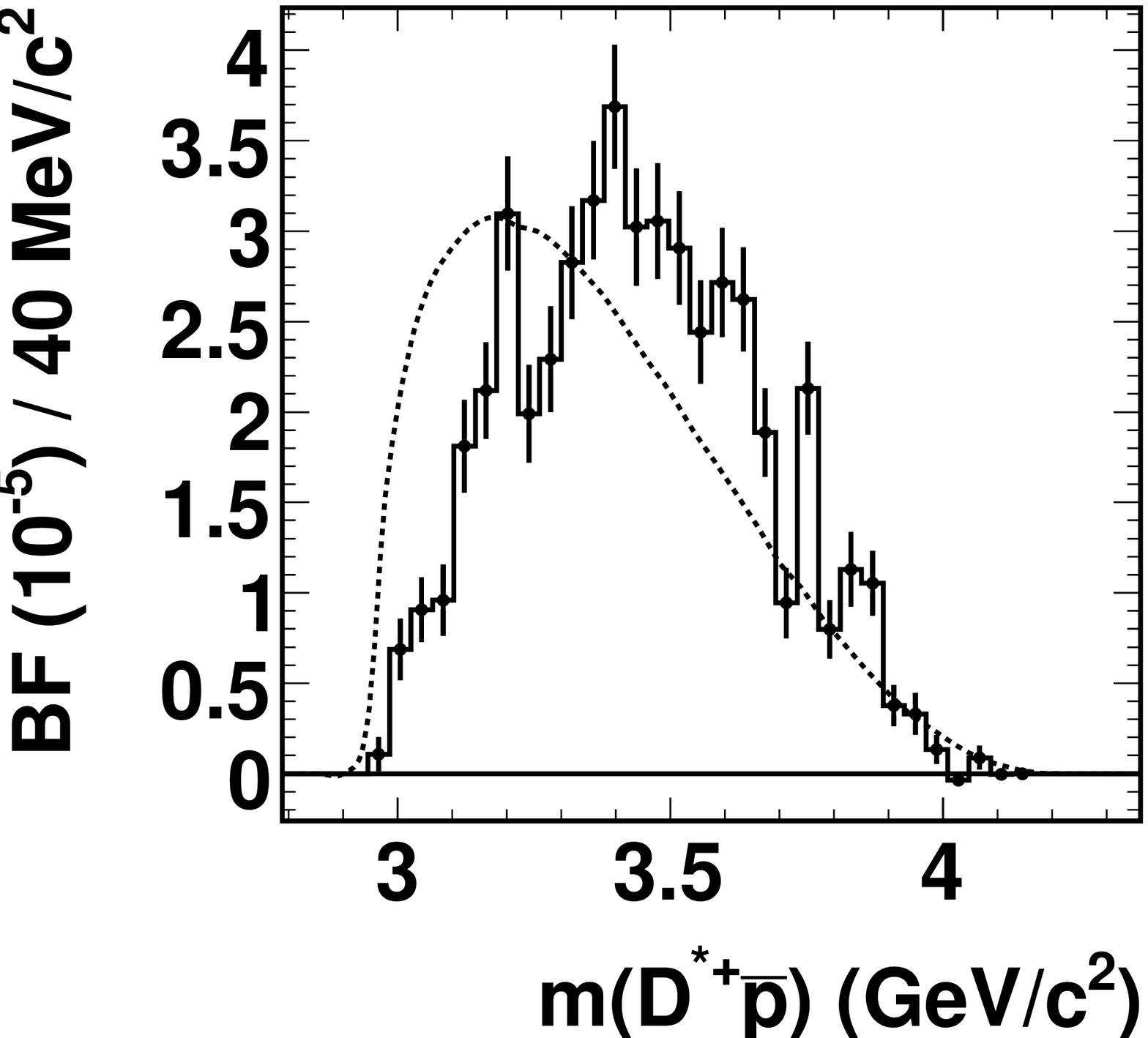}
    \put(66,74){\small{\babar}}
    \put(66,66){\small{prelim.}}
  \end{overpic}
}%
\subfloat[$\Dstarp\ppbar\pi$, $m(\Dp{p})$]{
  \hspace{-0.017\textwidth}%
  \begin{overpic}[width=0.25\textwidth]{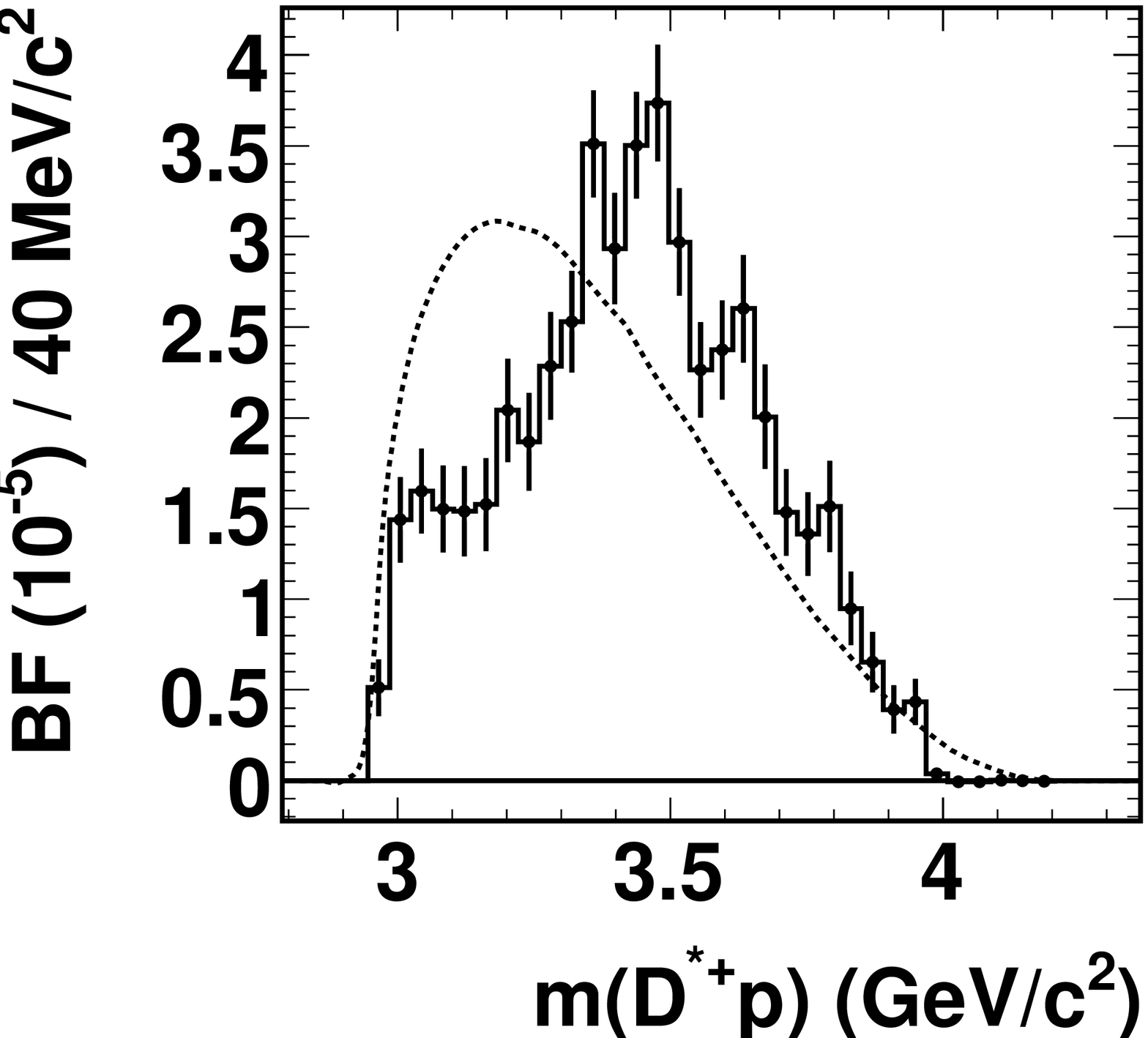}
    \put(66,74){\small{\babar}}
    \put(66,66){\small{prelim.}}
  \end{overpic}
}%
\subfloat[$\Dstarp\ppbar\pi$, $m(p\pim)$]{
  \hspace{-0.017\textwidth}%
  \begin{overpic}[width=0.25\textwidth]{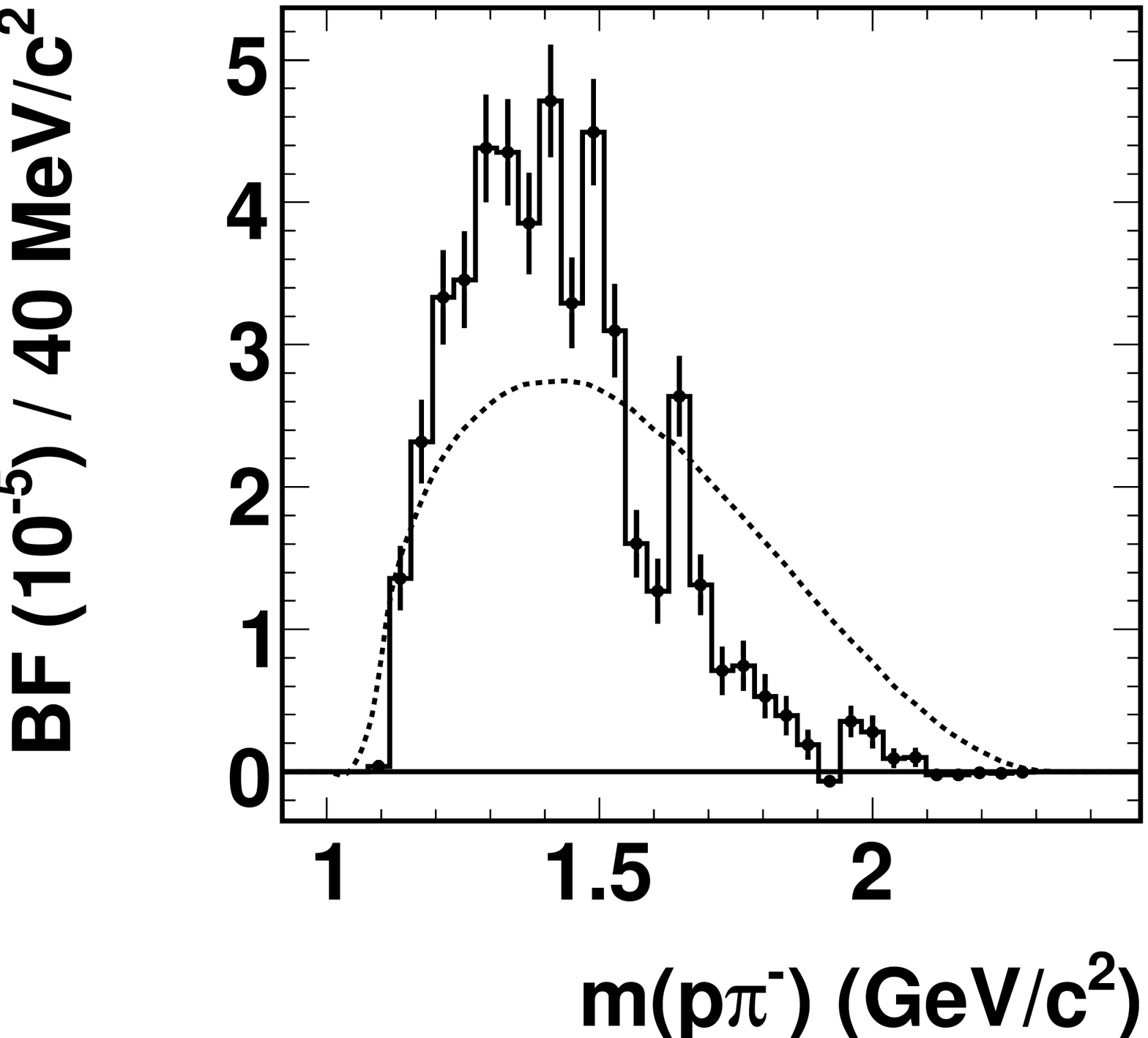}
    \put(66,74){\small{\babar}}
    \put(66,66){\small{prelim.}}
  \end{overpic}
}%
\\
\subfloat[$\Dz\ppbar\pi$, $m(\ppbar)$]{
  \begin{overpic}[width=0.25\textwidth]{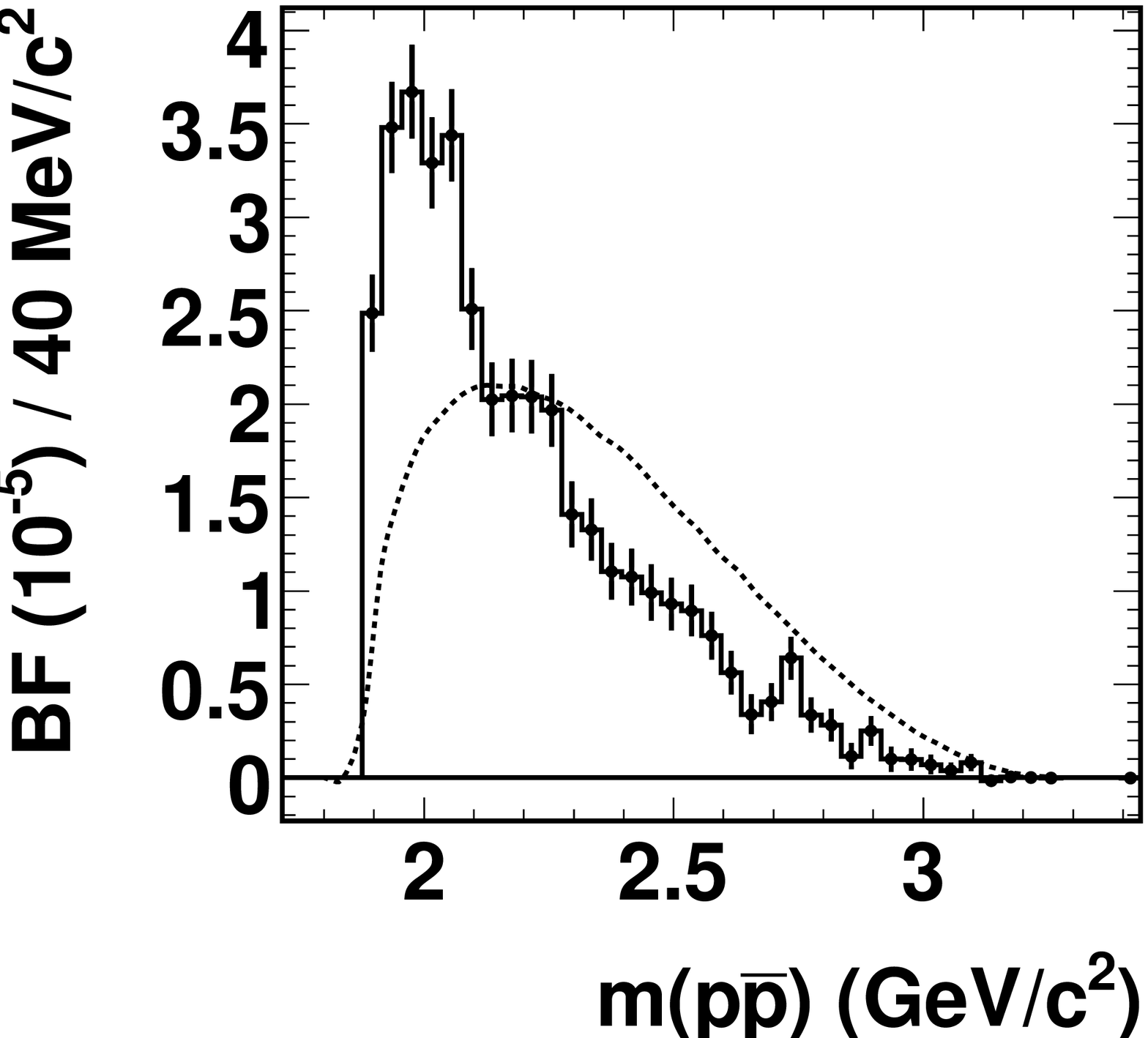}
    \put(66,74){\small{\babar}}
    \put(66,66){\small{prelim.}}
  \end{overpic}
}%
\subfloat[$\Dz\ppbar\pi$, $m(\Dp\pbar)$]{
  \hspace{-0.017\textwidth}%
  \begin{overpic}[width=0.25\textwidth]{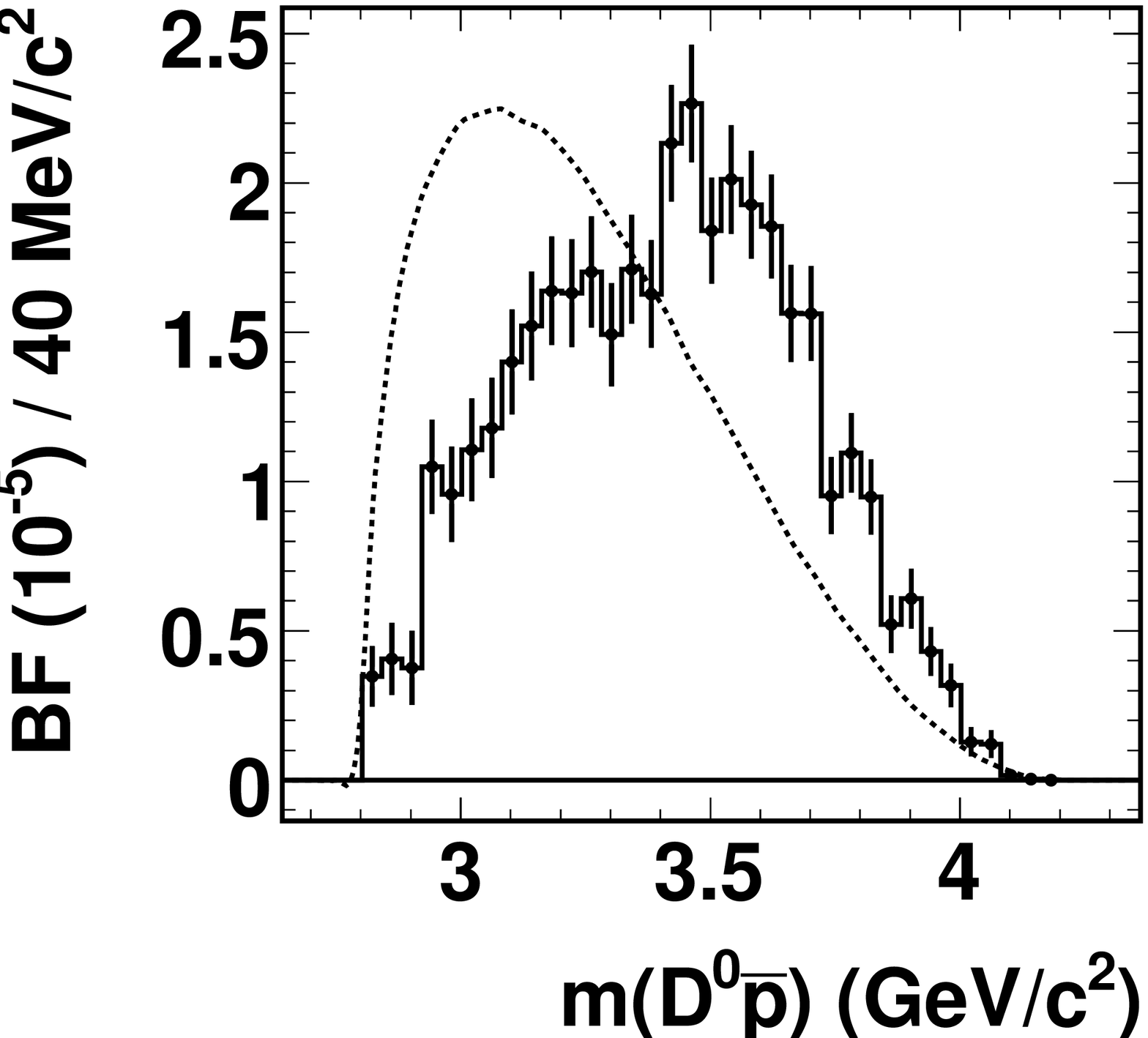}
    \put(66,74){\small{\babar}}
    \put(66,66){\small{prelim.}}
  \end{overpic}
}%
\subfloat[$\Dz\ppbar\pi$, $m(\Dp{p})$]{
  \hspace{-0.017\textwidth}%
  \begin{overpic}[width=0.25\textwidth]{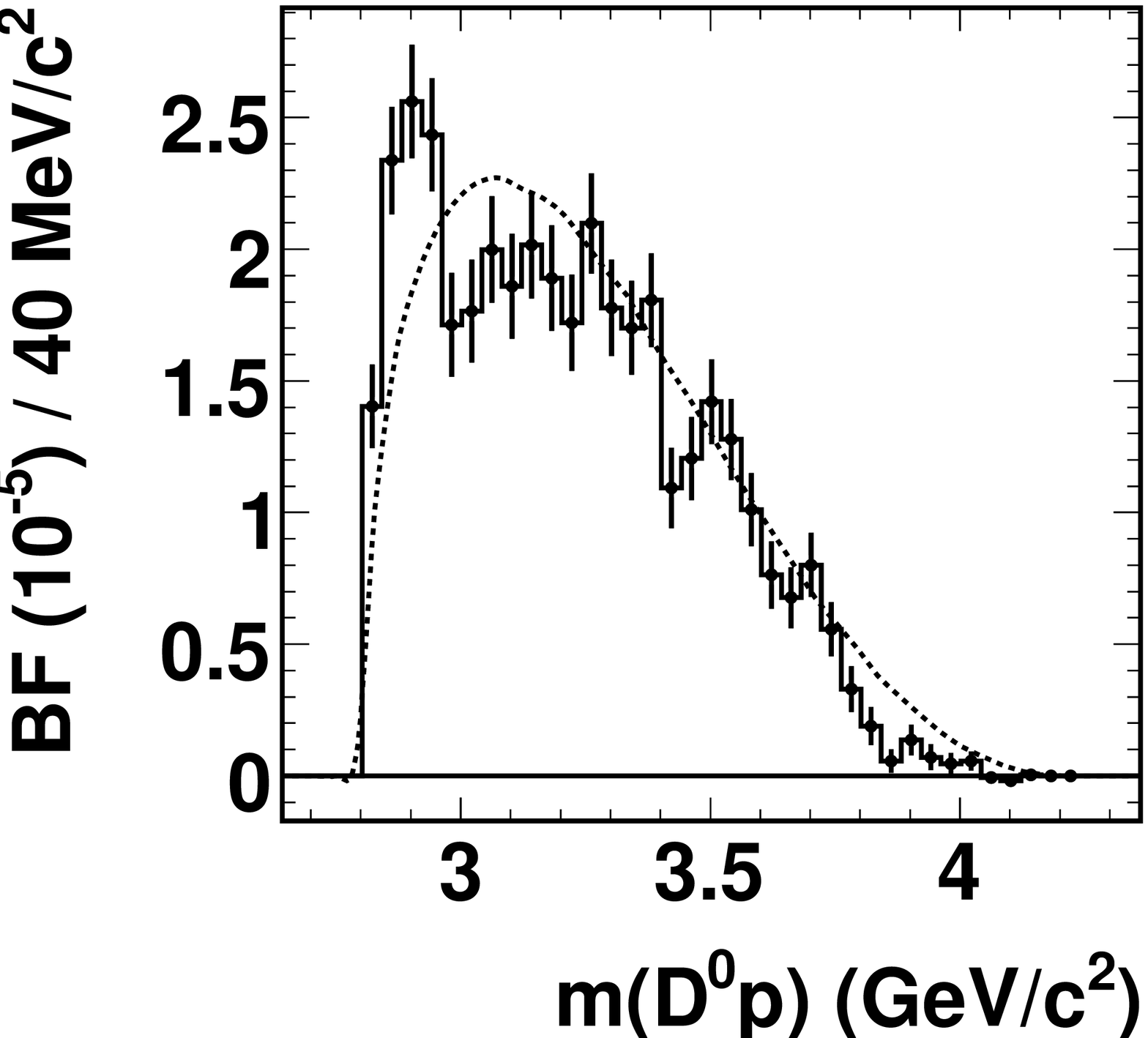}
    \put(66,74){\small{\babar}}
    \put(66,66){\small{prelim.}}
  \end{overpic}
}%
\subfloat[$\Dz\ppbar\pi$, $m(p\pim)$]{
  \hspace{-0.017\textwidth}%
  \begin{overpic}[width=0.25\textwidth]{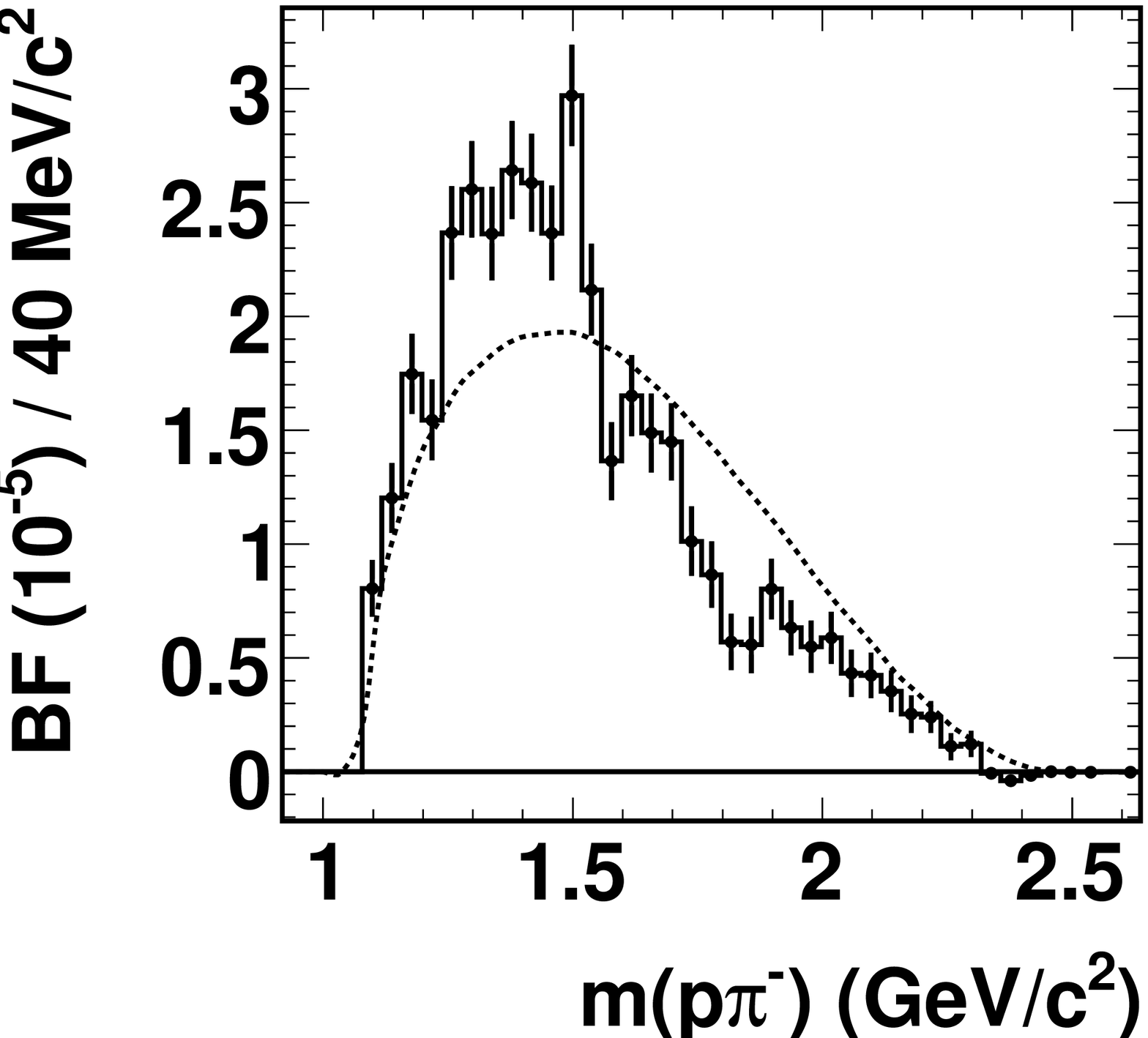}
    \put(66,74){\small{\babar}}
    \put(66,66){\small{prelim.}}
  \end{overpic}
}%
\\
\subfloat[$\Dstarz\ppbar\pi$, $m(\ppbar)$]{
  \begin{overpic}[width=0.25\textwidth]{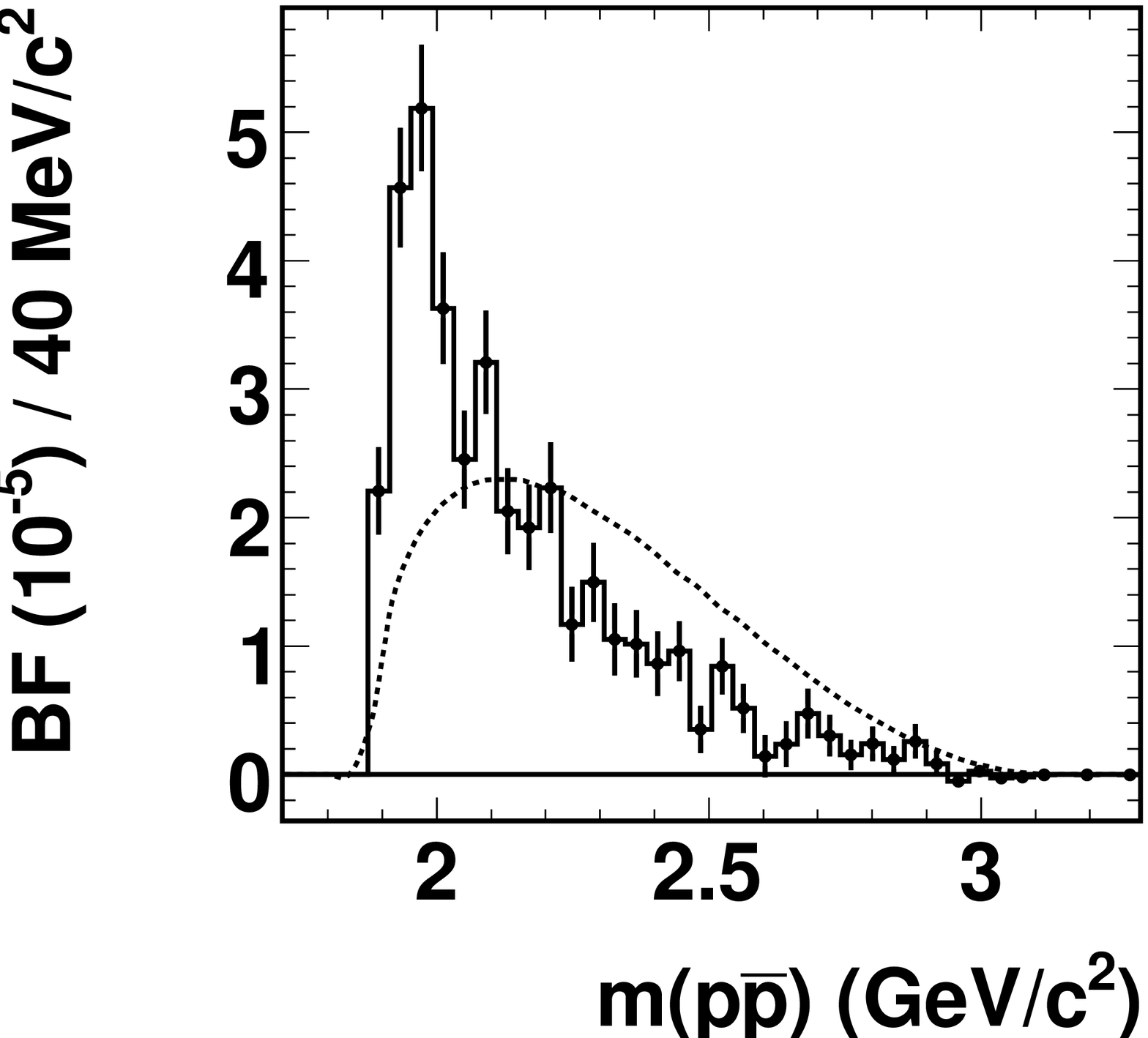}
    \put(66,74){\small{\babar}}
    \put(66,66){\small{prelim.}}
  \end{overpic}
}%
\subfloat[$\Dstarz\ppbar\pi$, $m(\Dp\pbar)$]{
  \hspace{-0.017\textwidth}%
  \begin{overpic}[width=0.25\textwidth]{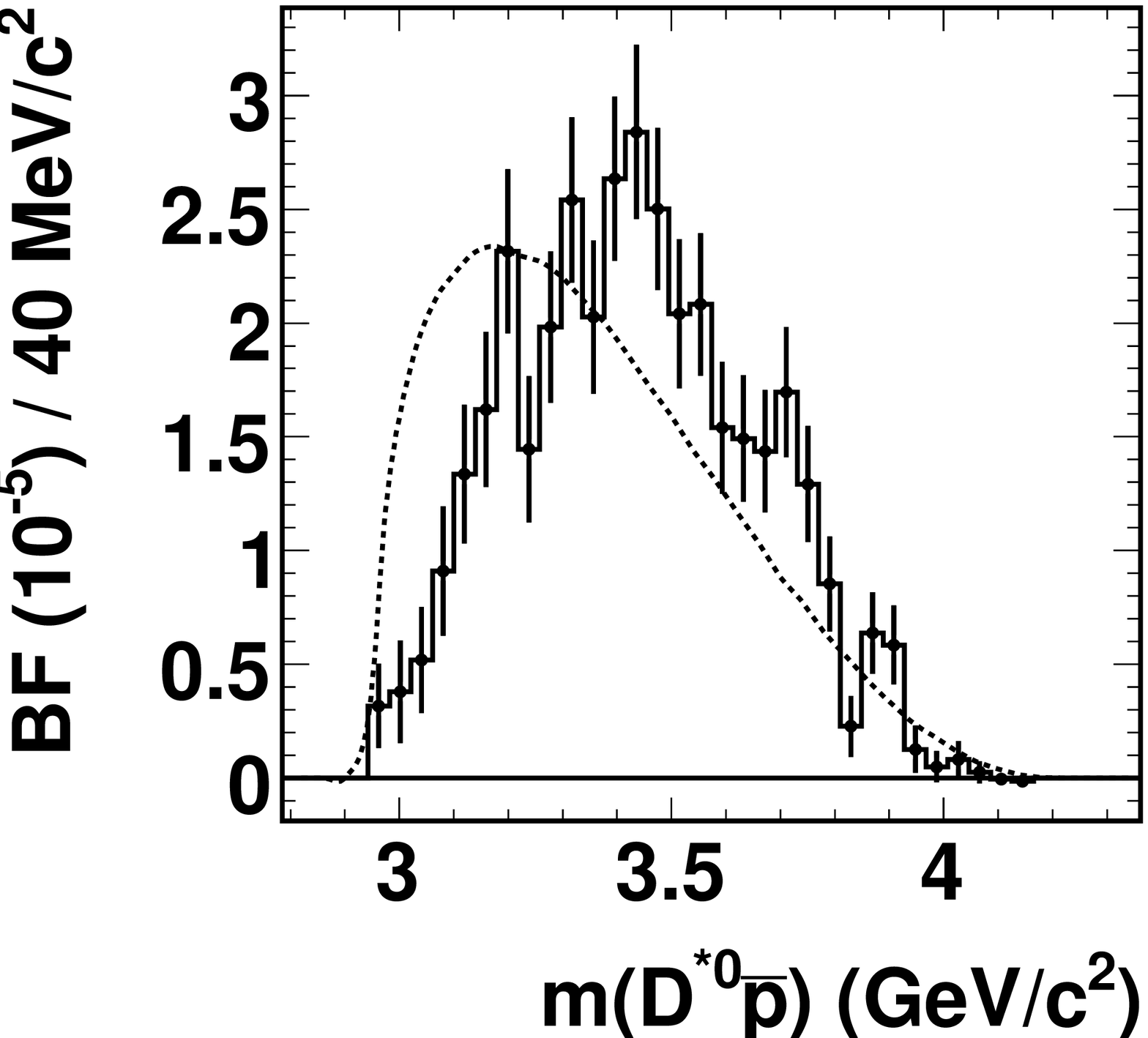}
    \put(66,74){\small{\babar}}
    \put(66,66){\small{prelim.}}
  \end{overpic}
}%
\subfloat[$\Dstarz\ppbar\pi$, $m(\Dp{p})$]{
  \hspace{-0.017\textwidth}%
  \begin{overpic}[width=0.25\textwidth]{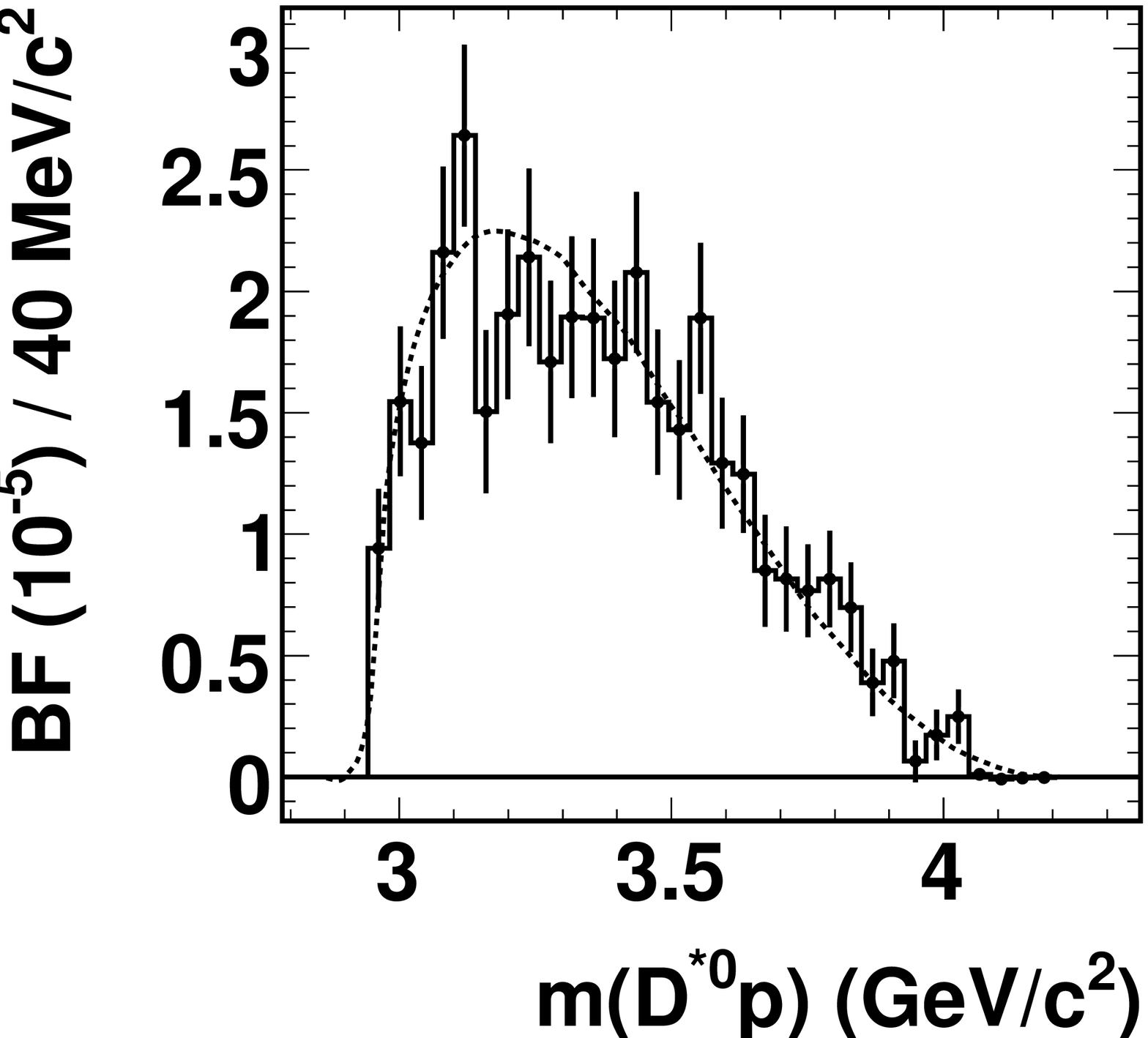}
    \put(66,74){\small{\babar}}
    \put(66,66){\small{prelim.}}
  \end{overpic}
}%
\subfloat[$\Dstarz\ppbar\pi$, $m(p\pim)$]{
  \hspace{-0.017\textwidth}%
  \begin{overpic}[width=0.25\textwidth]{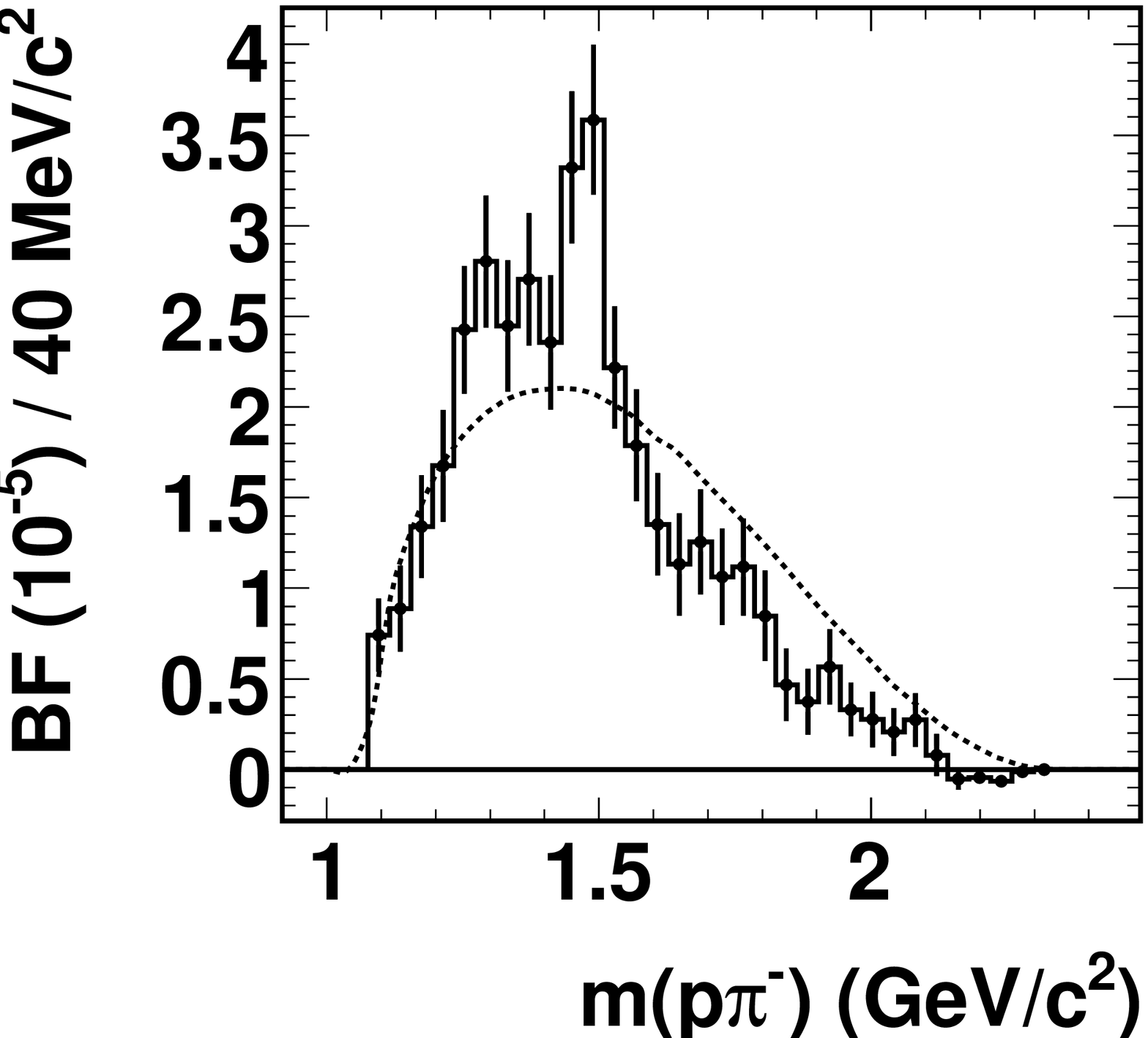}
    \put(66,74){\small{\babar}}
    \put(66,66){\small{prelim.}}
  \end{overpic}
}%
\caption{4-body $B$ decay differential branching fractions
  as functions of 
  $\ppbar$,
  $\Dmaybestar{p}$,
  $\Dmaybestar{\pbar}$, and
  $p\pim$
  in
  (abcd) $\Bzb\To\Dp\ppbar\pim$,
  (efgh) $\Bzb\To\Dstarp\ppbar\pim$,
  (ijkl) $\Bm\To\Dz\ppbar\pim$, and
  (mnop) $\Bm\To\Dstarz\ppbar\pim$, respectively.
  The smooth curve represents decays following the uniform phase space
  model.
}
\label{fig:4body_mass}
\end{figure}

Figure~\ref{fig:4body_mass} shows the differential branching fractions
for four mass variables for 4-body $B$ decays: $\ppbar$,
$\Dmaybestar{p}$, $\Dmaybestar{\pbar}$, and $p\pim$.  In $\ppbar$,
Figs.~\ref{fig:4body_mass}aeim,  we see an enhancement production near
threshold compared to the phase space expectations.  In
$\Dmaybestar{\pbar}$, Figs.~\ref{fig:4body_mass}bfjn, we see no obvious
hints of a narrow exotic 5-quark state around $3.1\gevcc$
\cite{Aktas:2004qf}.  In $\Dmaybestar{p}$,
Figs.~\ref{fig:4body_mass}egko, the broad threshold structure seen
3-body decays seems to be absent, except perhaps in
Fig.~\ref{fig:4body_mass}k for $\Bm\To\Dz\ppbar\pim$.  In $p\pim$,
Figs.~\ref{fig:4body_mass}dhlp, we observe a narrow structure near
$1.5\gevcc$, especially in Fig.~\ref{fig:4body_mass}d for
$\Bzb\To\Dp\ppbar\pim$, which refer to as $X$ in subsequent
discussions.

\begin{figure}[bp!]
\centering
\subfloat[$m(p\pim)$ for $\Bzb\To\Dp\ppbar\pim$ with pdfs from the
	top: Breit-Wigner for $X$ near $1.5\gevcc$ (top, blue);
	background pdf from (b) (bottom, red).]{
  \hspace{-0.017\textwidth}%
  \begin{overpic}[width=0.50\textwidth]{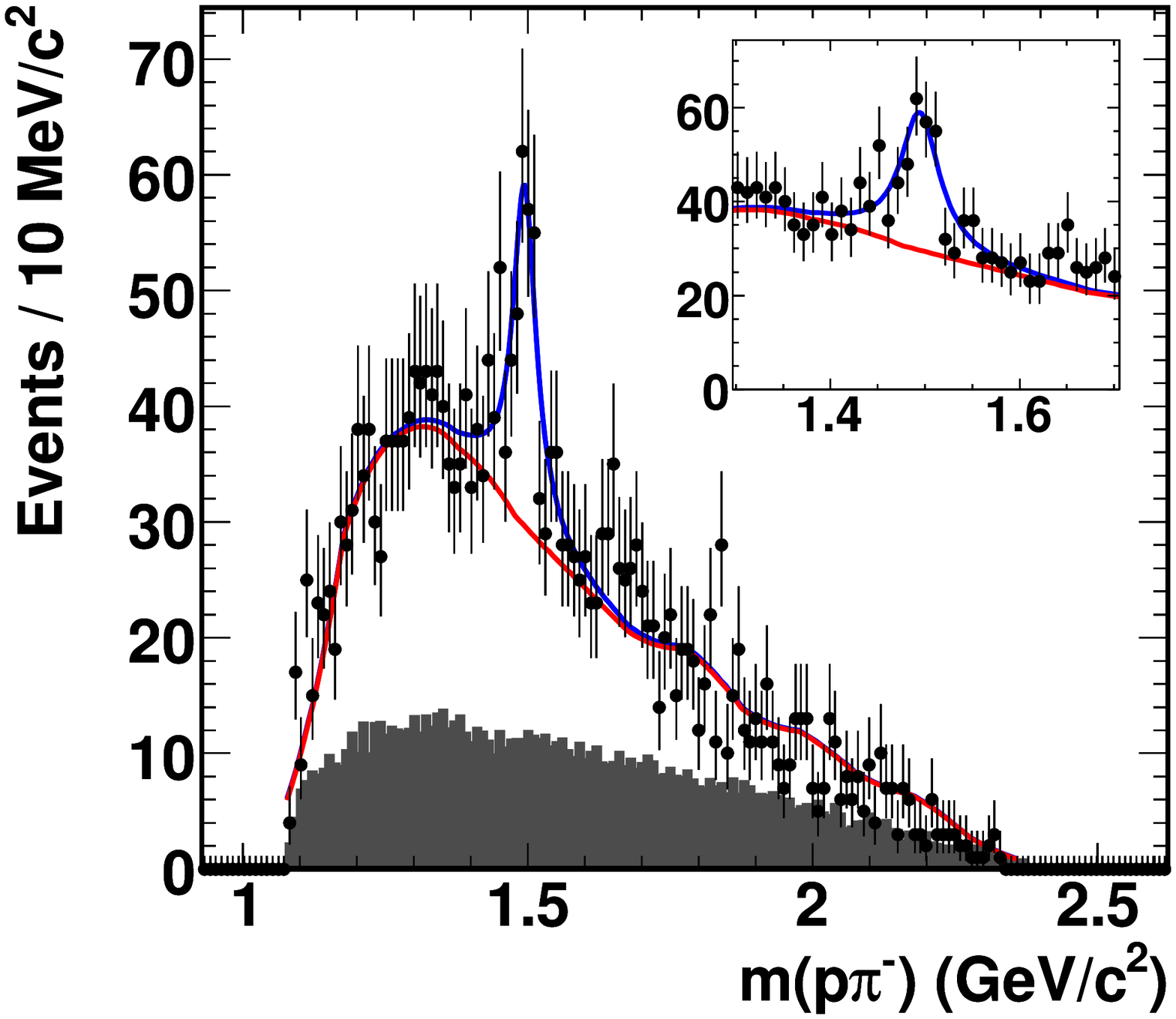}
    \put(76,43){{\babar}}
    \put(76,35){{prelim.}}
  \end{overpic}
}%
\subfloat[$m(\pbar\pim)$ for $\Bzb\To\Dp\ppbar\pim$ with the smoothed
	histogram pdf used as background in (a).]{
  \begin{overpic}[width=0.50\textwidth]{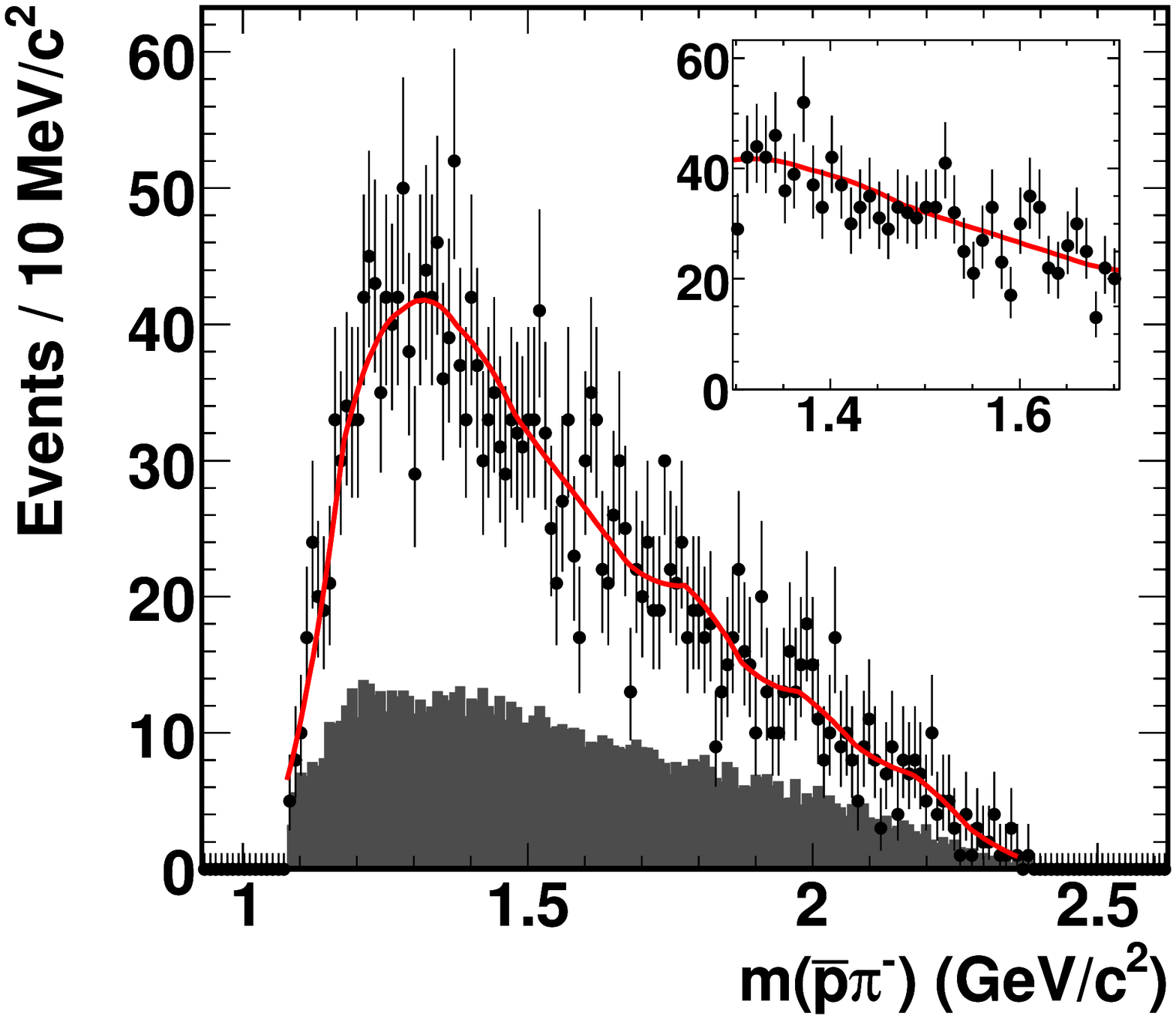}
    \put(76,43){{\babar}}
    \put(76,35){{prelim.}}
  \end{overpic}
}%
\\
\subfloat[$m(p\pim)$ for $\Bzb\To\Dstarp\ppbar\pim$;
	see (a) for descriptions.]{
  \hspace{-0.017\textwidth}%
  \begin{overpic}[width=0.50\textwidth]{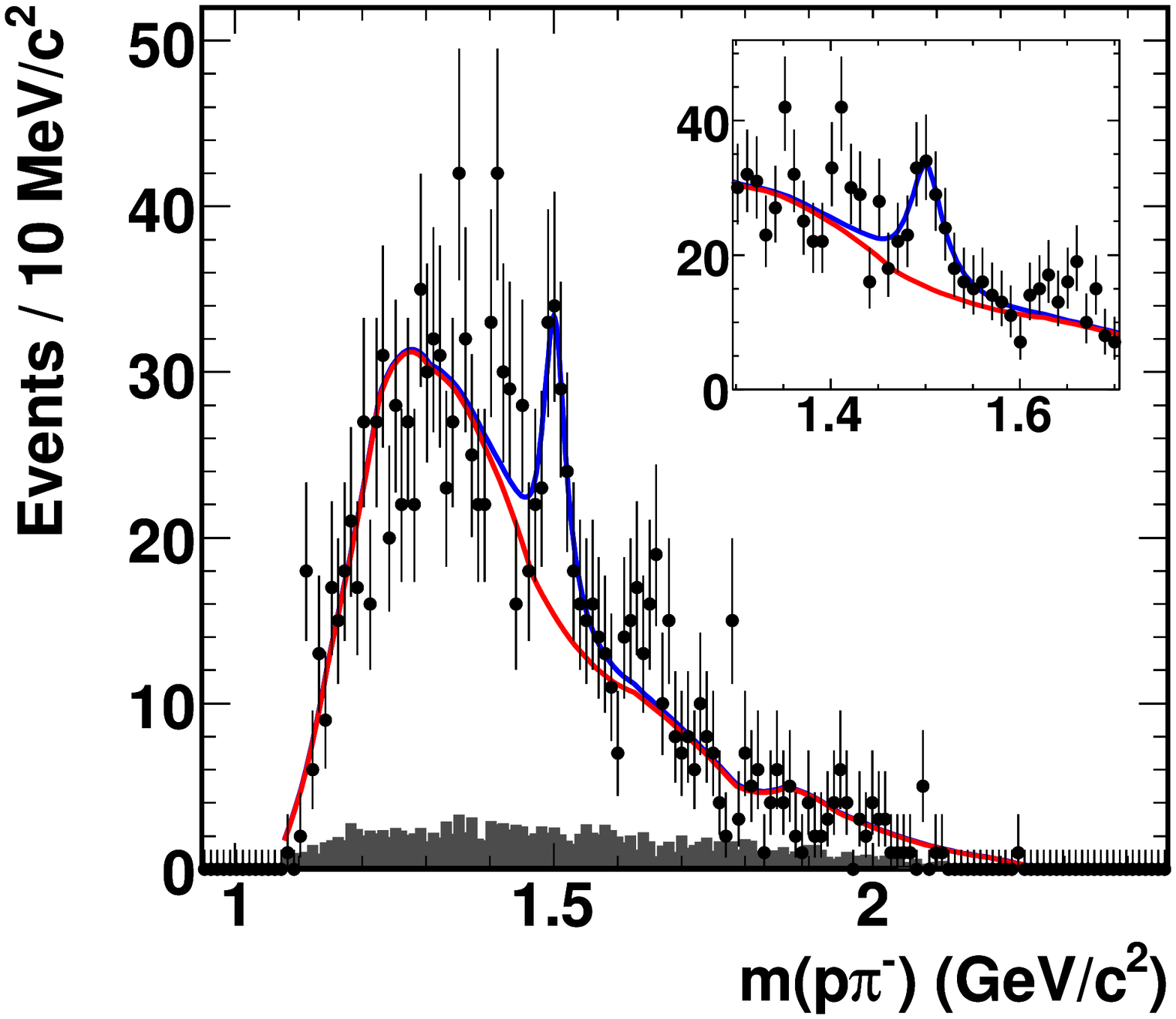}
    \put(76,43){{\babar}}
    \put(76,35){{prelim.}}
  \end{overpic}
}%
\subfloat[$m(\pbar\pim)$ for $\Bzb\To\Dstarp\ppbar\pim$;
	see (b) for descriptions.]{
  \begin{overpic}[width=0.50\textwidth]{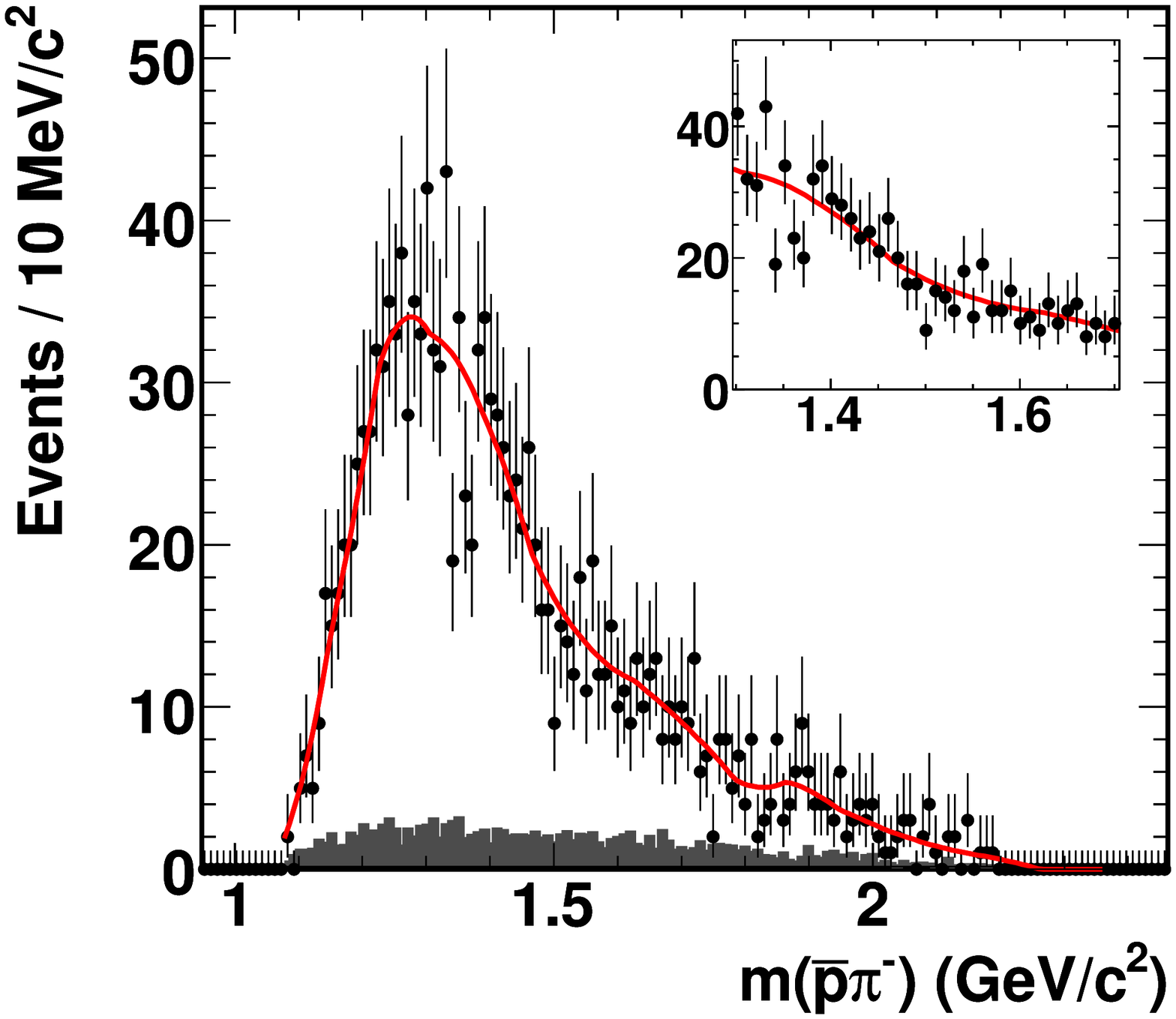}
    \put(76,43){{\babar}}
    \put(76,35){{prelim.}}
  \end{overpic}
}
\caption{Fits of $m(p\pi)$ in two neutral $B$ decays: (ab)
  $\Bzb\To\Dp\ppbar\pim$ and (cd) $\Bzb\To\Dstarp\ppbar\pim$.  The
  plotted sample are events in the signal box of \mes-\DeltaE\ within
  $2.5\sigma$ of the mean; the grey histograms are the scaled
  sidebands.  The in-set binning is the same as in the main figure.
}
\label{fig:ppi_bmodeAB}
\end{figure}

\begin{figure}[bp!]
\centering
\subfloat[$m(p\pim)$ for $\Bm\To\Dz\ppbar\pim$ with pdfs from the top:
	Breit-Wigner for $X$ near $1.5\gevcc$ (solid, blue);
	background from $m(\pbar\pim)$ in (b) (dashed, red).]{
  \hspace{-0.017\textwidth}%
  \begin{overpic}[width=0.50\textwidth]{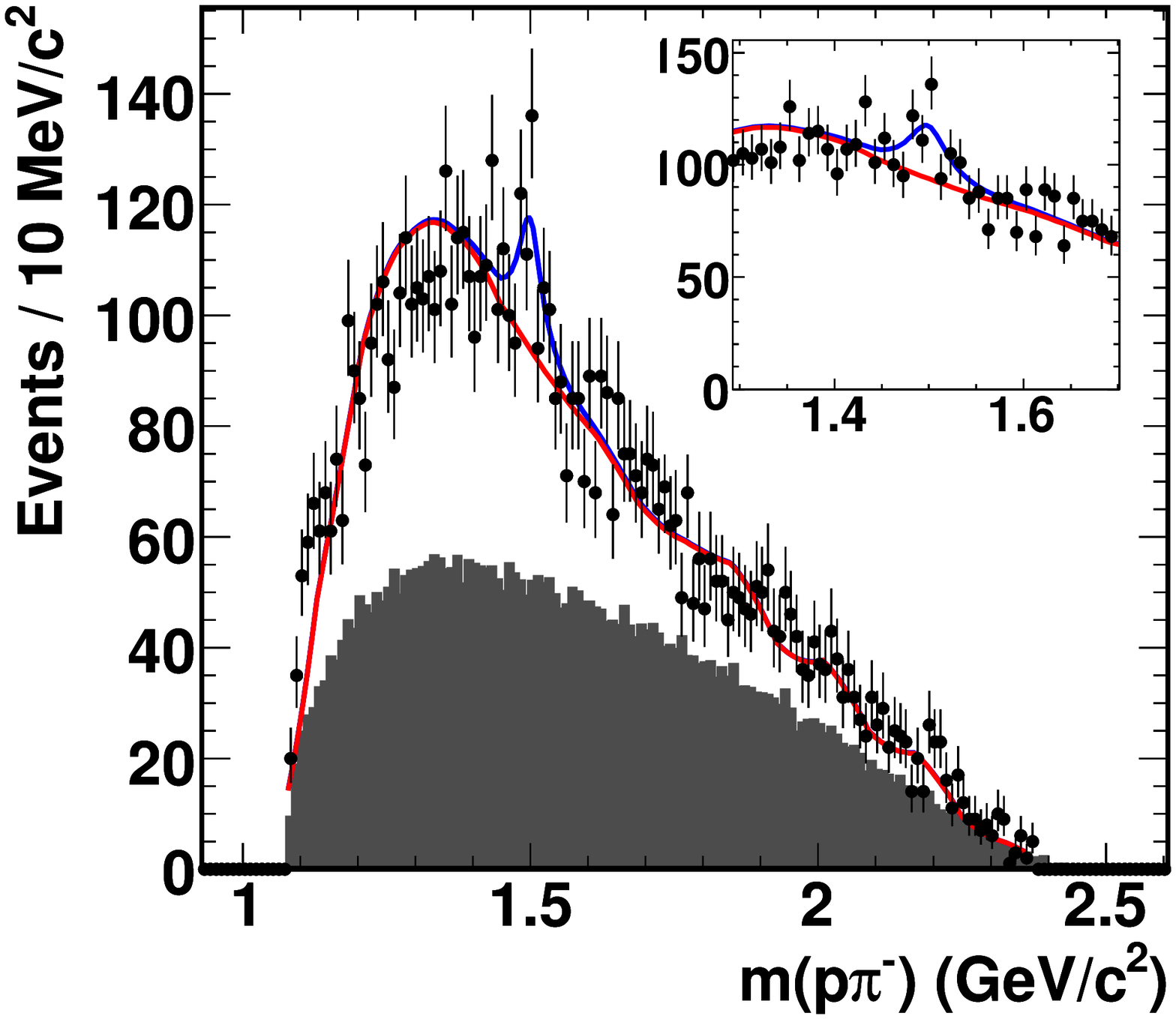}
    \put(76,43){{\babar}}
    \put(76,35){{prelim.}}
  \end{overpic}
}%
\subfloat[$m(\pbar\pim)$ for $\Bm\To\Dz\ppbar\pim$ with
	the smoothed histogram pdf used as background in (a).]{
  \begin{overpic}[width=0.50\textwidth]{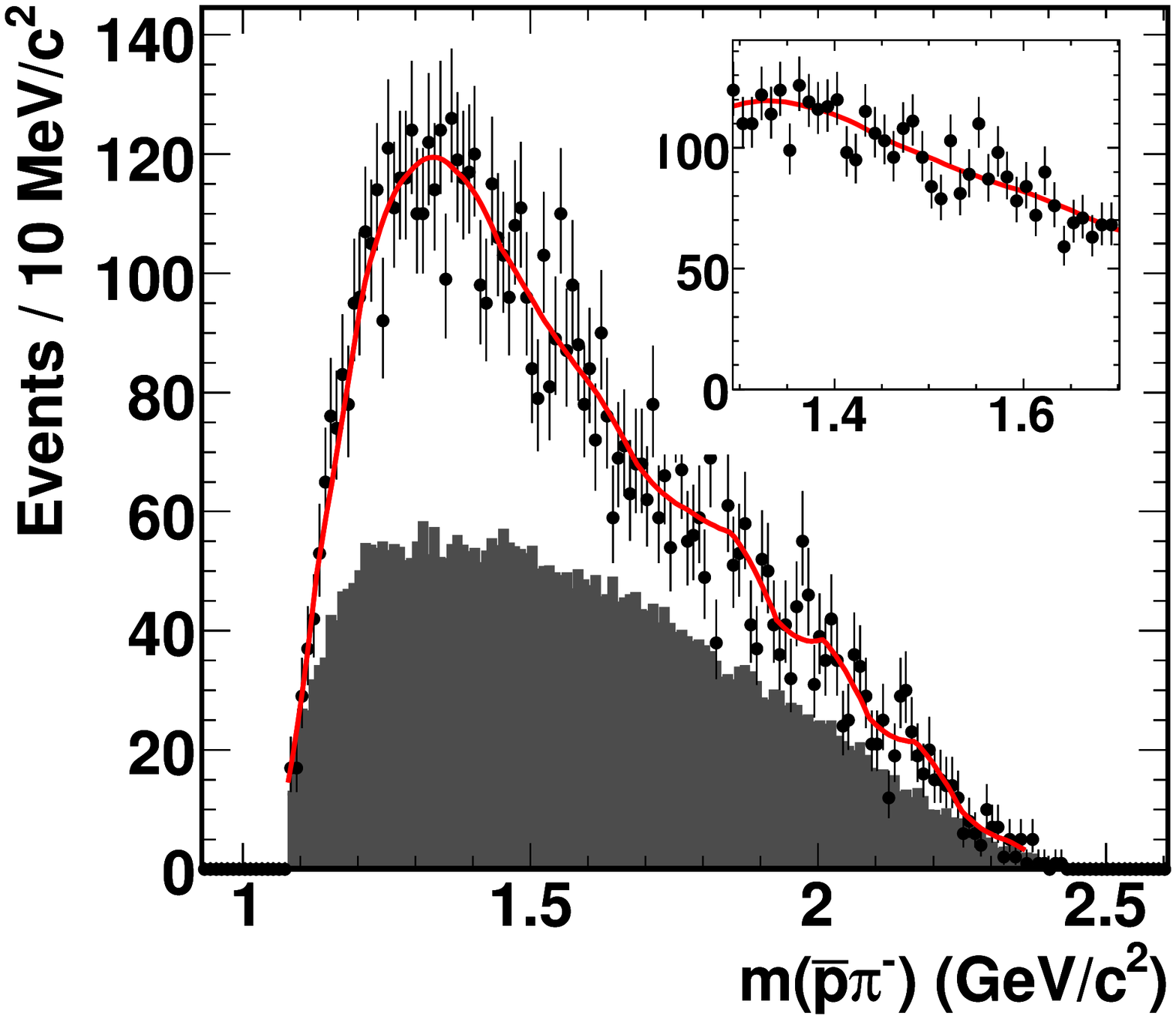}
    \put(76,43){{\babar}}
    \put(76,35){{prelim.}}
  \end{overpic}
}%
\\
\subfloat[$m(p\pim)$ for $\Bm\To\Dstarz\ppbar\pim$;
	see (a) for descriptions.]{
  \hspace{-0.017\textwidth}%
  \begin{overpic}[width=0.50\textwidth]{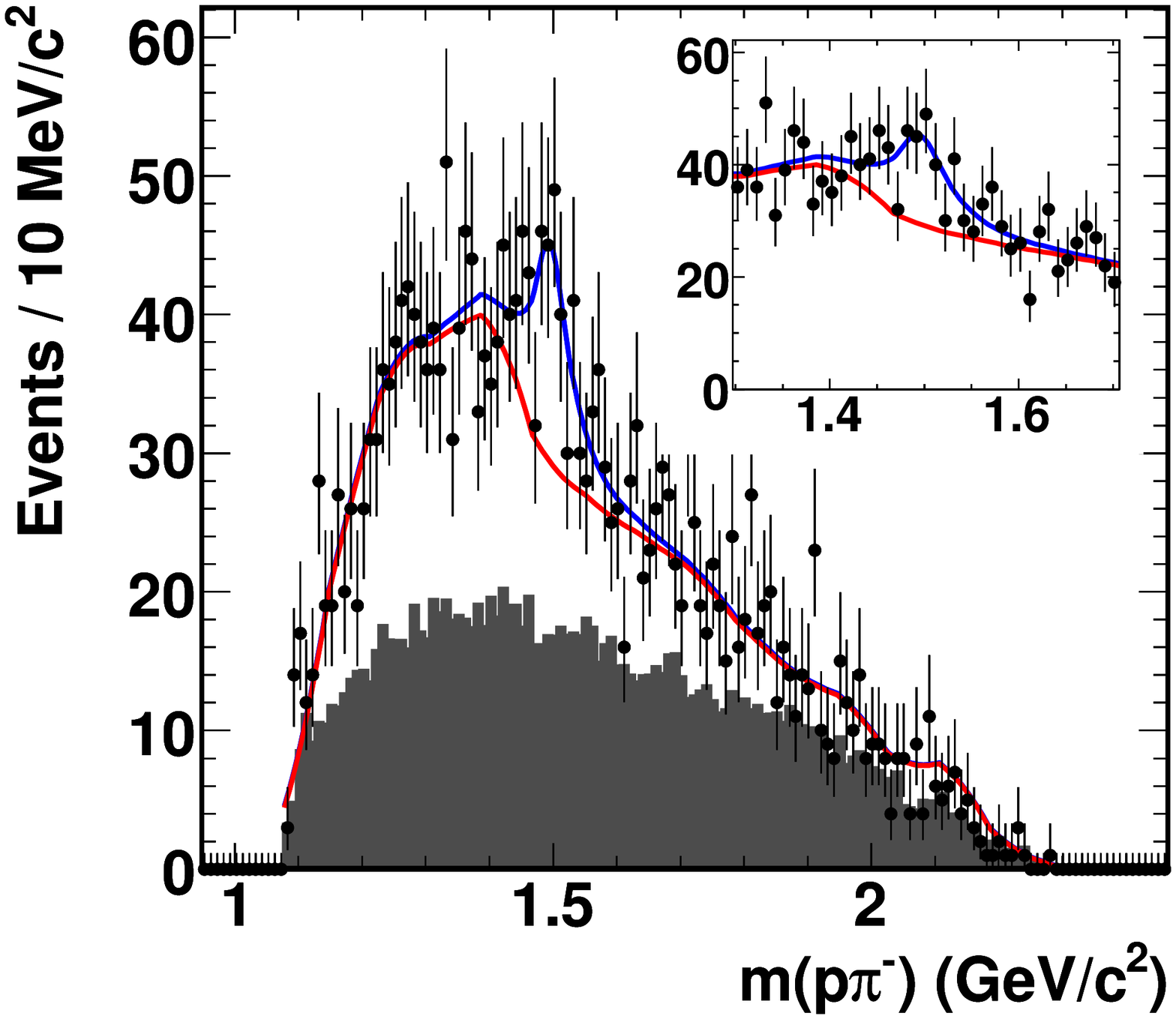}
    \put(76,43){{\babar}}
    \put(76,35){{prelim.}}
  \end{overpic}
}%
\subfloat[$m(\pbar\pim)$ for $\Bm\To\Dstarz\ppbar\pim$;
	see (b) for descriptions.]{
  \begin{overpic}[width=0.50\textwidth]{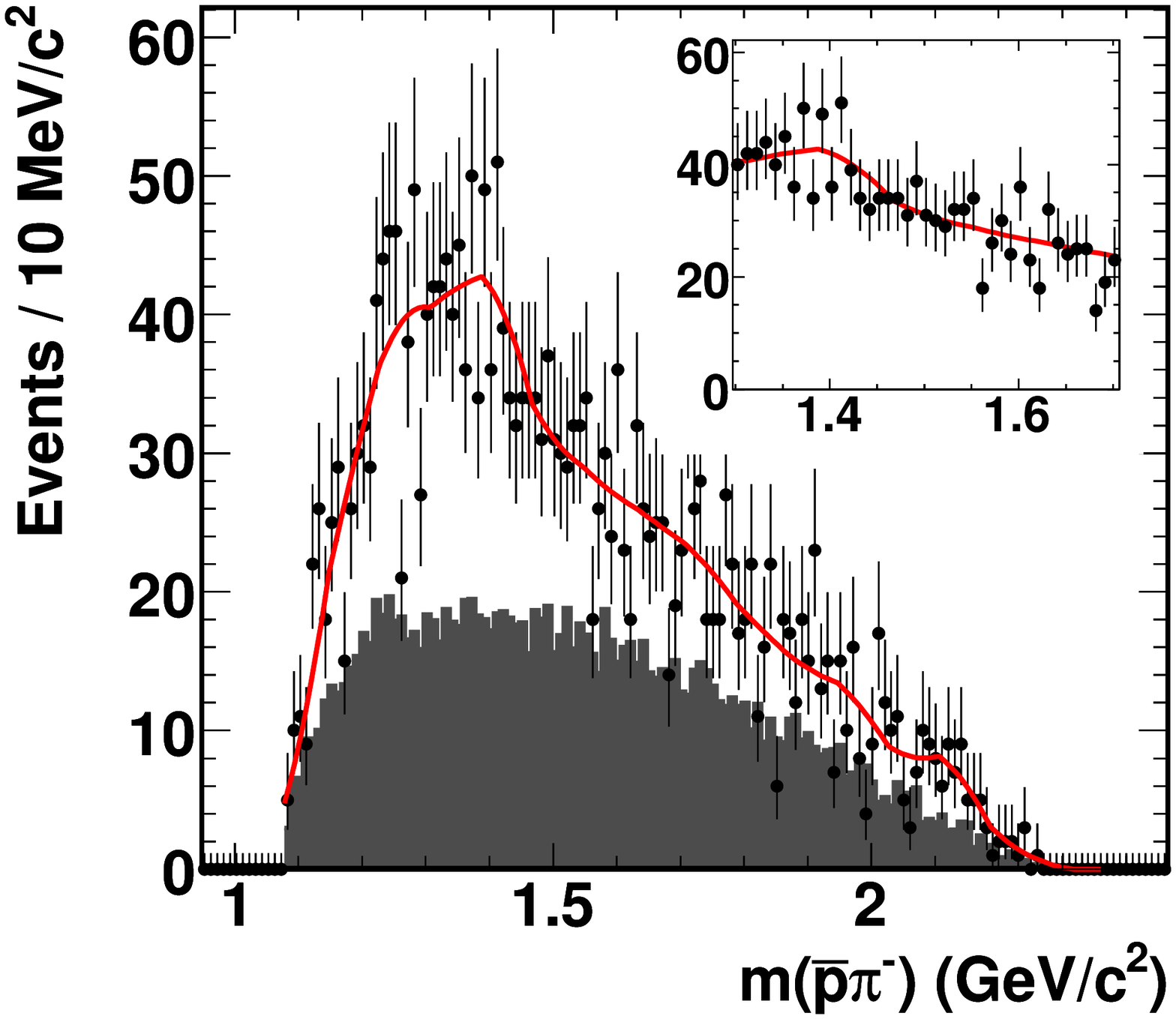}
    \put(76,43){{\babar}}
    \put(76,35){{prelim.}}
  \end{overpic}
}
\caption{Fits of $m(p\pi)$ in two charged $B$ decays:
	(ab) $\Bm\To\Dz\ppbar\pim$ and (cd) $\Bm\To\Dstarz\ppbar\pim$.
	The plotted sample are events in the signal box of
	\mes-\DeltaE\ within $2.5\sigma$ of the mean; the grey
	histograms are the scaled sidebands.  The in-set binning is
	the same as in the main figure.  Caveat emptor: The fit for
	(a) has the width fixed to $51\mevcc$ found in
	Fig.~\ref{fig:ppi_bmodeAB}a.
}
\label{fig:ppi_bmodeGH}
\end{figure}

\begin{figure}[bp!]
\centering
\subfloat[$m(p\pim)$ fit with pdfs from top to bottom: known $N^\ast$
	resonances at $1440$, $1520$, $1535$, $1650$ \mevcc\ (dashed,
	blue); background from $m(\pbar\pim)$ in $\Dp\ppbar\pim$
	(solid, red).]{
  \hspace{-0.017\textwidth}%
  \begin{overpic}[width=0.50\textwidth]{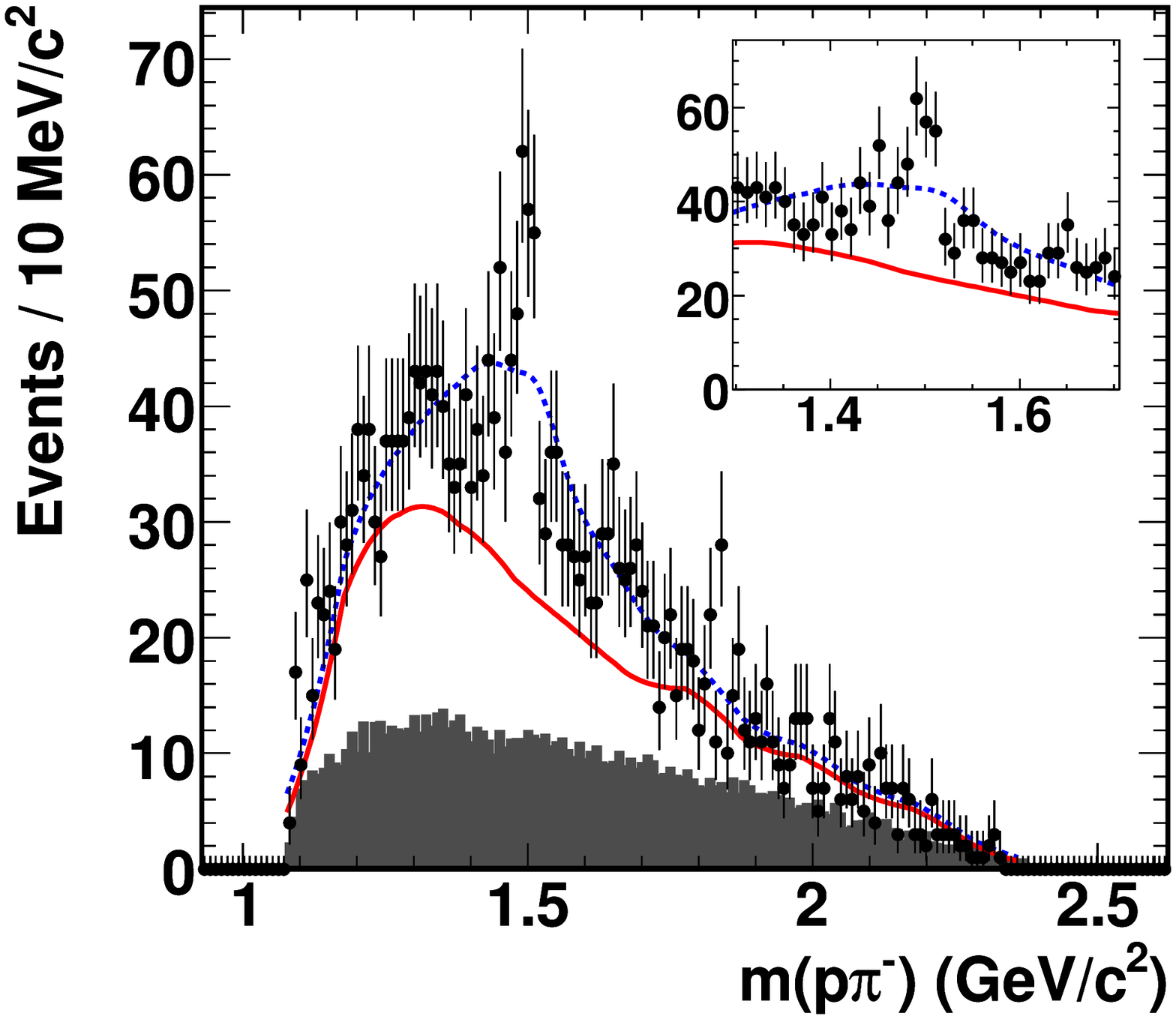}
    \put(76,43){{\babar}}
    \put(76,35){{prelim.}}
  \end{overpic}
}%
\subfloat[$m(p\pim)$ fit with pdfs from top to bottom: Breit-Wigner
	for $X$ (solid, blue); alternate background from
	$m(\pbar\pim)$ in $\Dstarp\ppbar\pim$ (dashed, red);
	background from $m(\pbar\pim)$ in $\Dp\ppbar\pim$ (solid,
	red).]{
  \begin{overpic}[width=0.50\textwidth]{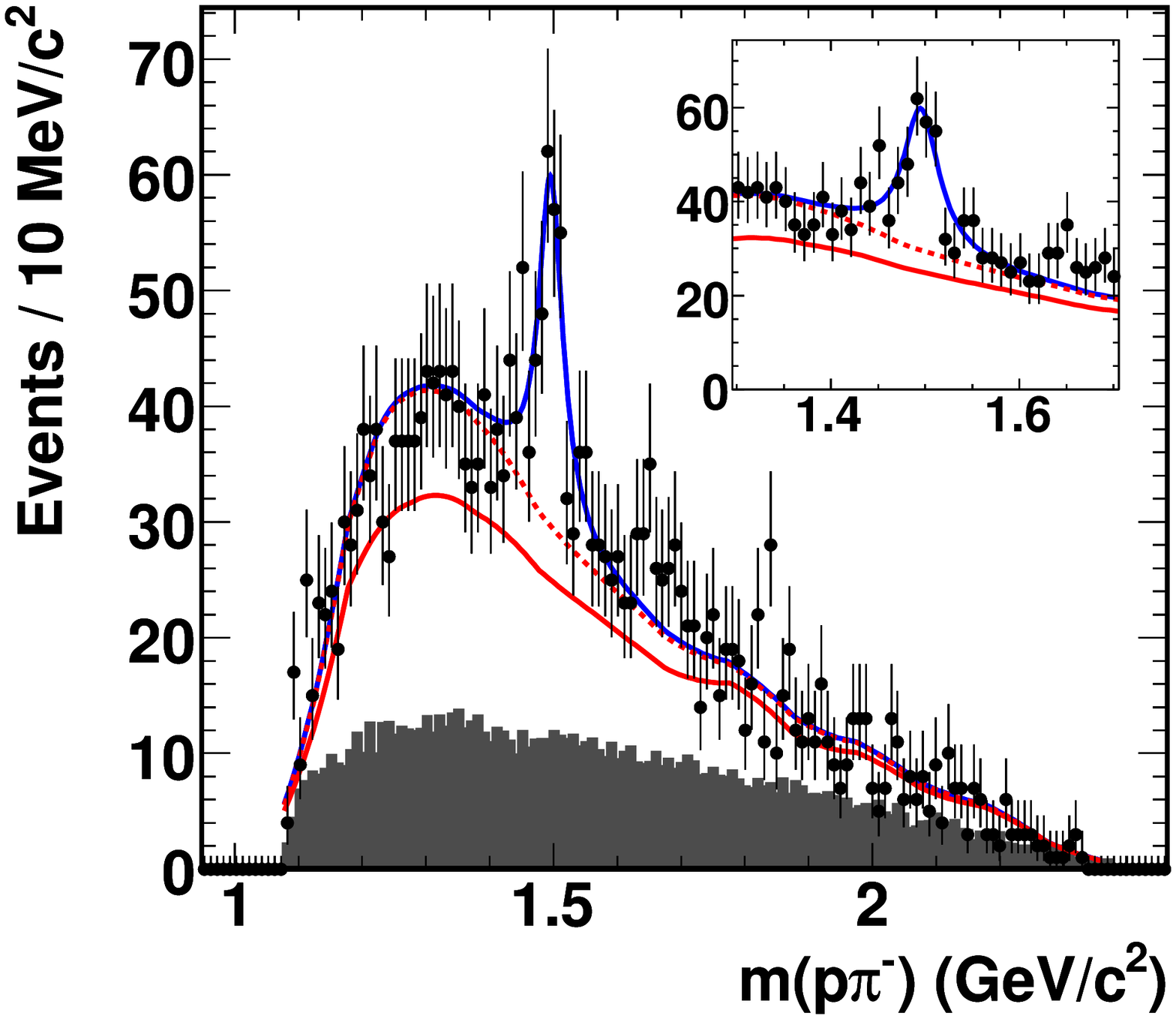}
    \put(76,43){{\babar}}
    \put(76,35){{prelim.}}
  \end{overpic}
}
\caption{Alternate fits of $m(p\pim)$ in $\Bzb\To\Dp\ppbar\pim$.  
	Known $N^\ast$ resonances are used in (a) and an alternate
	background shape from $\Bzb\To\Dstarp\ppbar\pim$ is used in
	(b) as a systematic study.  The in-set binning is the same as
	in the main figure.  The grey histograms are the scaled
	\mes-\DeltaE\ sidebands.
}
\label{fig:ppi_bmodeBalt}
\end{figure}

To investigate $X$ further, we show opposite-sign and same-sign $p\pi$
in finer bins of $10\mevcc$ in Figs.~\ref{fig:ppi_bmodeAB} and
\ref{fig:ppi_bmodeGH} for the two neutral and charged $B$ decay modes,
respectively.  (The detector resolution around $1.5\gevcc$ is less
than $4\mevcc$.) In Figs.~\ref{fig:ppi_bmodeAB}ac and
\ref{fig:ppi_bmodeGH}ac showing opposite-sign $p\pim$, the points with
error bars show the total number of events in the \mes-\DeltaE\ signal
box, which includes of three components: the $B$ signal without $X$,
the $B$ signal with $X$, and the non-signal events.  Although we do
not have any way a priori to determine the first component ($B$ signal
without $X$) a reasonable procedure, which we adopt, is to take the
smoothed histogram pdf from the same-sign mass combination of
$\pbar\pim$ in the same $B$ decay mode written as $P_{ss}$.  This
procedure has a systematic uncertainty that is difficult to quantify,
because the formation of $p$ and $\pbar$ is not neccessarily symmetric
with respect to the $\pim$ in the $B\To\Dmaybestar\ppbar\pim$ decays.
The distribution of $\pbar\pim$ events as well as the histogram pdf
are shown in Figs.~\ref{fig:ppi_bmodeAB}bd and
\ref{fig:ppi_bmodeGH}bd.  The second component (the $B$ signal with
$X$) is described by a Breit-Wigner line shape written as
$P_{bw}$.  The third component (the non-signal background) can be
determined from the \mes-\DeltaE\ sidebands scaled to the number of
non-signal events in the signal box.  The scaled sideband event
samples is shown in
Figs.~\ref{fig:ppi_bmodeAB}--\ref{fig:ppi_bmodeBalt} as a grey-shaded
histogram and we note that these are smooth across region around
$1.5\gevcc$.

Figures~\ref{fig:ppi_bmodeAB}a and \ref{fig:ppi_bmodeAB}c show the fits
of $p\pim$ for \nolbreaks{$\Bzb\To\Dp\ppbar\pim$} and
\nolbreaks{$\Bzb\To\Dstarp\ppbar\pim$}, respectively, using $P_{bw}$
and $P_{ss}$.  $P_{ss}$ is relatively flat across $X$ in Figs.
\ref{fig:ppi_bmodeAB}b and provides a good description of non-$X$ in
\ref{fig:ppi_bmodeAB}a.  However, $P_{ss}$ from Fig.
\ref{fig:ppi_bmodeAB}d is not as smooth around the signal region and
shows a dip just below $1.5\gevcc$.  We have not yet tried to quantify
the uncertainty on this shape.  The fit results for
\nolbreaks{$\Dp\ppbar\pim$} and \nolbreaks{$\Dstarp\ppbar\pim$}
converged to a mean (width) of $1494.4\pm4.1$ ($51\pm18$) \mevcc\ and
$1500.8\pm4.4$ ($43\pm17$) \mevcc, respectively, where the errors are
statistical.  We note a small excess of events above $1.6\gevcc$ with
respect to $P_{ss}$, but have not attempted to fit it.  The
statistical significance of $X$ in \nolbreaks{$\Dp\ppbar\pim$} and
\nolbreaks{$\Dstarp\ppbar\pim$} is found by comparing the likelihood
values of our nominal fit to the background-only hypothesis,
\nolbreaks{$\sqrt{2\ (\ln\mathcal{L}_{bw+ss}-\ln\mathcal{L}_{ss})}$},
which is $8.6$ and $6.9$ standard deviations, respectively, under
these background assumptions.

Figures~\ref{fig:ppi_bmodeGH}a and \ref{fig:ppi_bmodeGH}c show the fit
to $p\pim$ for \nolbreaks{$\Bm\To\Dz\ppbar\pim$} and
\nolbreaks{$\Bm\To\Dstarz\ppbar\pim$}, respectively.  The above
analysis is repeated, but the signals for $X$ in these modes are not
as clear compared to the neutral $B$ decays.  Moreover, the fit did
not converge for $\Bm\To\Dz\ppbar\pim$ following the procedure, so we
re-fit with the $X$ width fixed to the value found for
$\Bzb\To\Dp\ppbar\pim$.  The fit results are consistent with the
values found for neutral $B$ decays, but we leave them out in the
average.

Figure~\ref{fig:ppi_bmodeBalt} shows alternate fits to $p\pim$ in
\nolbreaks{$\Bzb\To\Dp\ppbar\pim$}.  The first,
Fig.~\ref{fig:ppi_bmodeBalt}a, uses known excited nucleon resonances
at $1440$, $1520$, $1535$, and $1650$ \mevcc.  The Breit-Wigner
parameters are fixed to their nominal values and are written as
$P_{N^\ast}$.  The fit with $P_{ss}$ and $P_{N^\ast}$ does not well
describe the data around $1.5\gevcc$.  The second,
Fig.~\ref{fig:ppi_bmodeBalt}b, uses $P_{bw}$, $P_{ss}$, and
$P_{ss}^\prime$, where the last component is the same-sign $\pbar\pim$
distribution from $\Bzb\To\Dstarp\ppbar\pim$.  This is used to
determine the systematic uncertainty for our choice of the background
pdf, which shifted the mean (width) by $0.8$ ($4$) \mevcc.  Lastly,
for the mass uncertainty, we add another $0.5\mevcc$ to the absolute
uncertainty due to the variation in the assumed magnetic field and the
detector material based on the study of $\Lambda_c$
\cite{Aubert:2005gt}.

Table~\ref{tab:oppsign} summarizes the results for $X$, which can be
described by a Breit-Wigner with
\begin{eqnarray}
m_X      &=& (1497.4\pm 3.0\pm 0.9) \mevcc\\ \nonumber
\Gamma_X &=& (\phantom{00.0}47\pm 12\phantom{.}\pm 4\phantom{.0}) \mevcc,
\end{eqnarray}
where the errors are statistical and systematic, respectively.  The
$X$ yields can be compared with the total $B$ yields and their ratios
range from \nolbreaks{$5$--$15$\percent}.

\begin{table}[bp!]
\centering
\caption{Fit results of $X$ near $1.5\gevcc$ in $m(p\pim)$ for 4-body
  $B$ decays.  The $X$ yield is $n_X$; ratio is $r_{X}{\eq}n_{X}/n_S$
  where the $B$ yield is from Table~\ref{tab:bfchain}.  The fit
  assumes the same-sign $m(\pbar\pim)$ combination as the background
  pdf.  All errors are statistical.
}
\label{tab:oppsign}
{\small
\begin{tabular}{l r@{$\pm$}l r@{$\pm$}l c r@{$\pm$}l r@{$\pm$}l }
\dbline
Decay 
& \multicolumn{2}{l}{Mean (\mevcc)}
& \multicolumn{2}{l}{Full width (\mevcc)}
& $\sqrt{2\Delta{\ln\mathcal{L}}}$
& \multicolumn{2}{l}{Yield $n_X$}
& \multicolumn{2}{l}{Ratio $r_X$ (\percent)} \\
\sgline
(a) $\Bzb\To\Dp\ppbar\pim$     	& $1494.4$ & $4.1$ & $51$ & $18$ & $\phantom{0}8.6$ & $227$ & $51$ & $12.5$ & $ 2.8$ \\
(b) $\Bzb\To\Dstarp\ppbar\pim$ 	& $1500.8$ & $4.4$ & $43$ & $17$ & $\phantom{0}6.9$ & $120$ & $31$ & $\phantom{0}8.8$ & $2.3$ \\
(c) $\Bm\To\Dz\ppbar\pim$      	& $1498.8$ & $6.2$ & \multicolumn{2}{l}{$51$, fixed to (a)} 
				& $\phantom{0}4.2$ & $183$ & $45$ & $\phantom{0}5.2$ & $1.3$ \\
(d) $\Bm\To\Dstarz\ppbar\pim$  	& $1495.2$ & $8.1$ & $71$ & $34$ &
$\phantom{0}5.5$ & $169$ & $59$ & $14.7$ & $5.1$ \\
Stat. average of (a) and (b)         & $1497.4$ & $3.0$ & $47$ & $12$
& $11\phantom{.0}$ & \multicolumn{2}{c}{-} & \multicolumn{2}{l}{-}  \\
\dbline
\end{tabular}
}
\end{table}

\subsection{\boldmath Five-body decays $B\To\Dmaybestar\ppbar\pi\pi$}

\begin{figure}[bp!]
\centering
\subfloat[$\Dz\ppbar\pi\pi$, $m(\ppbar)$]{
  \begin{overpic}[width=0.25\textwidth]{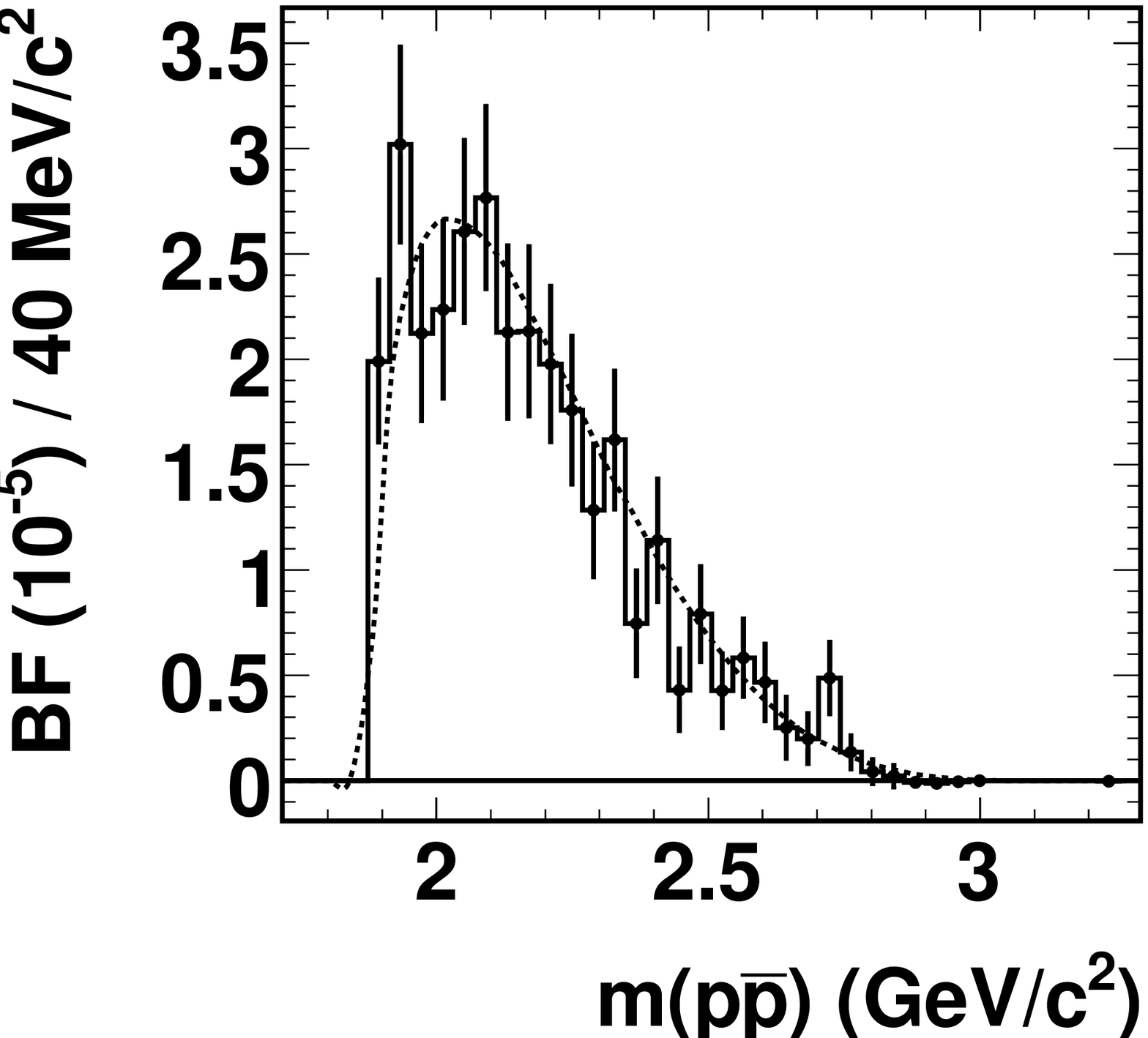}
    \put(66,74){\small{\babar}}
    \put(66,66){\small{prelim.}}
  \end{overpic}
}%
\subfloat[$\Dz\ppbar\pi\pi$, $m(\Dp\pbar)$]{
  \hspace{-0.017\textwidth}%
  \begin{overpic}[width=0.25\textwidth]{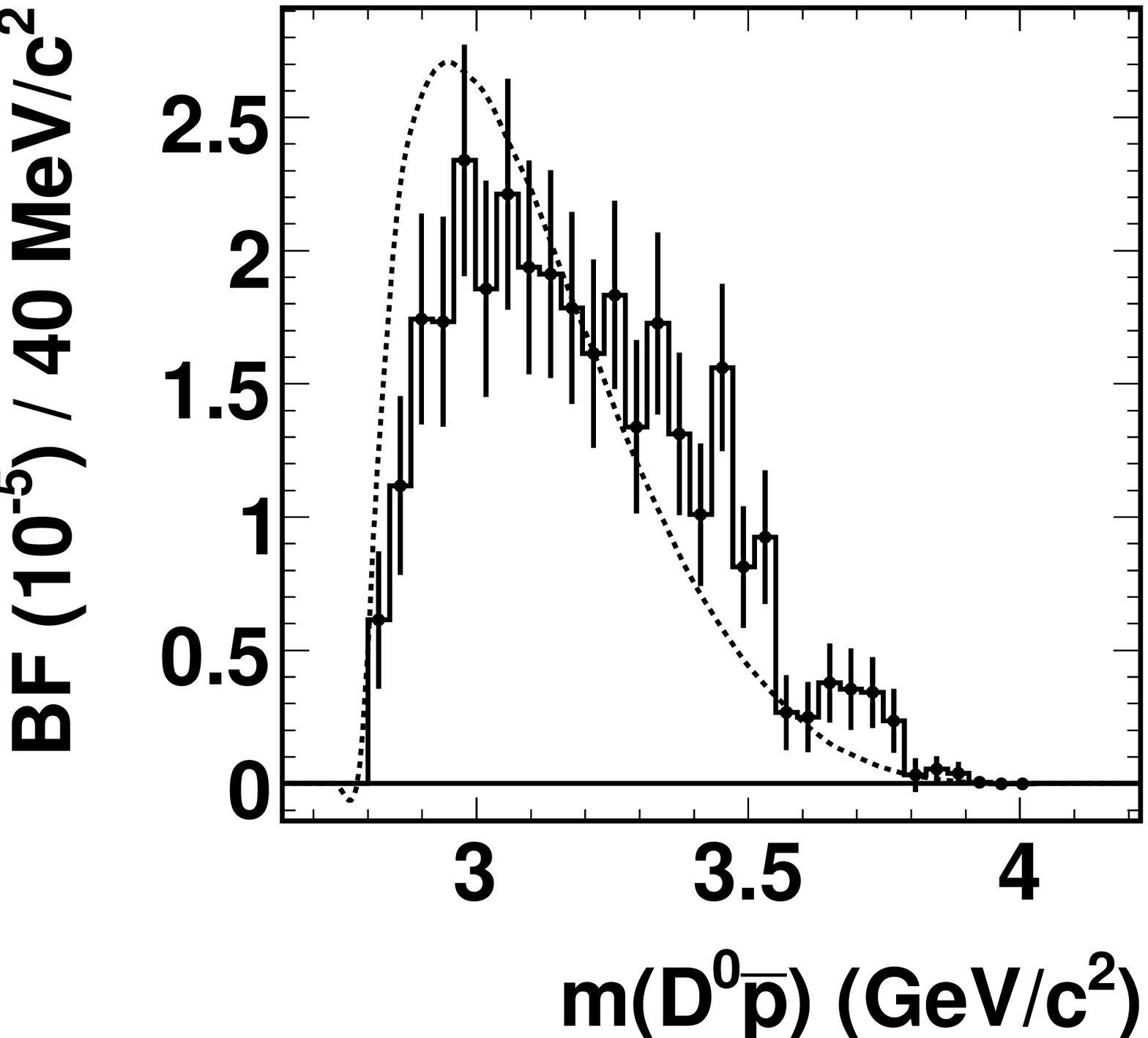}
    \put(66,74){\small{\babar}}
    \put(66,66){\small{prelim.}}
  \end{overpic}
}%
\subfloat[$\Dz\ppbar\pi\pi$, $m(\Dp{p})$]{
  \hspace{-0.017\textwidth}%
  \begin{overpic}[width=0.25\textwidth]{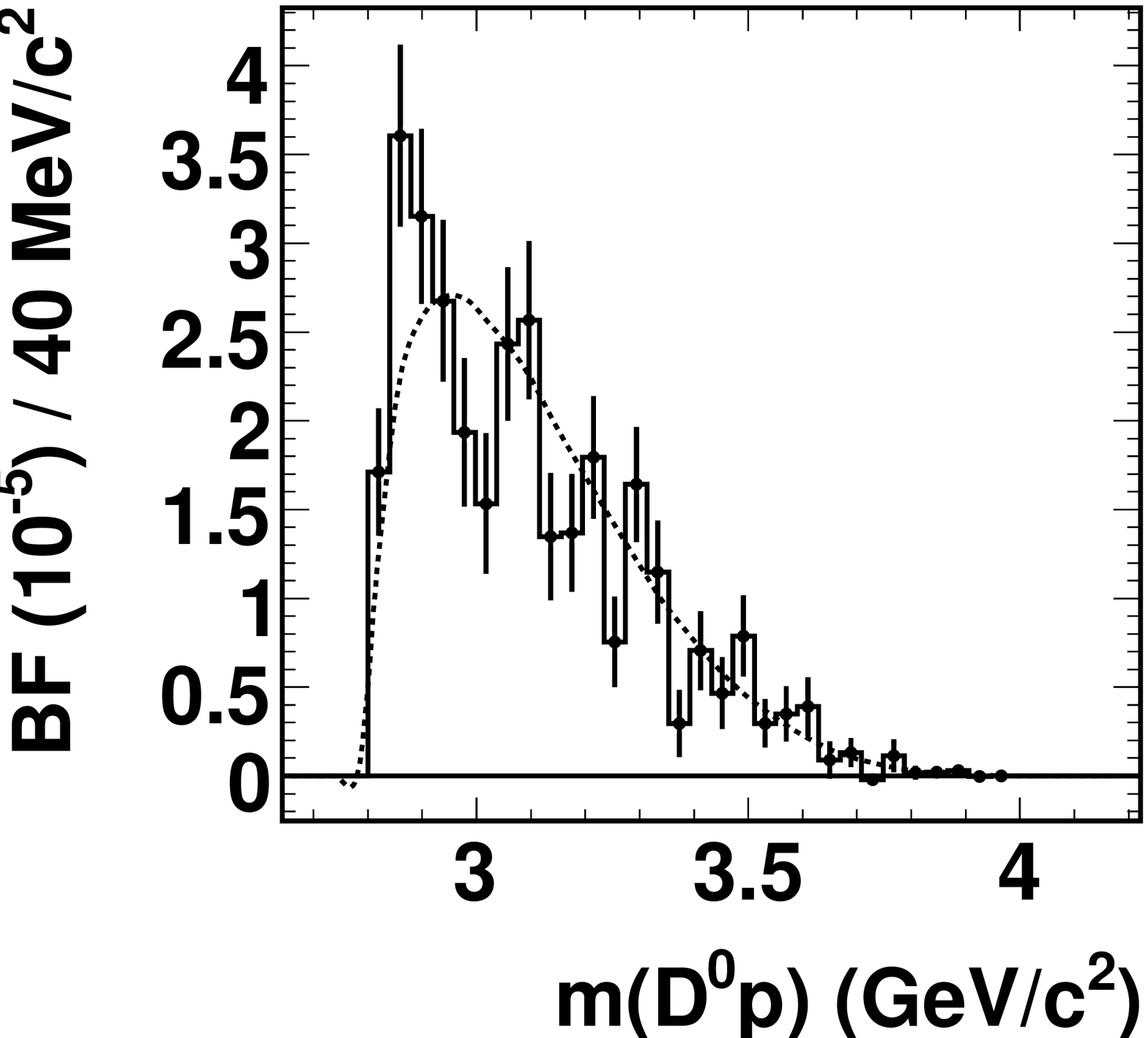}
    \put(66,74){\small{\babar}}
    \put(66,66){\small{prelim.}}
  \end{overpic}
}%
\subfloat[$\Dz\ppbar\pi\pi$, $m(p\pim)$]{
  \hspace{-0.017\textwidth}%
  \begin{overpic}[width=0.25\textwidth]{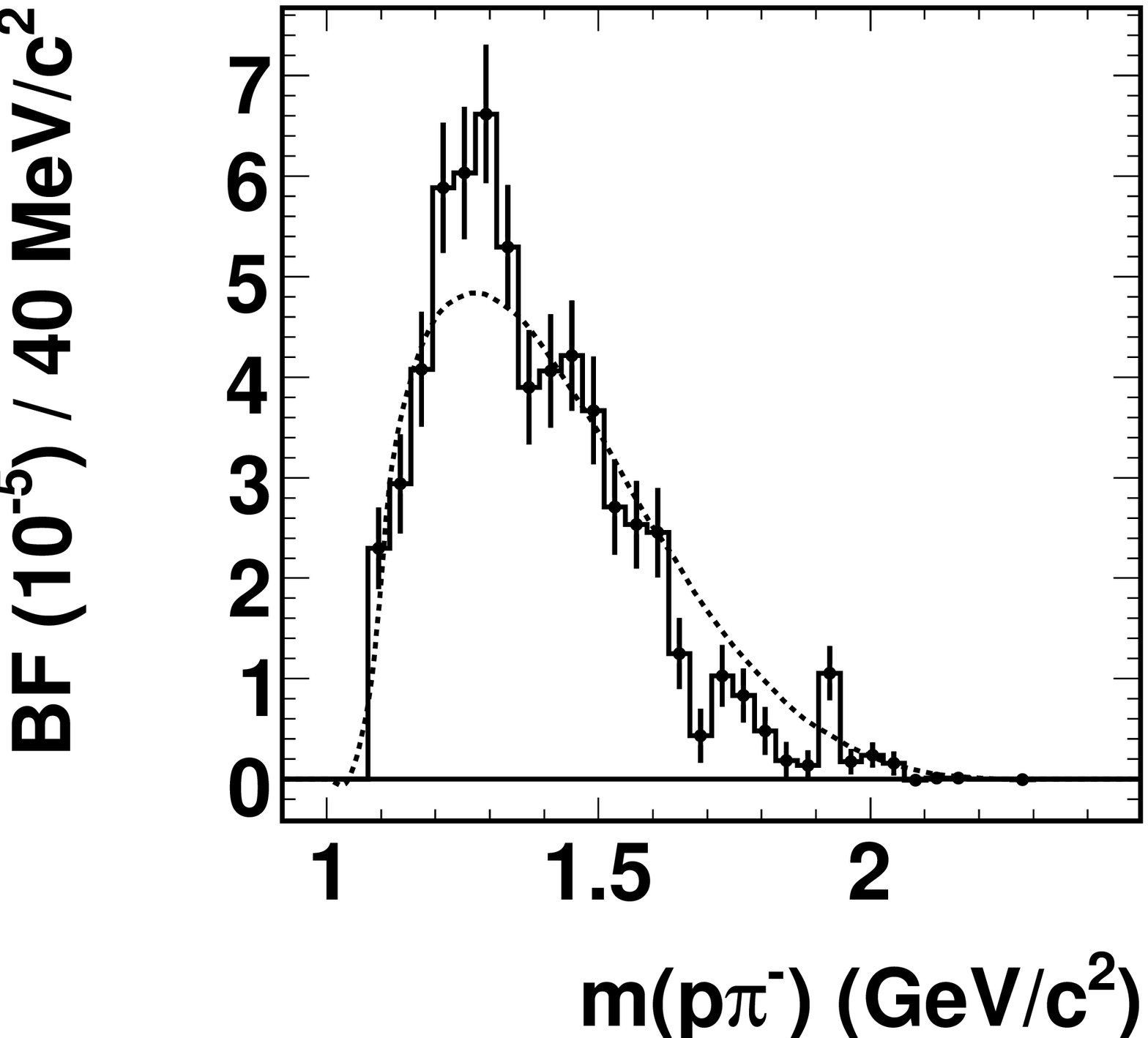}
    \put(66,74){\small{\babar}}
    \put(66,66){\small{prelim.}}
  \end{overpic}
}%
\\
\subfloat[$\Dstarz\ppbar\pi\pi$, $m(\ppbar)$]{
  \begin{overpic}[width=0.25\textwidth]{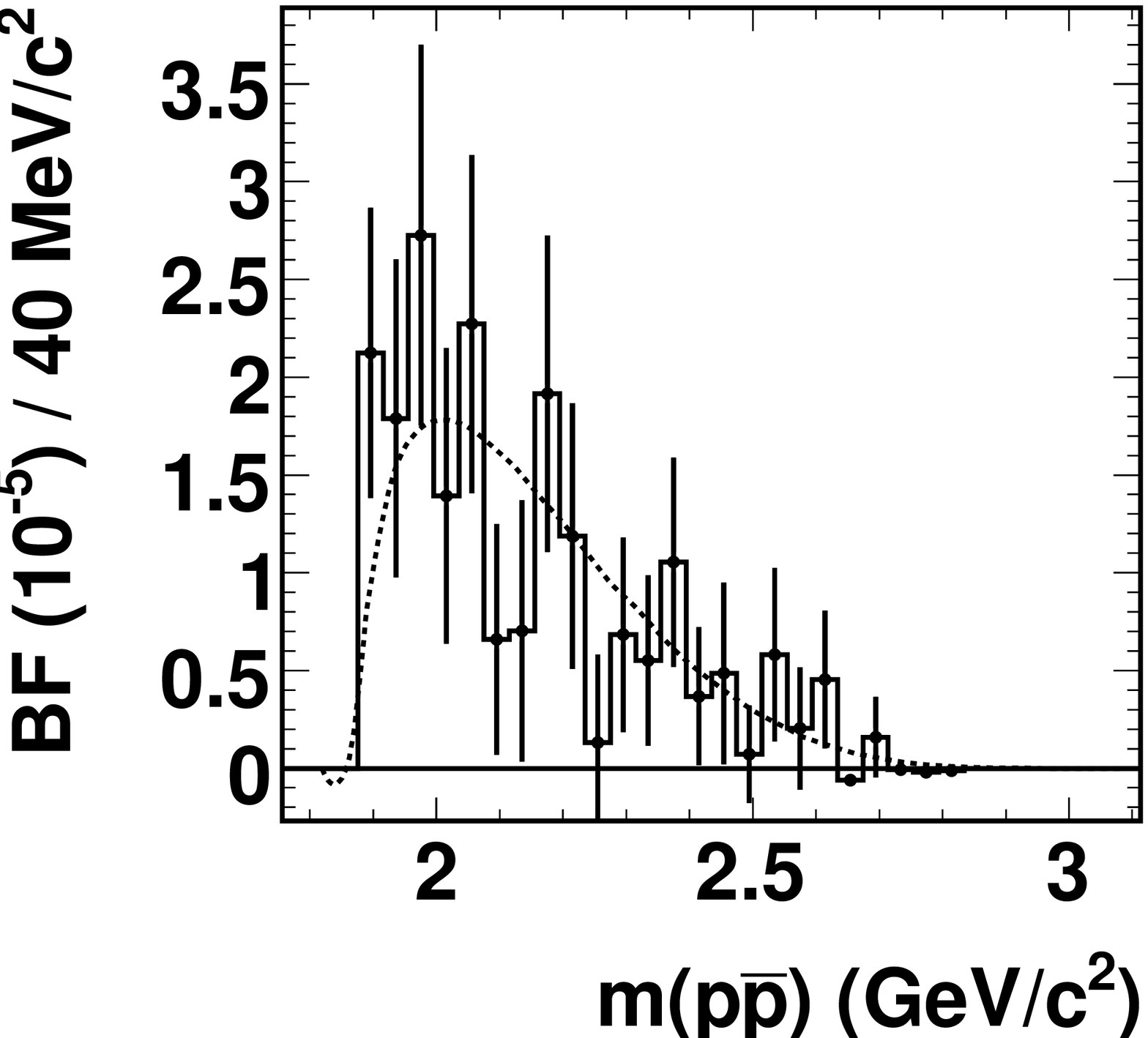}
    \put(66,74){\small{\babar}}
    \put(66,66){\small{prelim.}}
  \end{overpic}
}%
\subfloat[$\Dstarz\ppbar\pi\pi$, $m(\Dp\pbar)$]{
  \hspace{-0.017\textwidth}%
  \begin{overpic}[width=0.25\textwidth]{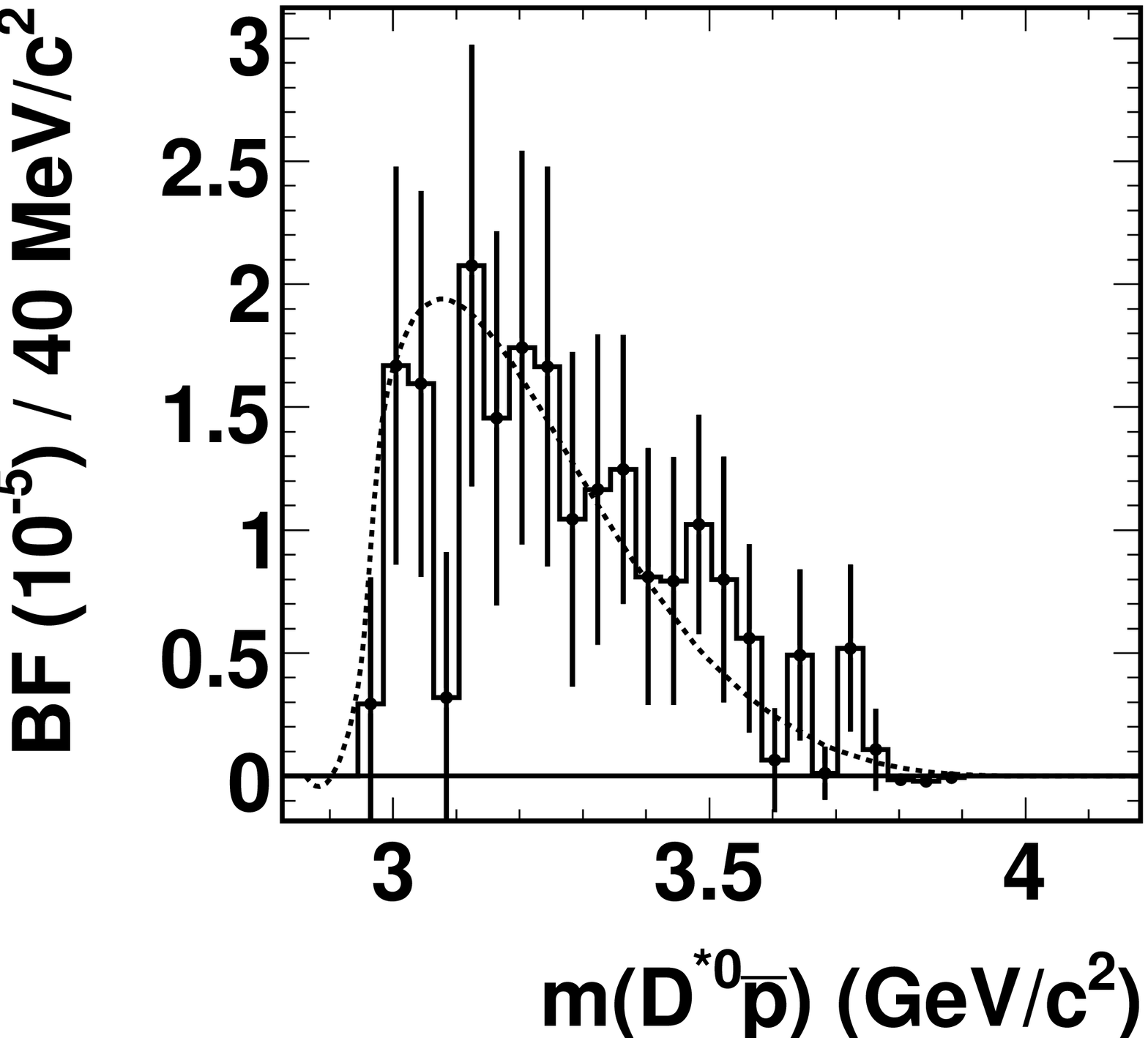}
    \put(66,74){\small{\babar}}
    \put(66,66){\small{prelim.}}
  \end{overpic}
}%
\subfloat[$\Dstarz\ppbar\pi\pi$, $m(\Dp{p})$]{
  \hspace{-0.017\textwidth}%
  \begin{overpic}[width=0.25\textwidth]{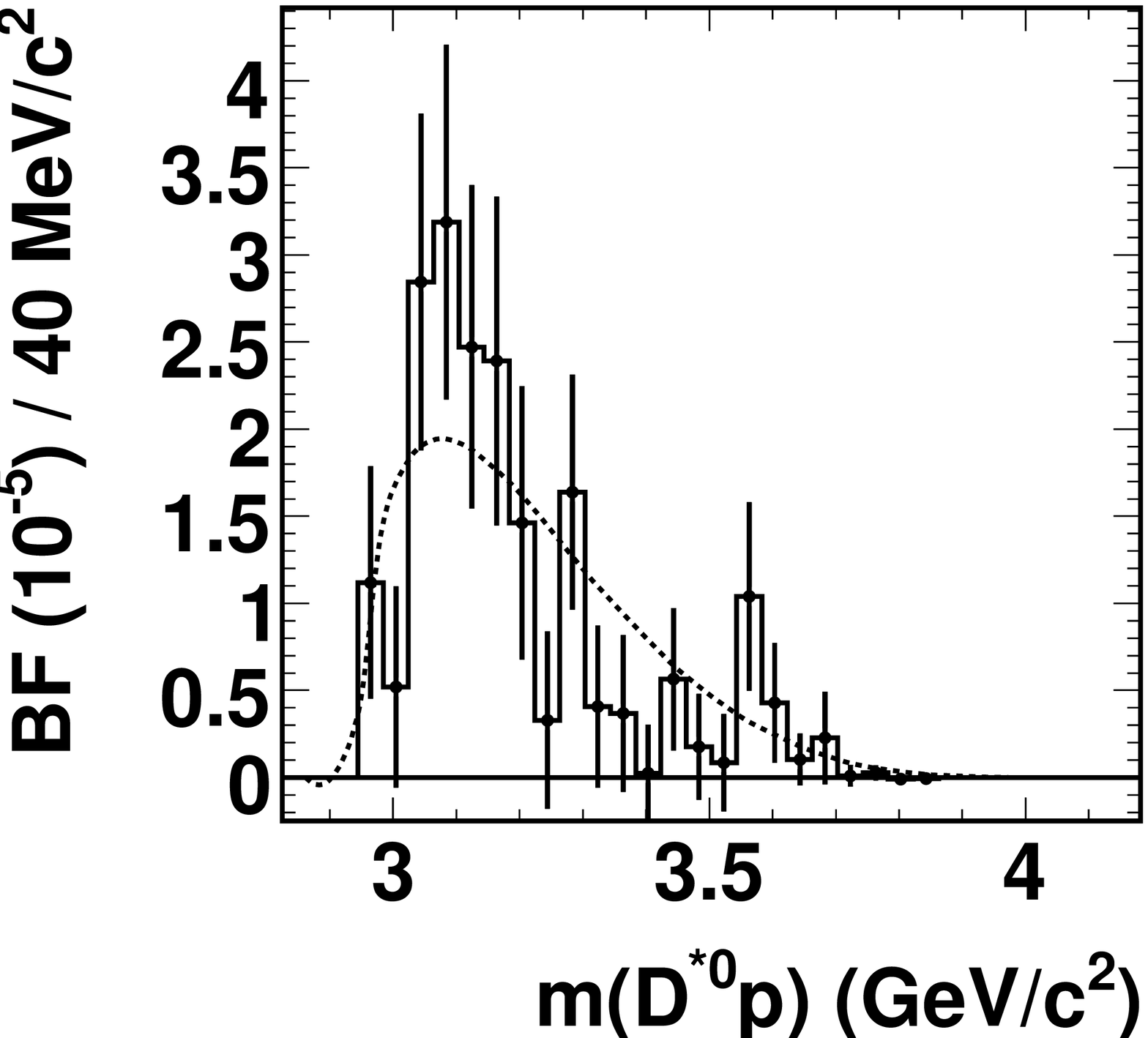}
    \put(66,74){\small{\babar}}
    \put(66,66){\small{prelim.}}
  \end{overpic}
}%
\subfloat[$\Dstarz\ppbar\pi\pi$, $m(p\pim)$]{
  \hspace{-0.017\textwidth}%
  \begin{overpic}[width=0.25\textwidth]{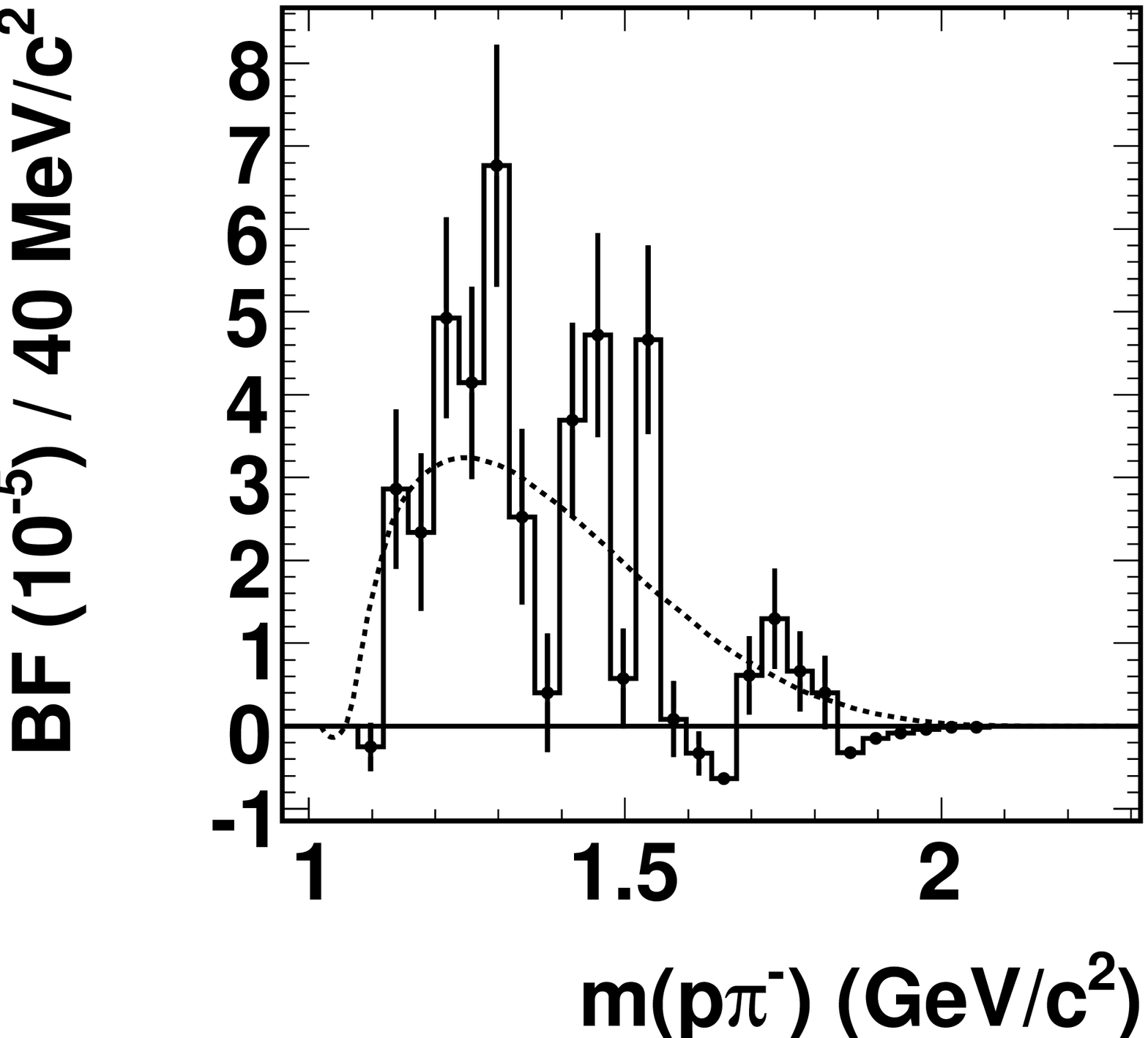}
    \put(66,74){\small{\babar}}
    \put(66,66){\small{prelim.}}
  \end{overpic}
}%
\\
\subfloat[$\Dp\ppbar\pi\pi$, $m(\ppbar)$]{
  \begin{overpic}[width=0.25\textwidth]{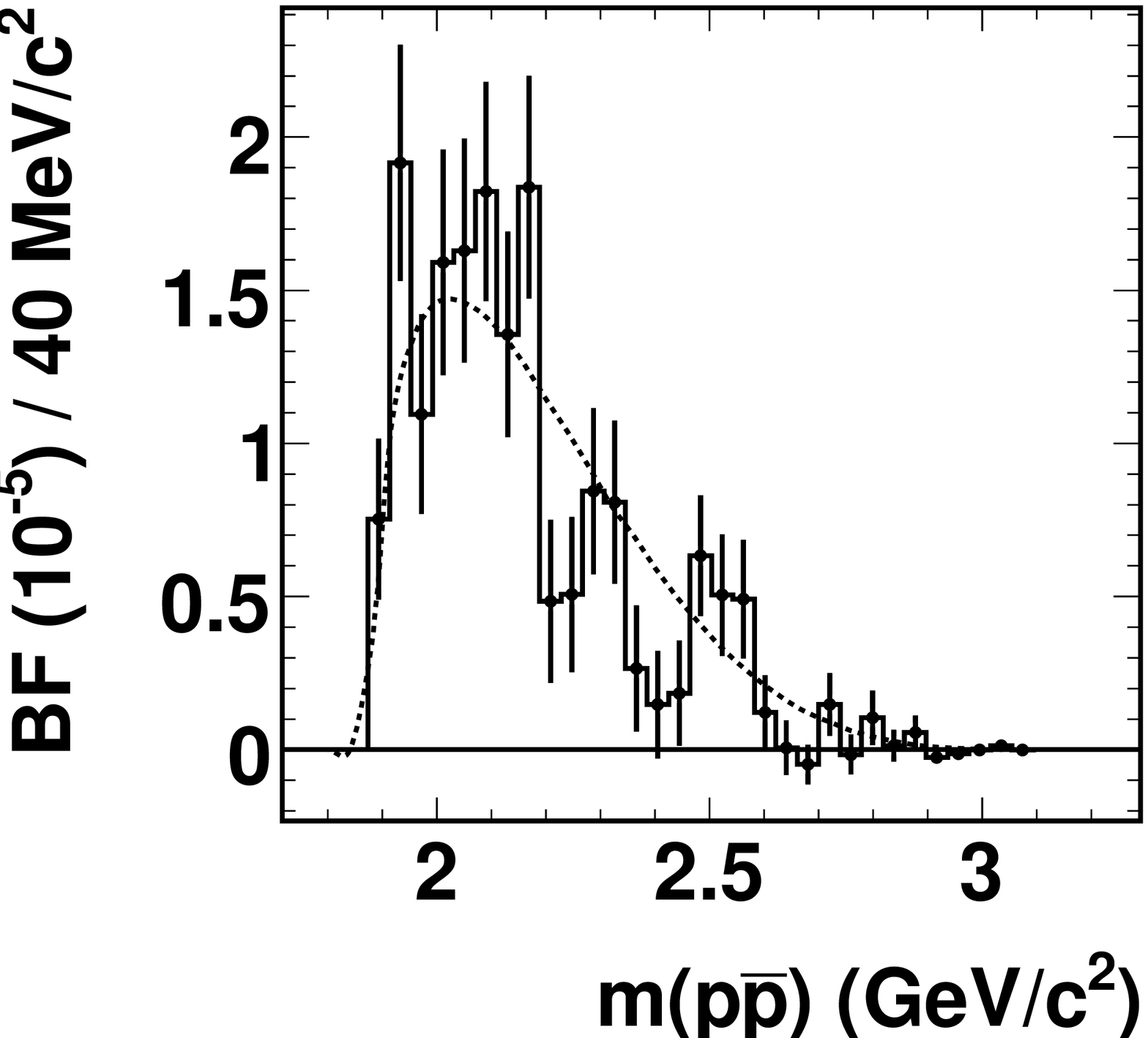}
    \put(66,74){\small{\babar}}
    \put(66,66){\small{prelim.}}
  \end{overpic}
}%
\subfloat[$\Dp\ppbar\pi\pi$, $m(\Dp\pbar)$]{
  \hspace{-0.017\textwidth}%
  \begin{overpic}[width=0.25\textwidth]{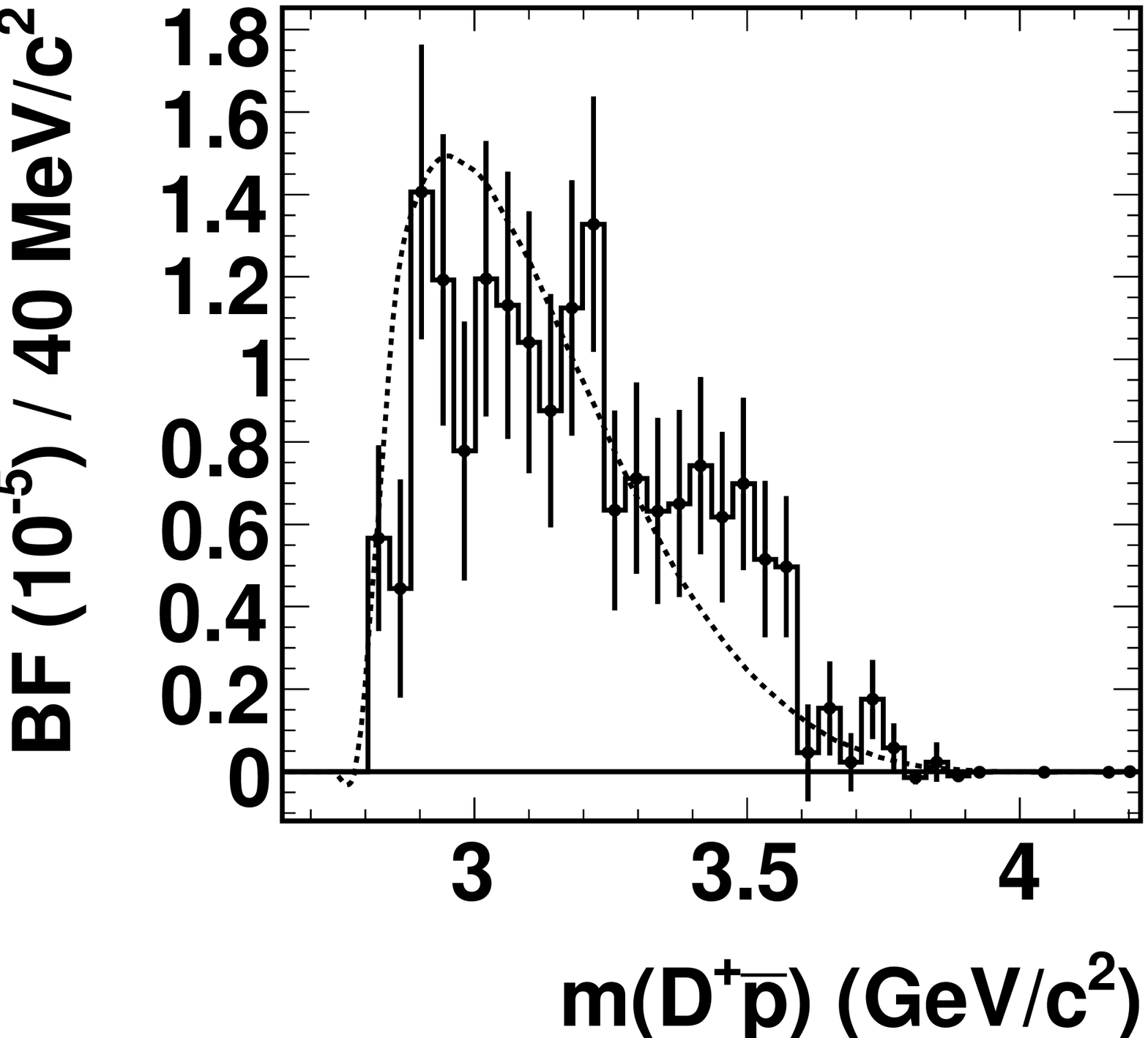}
    \put(66,74){\small{\babar}}
    \put(66,66){\small{prelim.}}
  \end{overpic}
}%
\subfloat[$\Dp\ppbar\pi\pi$, $m(\Dp{ p})$]{
  \hspace{-0.017\textwidth}%
  \begin{overpic}[width=0.25\textwidth]{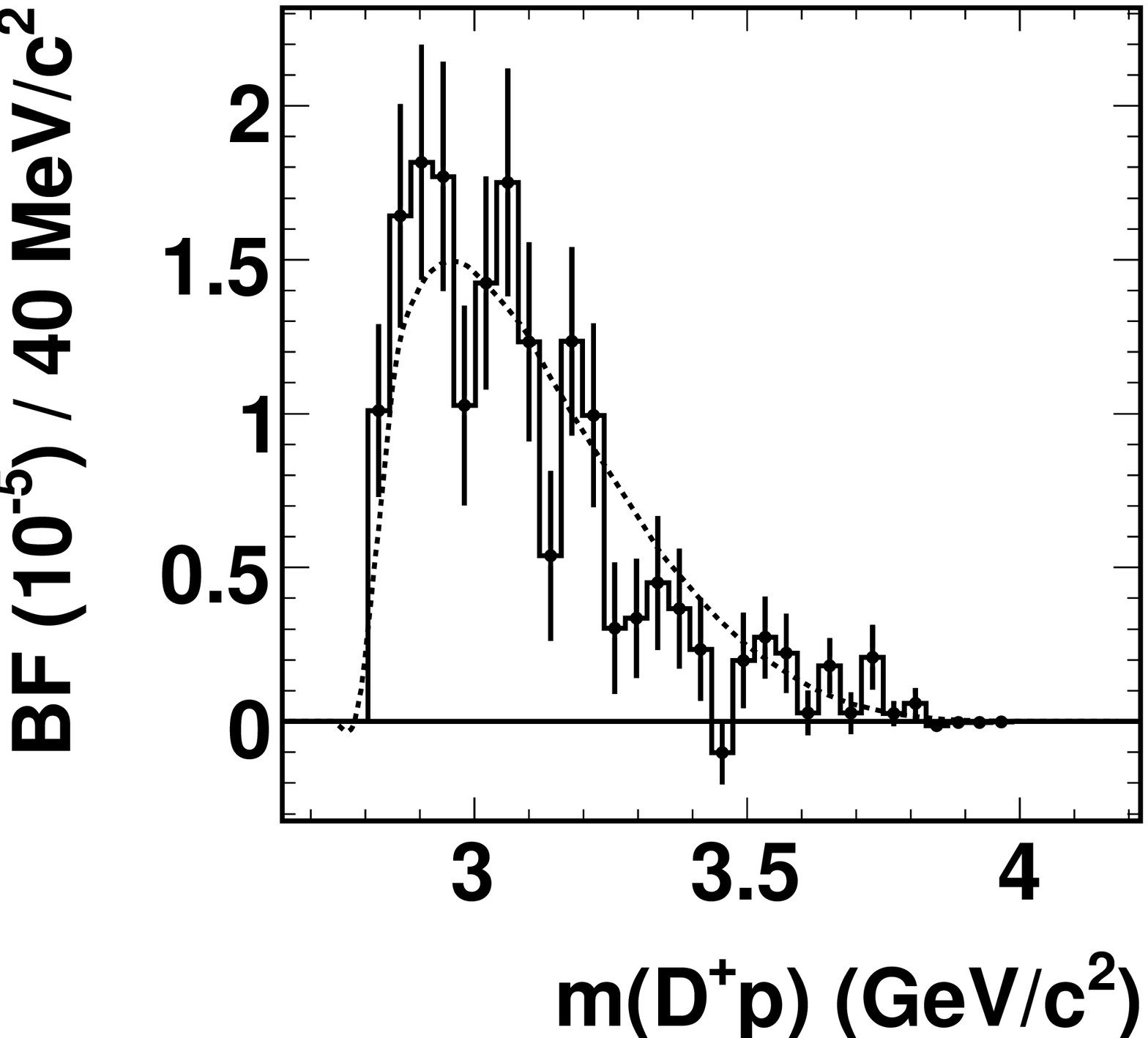}
    \put(66,74){\small{\babar}}
    \put(66,66){\small{prelim.}}
  \end{overpic}
}%
\subfloat[$\Dp\ppbar\pi\pi$, $m(p\pim)$]{
  \hspace{-0.017\textwidth}%
  \begin{overpic}[width=0.25\textwidth]{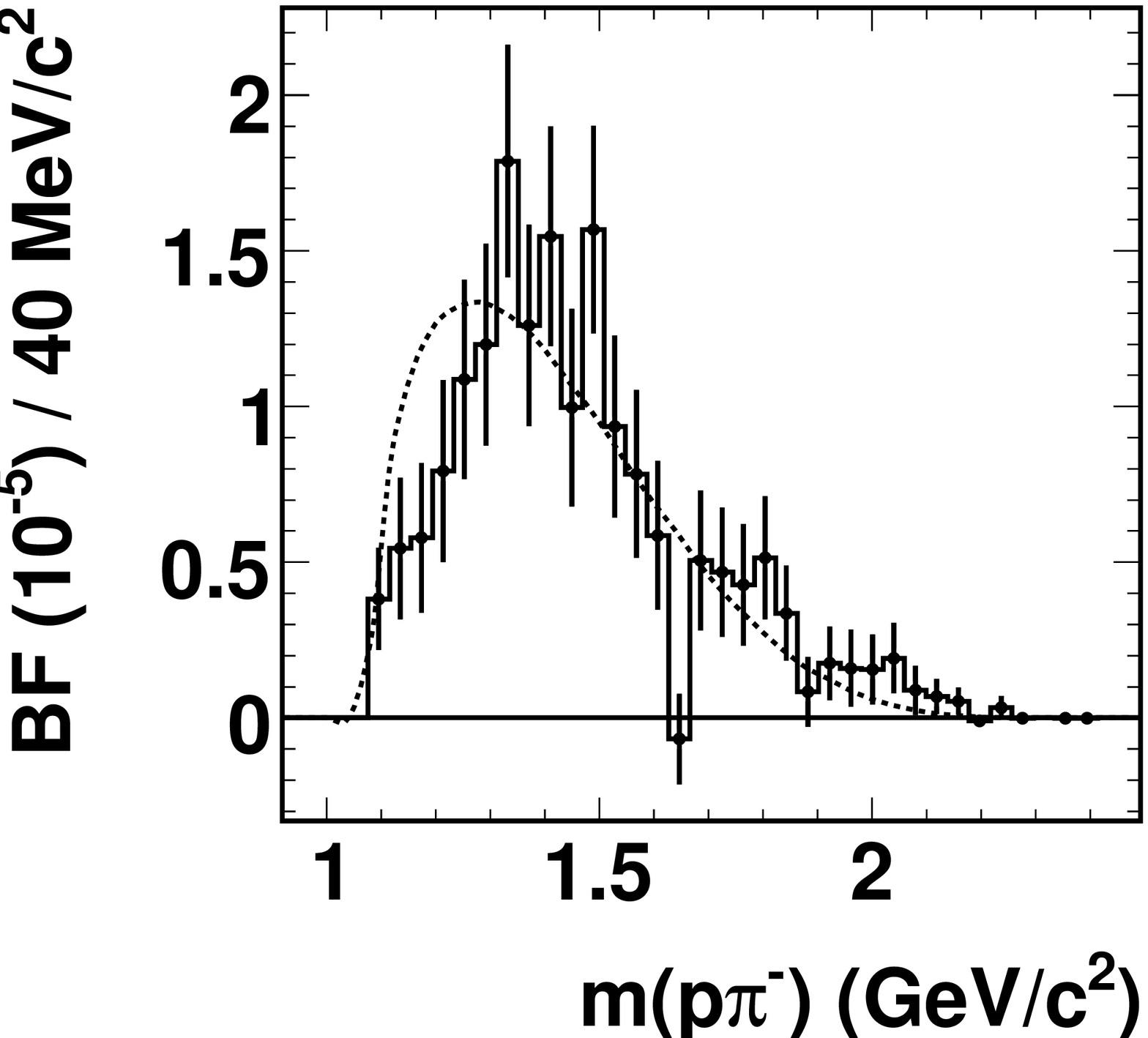}
    \put(66,74){\small{\babar}}
    \put(66,66){\small{prelim.}}
  \end{overpic}
}%
\\
\subfloat[$\Dstarp\ppbar\pi\pi$, $m(\ppbar)$]{
  \begin{overpic}[width=0.25\textwidth]{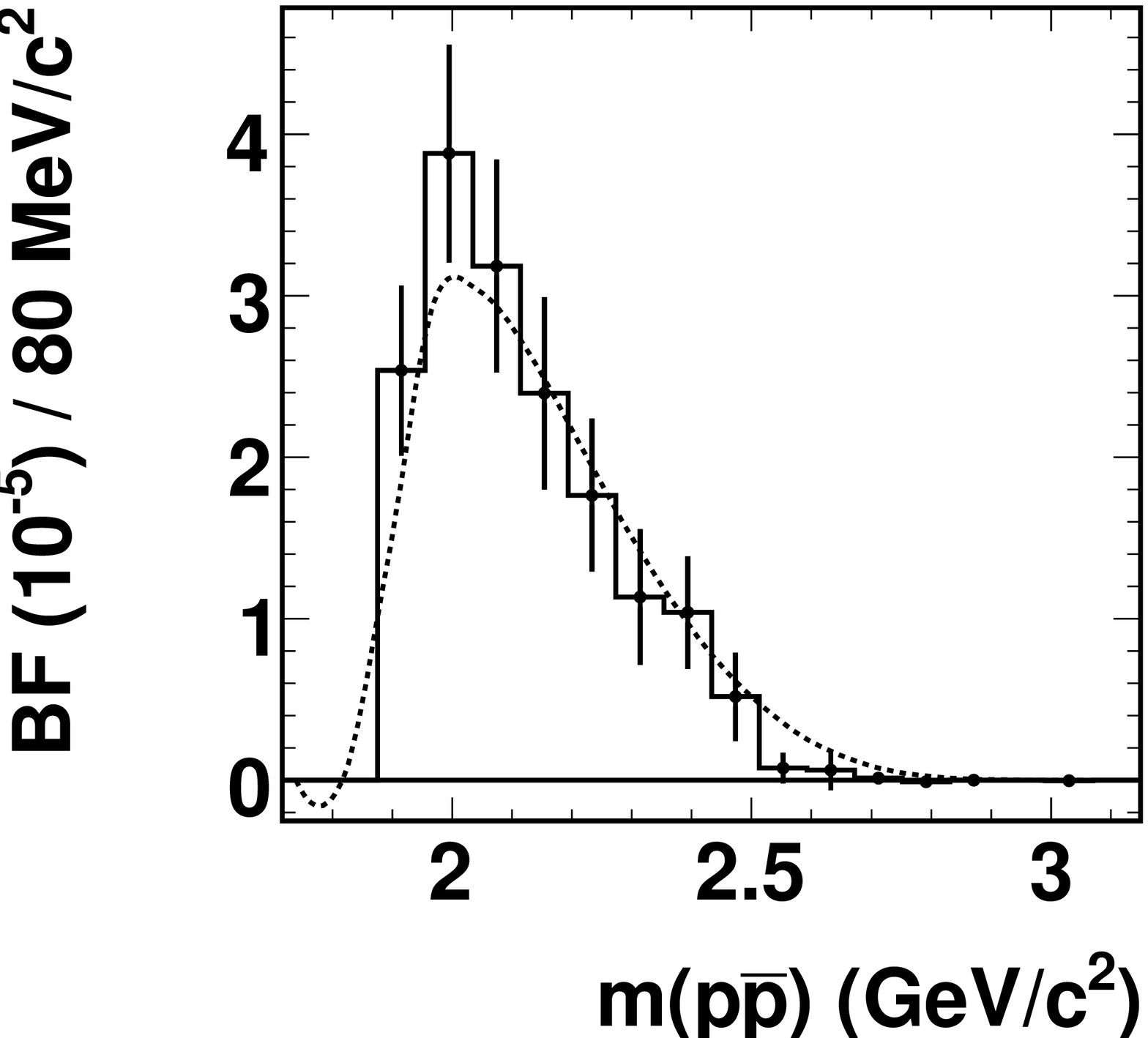}
    \put(66,74){\small{\babar}}
    \put(66,66){\small{prelim.}}
  \end{overpic}
}%
\subfloat[$\Dstarp\ppbar\pi\pi$,\! $m(\Dp\pbar)$]{
  \hspace{-0.017\textwidth}%
  \begin{overpic}[width=0.25\textwidth]{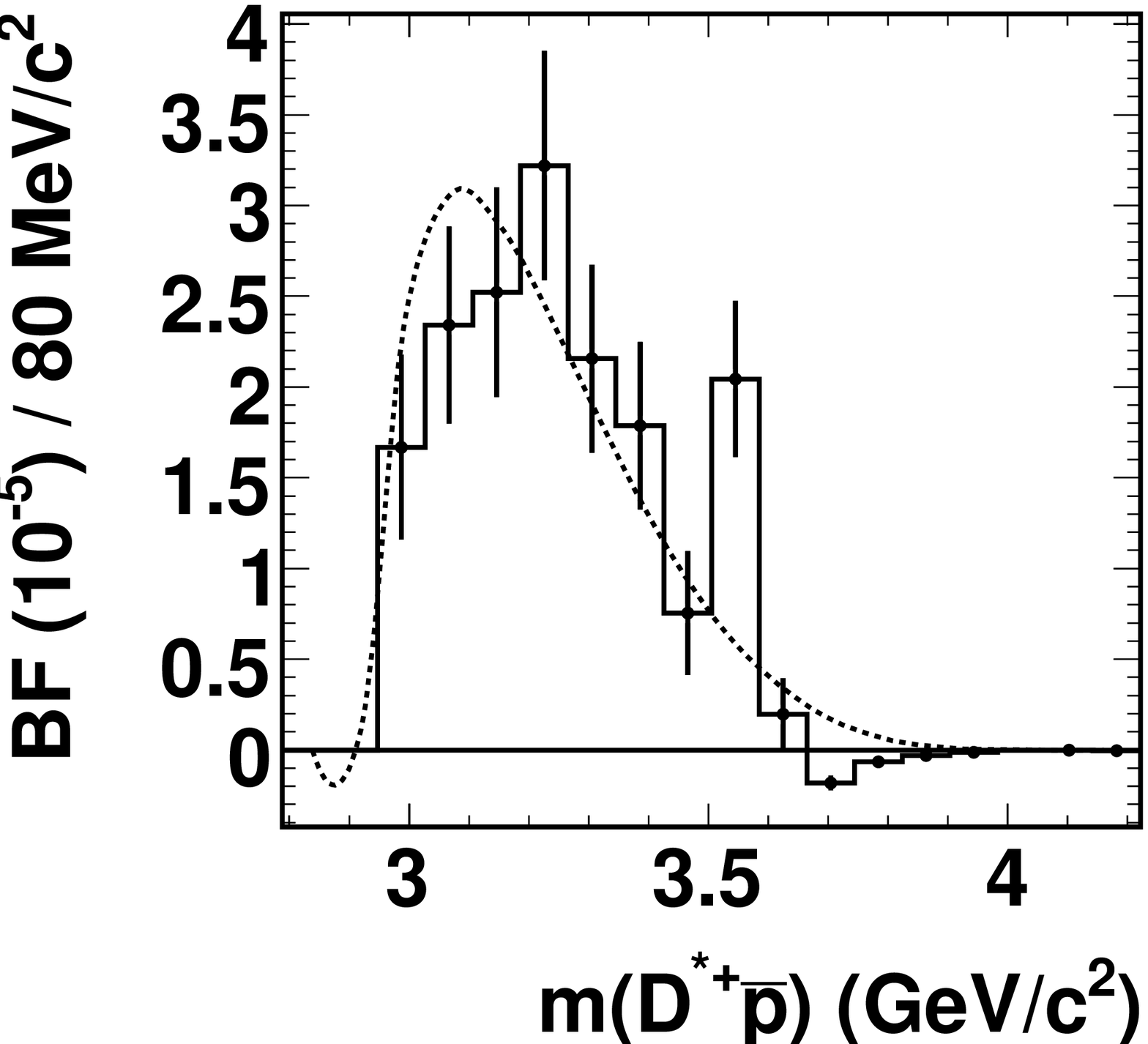}
    \put(66,74){\small{\babar}}
    \put(66,66){\small{prelim.}}
  \end{overpic}
}%
\subfloat[$\Dstarp\ppbar\pi\pi$, $m(\Dp{p})$]{
  \hspace{-0.017\textwidth}%
  \begin{overpic}[width=0.25\textwidth]{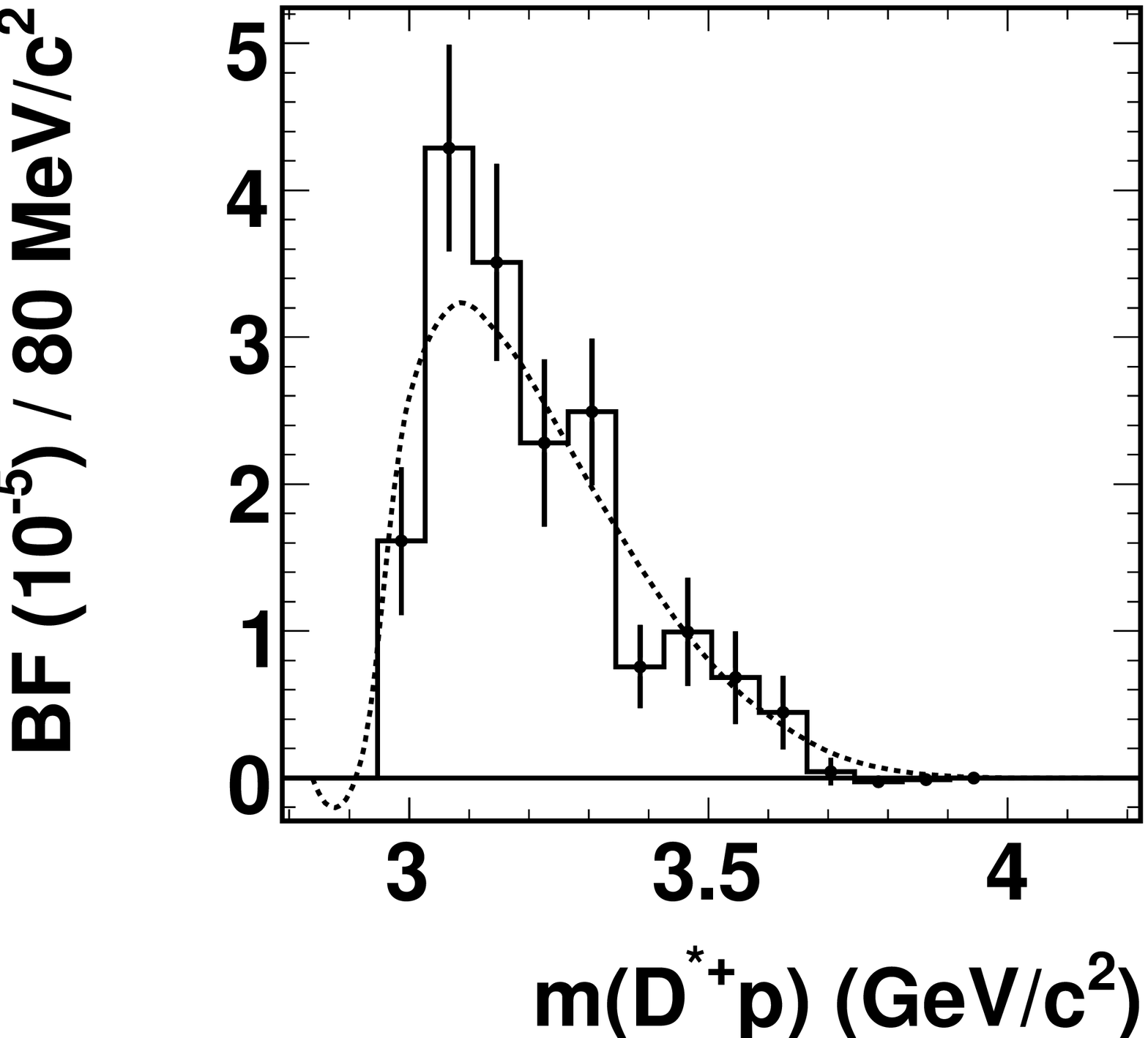}
    \put(66,74){\small{\babar}}
    \put(66,66){\small{prelim.}}
 \end{overpic}
}%
\subfloat[$\Dstarp\ppbar\pi\pi$, $m(p\pim)$]{
  \hspace{-0.017\textwidth}%
  \begin{overpic}[width=0.25\textwidth]{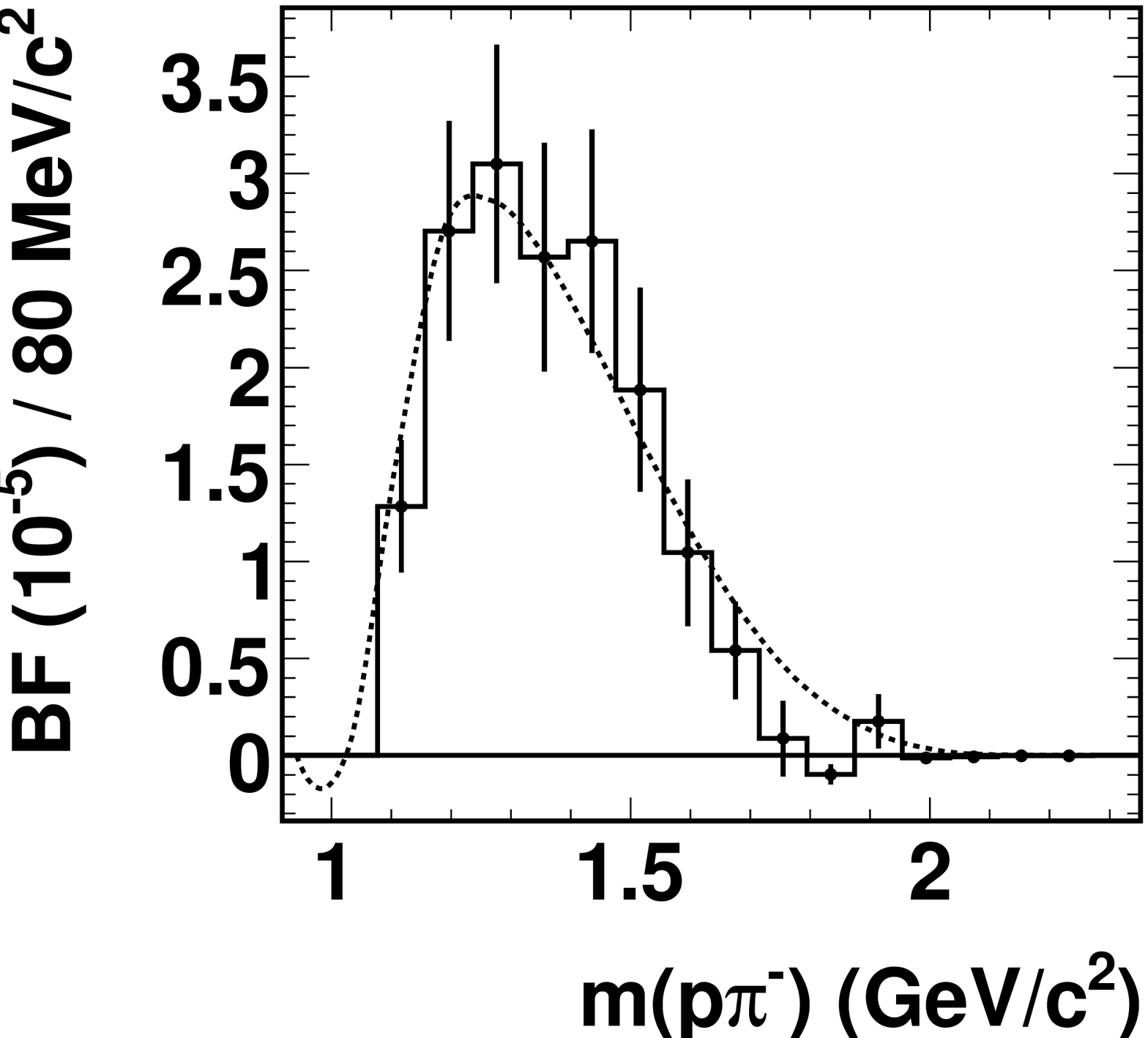}
    \put(66,74){\small{\babar}}
    \put(66,66){\small{prelim.}}
  \end{overpic}
}
\caption{
  5-body $B$ decay differential branching
  fractions as functions of 
  $\ppbar$,
  $\Dmaybestar{p}$,
  $\Dmaybestar{\pbar}$, and
  $p\pim$
  in
  (abcd) $\Bzb\To\Dz\ppbar\pim\pip$,
  (efgh) $\Bzb\To\Dstarz\ppbar\pim\pip$,
  (ijkl) $\Bm\To\Dp\ppbar\pim\pim$, and
  (mnop) $\Bm\To\Dstarp\ppbar\pim\pim$, respectively.  The smooth
  curve represents decays following the uniform phase space model.
}
\label{fig:5body_mass}
\end{figure}

Figure~\ref{fig:5body_mass} shows the differential branching fractions
for a selection of four mass variables for 5-body $B$ decays: $\ppbar$,
$\Dmaybestar{p}$, $\Dmaybestar{\pbar}$, and $p\pim$.  In
contrast to 3- and 4-body distributions, we see do not see wide
disagreement with phase space expectations.

\section{CONCLUSIONS}
\label{sec:conclusions}

Using \babar's data set of $455\times10^6$ \BB\ pairs, we present 
the observation and
study of ten baryonic $B$-meson decays to a $\Dmaybestar$,
a proton-antiproton pair, and a system of up to two pions,
of which six
    (\nolbreaks{$\Bm\To\Dz\ppbar\pim$},
     \nolbreaks{$\Bm\To\Dstarz\ppbar\pim$},
     \nolbreaks{$\Bzb\To\Dz\ppbar\pim\pip$},
     \nolbreaks{$\Bzb\To\Dstarz\ppbar\pim\pip$},
     \nolbreaks{$\Bm\To\Dp\ppbar\pim\pim$},
     \nolbreaks{$\Bm\To\Dstarp\ppbar\pim\pim$})
are first observations.  The branching fractions for 3- and 5-body
decays are suppressed compared to 4-body decays with the hierarchy
\nolbreaks{$\mathcal{B}_\textrm{3-body} < \mathcal{B}_\textrm{5-body}
< \mathcal{B}_\textrm{4-body}$}.  Branching fraction ratios
for modes related by exchange of $D$ mesons are of order unity.

The kinematic distributions show a number of peculiar features in the
$B$ sample.  For 3-body decays, non-overlapping threshold enhancements
are seen in $m(\ppbar)$ and $m(\Dmaybestarz{p})$.  For 4-body decays,
$m(p\pim)$ distribution shows a narrow peak with mass of
($1497.4\pm3.0\pm0.9$)\mevcc and full width of ($47\pm12\pm4$)\mevcc,
where the first (second) errors are statistical (systematic).  For
5-body decays, in contrast to 3- and 4-body decays, mass projections
are similar with phase space expectations.

All results are preliminary. 

\section{ACKNOWLEDGMENTS}
\label{sec:Acknowledgments}

We are grateful for the 
extraordinary contributions of our \pep2\ colleagues in
achieving the excellent luminosity and machine conditions
that have made this work possible.
The success of this project also relies critically on the 
expertise and dedication of the computing organizations that 
support \babar.
The collaborating institutions wish to thank 
SLAC for its support and the kind hospitality extended to them. 
This work is supported by the
US Department of Energy
and National Science Foundation, the
Natural Sciences and Engineering Research Council (Canada),
the Commissariat \`a l'Energie Atomique and
Institut National de Physique Nucl\'eaire et de Physique des Particules
(France), the
Bundesministerium f\"ur Bildung und Forschung and
Deutsche Forschungsgemeinschaft
(Germany), the
Istituto Nazionale di Fisica Nucleare (Italy),
the Foundation for Fundamental Research on Matter (The Netherlands),
the Research Council of Norway, the
Ministry of Education and Science of the Russian Federation, 
Ministerio de Educaci\'on y Ciencia (Spain), and the
Science and Technology Facilities Council (United Kingdom).
Individuals have received support from 
the Marie-Curie IEF program (European Union) and
the A. P. Sloan Foundation.

\end{document}